\documentclass[a4paper, 11pt]{article}
\usepackage[english]{babel}
\makeatletter

\usepackage{geometry,graphicx,color}
\usepackage{amsfonts, amsmath, amssymb, slashed,dsfont}
\usepackage{tensor}
\usepackage[pdftex]{hyperref}
\usepackage{epsfig}
\usepackage{pifont}
\usepackage{enumitem}
\usepackage{bm}
\usepackage{mathrsfs} % Allows \mathscr
\usepackage[hyperref,thmmarks,amsmath]{ntheorem}
\usepackage{amsmath,amssymb,slashed}
\usepackage{tikz}
\usetikzlibrary{positioning}

\numberwithin{equation}{section}
\DeclareMathOperator{\actr}{\triangleleft}
\DeclareMathOperator{\actl}{\triangleright}
\DeclareMathOperator{\coactr}{\blacktriangleleft}

\DeclareMathOperator{\bicros}{\actl\!\!\!\blacktriangleleft}

\newcommand\bbone{\mathbb{I}}

\newcommand\caZ{{\mathcal Z}}

\newcommand\algA{{\mathbb{A}}}
\newcommand\modM{{\mathbb{E}}}
\newcommand\algH{{\mathscr{H}}}
\newcommand\twiF{\mathscr{F}}

\newcommand\hR{{F}}
\newcommand\adrep{\mathrm{Ad}}

\newcommand\kS{{\mathfrak S}}

\DeclareMathOperator{\tr}{Tr} 
\newcommand{\sign}{\mathrm{sign}}
\DeclareMathOperator{\End}{\mathrm{End}}
\DeclareMathOperator{\Hom}{\mathrm{Hom}}
\DeclareMathOperator{\Aut}{\mathrm{Aut}}

\newcommand\Der{\mathrm{Der}}
\newcommand\Int{\mathrm{Int}}
\newcommand\Out{\mathrm{Out}}
\newcommand\kX{X}
\newcommand\kY{Y}
\newcommand{\ham}{\mathrm{Ham}}

\newcommand\dd{\mathrm{d}}

\newcommand\id{\mathrm{id}}

\newcommand\inv{\mathrm{inv}}

\DeclareMathOperator{\tcvp}{\hat{\circ}}

\newcommand\sstar{\text{\ding{86}}}
\newcommand{\kbar}{\mathchar'26\mkern-9mu k}

\newcommand{\wign}[6]{
\left(
  \begin{array}{ccc}
  #1 & #3 & #5 \\
  #2 & #4 & #6
  \end{array}
  \right)
}

\newcommand{\tops}[2]{\texorpdfstring{#1}{#2}}

\newcommand{\institute}[1]{\newcommand{\@institute}{#1}}
\renewcommand{\maketitle}{
\vspace*{0.5\baselineskip}
{% title
\center\LARGE\noindent\@title\par
}%
\vspace{1.5\baselineskip}
{% author
\center\normalsize\noindent\ignorespaces\@author\par
}%
\vspace{0.5\baselineskip}
{% institute
\center\normalsize\ignorespaces\@institute\par
}%
\vspace{2\baselineskip}
}%

\begin{document}

% -------------- Title -----------------
\title{Gauge theories on quantum spaces}
\author{Kilian Hersent$^a$, Philippe Mathieu$^b$, Jean-Christophe Wallet$^a$}
\institute{%
\textit{$^a$IJCLab, Universit\'e Paris-Saclay, CNRS/IN2P3, 91405 Orsay, France\\
$^b$Institut f\"ur Mathematik, Universit\"at Z\"urich, Winterthurerstrasse 190, CH-8057 Z\"urich.
}\\%
\bigskip
e-mail:
\href{mailto:kilian.hersent@universite-paris-saclay.fr}{\texttt{kilian.hersent@universite-paris-saclay.fr}}, 
\href{mailto:philippe.mathieu@math.uzh.ch}{\texttt{philippe.mathieu@math.uzh.ch}}, 
\href{mailto:jean-christophe.wallet@universite-paris-saclay.fr}{\texttt{jean-christophe.wallet@universite-paris-saclay.fr}}
}%
\maketitle

%---------------------- Abstract --------------------
\begin{abstract} 
We review the present status of gauge theories built on various quantum space-times described by noncommutative space-times. The mathematical tools and notions underlying their construction are given. Different formulations of gauge theory models on Moyal spaces as well as on quantum spaces whose coordinates form a Lie algebra are covered, with particular emphasis on some explored quantum properties. Recent attempts aiming to include gravity dynamics within a noncommutative framework are also considered.

\end{abstract}

\tableofcontents
\newpage

%--------------------- Main matter --------------------
\section{Introduction}
\label{sec:intro}
\paragraph{}
Quantum space-times, which are intended to substitute to the ordinary space-time at an energy scale currently identified with the Planck mass, where the notion of (differentiable) manifold becomes meaningless, can be related to coordinates which no longer commute. As such, quantum space-times are conveniently described by exploiting the tools and concepts of noncommutative geometry \cite{connes1}, so that they are often called noncommutative spaces. In what follows, we will use indifferently both terminologies. Recall that a quantum (noncommutative) space-time is modeled by a noncommutative but still associative algebra, its derivations appearing to be natural noncommutative analogs of the usual vector fields. Other natural correspondences between usual (commutative) objects of differential geometry and noncommutative entities will be detailed through this paper and in particular in section \ref{sec:nc_diff_calc}.

\paragraph{}
In quantum space-times, the space-time coordinates can be identified with operators whose spectra (eigenvalues) provide the possible space-time localizations of a particle (event). This natural identification even permits one to get a pictorial flavor of the quantum space whenever the above spectra are discrete.

An instructive example is provided by the noncommutative space $\mathbb{R}^3_\lambda$ \cite{Pepe-vit} (covered in section \ref{sec:R3L}) which, up to some technical details, can be identified with $\mathbb{R}^3_\lambda = \oplus_{n=1}^\infty \mathbb{M}_n(\mathbb{C})$ where $ \mathbb{M}_n(\mathbb{C})$ is the algebra of $n \times n$ complex matrices. Its corresponding coordinate algebra is $[x^\mu, x^\nu] = i \lambda \tensor{\varepsilon}{^{\mu\nu}_\rho}x^\rho$ (where $\lambda$ is a constant with length dimension) , that is the Lie algebra $\mathfrak{su}(2)$. Each of the $x^\mu$'s has a spectrum involving only integers which is also the case for the operator $\mathfrak{X}=X_\mu x^\mu$ when $ X^\mu X_\mu=1$. Assuming that a quantum mechanical framework holds, the operator $\mathfrak{X}$ permits one to measure one coordinate in the direction defined by the vector $X_\mu$. According to the property of its spectrum, the result of any measurement will be integer, thus exhibiting the discrete nature of this quantum space, which also inherits a continuous $SO(3)$ symmetry stemming from the action of $SU(2)$ on $\mathfrak{su}(2)$ by inner automorphisms as $SU(2)/\mathbb{Z}_2 \simeq SO(3)$.

While it is not possible to measure a position in different directions simultaneously, since the above operators do not commute, it is however possible to perform simultaneously another measurement by using the central operator $x^0 = \sum_{n=1}^\infty(n-1) P_n$, equation \eqref{eq:r3l_center_gen}, where $P_n$ is the orthogonal projector on $\mathbb{M}_n(\mathbb{C}) $. Its spectrum involves only positive integers. This latter operator, called the radius operator, permits one to measure the distance from the origin. In this description, $\mathbb{M}_n(\mathbb{C})$ looks like a kind of sphere of radius $n-1$, which actually is the algebra modeling the so called fuzzy sphere, while, by using the eigenvalues of the operator $\mathfrak{X}$ in the representation of $\mathbb{M}_n(\mathbb{C})$, one infers that the coordinate in the direction defined by $ X_\mu$ can only take the $n$ values $n-1$, $n-3$, $\dots$, $-n+1$.

\paragraph{}
There is a kind of consensus, for a rather long time, that the notion of manifold should not be valid at ultra short distance when the strength of the gravitational interaction becomes at the same order of magnitude than the one of the other interactions. In this regime, it is known that an immediate problem arises for the exact localization of events when combining quantum mechanics to general relativity. Indeed, according to the uncertainity principle in quantum mechanics, any measurement of the position of a coordinate $x$ to a given accuracy $\Delta x$ can be done provided an amount of energy $\Delta E \sim \mathcal{O}(\frac{1}{\Delta x})$. This energy is then brought to the volume element $(\Delta x)^3$, resulting in a huge energy density whenever $\Delta x$ is required to be very small. But according to the Einstein equations, this triggers the appearence of a black hole with Schwarzschild radius $R \sim \frac{E}{M_P}\ell_P$, where $\ell_P$ and $M_P$ are respectively the Planck length ($\ell_P\sim\mathcal{O}(10^{-35})\ \text{cm}) $ and the Planck mass ($M_P\sim\mathcal{O}(10^{19})\ \text{GeV}$). This radius therefore corresponds to the smallest length scale which can be probed.

\paragraph{}
To escape this difficulty, the authors of \cite{DFR} postulated modified uncertainity relations $ \Delta x^\mu \Delta x^\nu \geqslant \frac{1}{2}|\theta^{\mu\nu}|$, therefore generating a minimal length scale. This uncertainity relation stems from the following noncommutativity of the coordinates: $[x^\mu, x^\nu] = i \frac{\theta^{\mu\nu}}{M_P^2}$. This is the commutation relation characterizing the by-now popular Moyal space $\mathbb{R}^4_\theta$, see e.g.\ \cite{graciavar1}, which will be considered in section \ref{sec:moyal}. Further considerations on D-branes in \cite{Seiberg_1999} gave rise to extended commutation relations of the form
\begin{equation}
    [x^\mu, x^\nu]
    = i \frac{\theta^{\mu\nu(x)}}{M_P^2}
    = \frac{i}{M_P^2} \theta^{\mu\nu} + \frac{i}{M_P} \theta^{\mu\nu}_\rho x^\rho + \dots
    \label{premiereformule}.
\end{equation}
The second term in the right hand side of \eqref{premiereformule} corresponds to what is sometimes called in the physics literature a ``noncommutativity of Lie algebra type". By-now popular related quantum spaces are $\mathbb{R}^3_\lambda$, a deformation of $\mathbb{R}^3$ mentioned above and considered in section \ref{sec:R3L} and the deformation of the usual Minkowski space-time, called $\kappa$-Minkowski space-time introduced a long ago \cite{Lukierski_2017, Lukierski_1991, MR1994}. 

The latter space will be considered in section \ref{sec:kappa}. The corresponding coordinate algebra is $[x^0, x^j]= \frac{i}{\kappa}x^j$, $[x^j, x^k]= 0$ ($x^0$, $x^j$ are respectively time and space coordinates) where $\kappa$ is usually identified with the Planck mass. In this quantum space, the spatial part stays commutative while the time becomes ``noncommutative'' generating modified Lorentz boosts as well as modified relativistic dynamics, which produce sizable changes as energy scale approaches the Planck mass scale. This noncommutative space is regarded as phenomenologicaly promising.

\paragraph{}
A few years after, noncommutative structures have also shown up within Group Field Theory, see \cite{Oriti_2009} for a short review. These discoveries, performed within these different approaches to quantum gravity, have triggered a large interest for the study of field theories built on noncommutative (quantum) spaces, either as kind of prototypes or claimed to describe effective regimes of a more fundamental theory of quantum gravity. These approaches are attempts to characterize potentially observable effects from quantum gravity. For a recent review on the phenomenology of quantum gravity, see \cite{Addazi_2022}.

These mostly non local field theories are generically called noncommutative field theories. For reviews related to the early time of noncommutative field theories see \cite{dnsw-rev}. Compared to the usual (commutative) field theories, noncommutative field theories have somehow different features. Many efforts have been focused on the exploration of their quantum and renormalisation properties. This latter aspect is known to be often a difficult task in noncommutative field theories, due in particular to their non-local character, thus precluding the use of the standard (perturbative) machinery used for the ordinary local field theories. This may even be complicated by the possible appearance of the UV/IR mixing \cite{IRUVmix1}, a typical phenomenon of the noncommutative field theories spoiling renormalisability.

\paragraph{}
Gauge theories on quantum (noncommutative) spaces have also been the subject of an intense activity. These inherit of all the difficulties inherent to the noncommutative field theories. These difficulties are even supplemented by additional ones of the gauge theory context, including at least the construction of a suitable noncommutative differential calculus together with a reasonable extension of the notion of connection and its related gauge transformations. The purpose of this paper is to present an overview of the state of the art on the developments on gauge theories defined on quantum spaces which were carried out in physics during the last two decades.

\paragraph{}
The present review is organised as follows.

The section \ref{sec:star_prod} collects useful properties underlying the construction of various classes of associative (but noncommutative) products, called star-products, which equip the algebras modeling quantum spaces. There is an abundant mathematical literature devoted to the definition and construction of star-products. A part of those appearing in the physics literature pertains to the area of deformations, such as deformation quantization or formal deformation whereas the other part is based on twist deformation and makes use of Hopf algebra structures, in the vein of Drinfeld twists. Relevant references are for instance \cite{flato1, lecomte, Rieff, Kons1, Drinfeld_1990}.

\paragraph{}
The section \ref{sec:nc_diff_calc} presents different types of noncommutative differential calculi and different notions of noncommutative connections which will appear thorough this review. A special focus will be made on the derivation-based differential calculus and on the differential calculus obtained from twist deformation, which underlie the gauge theory models considered here. We will collect usefull properties of two noncommutative generalizations of the notion of connection. One, often used in the physics literature, exploits the structure of a right (or left) module over the algebra, see e.g.\ \cite{mdv99}. The other one uses a bimodule structure and can be viewed as a noncommutative generalisation of the notion of linear connection \cite{Dubois_Violette_1996, Madore_1997}.

\paragraph{}
The section \ref{sec:moyal} reviews the developments related to the gauge theories on Moyal spaces. These quantum spaces are physically characterized by the following commutation relation between coordinates (in obvious notations) $[x^\mu, x^\nu] = \theta^{\mu\nu}$, where $\theta^{\mu\nu}$ is a constant, which corresponds to the simplest ``noncommutativity'' among the coordinates. The gauge theories on Moyal spaces have received a lot of interest. They can be viewed and have been actually formulated, either as mere noncommutative analogs of a Yang-Mills theory or alternatively as matrix models. The various attempts to neutralize the UV/IR mixing, showing up in particular in the Yang-Mills-type formulation are presented, together with the complexity and possible quantum instability of the vacuum structure occurring in the gauge matrix formulation.

\paragraph{}
The section \ref{sec:R3L} deals with gauge theories on $\mathbb{R}^3_\lambda$, which, as the gauge theories on Moyal spaces, can be described, either as a noncommutative analog of a Yang-Mills theory or as a gauge matrix model. The quantum space $\mathbb{R}^3_\lambda$ corresponds to a $\mathfrak{su}(2)$ Lie algebra noncommutativity for the coordinates, \textit{i.e.}\ $[x^\mu, x^\nu] = i \lambda \tensor{\varepsilon}{^{\mu\nu}_\rho}x^\rho$ and inherits of specific properties stemming from the structure of the group algebra of $SU(2)$ and the Peter-Weyl theorem. The latter theorem renders $\mathbb{R}^3_\lambda$ ``close'', but not identical, to the fuzzy sphere, and allows again to relate this quantum space to the matrix algebra. Classical and quantum properties of gauge theories on $\mathbb{R}^3_\lambda$ are reviewed, among them possible quantum instabilities in Yang-Mills type formulation as well as the all order finitude of a class of matrix gauge theories. Relations with brane models together with group field theory models are exhibited.

\paragraph{}
The section \ref{sec:kappa} focuses on developments around the gauge theories on $\kappa$-Minkowski space-time, a deformation of the usual Minkowski space-time introduced a long time ago, see \cite{Lukierski_2017} for a general review. $\kappa$-Minkowski space is rigidly linked to a deformation of the Poincar\'e algebra, called the $\kappa$-Poincar\'e algebra, which can be viewed as the quantum analog of its algebra of symmetries, see \cite{Lukierski_1991, MR1994} for pioneering works. The $\kappa$-Minkowski space-time is often considered as a phenomenologically promising quantum space. The construction of gauge theories on this space is however more delicate to achieve and this section reviews various related developments based on the use of twists or exploiting invariance under $\kappa$-Poincar\'e symmetry.

\paragraph{}
The section \ref{sec:beyond_YM} collects various attempts to incorporate gravity in a noncommutative setting. These are not necessarily presented in a chronological order. It includes in particular some attempts using connexion on central bimodule, summarized in subsection \ref{subsec:nc_central_bimod}, the recent approach based on braided geometry or the emergent gravity proposal exploiting the matrix model formulation within Moyal spaces.

\paragraph{}
Finally, the section \ref{sec:conc} concludes and lists some open questions and outlook.

\paragraph{}
Notice that we will omit in this review the developments of the formulation of the standar model based on the use of spectral triples as well as those applied to Hall systems in condensed matter physics. Besides, we will omit the developments related to the gauge theories on the noncommutative torus and to noncommutative instantons.

\paragraph{}
Throughout this paper, the Einstein summation over repeated indices is applied, unless otherwise stated. Moreover, the notation of covariant and contravariant index notation is mostly used. One can lower or rise indices using the background metric $g^\mathrm{B}$, explicitly $x_\mu = g^\mathrm{B}_{\mu\nu} x^\nu$. Otherwise stated, this background metric is considered to be Euclidean, that is
\begin{equation}
    g^\mathrm{B}_{\mu\nu} = \delta_{\mu\nu}.
    \label{eq:background_metric}
\end{equation}
However, many results are put in a covariant way so that they do not depend on the background metric.

\newpage
\section{Building the star-products.}
\label{sec:star_prod}

\subsection{Basic properties.}
\label{subsec:star_prod_basic}
\paragraph{}
In this paper, we will consider different quantum (\textit{i.e.}\ noncommutative) spaces. Recall that one way to model a quantum version of a finite dimensional manifold, e.g.\ $\mathbb{R}^n$, is to equip a suitable space of functions on this manifold, denoted by $\mathcal{A}$, e.g.\ $\mathcal{C}^\infty(\mathbb{R}^n)$, with an associative but noncommutative product denoted generically by $\star$, called the star-product, such that $\algA = (\mathcal{A}, \star)$  becomes an associative algebra. This algebra\footnote{In the following, all the algebras considered will be unital. Moreover, the notation of $\mathcal{A}$ for the classical algebra and $\algA$ for the quantum one is kept throughout this section.} can then be viewed as a quantum analog of the above finite dimensional space corresponding to the chosen star-product. Whenever a notion of charge, and thus of charge conjugation, comes into play, which is needed in physics, the algebra must be enlarged with an involution, \textit{i.e.}\ a map 
\begin{align}
    & {}^\dag: \algA \to \algA, &
    (f \star g)^\dag
    &= g^\dag \star f^\dag, &
    (f^\dag)^\dag &= f,
    \label{eq:involution}
\end{align}
for any $f, g \in \algA$. Note that this involution is not necessarily the usual conjugation but may come from some algebraic structure underlying the noncommutative algebra, as we will see in the sequel.

\paragraph{}
There is a huge mathematical literature on star-products considered as deformations of pointwise products of functions in relation with deformation quantization of manifolds \cite{flato1}. The existence of star-products was proven for symplectic manifolds in \cite{lecomte} and for general Poisson manifold in \cite{Rieff, Kons1}. It appears that many star-products used in the physics literature pertain to the area of formal deformations where the product $f \star g$ is a formal power serie in some deformation parameter, usually identified with a large mass scale, with functions as coefficients of the expansion. This is the case which we will consider in this paper.

\paragraph{}
We find convenient to adopt a unified presentation of the star-products considered in this paper. To exhibit the related underlying properties, it is instructive to sketch briefly the old construction of the Moyal product. The detailed construction is carried out in the section \ref{subsubsec:moyal_prod}. The former old construction is based on the central notion of twisted convolution, introduced almost one century ago by von Neumann \cite{vonNeum}. The starting point was the first attempt by Weyl \cite{Weyl}, known as the Weyl quantization, realizing the correspondence between classical and quantum observables. It aims at mapping a classical phase space $\mathbb{R}^2$ with coordinates $(p, q)$ to a space of operators acting on some suitable Hilbert space generated by the operators $P$ and $Q$ satisfying the commutation relations $[P, Q] = i \hbar$ which defines the Heisenberg algebra. This is achieved by introducing a map given by \cite{Weyl}
\begin{equation}
    W(F)
    = \int \dd p \dd q \dd u \dd v\ F(p,q) e^{i ( u(P - p) + v(Q - q) )}
    = \int \dd u \dd v\ (\mathcal{F}F)(u,v) e^{i(u P + v Q)},
    \label{eq:weyl_pseudo_rep}
\end{equation}
where $u$ and $v$ run on all $\mathbb{R}$, which is well defined for any $F \in L^1(\mathbb{R}^2)$, in which $(\mathcal{F}F)$ denotes the Fourier transform\footnote{
Our convention for the Fourier transform is $(\mathcal{F}f)(s) = \int \dd^d x\ f(x) e^{-isx}$.
}
. Then, a product $\tcvp$ between two functions $F, G\in L^1(\mathbb{R}^2)$ can be defined by requiring that 
\begin{equation}
    W(F \tcvp G) = W(F) W(G)
    \label{eq:twist_conv_def}
\end{equation}
holds true. Equation \eqref{eq:twist_conv_def} gives rise to the so called twisted convolution \cite{vonNeum} which will be presented in section \ref{subsec:star_prod_conv_alg}.

\paragraph{}
To make contact with the deformation quantization framework, it is convenient to introduce a quantization map, denoted hereafter by $Q$, defined by
\begin{equation}
    Q(f) = W(\mathcal{F} f),
    \label{eq:weyl_map}
\end{equation}
which is required to satisfy
\begin{equation}
    Q(f \star g) = Q(f) Q(g)
    \label{eq:star_prod_quant},
\end{equation}
for any $h, k \in L^1(\mathbb{R}^2)$, where the (associative) product $\star$ hereby defined is the Moyal product. The map $Q$ is known as the Weyl quantization map.

Then, setting $F = \mathcal{F}f$, $G = \mathcal{F}g$ in \eqref{eq:twist_conv_def} and using \eqref{eq:weyl_map}, one infers that
\begin{equation}
    f\star g=\mathcal{F}^{-1}(\mathcal{F}f\tcvp\mathcal{F}g)
    \label{eq:star_prod_def},
\end{equation}
thus exhibiting the relationship between the Moyal product and the twisted convolution introduced in \cite{vonNeum}.

\subsection{Convolution algebras.}
\label{subsec:star_prod_conv_alg}
\paragraph{}
The above sketchy discussion exhibits clearly the main ingredients entering the construction. From a space of function $\mathcal{A}$ on a manifold, we extract a $*$-algebra structure on the multiplier space $\algA$ by deriving the product $\star$ and the involution ${}^\dagger$. The multiplier space $\algA$ is a subspace of the convolution algebra, sometimes strictly smaller for convergence reasons. The constructions of $\star$ and ${}^\dagger$ can generically follow the steps bellow.

\begin{enumerate}
	\item Start from a \textsc{Lie} algebra $\mathfrak{g}$ with a locally compact \textsc{Lie} group $G$ and derive operation laws on the group.
	
	    In the previous case, $\mathfrak{g}$ was the Heisenberg algebra and $G = \mathbb{H}_2$ was the Heisenberg group.
	
	\item Derive its Haar mesure(s) $\dd \nu$ and its convolution algebra, noted $\mathbb{C}(G) = (L^1(G), \hat{\circ})$.
	
	    In the previous case, the Haar measure of $\mathbb{H}_2$ is the Lebesgue measure on $\mathbb{R}^2$ and its convolution algebra is $(L^1(\mathbb{R}^2), \hat{\circ})$.
	
	\item Study the representations of the convolution algebra and characterize them.
	
	    In the previous case, the pseudo-representation \eqref{eq:weyl_pseudo_rep} was used. In fact, it appears more convenient to use the Schr\"odinger representation \eqref{eq:schroding_rep} and use the Stone-von Neumann theorem. 
	
	\item Define the quantization map. Here, we will use $Q(f) = \pi(\mathcal{F}f)$, where $f$ is an element of the convolution algebra $\mathbb{C}(G)$, $\pi:\mathbb{C}(G) \to \mathcal{B(H)}$ is a representation of $\mathbb{C}(G)$ on a Hilbert space $\mathcal{H}$ and $\mathcal{F}f$ denotes the Fourier transform of $f$.
	
	    In the previous case, $\pi$ was defined by a (kind of unitary) representation \eqref{eq:weyl_pseudo_rep} and so $Q$ was the Weyl quantization map \eqref{eq:weyl_map}. The Hilbert space was $L^2(\mathbb{R})$.
	
	\item Define the star-product $\star$ as $Q(f \star g) = Q(f) Q(g)$ and the involution ${}^\dagger$ as $Q(f^\dagger) = (Q(f))^*$, where $*$ is the involution of $\pi(\mathcal{B(H)})$. Note that this star-product does not depend on the chosen representation $\pi$, unlike the quantization map $Q$.
	
	    In the previous case, the star-product was obtained similarly through \eqref{eq:star_prod_quant}, but the involution was not defined.
\end{enumerate}

\paragraph{}
Note that, up to technical precautions, a convenient way to obtain a (non-degenerate) representation $\pi:\mathbb{C}(G) \to \mathcal{B}(\mathcal{H})$ of $\mathbb{C}(G)$ is to first determine $\pi_U$, a unitary representation of $G$, and to set 
\begin{equation}
    \pi(F) = \int_{G} \dd\nu(s)F(s)\pi_U(s),
    \label{eq:reps_conv_alg}
\end{equation}
in which $\dd\nu$ is a Haar measure on $G$.

\paragraph{}
Since we will have to use involutions, as mentioned at the beginning of this section, we will use representations respecting the involutive structure of the algebras, \textit{i.e.}\ $*$-representations, namely such that 
\begin{equation}
    (\pi(F))^* = \pi(F^\dag),
    \label{eq:star_rep_def}
\end{equation}
where ${}^\dag$ (resp.\ ${}^*$) is the involution equipping $\mathbb{C}(G)$ (resp.\ the image by $\pi$ of $\mathbb{C}(G)$, a subset of $\mathcal{B}(\mathcal{H})$). 

\paragraph{}
Finally, due to the structure of the quantization map and the fact that the Fourier transform is invertible, one obtains explicit expression of the star-product and the involution, in the general case, as follows
\begin{align}
    f \star g
    &= \mathcal{F}^{-1}( \mathcal{F}f \ \hat{\circ}\ \mathcal{F}g ), &
    f^\dagger
    &= \mathcal{F}^{-1}( (\mathcal{F}f)^* ).
    \label{eq:star_prod_inv_def}
\end{align}

\paragraph{}
Examples of star-products using this construction are done in this paper: the Moyal product is constructed in section \ref{subsubsec:moyal_prod}, the star-products for deformations of $\mathbb{R}^3$ are discussed in section \ref{subsec:R3L_star_prod} and the star-products of $\kappa$-Minkowski are built in section \ref{subsubsec:kM_star_prod_affine}.

\subsection{Twist deformations.}
\label{subsec:star_prod_twists}
\paragraph{}
Star-products can be built from other ways, sometimes equivalently to the previous one. One of the most renowned approach is based on twist deformation of the commutative pointwise product of functions in $\mathcal{A}$. This notion requires to handle quantum groups, also known as Hopf algebras. We here recall what Hopf algebras and their twists are. We refer the reader to textbooks like \cite{Klimyk_1997, Majid_1995} for more details.

\subsubsection{Hopf algebras.}
\label{subsubsec:star_prod_hopf_alg}

\begin{figure}[h]
    \centering
    \begin{tabular}{cc}
	\begin{tikzpicture}
		%Nodes
		\node (HHH) {$\algH \otimes \algH \otimes \algH$};
		\node (HHu) [node distance=1.4cm, right=of HHH] {$\algH \otimes \algH$};
		\node (HHd) [below=of HHH] {$\algH \otimes \algH$};
		\node (H)   [below=of HHu] {$\algH \!\!\!\!\!\phantom{\otimes}$};
		%Lines
		\draw[thick, <-] (HHu.west) -- 
		    node[anchor=south]{$\mu \otimes \id$} (HHH.east);
		\draw[thick, <-] (HHd.north) -- 
		    node[anchor=west]{$\id \otimes \mu$} (HHH.south);
		\draw[thick, <-] (H.north) --
		    node[anchor=west]{$\mu$} (HHu.south);
		\draw[thick, <-]  (H.west) -- 
		    node[anchor=south]{$\mu$} (HHd.east);
	\end{tikzpicture}%
	&
	\begin{tikzpicture}
	    %Nodes
		\node (HH) {$\algH \otimes \algH$};
		\node (HC) [node distance=1.2cm, right=of HH] {$\algH \otimes \mathbb{C}$};
		\node (CH) [node distance=1.2cm, left=of HH] {$\mathbb{C} \otimes \algH$};
		\node (H) [below=of HH] {$\algH$};
		%Lines
		\draw[thick, <-] (HH.west) -- 
		    node[anchor=south]{$\eta \otimes \id$} (CH.east);
		\draw[thick, <-] (HH.east) -- 
		    node[anchor=south]{$\id \otimes \eta$} (HC.west);
		\draw[thick, <-]  (H.north west) --
		    node[anchor=north east]{$\id$} (CH.south);
		\draw[thick, <-] (H.north east) -- 
		    node[anchor=north west]{$\id$} (HC.south);
		\draw[thick, <-] (H.north) --
		    node[anchor=west]{$\mu$} (HH.south);
	\end{tikzpicture}
	\end{tabular}
    \caption{Commutative diagram corresponding to the associativity \eqref{eq:alg_alg_def_assos} property (left) and to the unity \eqref{eq:alg_alg_def_unit} property (right) of the algebra $(\algH, \mu, \eta)$.}
    \label{fig:alg_com_diag}
\end{figure}

\begin{figure}[ht]
    \centering
    \begin{tabular}{cc}
	\begin{tikzpicture}
		%Nodes
		\node (HHH) {$\algH \otimes \algH \otimes \algH$};
		\node (HHu) [node distance=1.4cm, right=of HHH] {$\algH \otimes \algH$};
		\node (HHd) [below=of HHH] {$\algH \otimes \algH$};
		\node (H)   [below=of HHu] {$\algH \!\!\!\!\!\phantom{\otimes}$};
		%Lines
		\draw[thick, ->] (HHu.west) -- 
		    node[anchor=south]{$\Delta \otimes \id$} (HHH.east);
		\draw[thick, ->] (HHd.north) -- 
		    node[anchor=west]{$\id \otimes \Delta$} (HHH.south);
		\draw[thick, ->] (H.north) --
		    node[anchor=west]{$\Delta$} (HHu.south);
		\draw[thick, ->] (H.west) -- 
		    node[anchor=south]{$\Delta$} (HHd.east);
	\end{tikzpicture}%
	&
	\begin{tikzpicture}
	    %Nodes
		\node (HH) {$\algH \otimes \algH$};
		\node (HC) [node distance=1.2cm, right=of HH] {$\algH \otimes \mathbb{C}$};
		\node (CH) [node distance=1.2cm, left=of HH] {$\mathbb{C} \otimes \algH$};
		\node (H) [below=of HH] {$\algH$};
		%Lines
		\draw[thick, ->] (HH.west) -- 
		    node[anchor=south]{$\varepsilon \otimes \id$} (CH.east);
		\draw[thick, ->] (HH.east) -- 
		    node[anchor=south]{$\id \otimes \varepsilon$} (HC.west);
		\draw[thick, ->] (H.north west) --
		    node[anchor=north east]{$\id$} (CH.south);
		\draw[thick, ->] (H.north east) -- 
		    node[anchor=north west]{$\id$} (HC.south);
		\draw[thick, ->] (H.north) --
		    node[anchor=west]{$\Delta$} (HH.south);
	\end{tikzpicture}
	\end{tabular}
    \caption{Commutative diagram corresponding to the coassociativity \eqref{eq:coalg_def_coass} property (left) and to the counity \eqref{eq:coalg_def_counit} property (right) of the coalgebra $(\algH, \Delta, \varepsilon)$.}
    \label{fig:coalg_com_diag}
\end{figure}

\paragraph{}
First, an algebraic definition of an algebra $(\algH, \mu, \eta)$ is to say that it is a vector space $\algH$ with two linear maps: a product $\mu: \algH \otimes \algH \to \algH$ and a unit $\eta: \mathbb{C} \to \algH$. They must satisfy the associativity and unit properties
\begin{subequations}
\begin{align}
    \mu \circ (\mu \otimes \id)
    &= \mu \circ (\id \otimes \mu), &
    \text{(associativity)}
    \label{eq:alg_alg_def_assos} \\
    \mu \circ (\eta \otimes \id)
    &= \mu \circ (\id \otimes \eta)
    = \id. &
    \text{(unit property)}
    \label{eq:alg_alg_def_unit}
\end{align}
    \label{eq:alg_alg_def}
\end{subequations}
These relations can be summarized in the commutation of the diagrams of Figure \ref{fig:alg_com_diag}. They are related to the usual definition of an algebra through $hg = \mu(h \otimes g)$ and $1 = \eta(1)$, for any $h,g \in \algH$.

A coalgebra can be seen as the dualised version of an algebra so that one just has to ``reverse the arrows'' of the previous definition. Thus, a coalgebra $(\algH, \Delta, \varepsilon)$ is made of a vector space $\algH$ and two linear maps: a coproduct $\Delta: \algH \to \algH \otimes \algH$ and a counit $\varepsilon: \algH \to \mathbb{C}$. As above they must satisfy some properties that are called coassociativity and the counit property:
\begin{subequations}
\begin{align}
    (\Delta \otimes \id) \circ \Delta
    &= (\id \otimes \Delta) \circ \Delta, &
    \text{(coassociativity)}
    \label{eq:coalg_def_coass} \\
    (\varepsilon \otimes \id) \circ \Delta 
    &= (\id \otimes \varepsilon) \circ \Delta 
    = \id. &
    \text{(counit property)}
    \label{eq:coalg_def_counit}
\end{align}
    \label{eq:coalg_def}
\end{subequations}
These relations can be summarized in the commutation of the diagrams of Figure \ref{fig:coalg_com_diag}.

Now considering a vector space $\algH$, one says that $(\algH, \mu, \eta, \Delta, \varepsilon)$ is a bialgebra if it satisfies both \eqref{eq:alg_alg_def} and \eqref{eq:coalg_def} and if
\begin{align}
    \Delta(hg) &= \Delta(h) \Delta(g), &
    \varepsilon(hg) &= \varepsilon(h) \varepsilon(g), &
    \Delta(1) &= 1 \otimes 1, &
    \varepsilon(1) &= 1
    \label{eq:bialg_def}
\end{align}
is satisfied for any $h,g \in \algH$.

Finally, to obtain a Hopf algebra one has to add an invertible linear map $S: \algH \to \algH$ that is called the antipode, or coinverse. Finally, one says that $(\algH, \mu, \eta, \Delta, \varepsilon, S)$ is a Hopf algebra\footnote{
Generally, the notations are simplified to ``$\algH$ is a Hopf algebra'' without mentioning the maps.
}
if it satisfies
\begin{equation}
	\mu \circ (S \otimes \id) \circ \Delta
	= \mu \circ(\id \otimes S) \circ \Delta 
	= \eta \circ \varepsilon.
	\label{eq:hopf_def}
\end{equation}
This antipode has several properties that are, for any $h,g \in \algH$,
\begin{subequations}
\begin{align}
	S(hg) &= S(g) S(h) &
	S(1) &= 1, 
	\label{eq:prop_antipode_1} \\
	\Delta \circ S 
	&= \tau \circ (S \otimes S) \circ \Delta &
	\varepsilon \circ S &= \varepsilon,
	\label{eq:prop_antipode_2}
\end{align}
	\label{eq:prop_antipode}
\end{subequations}
where $\tau$ is the flip map, that is $\tau(h \otimes g) = g \otimes h$.

Note that one can also write the commutativity and cocommutativity properties with these notations: $\mu \circ \tau = \mu$ and $\tau \circ \Delta = \Delta$ respectively. Those will not be enforced here.

\paragraph{}
One can define the notion of dual Hopf algebra for a finite dimensional Hopf algebra. Explicitly, for a finite dimensional Hopf algbera $(\algH, \mu, 1, \Delta, \varepsilon, S)$, one defines $\algH'$ as its dual in the sens of vector spaces. Then, one defines an algebra structure $(\algH', \mu', 1)$ as
\begin{align}
    \mu'(f_1 \otimes f_2)(h) = (f_1 \otimes f_2) \Delta(h) = f_1(h_{(1)}) f_2(h_{(2)}), &&
    1(h) = \varepsilon(h).
    \label{eq:hopf_alg_dual_alg}
\end{align}
for any $f_1, f_2 \in \algH'$ and $h \in \algH$, where we used the Sweedler notations $\Delta(h) = h_{(1)} \otimes h_{(2)}$. The coalgebra sector $(\algH', \Delta', \varepsilon')$ is defined as
\begin{align}
    \Delta'(f)(g \otimes h) = f \circ \mu (g \otimes h) = f(gh) &&
    \varepsilon'(f) = f(1).
    \label{eq:hopf_alg_dual_coalg}
\end{align}
for any $f \in \algH'$ and $g, h \in \algH$. Finally, one can define an antipode $S'$ for $\algH'$, through
\begin{align}
    S'(f)(h) = f \big( S(h) \big).
    \label{eq:hopf_alg_dual_ant}
\end{align}
Finally, $(\algH', \mu', 1, \Delta', \varepsilon', S')$ is a Hopf algebra, which is dual to $\algH$.

An important thing to note in this definition is that the algebra sector of $\algH'$ is linked to the coalgebra sector of $\algH$, as $\mu'$ and $1$ are defined thanks to $\Delta$ and $\varepsilon$ in \eqref{eq:hopf_alg_dual_alg}, and conversely, the coalgebra sector of $\algH'$ is linked to the algebra sector of $\algH$, as $\Delta'$ and $\varepsilon'$ are defined thanks to $\mu$ and $1$ in \eqref{eq:hopf_alg_dual_coalg}. One can go deeper in identifying both structures by constructing a dual pairing of Hopf algebra. Still, this observation hints for the following property: a Hopf algebra is commutative (resp.\ cocommutative) if and only if its dual Hopf algebra is cocommutative (resp.\ commutative). This pairing shed some light on properties of the $\kappa$-Minkowski space as detailed in section \ref{sec:kappa}.

\subsubsection{Drinfeld twists.}
\label{subsubsec:star_prod_Drinfeld}
\paragraph{}
One can deform a Hopf algebra $\algH$ into another Hopf algebra, noted $\algH^\twiF$, using a Drinfeld twist. The latter is an invertible element $\twiF \in \algH \otimes \algH$ defined through two conditions:
\begin{subequations}
\begin{align}
    (\twiF \otimes 1) (\Delta \otimes \id) (\twiF)
    &= (1 \otimes \twiF) (\id \otimes \Delta) (\twiF), &
    \text{($2$-cocycle condition)}
    \label{eq:twist_def_cocycle} \\
    (\id \otimes \varepsilon) (\twiF)
    &= (\varepsilon \otimes \id) (\twiF)
    = 1 \otimes 1. &
    \text{(normalisation)}
    \label{eq:twist_def_norm}
\end{align}
    \label{eq:twist_def}
\end{subequations}
An additional condition is sometimes required in the context of deformation quantization stating $\twiF = 1 \otimes 1 + \mathcal{O}(\theta)$, where $\theta$ is the deformation parameter. This condition forces the twist to disappear at the commutative limit $\theta \to 0$.

Then, one can show (see \cite{Majid_1995} Theorem 2.3.4) that $\algH^\twiF$ defined by $(\algH, \mu, \eta, \Delta^\twiF, \varepsilon, S^\twiF)$ is also a Hopf algebra, where
\begin{align}
    \Delta^\twiF(h)
    &= \twiF \Delta(h) \twiF^{-1}, &
    S^\twiF(h)
    &= \chi S(h) \chi^{-1}
\end{align}
with $\chi = \twiF_1 S(\twiF_2)$ writing $\twiF = \twiF_1 \otimes \twiF_2$ and $h \in \algH$. One sees from the definition of the twisted algebra $\algH^\twiF$ that only the coproduct and the antipode are deformed, the algebra structure and the counit are untouched.

The Drinfeld twist were first put forward in \cite{Drinfeld_1990}, see also \cite{Oeckl_2000}.

\paragraph{}
The precise purpose of twists and Hopf algebra appears when one considers the symmetries of the space-time. Classically, One enforces symmetries on a space-time as a group $G$ acting on the functions $\mathcal{A}$ of this space. This action allows to see how quantities behave with respect to these symmetries.

One can enlarge this vision to quantum symmetries by making the quantum group (Hopf algebra) $\algH$ act on the space-time algebra $\mathcal{A}$. One turns $(\mathcal{A}, \cdot, \bbone)$ into $\algH$-module algebra\footnote{
The following definition is written for a left action $\actl$, however, it could equivalently be written for a right action $\actr$.
}, that is it must satisfy
\begin{subequations}
\begin{align}
    (hg) \actl a &= h \actl (g \actl a), &
    1 \actl a &= a,
    \label{eq:hopf_module_alg_module} \\
    h \actl (a \cdot b) &= (h_{(1)} \actl a) \cdot (h_{(2)} \actl b), &
    h \actl \bbone &= \varepsilon(h) \bbone,
    \label{eq:hopf_module_alg_alg}
\end{align}
    \label{eq:hopf_module_alg}
\end{subequations}
where $\actl: \algH \otimes \mathcal{A} \to \mathcal{A}$ is the action of $\algH$ on $\mathcal{A}$, $a,b \in \mathcal{A}$ and $h, g \in \algH$. We also used here the Sweedler notations $\Delta(h) = h_{(1)} \otimes h_{(2)}$. Equations \eqref{eq:hopf_module_alg_module} are consistency equations for the action, and equations \eqref{eq:hopf_module_alg_alg} are matching equations between the product of $\mathcal{A}$ and the coproduct of $\algH$.

Then, one can show (see for e.g.\ \cite{Aschieri_2012a} Theorem 3.4) that the twisted Hopf algebra $\algH^\twiF$ acts on an algebra $\algA$ defined by $(\mathcal{A}, \star, \bbone)$, where the new product is defined by
\begin{align}
    a \star b
    &= \cdot \circ \twiF^{-1} \actl (a \otimes b)
    = (\twiF^{-1}_1 \actl a) \cdot (\twiF^{-1}_2 \actl b).
    \label{eq:star_prod_twist}
\end{align}
Thus, through the deformation of its symmetries, the space-time is itself deformed, usually to a noncommutative space-time. More precisely, a deformation of the classical product $\cdot$ to the star-product $\star$ is due to the deformation of the coproduct $\Delta$ to $\Delta^{\twiF}$ through \eqref{eq:hopf_module_alg_alg}.

\subsubsection{Constructions of twists.}
\label{subsubsec:star_prod_eg}
\paragraph{}
One of the first applications of the latter formalism is the universal enveloping algebra $U(\mathfrak{g})$ of a Lie algebra $\mathfrak{g}$. At first, the enveloping algebra is the tensor algebra made out of generators of $\mathfrak{g}$, but it can be turned into a full Hopf algebra by considering
\begin{align}
    \Delta(x) &= x \otimes 1 + 1 \otimes x, &
    \varepsilon(x) &= 0, &
    S(x) &= -x,
\end{align}
for any generating element $x\in\mathfrak{g}$.

It was applied to $\mathfrak{g}$ being the Poincar\'{e} algebra \cite{Chaichian_2004, Chaichian_2005} to show that the Moyal space-time noncommutativity relation \eqref{eq:moyal_coord_nc}, despite breaking Lorentz invariance, is invariant under twisted Poincar\'{e} symmetry.

Another application was made to make the Moyal product invariant under deformed diffeomorphisms by considering $\mathfrak{g}$ to be the Lie algebra of vector fields. This will be more discussed in section \ref{subsubsec:braid_deformed_vf}.

\paragraph{}
The twist allows to built the Moyal product with an Abelian twist (see section \ref{subsubsec:moyal_prod}), a star-product on $\kappa$-Minkowski with an Abelian and a Jordanian twist (see section \ref{subsubsec:km_twist_star_prod}) and also a proposal for gravity formulation (see section \ref{subsubsec:braid_deformed_vf}).

\newpage
\section{Derivations and noncommutative differential calculus.}
\label{sec:nc_diff_calc}

\subsection{Trading vectors fields for derivations.}
\label{subsec:nc_diff_calc_der}
\paragraph{}
In this section, we collect the useful properties of the derivation-based differential calculus. This noncommutative differential calculus, introduced a long time ago \cite{DBV-1, mdv99}, underlies most of the physics literature dealing with noncommutative gauge theories. We will extensively use it to adopt a unified presentation of the materials used in the subsequent analysis. Other noncommutative differential calculi, dealing with particular classes of models will be introduced when necessary.

The derivation-based differential calculus is in some sense a noncommutative extension of the algebraic approach of the (usual) differential geometry by Koszul \cite{KOSZUL-T}, as it will be illustrated in a while on representative examples. In particular, the (commutative) manifold is replaced by an associative algebra modeling the ``noncommutative space''.  The role of the vector fields is played by the derivations of the algebra, while the main algebraic structures prevailing in the commutative set-up are preserved or extended consistently to the noncommutative setting. For a review, see e.g.\ \cite{mdv99}.

\paragraph{}
We start from an associative unital involutive algebra, with product $\star$ and center $\caZ(\algA)$. Recall that the center gathers the elements of $\algA$ that commute with any element of the algebra. The unit and involution are denoted by $\bbone$ and $a\mapsto a^\dag$ for any $a\in \algA$. Let $\mathrm{Der}(\algA) $ be the linear space of the derivations of $\algA$, that is, the linear maps $\kX : \algA \rightarrow \algA$, satisfying the Leibniz rule
\begin{equation}
    \kX(a \star b) 
    = \kX(a) \star b + a \star \kX(b),
    \label{eq:leibniz}
\end{equation}
for any $a,b\in \algA$. We will sometimes need to work with real derivations. Recall that a derivation is real if 
\begin{equation}
    X(a)^\dag
    = X(a^\dag),
    \label{eq:real_der}
\end{equation}
for any $a\in\algA$. Not all the derivations needed in noncommutative gauge theories are real, as we will see in section \ref{sec:kappa}.

One easily verifies that $\mathrm{Der}(\algA)$ inherits a Lie algebra structure from the bracket defined by $[\kX, \kY ](a) = \kX  (\kY (a)) - \kY (\kX (a))$, for any $a\in\algA$, $\kX,\kY \in \mathrm{Der}(\algA)$. Furthermore, defining the action $(z \actl \kX) (a) = z \star \kX (a)$, for any $z \in \caZ(\algA)$, $\kX \in \mathrm{Der}(\algA)$, $\mathrm{Der}(\algA)$ is a $\caZ(\algA)$-module for this action. The latter action will be simply denoted by the product: $z \star X \in \Der(\algA)$.

\paragraph{}
The linear subspace $\Int(\algA) \subset \mathrm{Der}(\algA)$ involving derivations such that
\begin{align}
    \adrep_a : b \mapsto [a,b], &&
    a \in \algA
\end{align}
is called the space of inner derivations. Inner derivations will play an important role in the following, in particular within the noncommutative gauge theories on Moyal space (see section \ref{sec:moyal}). It can be easily verified that $\Int(\algA)$ is a $\caZ(\algA)$-submodule. Finally, derivations in $\mathrm{Der}(\algA)$ which are not inner, \textit{i.e.}\ $\Out(\algA) = \mathrm{Der}(\algA) / \Int(\algA)$, are called outer derivations.

\paragraph{}
The derivation-based differential calculus is defined as follows. Consider the space of $\caZ(\algA)$-multilinear antisymmetric maps $\omega: \mathrm{Der}(\algA)^n\to\algA$, $n\in\mathbb{N}$, denoted by ${\Omega}^n(\algA)$, with ${\Omega}^0(\algA)=\algA$.

Set 
\begin{equation}
    \Omega^\bullet(\algA) 
    = \bigoplus_{n \geqslant 0} \Omega^n(\algA).
    \label{eq:ncdc_set_forms}
\end{equation}
Then, define the product $\wedge: \Omega^\bullet(\algA) \to \Omega^\bullet(\algA)$ by for any $\omega\in{\Omega}^p(\algA), \eta\in{\Omega}^q(\algA)$,
\begin{equation}
\begin{aligned}
    (\omega \wedge \eta) & (\kX_1, \dots, \kX_{p+q}) \\
    &= \frac{1}{p!q!} \sum_{\sigma\in \kS_{p+q}} (-1)^{{\sign}(\sigma)}
    \omega(\kX_{\sigma(1)}, \dots, \kX_{\sigma(p)}) \star \eta(\kX_{\sigma(p+1)}, \dots, \kX_{\sigma(p+q)}),
    \label{eq:form_prod}
\end{aligned}
\end{equation}
where ${\sign}(\sigma)$ is the signature of the permutation $\sigma$ and $\kS_{p+q}$ is the symmetric group of $p+q$ elements. If the product $\star$ is not commutative, one has $\omega \wedge \eta \ne (-1)^{|\omega|\, |\eta|} \eta \wedge \omega$ showing that $\Omega^\bullet(\algA)$ \eqref{eq:ncdc_set_forms} is not graded commutative.

The remaining element to define a differential calculus is the differential $\dd$, which is a linear map $\dd: \Omega^p(\algA) \to \Omega^{p+1}(\algA)$, for every form degree $p\in\mathbb{N}$, that satisfies
\begin{align}
\begin{aligned}
    \dd \omega(\kX_1, \dots, \kX_{p+1}) 
    &= \sum_{j = 1}^{p+1} (-1)^{j+1} \kX_j \big( \omega( \kX_1, \dots, \vee_j, \dots, \kX_{p+1}) \big) \\
    &+ \sum_{1 \leqslant j < k \leqslant p+1} (-1)^{j+k} \omega( [\kX_j, \kX_k], \dots, \vee_j, \dots, \vee_k, \dots, \kX_{p+1}),
    \label{eq:koszul}
\end{aligned}
\end{align}
for any $\omega\in{\Omega}^p(\algA)$, in which $\vee_j$ denotes the omission of the element $X_j$. A simple calculation yields\footnote{
Notice that the Lie algebra and ${\cal{Z}}(\algA)$-module structures for $\mathrm{Der}(\algA)$ are essential for in the construction.
} 
\begin{equation}
    \dd^2 = 0
    \label{eq:d2_0}.
\end{equation}

From \eqref{eq:leibniz}, \eqref{eq:form_prod} and \eqref{eq:koszul}, one can show that the differential satisfies the following expected graded Leibniz rule
\begin{equation}
    \dd(\omega \wedge \eta)
    = \dd\omega \wedge \eta + (-1)^{|\omega|} \omega \wedge \dd\eta,
    \label{eq:ncdc_leibniz}
\end{equation}
for any $\omega,\eta\in\Omega^\bullet(\algA)$ where $|\omega|$ denotes the degree of $\omega$. Then, one verifies that the triplet
\begin{equation}
    (\Omega^\bullet(\algA), \wedge, \dd)
    \label{eq:ncdiff-alg}
\end{equation}
is a (${\mathbb{N}}$-graded) differential algebra defining the derivation-based differential calculus.

\paragraph{}
Notice that any Lie subalgebra $\mathfrak{g}$ of the whole Lie algebra of derivations $\mathrm{Der}(\algA)$, which is still a $\caZ(\algA)$-submodule, can be used instead of $\mathrm{Der}(\algA)$ in the above construction. This will give rise to another derivation-based differential calculus, sometimes called $\mathfrak{g}$-restricted differential calculus. It appears that this type of restricted derivation-based differential calculus underlies most of the literature devoted to the gauge theories on the Moyal space which will be considered in section \ref{sec:moyal}.

\paragraph{}
As an illustrative example of the discussion, it is instructive to recover the usual de Rham differential calculus from the framework presented above. Indeed, set $\algA = C^\infty(\mathcal{M})$, the algebra of smooth functions on a smooth manifold $\mathcal{M}$. The algebra structure of the latter is given by the pointwise product. Then, $\mathrm{Der}(\algA) = \Gamma(\mathcal{M})$, the space of sections of the tangent space $T\mathcal{M}$, which is known to be the $C^\infty(\mathcal{M})$-module of smooth vector fields on $\mathcal{M}$, and inherits a Lie algebra structure for the usual bracket $[u,v](f) = u(f) v(f) - v(f) u(f)$ for any $u, v \in \Gamma(\mathcal{M})$, $f \in C^\infty(\mathcal{M})$. Observe that $\Int(\algA) = \{0\}$, due to the commutativity of the algebra. Observe also that, in the present case, the commutativity property of the pointwise product also makes the algebra coincide with its center, \textit{i.e.}\ $\caZ(\algA) = C^\infty(\mathcal{M}) = \algA$. However, notice that in a general noncommutative situation, the associative algebra and its center have their own (distinct) role. It is then a simple matter of algebra to check that \eqref{eq:form_prod} and \eqref{eq:koszul} coincides with the usual rules defining the de Rham differential calculus.

\subsection{Connections on a right-module.}
\label{subsec:nc_conn_right}
\paragraph{}
Various noncommutative generalizations of the notion of connection have appeared in the literature. We will mainly use a noncommutative generalization of ordinary connection which exploits the structure of a right module over the algebra. The latter is the most widely used noncommutative connection in the physics literature. Another but quite close generalization, based on bimodules, is presented in section \ref{subsec:nc_central_bimod}. Noncommutative connections on a right-module over the algebra can be viewed as a noncommutative extension of the notion of Koszul connection (linear connection) on a vector bundle \cite{KOSZUL-T}, \cite{KOSZUL-conn}. Recall that this latter can be defined as a map $\nabla: \Gamma(\mathcal{E}) \to \Gamma(\mathcal{E} \otimes T^*\mathcal{M})$ such that $\nabla(mf)=\nabla(m)f+m\otimes df$ for any $m\in\Gamma(\mathcal{E})$, $f\in C^\infty({\mathcal{M}})$ where $T^*\mathcal{M}$ is the cotangent bundle over a smooth manifold $\mathcal{M}$ and $\Gamma(\mathcal{E})$ is the space of sections of a vector bundle $\mathcal{E}$ over $\mathcal{M}$. The latter space of sections is a module over the smooth functions with an action given by the pointwise product. Therefore, $\Gamma(\mathcal{E})$ will be generalized by a module $\modM$ over the algebra $\algA$ in the following. From this, one can define for any $X\in \Gamma(\mathcal{M})$ the map $\nabla_X: \Gamma(\mathcal{E}) \to \Gamma(\mathcal{E})$ satisfying in particular $\nabla_X(m f) = m X(f) + \nabla_X(m) f$, which represents the covariant derivative.

\paragraph{}
One should emphasise that the constructions in section \ref{subsec:nc_conn_right} and \ref{subsec:pract_scheme} deal with right modules, but could be equivalently done for left modules. The case of bimodules is discussed in section \ref{subsec:nc_central_bimod}.

\paragraph{}
The noncommutative generalization of the above goes as follows. Let $\modM$ be a right $\algA$-module, \textit{i.e.}\ we provide $\modM$ with a right action of the algebra $\algA$ on $\modM$ denoted $\actr: \modM \otimes \algA \to \modM$ that should be linear and satisfy 
\begin{align}
    (m \actr a) \actr b
    &= m \actr (a \star b), &
    m \actr \bbone
    &= m
    \label{eq:module_def}
\end{align}
for any $a,b \in \algA$, $m \in \modM$. This action can be seen as a product.

The connection is then defined as a linear map 
$\nabla_\kX : \modM \to \modM$ satisfying
\begin{subequations}
\begin{align}
    \nabla_\kX (m \actr a) 
    &= m \actr \kX(a) + \nabla_\kX (m) \actr a,
    \label{eq:nc_conn_def_leibniz} \\
    \nabla_{f \star \kX + \kY}(m) 
    &= \nabla_\kX(m) \actr f + \nabla_\kY(m),
    \label{eq:nc_conn_def_lin}
\end{align}
    \label{eq:nc_conn_def}
\end{subequations}
for any $\kX,\kY \in \mathrm{Der}(\algA)$, $a \in \algA$, $m \in \modM$, $f \in \caZ(\algA)$. The relation \eqref{eq:nc_conn_def_leibniz} makes $\nabla$ extend to a map 
\begin{align}
    &\nabla: \modM \to \modM \otimes_\algA \Omega^1(\algA), &
    \nabla(m \actr a)
    &= \nabla(m) \actr a + m \otimes \dd a,
    \label{eq:nc_conn_form_def} 
\end{align}
for any $m\in\modM$, $a\in\algA$ which can be further extended to a map $\nabla: \modM \to \modM \otimes_\algA \Omega^\bullet(\algA)$.

The curvature of $\nabla$ represents the obstruction for $\nabla$ to be a homomorphism of right module. Recall that a homomorphism of right module is a linear map $\varphi: \modM \to \modM$ that satisfies 
\begin{equation}
    \varphi(m \actr a) = \varphi(m) \actr a,
    \label{eq:homomorph_right_mod_def}
\end{equation}
for any $m \in \modM$, $a \in \algA$. The curvature can be defined as the linear map $\hR(\kX, \kY) : \modM \to \modM$ such that
\begin{equation}
    \hR(\kX, \kY) (m) 
    = [\nabla_\kX, \nabla_\kY ] (m) - \nabla_{[\kX, \kY]}(m)
    \label{eq:nc_curv_def}
\end{equation}
for any  $m\in\modM$, $\kX, \kY \in \mathrm{Der}(\algA)$.

One can easily verify that $\hR(\kX, \kY)$ is a homomorphism of right module, namely one has $\hR(\kX, \kY)(m \actr a) = \hR(\kX, \kY)(m) \actr a$ for any 
$m\in\modM$, $a\in\algA$, $\kX, \kY \in \mathrm{Der}(\algA)$.

\paragraph{}
The gauge group of $\modM$ is defined as the group of automorphisms of $\modM$, denoted $\Aut(\modM)$, compatible with the structure of right $\algA$-module. Namely, elements of $\Aut(\modM)$ should satisfy \eqref{eq:homomorph_right_mod_def}. The action of a gauge transformation $\varphi \in \Aut(\modM)$ on a connection $\nabla$ is defined such that the map
\begin{align}
    &\nabla_X: \modM \to \modM, &
    \nabla^\varphi_\kX 
    &= \varphi^{-1} \circ \nabla_\kX \circ \varphi
    \label{eq:nc_conn_gauge}
\end{align}
is still a connection, that is satisfy \eqref{eq:nc_conn_def}. The corresponding gauge transformation of the curvature is
\begin{equation}
    \hR^\varphi(X,Y)
    = \varphi^{-1} \circ \hR(X,Y) \circ \varphi.
    \label{eq:nc_curv_gauge}
\end{equation}

\paragraph{}
Most of the noncommutative gauge theories considered in the literature use the notion of hermitian connection, simply defined as a connection compatible with a hermitian structure and a noncommutative analog of a ``unitary gauge group''.

First, recall that a hermitian structure is a sesquilinear map, $h: \modM \otimes \modM \to \algA$, such that 
\begin{align}
    h(m_1, m_2)^\dag
    &= h(m_2, m_1), &
    h(m_1 \actr a_1, m_2 \actr a_2)
    &= a_1^\dag \star h(m_1, m_2) \star a_2,
    \label{eq:nc_hermit_struct}
\end{align}
for any $m_1, m_2 \in \modM$, $a_1, a_2 \in \algA$. Then, a noncommutative connection as defined in \eqref{eq:nc_conn_def} is hermitian if it satisfies, for any $m_1, m_2 \in \modM$,
\begin{equation}
    X \big( h(m_1, m_2) \big)
    = h \big(\nabla_X(m_1), m_2 \big) + h \big(m_1, \nabla_X(m_2) \big),
    \label{eq:nc_hermit_conn}
\end{equation}
for any {\textit{real}} derivation $X\in\mathrm{Der}(\algA)$.

Then, the unitary gauge transformations are defined as the gauge transformations $\varphi$ compatible with the hermitian structure, that is verifying
\begin{equation}
    h(\varphi(m_1), \varphi(m_2))
    = h(m_1, m_2)
    \label{eq:nc_unit_gauge}.
\end{equation}
We denote by $\mathcal{U}(\modM)$ the group of unitary gauge transformations.

\subsection{Two practical schemes}
\label{subsec:pract_scheme}

\subsubsection{The module as a copy of the algebra}
\label{subsubsec:nc_mod_is_alg}
\paragraph{}
In many noncommutative gauge theories considered in the physics literature, one assumes that the module $\modM$ is {\textit{one copy}} of the algebra, \textit{i.e.}\ $\modM = \algA$ with the action given by the product, \textit{i.e.}\ $\actr = \star$. This scheme allows a generalization of $U(1)$ gauge theory. 

One takes the canonical hermitian structure 
\begin{equation}
    h_0(m_1, m_2)
    = m_1^\dag \star m_2.
    \label{eq:nc_A_canon_herm}
\end{equation}
Under these assumptions, one easily realizes that \eqref{eq:nc_conn_def_leibniz} (taking $m = \bbone$) implies 
\begin{equation}
    \nabla_\kX (a) = \kX(a) + \nabla_\kX (\bbone) \star a
    \label{eq:nc_A_conn_leibniz}, 
\end{equation}
for any $a \in \algA$, so that the noncommutative connection is entirely determined by $\nabla_X(\bbone)$, for any $X \in \mathrm{Der}(\algA)$. Set now 
\begin{equation}
    \nabla_X(\bbone)
    = i A_X
    \label{eq:nc_A_gauge_pot}.
\end{equation}
The 1-form connection $A \in {\Omega}^1(\algA)$ is defined accordingly by $A: X \to A_X = i \nabla_X(\bbone)$, for any $X\in\mathrm{Der}(\algA)$.

Furthermore, by combining \eqref{eq:nc_hermit_conn}, \eqref{eq:nc_A_canon_herm} and \eqref{eq:nc_A_conn_leibniz}, one easily verifies that a connection is hermitian provided that the associated $1$-form connection is, explicitly
\begin{equation}
    A_X^\dag = A_X
    \label{eq:nc_A_gauge_pot_real},
\end{equation}
for any {\textit{real}} derivation $X$.

\paragraph{}
As far as gauge transformations are concerned, since any gauge transformation $\varphi$ must satisfy \eqref{eq:homomorph_right_mod_def} and since $\modM = \algA$, one should have in particular $\varphi(a) = \varphi(\bbone) \star a$ for any $a \in \algA$ so that the gauge transformations are entirely determined by the elements of the module with the form $\varphi(\bbone) \in \modM = \algA$.

Set now $\varphi(\bbone) = g \in \algA$. It follows from \eqref{eq:nc_unit_gauge} and \eqref{eq:nc_A_canon_herm} that the unitary gauge group $\mathcal{U}(\modM)$ is determined by the elements of $\modM = \algA$ satisfying $g^\dag \star g = g \star g^\dag = \bbone$, \textit{i.e.}\ 
\begin{equation}
    \mathcal{U}_{\modM = \algA}(\modM)
    = \{g \in \algA,\ g^\dag \star g = g \star g^\dag = \bbone\}.
    \label{eq:nc_U1}
\end{equation}
In the following, the notation $\mathcal{U}(\algA) := \mathcal{U}_{\modM = \algA}(\modM)$ is used. The gauge transformations for the connection and curvature are easily obtained from a simple calculation. One gets
\begin{align}
    A^g_X
    &= g^\dag \star A_X \star g - i g^\dag \star X(g),
    \label{eq:nc_A_gauge_pot_trans} \\
    \hR(X,Y)^g 
    &= g^\dag \star \hR(X,Y) \star g,
    \label{eq:nc_A_curv_trans}
\end{align}
for any  $X, Y\in\mathrm{Der}(\algA)$ and any $g \in \mathcal{U}(\algA)$.

\paragraph{}
We will encounter a particular situation within gauge theories on the Moyal spaces (see section \ref{sec:moyal}) as well as on $\mathbb{R}^3_\lambda$ (see section \ref{sec:R3L}) which is specific to noncommutative spaces. It turns out that gauge invariant connections may appear. This is due to the possible existence of inner connections in the Lie algebra of the derivation of the associative algebra modeling the noncommutative space.

Assume that the differential $\dd$, defined by \eqref{eq:koszul}, can be written for any $a\in\algA$ as 
\begin{align}
    \dd a
    &= [i \xi, a]_\star
    = i(\xi \star a - a \star \xi), &
    \xi &\in {\Omega}^1(\algA)
    \label{eq:nc_inner_diff}
\end{align}
where $\xi$ is a given 1-form such that $\xi^\dag = \xi$. Then, it is a simple matter of calculation to verify that the connection defined by
\begin{equation}
    \nabla^{\mathrm{inv}}(a)
    = \dd a - i \xi \star a,
    \label{eq:nc_A_cano_conn}
\end{equation}
for any $a \in \algA$, is gauge invariant, that is
\begin{equation}
    (\nabla^{\mathrm{inv}})^g(a)
    = \nabla^{\mathrm{inv}}(a)
    \label{eq:nc_A_cano_conn_invar}
\end{equation}
for any $g\in\mathcal{U}(\algA)$ and any $a \in \algA$. It is then natural to consider the following difference of two connections $\nabla - \nabla^{\mathrm{inv}}$, from which one readily defines a tensor 1-form $\mathcal{A} \in {\Omega}^1(\algA)$ given by
\begin{equation}
    \mathcal{A}
    = i (A + \xi)
    \label{eq:nc_A_covar_coord},
\end{equation}
and, as such, transforms covariantly under $\mathcal{U}(\algA)$, namely 
\begin{equation}
    \mathcal{A}^g
    = g^\dag \star \mathcal{A} \star g
    \label{eq:nc_A_covar_coord_trans}.
\end{equation}
Note that $\mathcal{A}_X$, for $X\in\mathrm{Der}(\algA)$, is sometimes called the covariant coordinate in the physics literature. This object will play a special role in the gauge theories on Moyal spaces as it will be shown in the section \ref{sec:moyal}.

\subsubsection{The module as \tops{$N$}{N} copies of the algebra}
\label{subsubsec:nc_mod_is_N_alg}
\paragraph{}
Another case considered in the physics literature is the one that assumes that the module $\modM$ is $N$ {\textit{copies}} of the algebra, \textit{i.e.}\ $\modM = \algA \times \cdots \times \algA = \algA^N$, with $N \in \mathbb{N}$. This case is very similar to the previous one except a matrix notation \eqref{eq:nc_AN_gauge_pot_matrix} is adopted. 

One can exemplify this matrix notation with the definition of the action. Explicitly, one defines the canonical basis of $\modM = \algA^N$ as $e_j = (0, \dots, 0, \overset{(j)}{\bbone}, 0, \dots, 0)$, where the only non-zero entry is at the $j$-th place, for $j = 1, \dots, N$. Then, for any $m \in \modM$, there are $m_j \in \algA$, for every $j = 1, \dots, N$, such that $m = \sum_{j=1}^N e_j \star m^j$. One could write equivalently $m = \big(m^1, \dots, m^N \big)$. The action is given by the product to all algebra factor, that is $m \actr a = \sum_{j = 1}^N e_j \star m^j \star a$, \textit{i.e.}\ $m \actr a = \big(m^1 \star a, \dots, m^N \star a \big)$. This scheme allows a generalization of $U(N)$ gauge theory. 

As before, one considers the canonical hermitian structure 
\begin{equation}
    h_0(m_1, m_2)
    = \sum_{j = 1}^N (m_1^j)^\dag \star m_2^j.
    \label{eq:nc_AN_canon_herm}
\end{equation}
The connection can again be determined by few elements considering \eqref{eq:nc_conn_def_leibniz} (taking $m = e_j$ for any $j = 1, \dots, N$) which implies 
\begin{equation}
    \nabla_\kX \left( \sum_{j = 1}^N e_j \star m^j \right) 
    = \sum_{j = 1}^N e_j \star \kX(m^j) + \sum_{j = 1}^N \nabla_\kX (e_j) \star m^j
    \label{eq:nc_AN_conn_leibniz}, 
\end{equation}
for any $m = \sum_{j = 1}^N e_j \star m^j \in \modM$, so that the noncommutative connection is entirely determined by the elements $(A_X)^k_j$ defined by
\begin{align}
    \nabla_X(e_j)
    &= i \sum_{k = 1}^N e_k \star (A_X)_j^k
    \label{eq:nc_AN_gauge_pot}
\end{align}
for any $j = 1, \dots, N$ and $X \in \mathrm{Der}(\algA)$. 

All these components can be gathered using the matrix notation. Let $\mathbb{M}_N(\mathbb{C})$ denotes the usual algebra of complex-valued $N \times N$ matrices with $N^2$ canonical basis elements\footnote{
These are matrices $\{E_j^k\}_{j,k= 1, \dots, N}$ with all zero entries except the entry at the $j$-th line and $k$-th row equal to $1$.
}
$E_j^k$, $j, k = 1, \dots, N$, verifying the relation $E_j^k E_l^m = \delta_l^k E_j^m$. Then, one defines $A_X \in \algA \otimes \mathbb{M}_N(\mathbb{C})$ given by
\begin{align}
    A_X
    &= \sum_{j,k = 1}^N E_j^k (A_X)_k^j
    = \begin{pmatrix}
        (A_X)^1_1 & \cdots & (A_X)^1_N \\
        \vdots    & \ddots & \vdots    \\
        (A_X)^N_1 & \cdots & (A_X)^N_N
    \end{pmatrix}.
    \label{eq:nc_AN_gauge_pot_matrix}
\end{align}
The 1-form connection $A \in \Omega^1(\algA) \otimes \mathbb{M}_N(\mathbb{C})$ is defined as above.

Furthermore, by combining \eqref{eq:nc_hermit_conn}, \eqref{eq:nc_AN_canon_herm} and \eqref{eq:nc_AN_conn_leibniz}, one easily verifies that a connection is hermitian provided that $((A_X)^j_k)^\dag = (A_X)^k_j$ for any $j,k = 1, \dots, N$ and $X \in \Der(\algA)$, a \textit{real} derivation. Using the matrix notation \eqref{eq:nc_AN_gauge_pot_matrix}, the latter condition writes
\begin{equation}
    A_X^\ddag = A_X
    \label{eq:nc_AN_gauge_pot_real},
\end{equation}
for any {\textit{real}} derivation $X$, where here ${}^\ddag$ stands for the composition of the involution of $\algA$ and the transposition in $\mathbb{M}_N(\mathbb{C})$. Explicitly, one has ${}^\ddag = {}^\dag \otimes {}^t$ and ${}^\ddag$ corresponds to an involution in $\algA \otimes \mathbb{M}_N(\mathbb{C})$. One can write this hermiticity condition as $iA_X \in \algA \otimes \mathfrak{u}(N)$.

\paragraph{}
As in the previous case, any gauge transformation $\varphi$ must satisfy \eqref{eq:homomorph_right_mod_def} and since $\modM = \algA^N$, one should have in particular $\varphi(m) = \sum_{j = 1}^N \varphi(e_j) \star m^j$ for any $m = \sum_{j = 1}^N e_j \star m^j \in \algA^N$ so that the gauge transformations are entirely determined by the elements of the module with the form $\varphi(e_j) \in \modM = \algA^N$, for all $j = 1, \dots, N$.

Set now $\varphi(e_j) = \sum_{k = 1}^N e_k \star U_j^k \in \algA^N$. Using the same matrix notation as in \eqref{eq:nc_AN_gauge_pot_matrix}, one can write $U \in \algA \otimes \mathbb{M}_N(\mathbb{C})$, which completely determines the gauge transformation $\varphi$. It follows from \eqref{eq:nc_unit_gauge} and \eqref{eq:nc_AN_canon_herm} that the unitary gauge group $\mathcal{U}(\modM)$ is determined by the elements of $\modM = \algA^N$ satisfying $\sum_{l = 1}^N (U_j^l)^\dag \star U_k^l = \delta_{jk} \bbone$. This equation can be written in matrix notations as $U^\ddag \star U = U \star U^\ddag = \bbone_N$, where ${}^\ddag$ is the involution in $\algA \otimes \mathbb{M}_N(\mathbb{C})$ defined in \eqref{eq:nc_AN_gauge_pot_real} and $\bbone_N = \mathrm{diag}(\bbone, \dotsc, \bbone)$ is the identity matrix of $\algA \otimes \mathbb{M}_N(\mathbb{C})$.

This can be summarized as
\begin{equation}
    \mathcal{U}_{\modM = \algA^N}(\modM)
    = \{U \in \algA \otimes \mathbb{M}_n(\mathbb{C}),\ U^\ddag \star U = U \star U^\ddag = \bbone_N\}.
    \label{eq:nc_UN}
\end{equation}
In the following, the notation $\mathcal{U}(\algA,N) := \mathcal{U}_{\modM = \algA^N}(\modM)$ is used. The gauge transformations for the connection and curvature are obtained as above and are given in matrix notations by 
\begin{align}
    A^U_X
    &= U^\ddag \star A_X \star U - i U^\ddag \star X(U),
    \label{eq:nc_AN_gauge_pot_trans} \\
    \hR(X,Y)^U 
    &= U^\ddag \star \hR(X,Y) \star U,
    \label{eq:nc_AN_curv_trans}
\end{align}
for any  $X, Y\in\mathrm{Der}(\algA)$ and any $U \in \mathcal{U}(\algA, N)$. Here, we used the notation $X(U) := \sum_{j, k = 1}^N E^k_j X(U_k^j)$.

\paragraph{}
The existence of gauge invariant connections also stands in this exemple.

Assuming that \eqref{eq:nc_inner_diff} is again valid, then the connection defined by
\begin{equation}
    \nabla^{\mathrm{inv}} \left( \sum_{j = 1}^N e_j \star m^j \right)
    = \sum_{j = 1}^N e_j \star \dd m^j - i \sum_{j = 1}^N e_j \star \xi \star m_j,
    \label{eq:nc_AN_cano_conn}
\end{equation}
for any $\sum_{j = 1}^N e_j \star m^j \in \algA^N$, is gauge invariant, that is
\begin{equation}
    (\nabla^{\mathrm{inv}})^U(m)
    = \nabla^{\mathrm{inv}}(m)
    \label{eq:nc_AN_cano_conn_invar}
\end{equation}
for any $U\in\mathcal{U}(\algA, N)$ and any $m \in \algA^N$. It is then natural to consider the following difference of two connections $\nabla - \nabla^{\mathrm{inv}}$, from which one readily defines a tensor 1-form $\mathcal{A} \in {\Omega}^1(\algA) \otimes \mathbb{M}_N(\mathbb{C})$ in matrix notations, given by
\begin{equation}
    \mathcal{A}
    = i (A + \xi \star \bbone_N)
    \label{eq:nc_AN_covar_coord},
\end{equation}
and, as such, transforms covariantly under $\mathcal{U}(\algA,N)$, namely 
\begin{equation}
    \mathcal{A}^U
    = U^\ddag \star \mathcal{A} \star U
    \label{eq:nc_AN_covar_coord_trans}.
\end{equation}
Again $\mathcal{A}_X$, for $X\in\mathrm{Der}(\algA)$, is sometimes called the covariant coordinate in the physics literature.

\subsection{Connections on central bimodules}
\label{subsec:nc_central_bimod}
\paragraph{}
Another generalization of the notion of sections on a vector bundle is the bimodule, that is a module with both left and right structures. Some authors \cite{Dubois_Violette_1996, Madore_1997} advocate that the bimodule structure is essential to recover the physical properties of a linear connection. The first authors put forward that, from a quantum mechanical point of view, the suitable generalization of the sections on a vector bundle is the hermitian elements of an involutive bimodule over an involutive algebra, as the generalization of functions is not an associative algebra but rather the hermitian elements of an involutive algebra. The second claims that a generalized connection must satisfy both left and right Leibniz rule (see \eqref{eq:bimodule_conn} for more details), otherwise neither the reality condition of a linear connection, nor Lagrangians could be built.

The central trait of bimodule is added in \cite{Dubois_Violette_1996, Dubois_Violette_1994} to be able to define a dual connection, as explained below. Furthermore, from the commutative point of view, bimodules and central bimodules are similar.

\paragraph{}
Let $\modM$ be an $\algA$-bimodule, that is a vector space with two linear maps, a left action $\actl : \algA \otimes \modM \to \modM$ and a right action $\actr : \modM \otimes \algA \to \modM$, which satisfy
\begin{subequations}
\begin{align}
    b \actl (a \actl m)
    &= (b \star a) \actl m, &
    \bbone \actl m
    &= m, \\
    (m \actr a) \actr b
    &= m \actr (a \star b), &
    m \actr \bbone
    &= m,
\end{align}
\end{subequations}
for any $a,b \in \algA$ and $m \in \modM$. This bimodule is further said central if its elements commute with the center of the algebra $\caZ(\algA)$. Explicitly, for any $m \in \modM$ and $f \in \caZ(\algA)$, $f \actl m = m \actr f$.

We consider $\modM$ to be a central bimodule in the following.

\paragraph{}
One can define the notion of a dual bimodule. The dual bimodule $\modM'^{{}_\algA}$ corresponds to the homomorphisms of bimodule from $\modM$ to $\algA$. Explicitly, $\mu \in \modM'^{{}_\algA}$ means that $\mu : \modM \to \algA$ is a linear map that satisfies
\begin{align}
    \mu(a_1 \actl m \actr a_2)
    &= a_1 \star \mu(m) \star a_2,
    & m \in \modM, a_1,a_2 \in \algA.
    \label{eq:bimod_homomorp_def}
\end{align}
Note that, here, $\algA$ has been granted the canonical $\algA$-bimodule structure given by the product: $\actr_\algA = \actl_\algA = \star$. One can show \cite{Dubois_Violette_1996} that if $\modM$ is a central $\algA$-bimodule, then $\modM'^{{}_\algA}$ is a $\caZ(\algA)$-bimodule and, conversely, if $\modM$ is a $\caZ(\algA)$-bimodule, then $\modM'^{{}_\algA}$ is a central $\algA$-bimodule.

\paragraph{}
The connection on a bimodule is defined \cite{Dubois_Violette_1996} as satisfying the right Leibniz rule \eqref{eq:nc_conn_def_leibniz} and the left one at the same time. Explicitly, the connection is a linear map $\nabla_\kX : \modM \to \modM$, such that
\begin{subequations}
\begin{align}
     \nabla_\kX (a_1 \actl m \actr a_2) 
    &= a_1 \actl m \actr \kX(a_2) + a_1 \actl \nabla_\kX (m) \actr a_2 + \kX(a_1) \actl m \actr a_2,
    \label{eq:nc_conn_cenbim_def_leibniz} \\
    \nabla_{f \star \kX + \kY}(m) 
    &= \nabla_\kX(m) \actr f + \nabla_\kY(m)
    = f \actl \nabla_\kX(m) + \nabla_\kY(m),
    \label{eq:nc_conn_ cenbim_def_lin}
\end{align}
    \label{eq:nc_conn_cenbim_def}
\end{subequations}
for any $a_1, a_2 \in \algA$, $m \in \modM$, $\kX, \kY \in \Der(\algA)$ and $f \in \caZ(\algA)$. Note that the second equality in \eqref{eq:nc_conn_ cenbim_def_lin} stands because $\modM$ is central. However, one can also define a connection on a bimodule through \eqref{eq:bimodule_conn}, as detailed in section \ref{subsec:central_bimodules}.

The duality between an $\algA$-bimodule and a $\mathcal{Z}(\algA)$-bimodule can be exploited to define a dual connection. Let $\nabla$ be a connection on $\modM$, taken to be a $\algA$-bimodule here\footnote{
The choice of $\modM$ to be a $\caZ(\algA)$-bimodule is also possible and triggers a dual connection over the $\algA$-bimodule $\modM'^{{}_\algA}$.
}, then one can define a connection $\nabla'$, over the $\caZ(\algA)$-bimodule $\modM'^{{}_\algA}$ by
\begin{align}
	\nabla'_X(\mu)(m) 
	&= X \big( \mu(m) \big) - \mu \big( \nabla_X(m) \big), &
	X \in \Der(\algA), m \in \modM, \mu \in \modM'^{{}_\algA}.
\label{eq:dual_conn_def}
\end{align}
This connection is unique.

\paragraph{}
The curvature of a connection on a bimodule is defined similarly to single modules \eqref{eq:nc_curv_def} and has the property of a homomorphism of bimodule, \textit{i.e.}\ $\hR(X, Y) (a_1 \actl m \actr a_2) = a_1 \actl \hR(X,Y)(m) \actr a_2$, for any $a_1, a_2 \in \algA$, $m \in \modM$ and $X, Y \in \Der(\algA)$.

\paragraph{}
Furthermore, the (unitary) gauge group, the hermitian structure and the hermiticity of the connection are defined as in section \ref{subsec:nc_conn_right}.

\paragraph{}
The central bimodule setting was built to generalize the notion of a linear connection, as discussed in section \ref{subsec:central_bimodules}. To do so, one would want to take the module to be the derivations, that is $\modM = \Der(\algA)$. Except $\Der(\algA)$ is not an $\algA$-module, but rather a $\caZ(\algA)$-module.

One can however consider $\modM = \Omega^1(\algA)$ which has a canonical bimodule structure given by the star-product. It is further central straightforwardly. Furthermore, one has $\modM'^{{}_\algA} = \Der(\algA)$, so that a connection on $\Omega^1(\algA)$ is uniquely linked to a connection on $\Der(\algA)$.

\subsection{Differential calculus from a twist deformation}
\label{subsec:diff_calc_twist}
\paragraph{}
Another way of defining the differential calculus in a consistent manner consists of using the twist deformation of a classical differential calculus \cite{Majid_1999, Sitarz_2001}. The twist deformation of a classical space of function was introduced in section \ref{subsec:star_prod_twists} to build the star-product, and thus to quantize the space. This method can also be applied to quantize the differential calculus.

Thus, let us consider an algebra of smooth functions $\mathcal{A}$ on a manifold, with the commutative pointwise product $\cdot$, and let $\Xi$ be the Lie algebra of vector fields. Suppose that a quantum group $\algH$ acts on $\mathcal{A}$ through \eqref{eq:hopf_module_alg}. Consider a twist $\twiF$ defined by \eqref{eq:twist_def}, that turns $\mathcal{A}$ into its quantum version $\algA$ with the product $\star$ in \eqref{eq:star_prod_twist} and the space of symmetries $\algH^\twiF$.

\paragraph{}
One can define a differential calculus on $\mathcal{A}$, as usual, with $n$-forms being alternating $\mathcal{A}$-multilinear maps from $\Xi \otimes_\mathcal{A} \cdots \otimes_\mathcal{A} \Xi$ ($n$ times) to $\mathcal{A}$. This set of forms is denoted $\Omega^n(\mathcal{A})$. One can then define the wedge product $\wedge_\mathcal{A}$ and the differential $\dd_\mathcal{A}$ through \eqref{eq:form_prod} and \eqref{eq:koszul} respectively. This differential calculus is covariant under the action of $\algH$.

Now, one can twist all this structure, starting by the tensor product $\otimes_\mathcal{A}$, through
\begin{equation}
    X \otimes_\algA Y
    = \twiF^{-1}_1 (X) \otimes_\mathcal{A} \twiF^{-1}_2(Y)
    \label{eq:tensor_prod_twist}
\end{equation}
for any $X, Y \in \Xi$. The tensor product $\otimes_\algA$ is often called star-tensor product. From this star-tensor product, one builds the $n$-forms of $\algA$ as the alternating $\algA$-multilinear maps from $\Xi \otimes_\algA \cdots \otimes_\algA \Xi$ ($n$ times) to $\algA$. This set of forms is denoted by $\Omega^n(\algA)$.

The twist allows to export the bimodule structure of forms on $\mathcal{A}$ to a bimodule structure for the forms on $\algA$. Explicitly, the structure of $\mathcal{A}$-bimodule of $\Omega^n(\mathcal{A})$, in which the action of $\mathcal{A}$ is defined through the product $\cdot$, turns $\Omega^n(\algA)$ into a $\algA$-bimodule, in which the action of $\algA$ is defined through the product $\star$. This description can be enlarged to any tensorial forms.

On the other hand, one can define a $1$-form on $\algA$, $\Omega^1(\algA)$ as a $\algA$-linear map from $\Xi$ to $\algA$ and then define a twisted wedge product of forms $\wedge_\algA$ by
\begin{equation}
    \omega \wedge_\algA \eta
    = \twiF^{-1}_1 (\omega) \wedge_\mathcal{A} \twiF^{-1}_2(\eta)
    \label{eq:wedge_prod_twist}
\end{equation}
for any $\omega, \eta \in \Omega^1(\algA)$. It appears that these two constructions are equivalent \cite{Aschieri_2006}. Finally, this $\wedge_\algA$ makes $\Omega^\bullet(\algA) = \bigoplus_{n \geqslant 0} \Omega^n(\algA)$ covariant under the action of $\algH^\twiF$.

The differential map of $\algA$ can be shown \cite{Sitarz_2001} to be the same as $\mathcal{A}$. Note that the twisted tensor product \eqref{eq:tensor_prod_twist} can be use to define general twisted tensor fields.

\paragraph{}
The Lie algebra of vector field $\Xi$ can also be twisted to have a quantum version of vector fields. This formalism can be pushed even further to define e.g.\ (twisted) covariant derivatives and (twisted) curvature. More details are given in section \ref{subsec:braided_geometry}.

\paragraph{}
For physical purpose, one should note that several twists can give rise to the same structures but give physically nonequivalent properties as put forward by \cite{Borowiec_2009}.

\newpage
\section{Gauge theories on Moyal spaces \tops{$\mathbb{R}^{2n}_\theta$}{R\^2n\_theta}.}
\label{sec:moyal}

\subsection{The Moyal product and the Moyal space \tops{$\mathbb{R}^{2n}_\theta$}{R\^2n\_theta}.}
\label{subsec:moyal_prod_space}

\subsubsection{The star-product.}
\label{subsubsec:moyal_prod}
\paragraph{}
In this section, the Moyal product is constructed in the 2-dimensional case for clarity. Its extension to the $2n$-dimensional case is straightforward. This construction follows the construction of the star-product in the Weyl quantization scheme developed in section \ref{subsec:star_prod_basic}. In other words, what follows is an explicit example of the construction \eqref{eq:weyl_pseudo_rep}$-$\eqref{eq:star_prod_def}.

Start from the algebra of noncommutative coordinates given by{\footnote{We adapt the notations of the subsection \ref{subsec:star_prod_basic} to a noncommutative space view, \textit{i.e.}\  we set $P=x^1$, $Q=x^2$.}}
\begin{equation}
    [x^1, x^2]
    = iZ,
    \label{eq:lie_heisenb}
\end{equation}
which therefore merely corresponds to a central extension of a commutative Lie algebra with central element $Z$ ($[x^1,Z]=[x^2,Z]=0$).

\paragraph{}
The Lie group associated to \eqref{eq:lie_heisenb}, called the Heisenberg group $\mathbb{H}_3$, can be conveniently described as the real (nilpotent) 3-dimensional group of elements $s(z, u, v)$ with $z, u, v \in \mathbb{R}$ such that
\begin{subequations}
\begin{gather}
    s(z, u, v) s(z', u', v')
    = s(z + z' + \frac{1}{2}(uv' - u'v), u + u', v + v'),
    \label{eq:multipli-heisenb} \\
    s^{-1}(z, u, v)
    = s(-z, -u, -v), \qquad
    s(0, 0, 0) 
    = \bbone.
    \label{eq:invers-heisenb}
\end{gather}
    \label{eq:heisenb}
\end{subequations}

Recall that $\mathbb{H}_3$ is unimodular with Haar measure given by the simple Lebesgue measure on $\mathbb{R}^3$, namely $\dd\mu(s) = \dd z \dd u \dd v$ for any $s(z,u,v)\in\mathbb{H}_3$, while the convolution product is given by
\begin{equation}
    (f\circ g)(t)
    = \int \dd\mu(s) f(s) g(s^{-1}t),
    \label{convol-def}
\end{equation}
for any $f,g\in L^1(\mathbb{H}_3)$ which, combined with \eqref{eq:heisenb}, can be rewritten as
\begin{equation}
    (f\circ g)(Z,U,V)
    = \int_{\mathbb{R}^3} \dd z \dd u \dd v\ f(z, u, v) g(Z - z + \frac{1}{2}(Uv - Vu), U - u, V - v),
    \label{eq:convol_heisenb}
\end{equation}
using the fact that functions on $\mathbb{H}_3$ can be viewed as functions on $\mathbb{R}^3$, \textit{i.e.}\  $f(s)=f(z,u,v)$ for $s(z,u,v)\in\mathbb{H}_3$.

\paragraph{}
From the unitary irreducible representation of $\mathbb{H}_3$ (\textit{i.e.}\  the Schr\"odinger representation) on $L^2(\mathbb{R})$ defined for any $s\in\mathbb{H}_3$, $\psi\in L^2(\mathbb{R})$ by
\begin{align}
    S_\theta: \mathbb{H}_3 \to \mathcal{B}(L^2(\mathbb{R})), &&
    (S_\theta[s(z, u, v)] \psi)(\xi)
    &= e^{-i 2 \pi \theta z} e^{i 2 \pi (\theta \frac{uv}{2} + u \xi)} \psi(\xi + \theta v),
    \label{eq:schroding_rep}
\end{align}
which is unique for a given $\theta > 0$, up to unitary equivalence. By the Stone-von Neumann theorem, one obtains the (non-degenerate) irreducible representation of the convolution algebra $\mathbb{C}(\mathbb{H}_3) = (L^1(\mathbb{H}_3), \circ)$,  
\begin{align}
    \pi: \mathbb{C}(\mathbb{H}_3) \to \mathcal{B}(L^2(\mathbb{R})), &&
    \pi(f)
    &= \int_{\mathbb{R}^3} \dd z \dd u \dd v\ f(z, u, v) S_\theta[s(z, u, v)]
    \label{eq:reps-convolalg}
\end{align}
which satisfies
\begin{equation}
    \pi(f \circ g)
    = \pi(f) \pi(g)
    \label{eq:pi-mapalg}.
\end{equation}

At this stage, we have followed all the steps of the construction summarized in the last paragraph of the previous section. What remains to do is to reduce the whole scheme to functions of $L^1(\mathbb{R}^2)$.

\paragraph{}
To achieve this, define a map $\#:L^1(\mathbb{R}^3) \to L^1(\mathbb{R}^2)$ by setting
\begin{equation}
    f^\#(u, v)
    = \int_\mathbb{R} \dd z\ f(z, u, v) e^{- i 2 \pi \theta z}
    \label{eq:diese-map}
\end{equation}
so that
\begin{equation}
    (\pi(f) \psi)(\xi)
    = \int_{\mathbb{R}^2} \dd u \dd v\ f^\#(u, v) e^{i 2 \pi(\theta \frac{uv}{2} + u \xi)} \psi(\xi + \theta v)
    \label{eq:diese-pi-f}.
\end{equation}

Then, consider the action of $\#$ on the convolution product \eqref{eq:convol_heisenb}. A standard calculation yields 
\begin{equation}
    (f\circ g)^\#(u,v)
    = \int_{\mathbb{R}^2} \dd u' \dd v'\ f^\#(u', v') g^\#(u - u', v - v') e^{i \pi \theta(u v' - u' v)}
    \label{eq:source-twisconv},
\end{equation}
which defines the twisted convolution $\tcvp$ \cite{vonNeum} as
\begin{equation}
    (f\circ g)^\#(u, v)
    = (f^\# \tcvp g^\#)(u, v)
    \label{eq:twisted-convol}.
\end{equation}

Finally, in order to make contact with the discussion of the section \ref{subsec:star_prod_basic}, simply set $f^\#(u, v) = \mathcal{F}f(u, v)$ and use \eqref{eq:star_prod_quant} to obtain
\begin{equation}
    f \star_\theta g
    = \mathcal{F}^{-1}(\mathcal{F}f \tcvp \mathcal{F}g)
    \label{eq:conv_moyal_prod}
\end{equation}
which shows that the Moyal product is nothing but (the inverse Fourier transform of) a twisted convolution defined in \eqref{eq:source-twisconv}. Note that the factor $\theta$ occurring in the exponential in \eqref{eq:source-twisconv} stems from the non-trivial action of the central element $Z$ of \eqref{eq:lie_heisenb} in the representation. Reinstalling the dimensions, one infers from \eqref{eq:lie_heisenb} that $\theta$ has mass dimension $-2$, which is the deformation parameter of the 2-dimensional space with the noncommutative coordinates $x_1,x_2$ introduced in \eqref{eq:lie_heisenb}.

\paragraph{}
A simple calculation starting from \eqref{eq:conv_moyal_prod} yields the popular expression for the Moyal product in 2 dimensions whose extension to $2n$ dimensions is straightforward. It is given by
\begin{equation}
    (f \star_\theta g)(x) 
    = \frac{1}{(\pi\theta)^{2n}} \int \dd^{2n}y\ \dd^{2n}z\ f(x+y) g(x+z) e^{- 2 i y \Theta^{-1} z}
    \label{eq:moyal_prod}
\end{equation}
for any $f,g\in\mathcal{S}(\mathbb{R}^{2n})$, the space of (complex valued) Schwartz on $\mathbb{R}^{2n}$, where we have set  $y \Theta^{-1} z = y^\mu \Theta^{-1}_{\mu\nu} z^\nu$, in obvious notations, in which $\Theta^{-1}$ is the invertible skew-symmetric matrix given by
\begin{align}
    \Theta^{-1}
    &= \theta\ \mathrm{diag}(J,\dots, J), &
    J 
    &= \begin{pmatrix} 0 & -1 \\ 1 & 0 \end{pmatrix}.
    \label{eq:Zebelleparam}
\end{align}
Recall that Einstein summation convention holds here, and that the background metric is chosen to be the Euclidean one \eqref{eq:background_metric}.

\paragraph{}
As anticipated in section \ref{subsubsec:star_prod_eg}, the previous Moyal product \eqref{eq:moyal_prod} can be derived through twist deformation. Consider first the Abelian twist for a given smooth manifold $\twiF = e^{- \frac{i}{2} \Theta^{\mu\nu} X_\mu \otimes X_\nu}$, for $X_\mu$ a family of independent vector fields. As we consider here a deformation of $\mathbb{R}^{2n}$, we can take $X_\mu = \partial_\mu$ so that
\begin{align}
    \twiF 
    &= e^{- \frac{i}{2} \Theta^{\mu\nu} \partial_\mu \otimes \partial_\nu}.
    \label{eq:moyal_abelain_twist}
\end{align}
Then one can compute that this twist indeed satisfies requirements \eqref{eq:twist_def}. Then, one can use \eqref{eq:star_prod_twist} to compute the star-product obtained by this twist and gets
\begin{align}
\begin{aligned}
    (f \star_\theta g)(x)
    &= \underset{y \to x}{\mathrm{lim}}\ e^{ \frac{i}{2} \Theta^{\mu\nu} \frac{\partial}{\partial x^\mu} \frac{\partial}{\partial y^\nu}} f(x) g(y) \\
    &= f(x)g(x) + \sum_{k = 1}^\infty \left(\frac{i}{2}\right)^{\! k} \frac{1}{k!} \Theta^{\mu_1\nu_1} \cdots \Theta^{\mu_k\nu_k} \partial_{\mu_1} \cdots \partial_{\mu_k}f(x)\ \partial_{\nu_1} \cdots \partial_{\nu_k} g(x).
\end{aligned}
    \label{eq:moyal_prod_twist}
\end{align}
One can check that the formal expansion of \eqref{eq:moyal_prod} in $\Theta$ gives the formula \eqref{eq:moyal_prod_twist}.

\paragraph{}
The Moyal product \eqref{eq:moyal_prod} can be extended to larger spaces than $\mathcal{S}(\mathbb{R}^{2n})$ pertaining to tempered distribution spaces, which are large enough to involve polynomial functions, plane waves, Dirac distribution and thus can be used in models for physics. The extension to the relevant multiplier space, denoted hereafter by $\mathcal{M}(\mathbb{R}^{2n})$, is presented in details in \cite{graciavar1}.

In the following, it will be sufficient for our purpose to describe the $2n$-dimensional Moyal space with the following associative algebra
\begin{equation}
    \mathbb{R}^{2n}_\theta
    = (\mathcal{M}(\mathbb{R}^{2n}), \star_\theta),
    \label{eq:moyal_space_def}
\end{equation}
which models the noncommutative space whose coordinates verify
\begin{equation}
    [x^\mu, x^\nu]_\theta
    = i \Theta^{\mu\nu}.
    \label{eq:moyal_coord_nc}
\end{equation}
The commutator here stands for $[x^\mu, x^\nu]_\theta = x^\mu \star_\theta x^\nu - x^\nu \star_\theta x^\mu$.

$\mathbb{R}^{2n}_\theta$ has a natural involution given by the usual complex conjugation $f \to f^\dag$ and one easily verifies that $(f \star_\theta g)^\dag = g^\dag \star_\theta f^\dag$. The latter equation is consistent with $Q(f^\dag) = Q(f)^\dag$ where the symbol $^\dag$ in the right-hand side corresponds to adjoint operator. There is a natural integral (trace) on $\mathbb{R}^{2n}_\theta$ given by the usual Lebesgue integral, \textit{i.e.}\  the Haar measure supported by the convolution algebra. It verifies
\begin{equation}
    \int \dd^{2n}x\ (f \star_\theta g)(x)
    = \int \dd^{2n}x\ (g \star_\theta f)(x)
    = \int \dd^{2n}x\ f(x)g(x).
    \label{eq:moyal_prod_closed}
\end{equation}

\subsubsection{The matrix base.}
\label{subsubsec:matrix_base}
\paragraph{}
It may be necessary for technical reasons to trade the associative algebra of functions \eqref{eq:moyal_space_def} for an algebra of operators. This can be achieved within the formalism of the Moyal ``matrix base''. It permits in particular to represent the action of a noncommutative field theory as a matrix model. Note that this has been successfully applied to the noncommutative $\phi^4_4$ theory with harmonic term \cite{Grosse:2003aj-pc} in order to carry out the proof of the renormalizability to all orders. Through this, the effect of the harmonic term on the propagator neutralizing the UV/IR mixing was exhibited, as we will recall in the sequel. In this section, we will consider the 2-dimensional case. The extension of $2n$ dimensions is straightforward.

\paragraph{}
A convenient presentation of the matrix base goes as follows. First, recall that the family of Wigner transition eigenfunctions of the 1-dimensional harmonic oscillator, denoted by $\{f_{mn}(x) \}_{m,n\in\mathbb{N}}$, is an orthogonal basis of $\mathcal{S}(\mathbb{R}^2)$, the space of complex-valued Schwartz functions on $\mathbb{R}^2$ \cite{graciavar1}. The following relations holds \cite{graciavar1}
\begin{subequations}
\begin{gather}
    f_{mn} \star_\theta f_{kl}
    = \delta_{nk} f_{ml}, \qquad
    f_{mn}^\dag
    = f_{nm},
    \label{eq:moyal_matbas_prod}\\
    \langle f_{mn}, f_{kl} \rangle_{L^2} 
    = \int \dd^2x\ (f_{mn}^\dag \star_\theta f_{kl})(x)
    = 2 \pi \theta \delta_{mk} \delta_{nl}
    \label{eq:moyal_matbas_scal}.
\end{gather}
    \label{eq:moyal_matbas}
\end{subequations}

The explicit expressions for the $f_{mn}$'s are not needed here. It follows that, for any function $a,b\in\mathcal{S}(\mathbb{R}^2)$, one has the expansions $a(x) = \sum_{m,n} a_{mn} f_{mn}(x)$, $b(x) = \sum_{m,n} b_{mn} f_{mn}(x)$, with $a_{mn},\ b_{mn}\in\mathbb{C}$, which combined with \eqref{eq:moyal_matbas_prod} yields 
\begin{equation}
    (a \star_\theta b)(x)
    = \sum_{m,n\in\mathbb{N}} \left( \sum_{k\in\mathbb{N}} a_{mk} b_{kn} \right) f_{mn}(x),
    \label{eq:moyal_mat_prod}
\end{equation}
where the coefficient in the expansion \eqref{eq:moyal_mat_prod} is \textit{formally} expressed as a product of matrices $(a_{mn})$ and $(b_{mn})$.

\paragraph{}
Next, it is known \cite{graciavar1} that the algebra $\algA_\theta = (\mathcal{S}(\mathbb{R}^2), \star_\theta)$ is isomorphic to the algebra $\mathbb{M}_\theta \subset \ell^2(\mathbb{N}^2)$ defined as 
\begin{equation}
    \mathbb{M}_\theta
    = \{(\phi_{m,n})_{m,n\in\mathbb{N}}, \ \rho_n(\phi)<\infty, \ \forall n\in\mathbb{N}\}
    \label{eq:moyal_mat_alg_def},
\end{equation}
with 
\begin{equation}
    \rho_n(\phi)
    = \sum_{p,k} \theta \left(p + \frac{1}{2} \right)^n \left(k + \frac{1}{2} \right)^n |\phi_{pk}|^2,
    \label{eq:moyal_mat_alg_rho}
\end{equation}
which is the subalgebra of $\ell^2(\mathbb{N}^2)$ involving rapid decay matrices\footnote{
Note that the $\rho_n$'s endow ${\mathbb{M}}_\theta$ with a Fr\'echet algebra structure so that the isomorphism \eqref{eq:isomor-frechet} is a Fr\'echet algebra isomorphism. Recall that $\mathcal{S}(\mathbb{R}^2)$ is a Fr\'echet algebra since any $f\in\mathcal{S}(\mathbb{R}^2)$ satisfies $|x_1^{\alpha_1} x_2^{\alpha_2} \partial_1^{\beta_1} \partial_2^{\beta_2} f(x)| < \infty$ for any $\alpha_i, \beta_i \in \mathbb{R}^+$.
}.
Note that it is a $*$-algebra isomorphism. The isomorphism is defined by
\begin{equation}
    \phi_{mn} 
    \mapsto \sum_{m,n} \phi_{mn} f_{mn} \in \mathcal{S}(\mathbb{R}^2),
    \label{eq:isomor-frechet}
\end{equation}
with inverse
\begin{equation}
    \phi \in \mathcal{S}(\mathbb{R}^2)
    \mapsto \frac{1}{2 \pi \theta} \langle \phi, f_{mn} \rangle_{L^2}
    \label{eq:isomor-invers-frechet}.
\end{equation}

\paragraph{}
Now, one observes that the algebra ${\mathbb{M}}_\theta$ has a natural faithful representation stemming from the representation of matrices as operators on $\ell^2(\mathbb{N})$. Indeed, this corresponds to the product of a matrix by a column vector.

To see that, denote by $(e_n)_{n\in\mathbb{N}}$, the canonical base of $\ell^2(\mathbb{N})$ with dual basis $(e'_n)_{n\in\mathbb{N}}$, thus obeying $e'_n(e_p) = \delta_{np}$.  For any $\phi \in {\cal{M}}_\theta$, $\Phi = \sum_{m,n} \phi_{mn} e_m \otimes e_n$, define the representation $\eta$ as
\begin{align}
    \eta: \ell^2(\mathbb{N}) \otimes \ell^2(\mathbb{N}) \to {\cal{B}}(\ell^2(\mathbb{N})), &&
    \eta(e_m \otimes e_n) 
    &= e_m \otimes e'_n, &
    \forall m,n\in\mathbb{N}.
    \label{eq:moyal_mat_rep}
\end{align}

But since $\eta$ is faithful, one has the isomorphism of algebra 
\begin{equation}
    \eta({\mathbb{M}}_\theta) \simeq {\mathbb{M}}_\theta,
    \label{eq:moyal_mat_isomor}
\end{equation}
so that one can merely identify ${\mathbb{M}}_\theta$ as the algebra involving elements of the form 
\begin{equation}
    \widehat{\phi}
    = \sum_{m,n} \phi_{mn} e_m \otimes e'_n.
\end{equation}
In order to make contact with standard physics notations, simply set 
\begin{equation}
    e_m \otimes e'_n
    := \widehat{f}_{mn}
    = |m\rangle \langle n|,
\end{equation}
so that
\begin{equation}
    \widehat{\phi}
    =\sum_{m,n} \phi_{mn} |m\rangle \langle n|.
\end{equation}

\paragraph{}
Summarizing the above, we have traded functions of $\algA_\theta$ given by $\phi(x) = \sum_{m,n} \phi_{mn} f_{mn}(x)$ for operators of $\mathbb{M}_\theta$ given by $\widehat{\phi} = \sum_{m,n} \phi_{mn} |m\rangle \langle n|$. In the operator language, the ${f}_{mn}(x)$'s are the symbols of the operators $\widehat{f}_{mn}$ and the star-product $\star_\theta$ represents the product between symbols which keeps track of the operator product.

Finally, by using \eqref{eq:moyal_matbas}, one easily realizes that
\begin{equation}
    \int \dd^2x\ \phi(x)
    = 2 \pi \theta \sum_m \phi_{mm}
    = 2 \pi \theta \tr(\widehat{\phi}).
\end{equation}
The above algebras can be enlarged to multiplier algebras as carried out in \cite{graciavar1} so as to include plane waves, delta distributions and unity.

\subsection{Differential calculus, connections and curvatures on \tops{$\mathbb{R}^{4}_\theta$}{R\^4\_theta}.}
\label{subsec:moyal_diff_calc}
\paragraph{}
A convenient differential calculus underlying most of the literature on gauge theories on Moyal spaces can be straightforwardly extracted from the general framework given in section \ref{subsec:nc_diff_calc_der}. It is generated by the simplest Abelian Lie algebra of derivations $\mathfrak{D}_1$ generated by the $\partial_\mu$'s. As $\mathfrak{D}_1$ is Abelian, the generators must satisfy
\begin{equation}
    [\partial_\mu,\partial_\nu] =0,
\end{equation}
$\mu, \nu = 1, \dots, 4$. Here, $\partial_\mu$ denotes the usual derivative w.r.t.\ $x^\mu$ and moreover is an inner derivation of the Moyal algebra since one has
\begin{align}
    \partial_\mu f
    &= [\xi_\mu, f]_\theta, &
    \xi_\mu 
    &= -i \Theta_{\mu\nu}^{-1} x^\nu,
    \label{eq:moyal_inner_der}
\end{align}
for any function $f \in \mathbb{R}^4_\theta$.

\paragraph{}
Therefore, according to section \ref{subsec:nc_diff_calc_der}, the relevant differential algebra is completely characterized by
\begin{equation}
    \left( {\Omega}^\bullet(\mathbb{R}^4_\theta) 
    = \bigoplus_{n=0}^4 {\Omega}^n(\mathbb{R}^4_\theta),\ \wedge_\theta,\ \dd \right),
\end{equation}
where ${\Omega}^n(\mathbb{R}^4_\theta)$ denotes the space of $\mathbb{C}$-linear antisymmetric maps\footnote{Recall that $\mathcal{Z}(\mathbb{R}^4_\theta) = \mathbb{C} \bbone_4$.}
$\omega: \mathfrak{D}_1^n \to \mathbb{R}^4_\theta$, $n = 0, \dots, 4$.
The product $\wedge_\theta: {\Omega}^\bullet(\mathbb{R}^4_\theta) \to {\Omega}^\bullet(\mathbb{R}^4_\theta)$ is defined by
\begin{equation}
\begin{aligned}
    (\omega \wedge_\theta \eta)&(\kX_1, \dots, \kX_{p+q}) \\
    &= \frac{1}{p!q!} \sum_{\sigma\in \kS_{p+q}} (-1)^{{\sign}(\sigma)} \omega(\kX_{\sigma(1)}, \dots, \kX_{\sigma(p)}) \star_\theta \eta(\kX_{\sigma(p+1)}, \dots, \kX_{\sigma(p+q)})
\end{aligned}
    \label{eq:form_prod-bis}
\end{equation}
with ${\sign}(\sigma)$ the signature of the permutation $\sigma$,  $\kS_{p+q}$ the symmetric group of $p+q$ elements. The differential is defined as
$\dd:{\Omega}^p(\mathbb{R}^4_\theta) \to {\Omega}^{p+1}(\mathbb{R}^4_\theta)$, for any $p\in\mathbb{N}$, and satisfies
\begin{equation}
    \dd \omega(\kX_1, \dots, \kX_{p+1}) 
    = \sum_{j = 1}^{p+1} (-1)^{j+1} \kX_j \Big( \omega(\kX_1,..\vee_j.., \kX_{p+1}) \Big)
    \label{eq:koszul-bis}
\end{equation}
for any $\omega \in {\Omega}^p(\mathbb{R}^4_\theta), \eta \in {\Omega}^q(\mathbb{R}^4_\theta)$, where $\vee_j$ again denotes the omission of the element $X_j$ and 
\begin{equation}
    X_j = \partial_{j}
\end{equation}
must be set in \eqref{eq:form_prod-bis} and \eqref{eq:koszul-bis}.

\paragraph{}
Some comments are in order here.

Observe first that the above differential calculus, \textit{albeit} noncommutative in essence, bears some similarities with the usual de Rham calculus of standard differential geometry. As an example, from \eqref{eq:koszul-bis} the action of $\dd$ on any function (zero-form) $f$ yields $\dd f(X_\mu) = \partial_\mu f$. Note, by the way, that differential forms in the present framework are defined through their ``components'' \textit{i.e.}\  their evaluation on the relevant number of derivations in $\mathcal{D}_1$. Pick for instance a one form $A \in \Omega^1(\mathbb{R}^4_\theta)$. Then, its evaluation on any element $\partial_\mu$ of the algebra $\mathfrak{D}_1$ is $A(\partial_\mu)$. We will use from now on the usual notation for differential forms and denote the evaluation of any $p$-forms as
\begin{equation}
    \omega(\partial_{\mu_1}, \partial_{\mu_2}, \dots, \partial_{\mu_p})
    := \omega_{{\mu_1}, {\mu_2}, \dots, {\mu_p}}.
    \label{eq:zenotation}
\end{equation}
This notation will prevail below when the relevant gauge connection and curvature will be introduced.

Next, it can be verified that 
\begin{equation}
    \omega \wedge_\theta \eta 
    \ne (-1)^{|\omega||\eta|} \eta \wedge_\theta \omega
\end{equation}
where $|\omega|$ denotes the degree of $\omega$ (same for $\eta$), so that the differential algebra is not graded commutative, contrary to the case of the de Rham calculus. This difference stems simply from the noncommutativity of the star-product.

Finally, going back to the Lie algebra of derivation $\Der(\mathbb{R}^4_\theta) $ of the algebra modeling the Moyal space $\mathbb{R}^4_\theta$, it appears that this latter is infinite dimensional so that a differential calculus based on it would give rise to gauge potentials having an infinite number of components. This could not be reconciled, as a physically acceptable commutative limit, with any standard (commutative) gauge theories. This feature is of course absent when using restricted differential calculus based on a finite dimensional Lie algebra of derivations, introduced in section \ref{subsec:nc_diff_calc_der}, as the algebra $\mathfrak{D}_1$ used above. Note that $\mathfrak{D}_1$ is one of the two algebras in $\Der(\mathbb{R}^4_\theta) $ whose elements can be interpreted as infinitesimal symplectomorphisms, \textit{i.e.}\  area-preserving diffeomorphisms. The other algebra $\mathfrak{D}_2$ is defined as the image of the space of polynomial functions of maximal degree $2$, denoted by $\mathcal{P}_2$, by the adjoint map $\mathrm{Ad}_P := [P,.]_\theta$, namely $\mathfrak{D}_2 = \{X_P = \mathrm{Ad}_P,\ P \in \mathcal{P}_2\}$.

The rest of section \ref{sec:moyal} deals essentially with differential structure based on $\mathfrak{D}_1$. For past works on noncommutative gauge theories based on differential calculus generated by $\mathfrak{D}_2$, see \cite{marmoetal, cawa1, cawa}.

\paragraph{}
The main part of the literature on gauge theories on Moyal spaces can be described using the notion of noncommutative connection on a right module $\modM$ as described in section \ref{subsec:nc_conn_right}. From this framework, the choice where $\modM \simeq \mathbb{R}^4_\theta$, \textit{i.e.}\  the module is one copy of the algebra, corresponds to a noncommutative analog of $U(1)$ gauge theories. The extension to a noncommutative analog of $U(N)$ gauge theories is obtained by taking the module to be the product of $N$ copies of $\mathbb{R}^4_\theta$, namely $\modM\simeq(\mathbb{R}^4_\theta)^N$.

\paragraph{}
When $\modM\simeq\mathbb{R}^4_\theta$, a mere application of the results presented in section \ref{subsec:nc_conn_right} leads to the hermitian connection defined by (recall we use the notation defined by \eqref{eq:zenotation})
\begin{align}
    \nabla_\mu(f)
    &= \partial_\mu f - i A_\mu \star_\theta f, &
    A_\mu
    &= \nabla_\mu(\bbone),
    \label{eq:connect-moyal}
\end{align}
for any $f\in\mathbb{R}^4_\theta$ with
\begin{equation}
    A_\mu^\dag
    = A_\mu
\end{equation}   
for the hermitian structure defined by \eqref{eq:nc_A_canon_herm}. The corresponding curvature is
\begin{equation}
    i F_{\mu\nu}
    = \partial_\mu A_\nu - \partial_\nu A_\mu - i [A_\mu, A_\nu]_\theta.
    \label{eq:courb-moyal}
\end{equation}
Since \eqref{eq:nc_inner_diff} holds true in view of \eqref{eq:moyal_inner_der}, it follows from section \ref{subsubsec:nc_mod_is_alg} that a noncommutative gauge invariant connection shows up. This latter is given by
\begin{equation}
    \nabla^{\inv}(f)
    = i f \star_\theta \Theta^{-1}_{\mu\nu} x^\nu
    \label{eq:moyal_inv_conn}
\end{equation}
for any $f\in\mathbb{R}^4_\theta$. This singles out a distinguished field variable as a covariant tensor form defined by \eqref{eq:nc_A_covar_coord}, whose expression in the present case is
\begin{equation}
    \mathcal{A}_\mu
    = - i (A_\mu + \Theta^{-1}_{\mu\nu} x^\nu).
    \label{eq:moyal_covar_coord}
\end{equation}
Expressing \eqref{eq:connect-moyal} and \eqref{eq:courb-moyal} in term of $\mathcal{A}_\mu$ yields
\begin{align}
    \nabla_\mu(f)
    &= i f \star_\theta \Theta^{-1}_{\mu\nu} x^\nu + \mathcal{A}_\mu, & 
    F_{\mu\nu}
    &= [\mathcal{A}_\mu, \mathcal{A}_\nu]_\theta - i \Theta^{-1}_{\mu\nu}.
    \label{eq:moyal_covar_curv}
\end{align}

\paragraph{}
According to section \ref{subsubsec:nc_mod_is_alg}, the unitary gauge group \eqref{eq:nc_U1} is defined by
\begin{equation}
    \mathcal{U}(\mathbb{R}^4_\theta)
    = \{g \in \modM \simeq \mathbb{R}^4_\theta,\ g^\dag \star_\theta g = g \star_\theta g^\dag =\bbone \}
    \label{eq:moyal_U1}.
\end{equation}
The corresponding gauge transformations are given by
\begin{subequations}
\begin{align}
    A^g_\mu 
    &= g^\dag \star_\theta A_\mu \star_\theta g + i g^\dag \star_\theta \partial_\mu g,
    \label{eq:moyal_gauge_trans_con}\\
    F^g_{\mu\nu}
    &= g^\dag \star_\theta F_{\mu\nu} \star_\theta g,
    \label{eq:moyal_gauge_trans_curv}\\
    \mathcal{A}^g_\mu
    &= g^\dag \star_\theta \mathcal{A}_\mu \star_\theta g,
    \label{eq:moyal_gauge_trans_cov}
\end{align}
    \label{eq:moyal_gauge_trans}
\end{subequations}
for any $g\in\mathcal{U}(\mathbb{R}^4_\theta)$, where the covariant trait of the tensor form $\mathcal{A}_\mu$ is apparent from \eqref{eq:moyal_gauge_trans_cov}. Note that this tensor form is often called ``covariant coordinate'' in the physics literature on string models and matrix models.

The curvature corresponding to the gauge invariant connection \eqref{eq:moyal_inv_conn} is
\begin{equation}
    i F^{\inv}_{\mu\nu}
    = \Theta^{-1}_{\mu\nu}
    \label{eq:Moyal_curv_inv_gauge}
\end{equation}
as it can be verified by a direct computation using the definition of the curvature, e.g.\ \eqref{eq:nc_curv_def}. The gauge invariance of \eqref{eq:moyal_inv_conn} can be verified in the same way.

\paragraph{}
The extension to the larger symmetry group can be achieved by considering, instead of $\modM\simeq\mathbb{R}^4_\theta$, the product of $N$ algebras $\mathbb{R}^4_\theta$, namely $\modM \simeq \mathbb{R}^4_\theta \times \mathbb{R}^4_\theta \times ... \times \mathbb{R}^4_\theta := (\mathbb{R}^4_\theta)^N$ as the relevant module. Owing to the linear structure of the module, any of its elements $m$ can thus be expressed as $m = \textstyle \sum_{j=1}^N m_j e_j$ with $m_j \in \mathbb{R}^4_\theta$, the $j$-th algebra factor in $\modM$, and $e_j = (0, \dots, 0, \overset{(j)}{1}, 0, \dots,0)$ where the only non-zero entry is at the $j$-th place. It is then a simple matter of algebra to realize that the resulting connection can be defined \cite{cawa} by an (anti)-hermitian matrix $A_\mu \in \mathbb{R}^4_\theta \otimes \mathbb{M}_N(\mathbb{C})$ given by
\begin{align}
    A_\mu
    &= - i E_j^k(A_\mu)_k^j, &
    j, k 
    &= 1, 2, \dots, N
    \label{eq:moyal_Ngaugepot}
\end{align}
(summation over $j, k$ understood) where $ (A_\mu)_k^j \in \mathbb{R}^4_\theta$ and $\mathbb{M}_N(\mathbb{C})$ denotes the usual algebra of complex-valued $N \times N$ matrices with $N^2$ canonical basis elements\footnote{
These are matrices $(E_j^k)_{j,k= 1, \dots, N}$ with all zero entries except the entry at the $j$-th line and $k$-th row equal to $1$.
}
$E_j^k$, $j, k = 1, \dots, N$, verifying the relation $E_j^kE_l^m = \delta_l^k E_j^m$. The hermitian structure underlying the present construction is a mere generalization of \eqref{eq:nc_A_canon_herm}, namely $h_0(m_1,m_2) = \sum_{j = 1}^N (m_1)_j^\dag (m_2)_j$ in obvious notations.

The related curvature is straightforwardly obtained from $F_{\mu\nu} = [\nabla_\mu, \nabla_\nu]_\theta - \nabla_{[\partial_\mu, \partial_\nu]_\theta}$ where the matrix product is understood. In the present case, the group of unitary gauge transformations involves the automorphisms of the module preserving the hermitian structure mentioned above. It reduces \cite{cawa1}, \cite{cawa} to the group of unitary matrices $N \times N$ of the form $E^j_k a^k_j$, with $a^j_k \in \mathbb{R}^4_\theta$, denoted by $\mathcal{U}(\mathbb{R}^4_\theta, N)$. The corresponding gauge transformations are (with matrix product understood)
\begin{align}
    A_\mu^U
    &= U^\dag \star_\theta A_\mu \star_\theta U + U^\dag \star_\theta \partial_\mu U, &
    F^U_{\mu\nu}
    &= U^\dag \star_\theta F_{\mu\nu} \star_\theta U
\end{align}
for any $U\in\mathcal{U}(\mathbb{R}^4_\theta,N)$.

\paragraph{}
This can be put into a more convenient form for physical purpose by using the $N^2$ generators of the $\mathfrak{u}(N)$ Lie algebra $T^a$ for $a=0, 1, \dots, N^2-1$, with
\begin{subequations}
    \begin{align}
        [T^a, T^b] 
        &= i f^{abc} T^c, &
        \{T^a,T^b\}
        &= \frac{1}{N}\delta^{ab} \bbone_N + d^{abc} T^c, &
        a,b,c &= 0, \dots, N^2-1
        \label{eq:uN_lie_alg_comm}
        \\
        T^0
        &= \frac{1}{2 \sqrt{N}} \bbone_N, &
        f^{0ab}
        &= 0, &
        d^{ab0} &=
        \sqrt{\frac{2}{N}}\delta^{ab}
        \label{eq:uN_lie_alg_0}
        \\
        \mathrm{Tr}(T^a)
        &= 0, &
        \tr(T^a T^b)
        &= \frac{1}{2} \delta^{ab}, &
        a,b &= 1, \dots, N^2-1,
        \label{eq:uN_lie_alg_tr}
    \end{align}
    \label{eq:uN_lie_alg}
\end{subequations}
where $f^{abc}$ (resp.\ $d^{abc}$) is as usual totally antisymmetric (resp.\ symmetric). In particular, one finds that \eqref{eq:moyal_Ngaugepot} describes a $\mathfrak{u}(N)$-valued gauge potential $A_\mu = A_\mu^a T^a$. Besides, a simple calculation leads to the expression for the corresponding curvature given by
\begin{equation}
    i F_{\mu\nu}
    = i F_{\mu\nu}^a T^a
    = \big( \partial_\mu A_\nu^a - \partial_\nu A_\mu^a + \frac{1}{2} f^{abc} \{A_\mu^b, A_\nu^c\}_\theta - \frac{i}{2} d^{abc} [A_\mu^b, A_\nu^c]_\theta \big) T^a.
    \label{eq:moyal_uN_curv}
    \end{equation}
where the anticommutator writes $\{A_\mu, A_\nu\}_\theta = A_\mu \star_\theta A_\nu + A_\nu \star_\theta A_\mu$.
    
The use of \eqref{eq:moyal_uN_curv} gives rise to an interesting generalization of the $\mathcal{U}(\mathbb{R}^4_\theta)$ gauge theories characterized by an action of the form
\begin{equation}
    S_{\mathfrak{u}(N)}
    = \int \dd^4x\ \tr_{\mathfrak{u}(N)}(F_{\mu\nu} \star_\theta F^{\mu\nu})  
    \label{eq:Moyal_action_un}
\end{equation}
where $\tr_{\mathfrak{u}(N)}$ is the trace on the Lie algebra $\mathfrak{u}(N)$. See \cite{moyun1, moyun2, moyun2bis, moyun3, moyun4}.

\paragraph{}
However, the formal commutative limit $\theta \to 0$ does not reproduce a usual $SU(2)\times U(1)$ Yang-Mills theory as it can be verified from $S_{\mathfrak{u}(N)}$ and \eqref{eq:moyal_uN_curv} by a simple computation. Moreover, the one-loop beta function is negative and independent of $\theta$. Taking the commutative limit therefore cannot reproduce the expected behaviour of the running $U(1)$ gauge coupling constant \cite{moyun2}, \cite{moyun2bis}. Hence, the physical relevance of the corresponding noncommutative theory is questionable. Nevertheless, this noncommutative gauge theory provides an interesting toy model to test various features of the UV/IR mixing which will be examined in the subsection \ref{subsec:moyal_YM}.

\paragraph{}
At this point, one remark is in order.

The existence of the gauge invariant connection \eqref{eq:moyal_inv_conn} mentioned above actually singles out two possible field variables, hence two different types of gauge invariant noncommutative theories which can be obtained:

The first one are characterized by an action functional $S(A_\mu)$, hence only based on the variable $A_\mu$, thus related to a 1-form gauge connection, in other words a (noncommutative) gauge potential. These noncommutative gauge theories represent a large part of the literature on the gauge theories on Moyal spaces and bear some (algebraic) similarities with the (commutative) Yang-Mills theories. This is especially apparent for the simplest noncommutative extension of Yang-Mills theory for which the interaction vertices only differ from their commutative counterpart by (momentum dependent) phase factors as we will see in a while. However, UV/IR mixing shows up in this gauge theory, which prevents \textit{a priori} its perturbative renormalizability to be achieved. This motivated the quest for solutions curing this mixing, obtained by adapting to a gauge theory context some available solutions of the UV/IR mixing proposed for noncommutative scalar field theories. These will be reviewed below.

The second type of gauge invariant noncommutative theories uses the covariant tensor-form $\mathcal{A}_\mu$ \eqref{eq:moyal_covar_coord} as field variable. Note that \eqref{eq:moyal_covar_coord} is also known in the literatre of string theory as ``covariant coordinates''. It turns out that the corresponding gauge invariant actions $S(\mathcal{A}_\mu)$ can be naturally viewed as describing matrix models which is reminiscent of the type IIB matrix models \cite{matrix1}. Indeed, one can realizes that the general form of suitable $\mathcal{U}(\mathbb{R}^{4}_\theta)$ gauge invariant action $S(\mathcal{A}_\mu)$ is
\begin{equation}
    S
    = \int \dd^4x \Big( \frac{1}{4} F_{\mu\nu}(\mathcal{A}) \star_\theta F^{\mu\nu}(\mathcal{A}) + \frac{\Omega^2}{4} \{{\cal{A}}_\mu, {\cal{A}}_\nu\}^2_\theta + \kappa {\cal{A}}_\mu \star_\theta {\cal{A}}^\mu \Big)
    \label{eq:moyal_mat_action}
\end{equation}
where the curvature $F_{\mu\nu}(\mathcal{A})$ is now expressed in term of $\mathcal{A}$ via the second relation in \eqref{eq:moyal_covar_curv} and $\Omega$ and $\kappa$ are some constants. 

Equation \eqref{eq:moyal_mat_action} is nothing but a matrix model action, a feature which becomes apparent whenever \eqref{eq:moyal_mat_action} is expressed in the matrix base for $\mathbb{R}^4_\theta$. This class of interesting matrix-model description of noncommutative gauge theories has received some attention from various viewpoints, e.g.\ \cite{Steinacker_2007, wess-madore, steinhacker2, grimwulk, wall-wulk1, wall-wulk2}.

For instance, it can be shown that gravity seems to be an intrinsic part of the matrix model formulation of $U(N)$ noncommutative gauge theories \cite{Steinacker_2007}, in the spirit of \cite{wess-madore}, and provides a possible interpretation of the UV/IR mixing in terms of an (induced) gravity action \cite{steinhacker2}. This, together with other approaches will be reviewed in section \ref{subsec:emergent_grav}.

\subsection{Gauge theories on \tops{$\mathbb{R}^{4}_\theta$}{R\^4\_theta} as Yang-Mills type models}
\label{subsec:moyal_YM}

\subsubsection{The simplest noncommutative Yang-Mills model}
\label{subsubsec:moyal_simplest_YM}
\paragraph{}
We first focus on noncommutative gauge theories with gauge group $\mathcal{U}(\mathbb{R}^4_\theta)$, \textit{i.e.}\  a noncommutative analog of $U(1)$. The corresponding action on $\mathbb{R}^4_\theta$ is simply
\begin{equation}
    S_{\mathrm{cl}}
    = \frac{1}{g^2} \int \dd^4x\ (F_{\mu\nu} \star_\theta F^{\mu\nu})(x),
    \label{eq:moyal_YM_action}
\end{equation}
with $F_{\mu\nu}$ given by \eqref{eq:courb-moyal}, which is obviously invariant under $\mathcal{U}(\mathbb{R}^4_\theta)$ \eqref{eq:moyal_U1} in view of \eqref{eq:moyal_gauge_trans}. Here, $g$ is a coupling constant. The mass dimensions of the parameter and fields are as in the commutative case, namely $[A_\mu]=1$, $[g^2]=0$.

\paragraph{}
A standard computation yields the cubic and quartic vertices, which in momentum space reads
\begin{equation}
    V^3_{\alpha\beta\gamma}(k_1, k_2, k_3)
    = - 2 i \sin \left(\frac{k_1 \wedge k_2}{2} \right)
    \big[ (k_2 - k_1)_\gamma \delta_{\alpha\beta} 
    + (k_1 - k_3)_\beta \delta_{\alpha\gamma} 
    + (k_3 - k_2)_\alpha \delta_{\beta\gamma} \big],
    \label{eq:moyal_cubic_vert}
\end{equation}
and
\begin{align}
\begin{aligned}
    V^4_{\alpha\beta\gamma\delta}(k_1,k_2,k_3,k_4)
    &= - 4 \bigg[(\delta_{\alpha\gamma} \delta_{\beta\delta} - \delta_{\alpha\delta} \delta_{\beta\gamma}) \sin \left(\frac{k_1 \wedge  k_2}{2} \right) \sin \left(\frac{k_3 \wedge k_4}{2} \right) \\ 
    &\phantom{=}
    + (\delta_{\alpha\beta} \delta_{\gamma\delta} - \delta_{\alpha\gamma} \delta_{\beta\delta}) \sin \left(\frac{k_1 \wedge k_4}{2} \right) \sin \left(\frac{k_2 \wedge  k_3}{2} \right) \\
    &\phantom{=}
    + (\delta_{\alpha\delta} \delta_{\beta\gamma} - \delta_{\alpha\beta} \delta_{\delta\gamma}) \sin \left(\frac{k_3 \wedge k_1}{2} \right) \sin \left(\frac{k_2 \wedge k_4}{2} \right) \bigg],
\end{aligned}
    \label{eq:moyal_quartic_vert}
\end{align}
in which 
\begin{equation}
   k \wedge p
   := \Theta^{\mu\nu} k_\mu p_\nu,  
\end{equation}
and all the momenta are entering the vertex. The overall delta function enforcing the momentum conservation is understood. One should note that the $\delta$'s in \eqref{eq:moyal_cubic_vert} and \eqref{eq:moyal_quartic_vert} arise because of the background metric choice \eqref{eq:background_metric}.

It can be verified that the purely Lorentzian structure of the kinetic operator of \eqref{eq:moyal_YM_action} coincides with the one of a commutative Abelian Yang-Mills theory, which is a mere consequence of \eqref{eq:courb-moyal} and \eqref{eq:moyal_prod_closed}. Besides, it can be noticed that the structure of the cubic and quartic vertices have some similarity with the ones of the commutative non-Abelian Yang-Mills vertices, up to phase factors of the form $\sin(p \wedge q)$. 

\paragraph{}
The gauge-fixing can be carried out using the BRST machinery in a way similar to what is done for commutative Yang-Mills theory. According to the BRST liturgy, this is achieved by adding to the classical action \eqref{eq:moyal_YM_action} a BRST-exact term. For instance, working in the gauge $\partial_\mu A^\mu = \lambda$ ($\lambda$ real constant), the gauge-fixed action reads
\begin{equation}
    S
    = S_{\mathrm{cl}} + s \int \dd^4x\ (\overline{C} \partial_\mu A^\mu + \frac{\lambda}{2} \overline{C} b)
    \label{eq:moyal_gauge_fix_act}
\end{equation}
where $s$ is the nilpotent BRST operation, sometimes called the Slavnov operation in the literature, defined by
\begin{align}
    s A_\mu
    &= \partial_\mu C - i [A_\mu, C]_\theta, &
    s C
    &= \frac{i}{2} [C,C]_\theta, &
    s\overline{C}
    &= b, &
    s b
    &= 0
    \label{eq:moyal_slavnov}.
\end{align}
From \eqref{eq:moyal_slavnov}, one can check by simple calculation that $s^2=0$. Recall that $s$ increases the ghost number of fields by $+1$, acting as a graded derivation with grading defined by the sum of the degree of differential forms and ghost number, modulo $2$. In \eqref{eq:moyal_gauge_fix_act}, $C$, $\overline{C}$ and $b$ are respectively the ghost, antighost and St\"uckelberg field, with respective ghost number $1$, $-1$ and $0$.

\paragraph{}
The above gauge-fixed theory has some interesting properties. In particular, unlike other noncommutative gauge theories which will be reviewed later on, \eqref{eq:moyal_gauge_fix_act} does not produce non-vanishing 1-point (tadpole) function at the one-loop order, as it is the case for the commutative Yang-Mills theory. This simply stems from the combination the algebraic structures of the interaction vertices and the $A_\mu$ and ghost propagators. In some sense, \eqref{eq:moyal_gauge_fix_act} , together with its $U(N)$ extension, has some flavor of a commutative Yang-Mills theory (remember however the remark below \eqref{eq:Moyal_action_un}). Roughly speaking, the former is obtained from the latter by simply trading the usual commutative product for the Moyal star-product.

\paragraph{}
The coupling to fermion has been examined in many works, see for instance \cite{hayakawa}. Even more phenomenological explorations have been carried out by considering noncommutative version of the standar model on the Moyal space, by taking advantage of the Seiberg-Witten map, thus considering $\theta$-expanded models. See section \ref{subsec:moyal_theta_exp} for more details. These models are interesting in themselves and in particular are free of UV/IR mixing. However, they are not renormalizable as analyzed in \cite{grimwulk}. We will anyway review them at the end of this section.

\paragraph{}
It is well known that \eqref{eq:moyal_gauge_fix_act} exhibits UV/IR mixing. This problematic phenomenon often (but not always) affects noncommutative field theories and noncommutative gauge theories. UV/IR mixing has been first evidenced in \cite{IRUVmix1} in the context of scalar theories on $\mathbb{R}^4_\theta$, see also \cite{matusis} where this observation has been extended to noncommutative gauge theories. There is a huge literature on the subject, essentially focused on scalar field theories. For reviews, see e.g.\ \cite{autro-revue}.

\paragraph{}
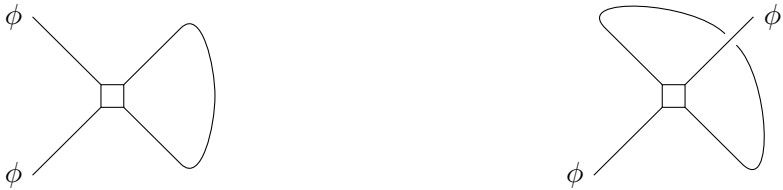
\begin{figure}[h]
    \begin{minipage}{.49\textwidth}
         \centering
    \begin{tikzpicture}[scale = 1.5]
        \draw[black] (-.1,-.1) rectangle (.1,.1);
        \draw[black] (-.7, -.7) node[anchor= east]{$\phi$} to (-.1, -.1);
        \draw[black] (-.7, .7) node[anchor= east]{$\phi$} to (-.1, .1);
        \draw[black] (.1,.1) to (.6,.6) to[out=45, in=90] (.9,0);
        \draw[black] (.1,-.1) to (.6,-.6) to[out=-45, in=-90] (.9,0);
    \end{tikzpicture}
    \end{minipage}
    \begin{minipage}{.49\textwidth}
         \centering
    \begin{tikzpicture}[scale = 1.5]
        \draw[black] (-.1,-.1) rectangle (.1,.1);
        \draw[black] (-.7, -.7) node[anchor= east]{$\phi$} to (-.1, -.1);
        \draw[black] (.7, .7) node[anchor= west]{$\phi$} to (.1, .1);
        \draw[black] (-.1,.1) to (-.6,.6) to[out=135, in=135] (.45,.55);
        \draw[black] (.1,-.1) to (.6,-.6) to[out=-45, in=-45] (.55,.45);
    \end{tikzpicture}
    \end{minipage}
    
    \caption{Feynman diagrams for $\phi^4$ noncommutative field theory. The left diagram shows a planar contraction, explicitly the diagram can be drawn on a plan. The right diagram pictures a non-planar diagram, that is a diagram that cannot be drawn on a plan. The latter diagram triggers IR singularities.}
    \label{fig:plan_non_planar}
\end{figure}

To synthesize the essential features of the UV/IR mixing, recall that it manifests itself at the one-loop level in the 2-point function as infrared singularities whose leading singularity is of polynomial type, \textit{i.e.}\ $\sim \frac{1}{p^\alpha}$ where $\alpha$ is a constant and $p$ is the external momentum. These IR singularities show up typically in contributions related to the so-called non-planar diagrams (see Figure \ref{fig:plan_non_planar}), which, by the way, are UV finite. As an illustrative example, a typical 1-loop contribution of a non-planar diagram to the 2-point function of a noncommutative $\phi^4$ theory takes the form
\begin{equation}
    \omega_2^{\mathrm{NP}}
    \sim \int \frac{\dd^4k}{(2 \pi)^4} \frac{e^{i p_\mu k_\nu \Theta^{\mu\nu}}}{k^2 + m^2}
    \label{eq:moyal_nonplanar_2pf}
\end{equation}
where $m$ is a mass parameter and $p$ the external momentum, which is UV finite for $p \ne 0$ thanks to the oscillating factor in the numerator. The integral over $k$ can be performed leading to
\begin{equation}
    \omega_2^{\mathrm{NP}}
    \sim \sqrt{\frac{m^2}{(\Theta p)^2}} K_1 \big( m^2(\Theta p)^2 \big)
    \underset{p \to 0}{\sim} \frac{1}{p^2}
    \label{eq:moyal_nonplanar_div}
\end{equation}
where $(\Theta p)^\mu = \Theta^{\mu\nu}p_\nu$ and $K_1(z)$ is the first Bessel function of second kind. From this, one easily finds that \eqref{eq:moyal_nonplanar_div} becomes singular when $p \to 0$.

At the two-loop level, some of the corresponding diagrams happen to become subdiagrams while the above external momentum $p$ becomes an internal momentum which, upon integration, generates UV divergences in the two-loop diagrams. In other words, UV and IR scales are linked in a non-trivial way.

\paragraph{}
The presence of these polynomial singularities, as leading singularities, sparks overwhelming problems to carry out a Wilson-type renormalization, thus motivating the quest for solutions to overcome the UV/IR mixing. Those will be reviewed below. Notice that other subleading IR singularities of logarithimic type are present. But these logarithmic (integrable) singularities alone could be accommodated for in the renormalization. However, quadratic and linear IR singularities definitely spoil the usual renormalization process. Basically, the subtraction of the new (UV) divergences generated by these polynomial IR singularities cannot be clearly related to parameters involved in the action.

\paragraph{}
As far as the noncommutative gauge theories on $\mathbb{R}^4_\theta$ are concerned, the IR singularities generating UV/IR mixing appear in the vacuum polarization tensor. Going back for the moment to the $U(N)$ gauge theory, a standard computation in the Lorentz gauge of the related vacuum polarization within the dimensional regularization scheme \cite{cawa} leads to the contribution to the IR limit given by ($d = 4 + \epsilon$)
\begin{equation}
    \omega^{ab}_{\mu\nu}(p)
    = \int \frac{\dd^d k}{(2 \pi)^d}
    \frac{\delta^{ab}}{k^2(k - p)^2} P_{\mu\nu}(k, p) \times (2 - \delta^{a0} \cos(k \wedge p)),
\end{equation}
where $P_{\mu\nu}(k,p)$ is a polynomial function whose explicit form is not needed here and $a,b$ are indices of $\mathfrak{u}(N)$. The second contribution $\sim\cos(k\wedge p)$ corresponds to the non-planar contribution involving the dangerous IR singularities, which obviously vanishes whenever $a \ne 0$.

In view of \eqref{eq:uN_lie_alg}, one therefore concludes [XXX] that the pure $SU(N)$ part of the gauge theory is free of UV/IR mixing. In other words, UV/IR mixing shows up only in the $U(1)$ gauge theory.

\paragraph{}
In the $U(1)$ case, the related leading IR singularity can be obtained after a standard but tedious computation. The IR limit of the vacuum polarization tensor is given by
\begin{equation}
    \omega^{\mu\nu}(p)
    = (d-2) \Gamma \left( \frac{d}{2} \right) \frac{\tilde{p}^\mu \tilde{p}^\nu}{\pi^{d/2} (\tilde{p}^2)^{d/2}} + \dots,
    \label{eq:moyal_poly_sing}
\end{equation}
where we set
\begin{equation}
    \tilde{p}^\mu
    = \Theta^{\mu\nu} p_\nu
    \label{eq:moyal_tilde_p}
\end{equation}
and $\Gamma(z)$ is the Euler gamma function. The polarization tensor verifies the transversality condition
\begin{equation}
    p_\mu \omega^{\mu\nu}(p)
    = 0,
\end{equation}
stemming from the Slavnov-Taylor identities. The IR singular expression \eqref{eq:moyal_poly_sing} signals the occurrence of UV/IR mixing in the gauge theory, unless $d = 2$, which is of very limited physical interest.

\paragraph{}
It appears that \eqref{eq:moyal_poly_sing} does not depend on the gauge choice. This interesting observation has been done in \cite{ruiz-ruiz} and \cite{austros} where different interpolating gauges have been considered, thus showing that the corresponding singularity is by the way not a gauge artifact. Besides, an early attempt to interpret the leading IR singularity in terms of a matrix model formulation of the gauge theory has been proposed in \cite{zondermeuuh}.

\paragraph{}
As far as renormalizability is concerned, both noncommutative $U(1)$ and $U(N)$ gauge theories are renormalizable up to one loop order, see e.g.\ \cite{armoni1} while consistency with the related Slavnov-Taylor identities stemming from the BRST invariance has been shown to hold \cite{cpmartin1}, \cite{slavnovjc}. Unfortunately, the appearance of UV/IR mixing likely prevents these gauge theories to be perturbatively renormalizable to all orders. Finally, we mention early work exploring gauge theories on the noncommutative torus \cite{torus1}, \cite{torus2}, \cite{torus3}

\subsubsection{\tops{$BF$}{BF} terms as a cure to the UV/IR mixing.}
\label{subsubsec:moyal_BF}
\paragraph{}
We now focus only on the $U(1)$ case.

The first attempt to get rid of the UV/IR mixing in gauge theories on $\mathbb{R}^4_\theta$ was proposed by Slavnov \cite{slavnov1}. This amounts to add to the classical action \eqref{eq:moyal_YM_action} a BF term of the form
\begin{equation}
    S_{\mathrm{bf}}
    = \frac{1}{2} \int \dd^4x\ \lambda \star_\theta \Theta^{\mu\nu} F_{\mu\nu}
    \label{eq:moyal_slavnov_act},
\end{equation}
with the assumption that $\Theta_{0j} = 0$, $j = 1, 2, 3$ which corresponds to a ``commutative time''. Here, $\lambda$ plays the role of the $B$-field of the $BF$ term, see e.g.\ in \cite{Loss1}, \cite{Loss2} for a discussion.

\paragraph{}
The polarization tensor of the resulting action has still a leading IR singularity of the form \eqref{eq:moyal_poly_sing} which is still independent of the choice of the gauge function \cite{danyblasch1}  while the free propagator $P_{\mu\nu}(p)$ of the gauge potential $A_\mu$ is modified to become transverse w.r.t $\tilde{p}^\mu$ defined in \eqref{eq:moyal_tilde_p}, \textit{i.e.}\ 
\begin{equation}
   \tilde{p}^\mu P_{\mu\nu}(p)=0.
\end{equation}
One then concludes that the leading IR singularity is neutralized whenever the polarization tensor is connected to a diagram by a propagator of the gauge potential.

\paragraph{}
This interesting attempt is appealing. However, the formal commutative limit does not reproduce a gauge theory, but a scalar field theory instead. Besides, possible new UV divergences may appear if the tensor $\Theta^{\mu\nu}$ is not of full rank \cite{danyblasch1}. In addition, notice that the $\lambda$-field propagates and is involved in interaction vertices with $A_\mu$ which therefore enlarges the diagrammatic of the model. 

\paragraph{}
Variations of the so-called Slavnov model have been considered and shown to be resilient to the dangerous IR singularities. The authors of \cite{Loss1} consider a modification of the Slavnov action given by
\begin{equation}
    S
    = \int \dd^4x\ -\frac{1}{4} F_{\mu\nu} \star_\theta  F^{\mu\nu} + \frac{\theta}{2} \lambda \star_\theta \epsilon^{jk} F_{jk} + b \star_\theta n^j A_j - \overline{C} \star_\theta n^j D_j C,
    \label{eq:moyal_slavnov_modif}
\end{equation}
assuming 
\begin{align}
    \Theta^{0j}
    &= 0, &
    \Theta^{jk}
    &= \epsilon^{jk}\theta,
    \label{eq:moyal_slavnov_modif_theta}
\end{align}
together with the constant vector $n^\mu$ belonging to the plane $(x^1,x^2)$, thus working with an axial gauge function. Here $\epsilon$ denotes the fully antisymmetric Levi-Civita tensor and $D$ is the covariant derivarive defined as usual as $D_\mu C = \partial_\mu C - i [A_\mu, C]_\theta$. This action is invariant under the following BRST symmetry
\begin{subequations}
\begin{align}
    s A_\mu
    &= \partial_\mu C - i[A_\mu, C]_\theta, &
    s \overline{C}
    &= b, &
    s b 
    &= 0 \\
    s C
    &= \frac{i}{2} [C,C]_\theta, &
    s \lambda
    &= - i [\lambda, C]_\theta, &&
\end{align}
    \label{eq:moyal_slavmod_slavnov}
\end{subequations}
and one has $s^2=0$.

\paragraph{}
More interestingly, the action has an additional linear vector supersymmetry which appears thanks to the particular choice for the parameters given above. This symmetry is defined by the following nilpotent operations $\delta_j$, $j = 1, 2$, $\delta_j^2 = 0$ lowering the ghost number by one unit:
\begin{align}
\begin{aligned}
    \delta_j A_\mu 
    &= 0, &
    \delta_j C &= A_j, &
    \delta_j \overline{C} &= 0 \\
    \delta_j b
    &= \partial_j \overline{C}, &
    \delta_j \lambda
    &= \frac{1}{\theta} \epsilon_{jk} n^k \overline{C}, &&
\end{aligned}
\end{align}
again acting as graded derivations.

\paragraph{}
The noticeable consequences of this linear vector supersymmetry have been analyzed in \cite{Loss1} using the machinery of Slavnov-Taylor identities. It has been shown in particular that the propagator $P_{\mu\nu}$ for $A_\mu$ is zero whenever at least one of the indices is equal to $1$ or $2$. This, together with the fact that the vertex $\lambda AA$ is proportional to $\Theta^{jk} = \theta \epsilon_{jk}$, by the very construction of the action \eqref{eq:moyal_slavnov_modif}, \eqref{eq:moyal_slavnov_modif_theta} leads to the conclusion that any loop diagram involving the $\lambda AA$ vertex vanishes. This includes in particular all the dangerous diagrams involving insertions of polarization tensor contributions related to the additional field $\lambda$ and its interaction vertex.

\paragraph{}
An extension of this gauge model has been proposed in \cite{danyblasch2} where the initial Slavnov $BF$ term \eqref{eq:moyal_slavnov_act} is replaced by
\begin{equation}
    S^\prime
    = \int \dd^4x\ \frac{1}{2} \epsilon^{jkl}F_{jk} \star_\theta \lambda_l,
\end{equation}
thus now introducing in some sense 3 fields $\lambda_j$, $j = 1, 2, 3$ which formally amounts to satisfy the initial Slavnov constraint related to \eqref{eq:moyal_slavnov_modif}, \eqref{eq:moyal_slavnov_modif_theta} given by 
\begin{equation}
    \Theta^{12} F_{12} + \Theta^{13} F_{13} + \Theta^{23} F_{23} = 0,
    \label{eq:moyal_slavmod_sum_const}
\end{equation}
by requiring that each term in  \eqref{eq:moyal_slavmod_sum_const} vanishes separately. The various symmetries of this model have been characterized in \cite{danyblasch2}, among which a linear vector supersymmetry arises when the classical action is gauge-fixed with a space-like axial gauge. Again, one gets rid of the dangerous IR sick diagrams.

\paragraph{}
Although the above Slavnov modification represents a progress to cure the UV/IR mixing, it appears that the constraints enforced by the introduction of $BF$ terms reduce the number of degrees of freedom the gauge models so that the commutative limit does not describe a photon.

\subsubsection{Harmonic term, IR damping and gauge invariance.}
\label{subsubsec:moyal_harmonic_term}

\paragraph{}
The authors of \cite{danyblasch3} have proposed a nice solution to the UV/IR mixing. It amounts to incorporate a harmonic oscillator term in the action \eqref{eq:moyal_YM_action} in a way compatible with gauge invariance, or more precisely with BRST invariance, as we will show in a while. Observe that a harmonic term involving the gauge potential is $\sim x^2 A_\mu \star_\theta A_\mu$ which is obviously not gauge invariant. The effect of this term is to generate a strong damping of the infrared region of the momentum space. It is nothing but an adaptation to a gauge theory context of the features characterizing the noncommutative $\phi^4$ model with harmonic term on $\mathbb{R}^4_\theta$.

\paragraph{}
Recall that the noncommutative $\phi^4$ model with harmonic term \cite{Grosse:2003aj-pc} is the first example of perturbatively renormalizable to all orders noncommutative field theory on Moyal space. The corresponding action is given by\footnote{The convention for $\Theta^{\mu\nu}$ are $\Theta^{\mu\nu} = \begin{pmatrix} \mathcal{J} & 0 \\ 0 & \mathcal{J} \end{pmatrix}$ with $\mathcal{J} = \begin{pmatrix} 0 & \theta \\ - \theta & 0 \end{pmatrix}$.}
\begin{equation}
    S(\phi; m, g, \Omega)
    =\int \dd^4 x\ \frac{1}{2} \partial_\mu \phi \star_\theta \partial^\mu \phi 
    + m^2 \phi \star_\theta \phi 
    + \Omega \tilde{x}^\mu \phi \star_\theta \tilde{x}_\mu \phi 
    + \frac{g}{4!} \phi \star_\theta \phi \star_\theta \phi \star_\theta \phi
    \label{eq:moyal_phi4_act}
\end{equation}
where $\tilde{x}^\mu = \Theta^{-1}_{\mu\nu} x^\nu$, $\Omega$ is a real {\textit{positive}} parameter and $\phi$ is real-valued. Recall that this model is covariant under the following transformation called the Langmann-Szabo duality \cite{langmanszabo}
\begin{equation}
    S(\phi; m, g, \omega)
    \mapsto \Omega^2 S\left(\phi; \frac{m}{\Omega}, \frac{g}{\Omega}, \Omega^{-1} \right).
\end{equation}
It has vanishing beta function to all orders \cite{dissertori} at the self-dual point $\Omega = 1$, and can be solved at this point \cite{grosswulk22}. However, the major drawback of the inclusion of the harmonic term is the explicit breaking of translation invariance{\footnote{See however \cite{restau-symet}.}}.

\paragraph{}
The IR damping effect of the harmonic term can be illustrated by considering the propagator of the scalar field in \eqref{eq:moyal_phi4_act} expressed with the space variables. This latter is known to be given by the Mehler kernel expressed as
\begin{equation}
    P_H(x,y)
    = \int_0^\infty \dd t\ 
    \frac{1}{4 \pi^2 \theta^2 \Omega^{-2} \sinh^2(t)} e^{- \frac{\Omega}{4 \theta} \left( (x-y)^2 \coth \left(\frac{t}{2} \right) + (x+y)^2\tanh \left( \frac{t}{2} \right) - \frac{m^2 \theta t}{\Omega} \right)}.
    \label{eq:moyal_mehler}
\end{equation}
Setting now $m = 0$ and $y = 0$ in \eqref{eq:moyal_mehler}, one can perform the integration over $t$ giving
\begin{equation}
    P_H(x,0)
    \sim \frac{e^{\frac{- x^2 \theta}{4 \Omega}}}{x^2},
    \label{eq:moyal_harmdamp}
\end{equation}
while a similar computation performed in the absence of the harmonic term would give rise to
\begin{equation}
    P(x,0)
    \sim \frac{1}{x^2}.
    \label{eq:moyal_noharmdamp}
\end{equation}
It is apparent from equation \eqref{eq:moyal_harmdamp} that the decay of $P_H(x,0)$ for large values of $x$, equivalent to small momentum hence to the IR region, is stronger than the one of $P(x,0)$ for which a non-harmonic term is present. In some sense, this latter acts as an IR cut-off which renders harmless the effect of the leading IR singularities.

\paragraph{}
An interesting attempt to extend the above feature to gauge theory has been proposed in \cite{danyblasch3}. In order to cope with gauge invariance, the harmonic term is introduced through a (necessarely BRST-exact) gauge-fixing term suitably chosen. The resulting gauge-fixed action is, by construction, invariant under a BRST symmetry, thus trading at this stage the gauge invariance for a BRST invariance. The corresponding action can be cast into the form \cite{danyblasch3}
\begin{equation}
  S_H
  = \int \dd^4x\ 
  \frac{1}{4} F_{\mu\nu} \star_\theta F^{\mu\nu}
  + s \left(\frac{\Omega^2}{8} \tilde{c}_\mu \star_\theta \mathcal{C}^\mu + \overline{C} \star_\theta \partial^\mu A_\mu - \frac{1}{2} \overline{C} \star_\theta b \right)
  \label{eq:moyal_harm_act_gauge}
\end{equation}
where $\Omega$ is a positive real constant while
\begin{equation}
    \mathcal{C}^\mu
    = \{ \{\tilde{x}^\mu, A^\nu\}_\theta, A_\nu\}_\theta
    + [\{\tilde{x}^\mu, \overline{C}\}_\theta, C]_\theta
    + [\overline{C}, \{\tilde{x}^\mu, C\}_\theta]_\theta
\end{equation}
with the nilpotent BRST operation $s$ defined by
\begin{subequations}
\begin{align}
    s A_\mu
    &= \partial_\mu C - i [A_\mu, C]_\theta, &
    s C
    &= \frac{i}{2} [C, C]_\theta, &
    s \overline{C}
    &= b, \\
    sb
    &= 0, &
    s \tilde{c}^\mu
    &= \tilde{x}^\mu, &
    s \tilde{x}^\mu 
    &= 0.
\end{align}
\end{subequations}

Simple algebraic manipulations yield the following $A_\mu$ and ghost propagators in momentum space
\begin{align}
    P_{\mu\nu}(p,k)
    &= \delta_{\mu\nu} P_H(p,k), &
    P_{\mathrm{ghost}}(p,k)
    &= P_H(p,k)
\end{align}
where $P_H(p,k)$ is the Mehler kernel in momentum space given by
\begin{equation}
    P_H(p,k)
    = \frac{\theta^3}{8 \pi^2 \Omega^3} \int_0^\infty \dd t\ 
    \frac{1}{\sinh^2(t)} e^{-\frac{\theta}{4\Omega} (p-k)^2 \coth\left(\frac{t}{2}\right) - \frac{\theta}{4\Omega} (p+k)^2 \tanh\left(\frac{t}{2}\right)}
\end{equation}
so that $S_H$ in \eqref{eq:moyal_harm_act_gauge} achieves in principle the IR damping improvement previously exhibited in the noncommutative field theory with harmonic term sketched above. This noticeable feature is however balanced by somewhat troublesome properties arising at the one-loop order as we will recall below.

\paragraph{}
By perusing the structure of the action \eqref{eq:moyal_harm_act_gauge}, one realizes \cite{Blaschke_2010} that the interaction vertex involving $\tilde{c}^\mu$ does not contribute to the loop corrections so that the net result is that the Feynman rules for \eqref{eq:moyal_harm_act_gauge} relevant for the loop computations reduces to those for the simplest gauge-fixed action \eqref{eq:moyal_gauge_fix_act}. An upper-bound for the degree of UV divergence of the diagrams derived in \cite{Blaschke_2010} is given (in 4-d) by $d \leqslant 4 - E_A - E_{\mathrm{ghosts}} - E_{\tilde{c}} - 2 E_b$ where $E_\phi$ denotes the number of  external lines for a field of type $\phi$.

\paragraph{}
One salient feature arising in the gauge-fixed theory is the loss of transversality for the 2-point function of the $A_\mu$, unlike the commutative case. This stems from the breaking of translation invariance originated in the choice of the function $\mathcal{C}^\mu$. Indeed, first introduce
\begin{align}
    \Gamma
    &= \Gamma_{\mathrm{eff}} + \Gamma_{\mathrm{sources}}, &
    \Gamma_{\mathrm{sources}}
    &= \int \dd^4x\ j^\mu sA_\mu + \gamma sc 
\end{align}
where $\Gamma_{\mathrm{eff}}$ is the generating functional of the vertices, \textit{i.e.}\ the effective action and $j^\mu, \gamma$ are sources linearly coupled to the BRST variations of the fields, with respective ghost number $0$ and $-2$, one easily derives the expression for the Slavnov-Taylor functional identity related to the BRST symmetry. It can be written as
\begin{equation}
    \int \dd^4x\ 
    \frac{\delta \Gamma}{\delta j^\mu} \star_\theta \frac{\delta \Gamma}{\delta A_\mu} 
    + \frac{\delta \Gamma}{\delta \gamma} \star_\theta \frac{\delta \Gamma}{\delta C}
    + b \star_\theta \frac{\delta \Gamma}{\delta \overline{C}}
    + \tilde{x}^\mu \star_\theta \frac{\delta \Gamma}{\delta \tilde{c}^\mu},
    \label{eq:moyal_func_ST}
\end{equation}
which {\textit{formally}} differs from the Slavnov-Taylor identity for the commutative Yang-Mills theory by the last term involving $\tilde{x}^\mu$, signaling the breaking of translational invariance. As a consequence, upon functionally deriving \eqref{eq:moyal_func_ST} w.r.t.\ $A_\mu$ and $C$, one obtains
\begin{equation}
    \partial^y_\mu \frac{\delta^2 \Gamma}{\delta A_\nu(x) \delta A_\mu(y)}
    = \int \dd^4z\ \tilde{z}^\mu \frac{\delta^3 \Gamma}{\delta C(y) \delta A_\nu(x) \delta\tilde{c}^\mu(z)}
    \ne 0.
\end{equation}

\paragraph{}
Explicit one-loop computations of diagrams have been performed and analyzed in \cite{Blaschke_2010}. The most troublesome feature is the appearance of a non-vanishing tadpole for $A_\mu$. This may be interpreted as a vacuum instability of the theory triggered by the quantum fluctuation. So far, no deep investigation of the consequences and/or physical interpretation of this non-vanishing tadpole have appeared in the literature.

Besides, the computation of the polarization tensor of $A_\mu$ exhibits a UV divergence which is more severe than its commutative counterpart. This can be traced back to the breaking of translational invariance which generates a loss of transversality for the polarization tensor.

\paragraph{}
To conclude this section, the appearance of a harmonic term in a gauge theory model on $\mathbb{R}^4_\theta$ can be achieved through the inclusion of a suitably chosen BRST-exact term. While the resulting propagator for the $A_\mu$ has a strong decay property in the IR regime, as in the $\phi^4$ scalar model with harmonic term, and thus can be expected to neutralize the bad effects of the UV/IR mixing, new unpleasant features appear already at the one-loop order, such as a non-vanishing tadpole for $A_\mu$, a loss of transversality for the polarization tensor and stronger UV divergences for it. All of these features do not \textit{a priori} favor possible renormalizability to all orders of the gauge model \eqref{eq:moyal_harm_act_gauge}. No corresponding study has been performed so far.

As a closing remark, the one-loop computation within this gauge model performed in \cite{Blaschke_2010} indicates that UV counterterms (not present in the classical action) {\textit{seem}} to fit with terms involved in the gauge invariant action \eqref{eq:moyal_mat_action} describing a gauge theory on $\mathbb{R}^4_\theta$ as a matrix model. This formulation is sometimes called ``induced gauge theory'' in the literature. This observation has led the authors of \cite{Blaschke_2010} to conjecture that the induced gauge theory has a more fundamental role and may be the classical action to be used as a starting point. This interesting conjecture lacks however of a definite proof. We will review the present status of the induced gauge theory on Moyal space in subsection \ref{subsec:moyal_induced_gauge}.

\subsubsection{\tops{$\frac{1}{p^2}$}{1/p\^3} gauge models and IR damping.}
\label{subsubsec:moyal_p2gauge_model}
\paragraph{}
Another attempt to cure the UV/IR mixing has been proposed and explored by adapting to a gauge model context the main property of the so-called translation-invariant renormalizable noncommutative $\phi^4$ scalar model on $\mathbb{R}^4_\theta$ \cite{rivas1}. As for the harmonic solution presented in the previous section, the neutralizing effect on the UV/IR mixing comes from a damping effect of the IR region of the momentum space. The definite advantage of the present attempt is that the translation invariance is unbroken, unlike the harmonic solution.

\paragraph{}
Recall that the translation-invariant noncommutative $\phi^4$ scalar model is obtained from the simple noncommutative $\phi^4$ action by supplementing it with a term 
\begin{equation}
    \sim - \frac{a^2}{\theta^2} \int \dd^4x\ \phi \star_\theta \frac{1}{\partial^2} \star_\theta \phi,
    \label{eq:moyal_contretermIR}
\end{equation}
where $a$ is a dimensionless constant, which can be viewed as a counterterm for the IR quadratic singularity expected to occur. The corresponding action can be more conveniently expressed in the momentum space as
\begin{equation}
    S(\phi; a)
    = \int \frac{\dd^4k}{(2 \pi)^4} \frac{1}{2} \phi(-k) \left(k^2 + m^2 + \frac{a^2}{\tilde{k}^2} \right) \phi(k) + \frac{g}{4!} \mathcal{F}(\phi \star_\theta \phi \star_\theta \phi \star_\theta \phi)
    \label{eq:moyal_transinv_act}
\end{equation}
in which $\mathcal{F}$ denotes the Fourier transform. It can be readily realized that the propagator for $\phi$ is given by
\begin{equation}
    P(k)
    = \frac{1}{k^2 + m^2 + \frac{a^2}{k^2}},
    \label{eq:moyal_prop_dampall}
\end{equation} 
thus exhibiting obviously a net decay in the IR region whenever $a \ne 0$, even in the massless case $m = 0$.

The model has been shown to be renormalizable to all orders \cite{rivas1}, thanks in particular to the IR damping effect of the propagator. This effect can be illustrated by considering typical (non-planar) potentially dangerous higher order contributions to the 2-point function $\Pi^{(n)}(p)$ for $\phi$, $p$ being the external momentum, obtained from $n$ insertions in a large loop of the 1-loop 2-point function whose leading IR singularity due to the UV/IR mixing is given by \eqref{eq:moyal_nonplanar_div}. The corresponding contributions behave as
\begin{equation}
    \Pi^{(n)}(p)
    \sim \int \dd^4 k \frac{e^{ i k_\mu \Theta^{\mu\nu} p_\nu}}{\tilde{k}^{2n} (k^2 + m^2 + \frac{a^2}{k^2})^{n+1}}
    \label{eq:moyal_highloop}
\end{equation}
which does not involve an IR divergence whenever $a \ne 0$, whereas for $a = 0$ the contribution $\Pi^{(n)}(p)$ \eqref{eq:moyal_highloop} has a manifest IR divergence: the initial IR (leading) 1-loop singularity $\frac{1}{\tilde{k}^2}$ grows now order by order as $\sim \int \dd^4 k\ \frac{1}{\tilde{k}^{2n} (k^2 + m^2)^{n+1}}$ in the absence of IR damping effect in the propagator.

Notice that the IR finitude still holds when $a \ne 0$ for the massless case $m = 0$ which was regarded, soon after the model \eqref{eq:moyal_transinv_act} was put forward, as an encouraging indication toward a possible extension to the gauge theory models.

\paragraph{}
However, this task is drastically complicated by the requirement of gauge invariance, as expected. Since one is ultimately looking for a perturbatively renormalizable gauge theory model (to all orders), it seems natural to trade gauge invariance for BRST invariance requirement of the (gauge-fixed) action with IR damping of the $A_\mu$ propagator similar to the one for the $\phi$ propagator of the above translation-invariant noncommutative $\phi^4$ model. The corresponding nilpotent BRST operation is
\begin{align}
    s A_\mu
    &= \partial_\mu C - i [A_\mu, C]_\theta
    = D_\mu C, &
    s C
    &= \frac{i}{2} [C, C]_\theta, &
    s \overline{C}
    &= b, &
    s b
    &= 0.
    \label{eq:moyal_brst_ir}
\end{align}
Now, BRST invariance requirement forbids obviously the use of the most straightforward gauge extension of the counterterm \eqref{eq:moyal_contretermIR} which would be of the form $\int \dd^4x\ A_\mu \star_\theta \frac{1}{\tilde{\partial}^2} \star_\theta A^\mu$. A plausible solution respecting BRST invariance has been finally proposed \cite{danyblasch5}
\begin{equation}
    S_{\mathrm{IR}}
    = \int \dd^4 x\ F_{\mu\nu} \star_\theta \frac{1}{D^2 \widetilde{D}^2} \star_\theta F^{\mu\nu}
    \label{eq:moyal_conterIR_act}
\end{equation}
with $\widetilde{D}^\mu = \Theta^{\mu\nu}D_\nu$ and $D_\mu$ is the covariant derivative defined e.g.\ in the 1st of equations \eqref{eq:moyal_brst_ir}. Other possible solutions built from $F_{\mu\nu}$ have been claimed to be eliminated \cite{danyblasch5}. It is important to notice that \eqref{eq:moyal_conterIR_act} must be understood as an infinite power in the gauge potential $A_\mu$ which will generate a problem as we will show in a while.

\paragraph{}
The resulting gauge-fixed action one would obtain from \eqref{eq:moyal_conterIR_act} can be cast into the form \cite{danyblasch6}
\begin{subequations}
\begin{align}
   S_{\mathrm{tot}}
   &= \frac{1}{4} \int \dd^4 x\ F_{\mu\nu} \star_\theta F^{\mu\nu} + F_{\mu\nu} \star_\theta \frac{1}{D^2 \widetilde{D}^2} \star_\theta F^{\mu\nu} + S_{\mathrm{GF}} \\
   S_{\mathrm{GF}}
   &= s \int \dd^4 x\ \overline{C} \star_\theta (1 + \frac{1}{\partial^2 \tilde{\partial}^2}) \partial^\mu A_\mu - \frac{\alpha}{2} \overline{C} \star_\theta b
\end{align}
    \label{eq:moyal_conterIR_act_tot}
\end{subequations}
where $\alpha$ is, as usual, the gauge parameter and $s$ defined by \eqref{eq:moyal_brst_ir}.

BRST invariance is achieved thanks to $s \left( \frac{1}{D^2 \widetilde{D}^2} \star_\theta F_{\mu\nu} \right) = i [C, \frac{1}{D^2 \widetilde{D}^2} \star_\theta F_{\mu\nu}]_\theta$ \cite{danyblasch6} together with $s^2 = 0$. A standard computation yields the propagators for the $A_\mu$ and the ghost fields given respectively by
\begin{subequations}
\begin{align}
    P_{\mu\nu}(k)
    &= \frac{1}{k^2 + \frac{1}{\tilde{k}^2}} \left(- \delta_{\mu\nu} + \frac{k_\mu k_\nu}{k^2} - \alpha \frac{k_\mu k_\nu}{k^2 + \frac{1}{\tilde{k}^2}} \right) \\
    P_{\mathrm{ghost}}(k)
    &= \frac{1}{k^2 + \frac{1}{\tilde{k}^2}},
\end{align}
\end{subequations}
which thus exhibit apparently the desired IR damping as for \eqref{eq:moyal_prop_dampall}.

\paragraph{}
As far as renormalization is concerned, the action \eqref{eq:moyal_conterIR_act_tot} is still not in a suitable form. Indeed, it appears that the counterterm \eqref{eq:moyal_conterIR_act} generates an infinite number of $A_\mu$ self-interactions which therefore would require to handle with an infinite number of parameters which thus does not fit with renormalization based on power counting. One somewhat standard way out is to introduce (``unphysical'') auxiliary fields acting as a kind of multiplicators and re-express \eqref{eq:moyal_conterIR_act} in order to make the factor $\frac{1}{D^2\widetilde{D}^2}$ disappear such that the modified expression reduces to \eqref{eq:moyal_conterIR_act} upon functionally integrating over these auxiliary fields.

\paragraph{}
A nice solution has finally been elaborated in \cite{danyblasch7} which deals with almost all the technical difficulties. It is partly based on \cite{vilar} where a clever adaptation of the Zwanziger method \cite{zwanziger} of localization of non-local action for QCD is applied to noncommutative gauge theory on Moyal space. Recall that in QCD, the restriction to the first Gribov horizon \cite{gribov} implies that the gluon propagator vanishes in the IR{\footnote{The existence of Gribov copies has been examined in \cite{gribov-ital}.}}. This is achieved by supplementing the action with the following operator
\begin{equation}
    \gamma^4 g_3^2 \int \dd^4x\ f^{abc} f^{dec} A_\mu^b (O^{-1})^{ad} (A^\mu)^e,
    \label{eq:moyal_termQCD}
\end{equation}
where $\gamma$, $g_3$, $f^{abc}$ are respectively the Gribov parameter, the QCD coupling constant and the structure constant of the Lie algebra $\mathfrak{su}(3)$. The operator $(O^{-1})^{ad}$ is the inverse of the Fadeev-Popov operator. This yields a gluon propagator of the form 
\begin{equation}
    G^{ab}_{\mu\nu}(k)
    = \frac{\delta^{ab}}{k^2 + \frac{\gamma^2}{k^2}} \left( \delta_{\mu\nu} - \frac{k_\mu k_\nu}{k^2} \right).
\end{equation}
This propagator vanishes in the IR regime, which is what would be needed as an IR damping. By a suitable introduction of auxiliary fields, each accompanied with its associated ghost and upon using standard functional integration methods, it appears that \eqref{eq:moyal_termQCD} can be rewritten as the sum of two local actions, one being BRST exact and thus presumably not physically relevant and the other one depending on the Gribov parameter and not BRST invariant. This latter action represents the soft breaking term in the Gribov-Zwanziger approach in QCD.

\paragraph{}
An improvement of the gauge model \cite{vilar} was considered in \cite{danyblasch8} and further analyzed in \cite{danyblasch9} showing that the gauge model \cite{vilar} together with the improved gauge model involve propagators, in particular in the auxiliary field sector, with no required damping, leading to the occurrence of diagrams having IR divergences growing with their loop order.\\
This drawback was overcome in \cite{danyblasch7} and the corresponding action takes the form
\begin{equation}
S_{\mathrm{fin}}
    = S_{\mathrm{cl}} + S_{\mathrm{GF}} + S_{\mathrm{aux}} + S_{\mathrm{break}}
    \label{eq:moyal_finl_act}
\end{equation}
with 
\begin{subequations}
\begin{align}
    S_{\mathrm{cl}}
    &= \frac{1}{4} \int \dd^4 x\ F_{\mu\nu} \star_\theta F^{\mu\nu}, \\  
    S_{\mathrm{GF}}
    &= s \int \dd^4 x\ \overline{C} \star_\theta \partial^\mu A_\mu, \\
    S_{\mathrm{aux}}
    &= s \int \dd^4 x\ \overline{\psi}^{\mu\nu} \star_\theta B_{\mu\nu}, \\
    S_{\mathrm{break}}
    &= s \int \dd^4 x\ \left( (\overline{Q}^{\mu\nu\alpha\beta} \star_\theta B_{\mu\nu} + {Q}^{\mu\nu\alpha\beta} \star_\theta \overline{B}_{\mu\nu}) \star_\theta \frac{1}{\tilde{\partial}^2} \left( f_{\alpha\beta} + \sigma \frac{\Theta_{\alpha\beta}}{2} \tilde{f} \right) \right)
\end{align}
\end{subequations}
where $f_{\alpha\beta} = \partial_\alpha A_\beta - \partial_\beta A_\alpha$, $\tilde{f} = \Theta^{\mu\nu} f_{\mu\nu}$, $\sigma$ is a parameter with mass dimension $2$ and the BRST operation is defined by \eqref{eq:moyal_brst_ir} and
\begin{subequations}
\begin{align}
   s B_{\mu\nu}
   &= \psi_{\mu\nu}, &
   s \psi_{\mu\nu} 
   &= 0, &
   s {Q}_{\mu\nu\alpha\beta}
   &= {J}_{\mu\nu\alpha\beta}, &
   s {J}_{\mu\nu\alpha\beta} 
   &= 0 \\
   s \overline{\psi}_{\mu\nu}
   &= \overline{B}_{\mu\nu}, &
   s \overline{B}_{\mu\nu}
   &= 0, &
   s \overline{Q}_{\mu\nu\alpha\beta}
   &= \overline{J}_{\mu\nu\alpha\beta}, &
   s \overline{J}_{\mu\nu\alpha\beta}
   &= 0.
\end{align}
\end{subequations}
$({Q}_{\mu\nu\alpha\beta},\overline{Q}_{\mu\nu\alpha\beta})$ and $({J}_{\mu\nu\alpha\beta},\overline{J}_{\mu\nu\alpha\beta})$ are sources introduced to restore the full BRST invariance of the action \eqref{eq:moyal_finl_act} in the UV regime whereas they satisfy in the IR limit
\begin{align}
  \overline{Q}_{\mu\nu\alpha\beta}
  &= {Q}_{\mu\nu\alpha\beta}
  = 0, &
  \overline{J}_{\mu\nu\alpha\beta}
  &= {J}_{\mu\nu\alpha\beta}
  = \frac{\gamma^2}{4} (\delta_{\mu\alpha} \delta_{\nu\beta} - \delta_{\mu\beta} \delta_{\nu\alpha}),
\end{align}
giving rise to a breaking term \cite{danyblasch7}.

\paragraph{}
This gauge model exhibits many nice properties. The propagator for $A_\mu$ has the expected IR damping and has finite limits both in the IR and UV limit 
\begin{subequations}
\begin{align}
    P_{\mu\nu} 
    & \underset{\tilde{k}^2 \to 0}{\sim}
    \frac{\tilde{k}^2}{\gamma^4} \left( \delta_{\mu\nu} - \frac{k_\mu  k_\nu}{k^2} - \frac{\sigma_1^4}{\sigma_1^4 + \gamma^4}  \frac{\tilde{k}_\mu \tilde{k}_\nu}{\tilde{k}^2} \right),
    \label{eq:moyal_prop_fin_IR} \\
    P_{\mu\nu}
    & \underset{{k}^2\to\infty}{\sim} 
    \frac{1}{k^2} \left( \delta_{\mu\nu} - \frac{k_\mu  k_\nu}{k^4} \right)
    \label{eq:moyal_prop_fin_UV},
\end{align}
    \label{eq:moyal_prop_fin}
\end{subequations}
where $\sigma_1$ is some constant unessential for our present purpose. One feature of the model is that the mixed propagators involving both $A_\mu$ and $B,\overline{B}$ as well as the ``diagonal propagators'' for the $\psi$- and $B$-type fields cannot be connected to the interactions vertices. Hence, these do not contribute to loop corrections. Besides, the superficial UV degree of divergence of any diagram is given by $d \leqslant 4-E_A-E_{\mathrm{ghosts}}$ where $E_\phi$ again denotes the number of  external lines for a field of type $\phi$.

The one-loop corrections have been carried out in \cite{danyblasch7}. The vacuum polarization can be cast into the form  
\begin{equation}
  \Pi_{\mu\nu}(k)
  = \frac{2 g^2}{\pi^2 \varepsilon^2} \frac{\tilde{k}_\mu \tilde{k}_\nu}{(\tilde{k}^2)^2} 
  + \frac{13 g^2}{3 (4 \pi)^2}(k^2 \delta_{\mu\nu} - k_\mu k_\nu) \ln(\Lambda)
  + \text{finite},
\end{equation}
exhibiting the quadratic IR singularity (1st term) and a logarithmic UV divergence. Transversality for the polarization tensor holds true, namely
\begin{equation}
    p^\mu \Pi_{\mu\nu}
    = 0
\end{equation}
stemming simply from $p_\mu \tilde{p}^\mu = 0$. One-loop corrections for the cubic and quartic vertices for $A_\mu$ agree, considering the power counting given just above, with a simple logarithmic UV divergence for the quartic vertex. The beta function is found to be negative, indicating asymptotic freedom and the absence of Landau ghost. This sign agrees with the sign of the beta function for the simplest gauge theory model presented in section \ref{subsubsec:moyal_simplest_YM}.

The main observation is that the gauge model is renormalizable at the one-loop order. It has been conjectured that this property may well extend to all orders. But to date, a full proof is still lacking.

\subsubsection{Braided \tops{$L_\infty$}{L-infinity}-algebra gauge theory}
\label{subsubsec:moyal_braided_l_infty}
\paragraph{}
Classical gauge theory was shown to be perfectly implemented in a $L_\infty$-algebra formalism \cite{HZ2017}, encoding both gauge transformations and dynamics. $L_\infty$-algebras correspond to a generalization of the Lie algebras with an infinite number of brackets and infinite number of corresponding Jacobi identities. A $L_\infty$-algebra is a graded vector space $V = \bigoplus_{k \in \mathbb{Z}} V_k$, with graded antisymmetric multilinear maps $\ell_n: \bigotimes_n V \to V$ called $n$-brackets. However, classical gauge theory can be recovered by only using $4$ degrees, that is $V = V_0 \oplus \cdots \oplus V_3$, where degree $0$ fields are gauge parameters, degree $1$ fields are gauge fields, degree $2$ fields encodes the equation of motions and degree $3$ fields encodes the Noether identities.

\paragraph{}
Very recently, the generalization of this scheme to noncommutative gauge theory was undertaken. It was applied to the $L_\infty$ Einstein-Cartan gravity, as detailed in subsection \ref{subsubsec:braided_l_infty_alg}, and to the Moyal space \cite{Ciric_2022}. The deformation of a Lie algebra is done via a twist deformation (see subsection \ref{subsec:star_prod_twists} for more details) and leads to a braided geometry as detailed in subsection \ref{subsubsec:braid_braid}. Generalisation of this braiding to the $L_\infty$-algebra allows to deform such a theory, starting with the $n$-brackets that becomes braided. The construction is then similar to the classical case, with the braided $n$-brackets.

\paragraph{}
For the case of the Moyal twist deformation \eqref{eq:moyal_abelain_twist}, a $U(1)$ gauge theory is undertook. The quantization method uses the BV-formalism. The photon propagator is unchanged compare to the undeformed case and there are no three photon or four photon interactions, contrary to \eqref{eq:moyal_cubic_vert} and \eqref{eq:moyal_quartic_vert}. However, the fermion-photon vertex is non-trivially deformed. Therefore, the first deformed diagram would be the photon self-energy one. It appears that this diagram is UV divergent triggering the UV/IR mixing (see subsection \ref{subsubsec:moyal_simplest_YM} for more details) of the theory, this time without non-planar diagrams. The authors of \cite{Ciric_2022} advocates that this UV/IR mixing comes from an unadapted quantization procedure for braided theories.

\subsection{From \tops{$\theta$}{theta}-expanded gauge models to phenomenological predictions}
\label{subsec:moyal_theta_exp}
\paragraph{}
The $\theta$-expanded approach to noncommutative gauge theories on Moyal spaces received a lot of attention in the early days of these theories and provided a first insight into various related properties.

\subsubsection{The Seiberg-Witten map}
\label{subsubsec:moyal_SW_map}
\paragraph{}
This approach is essentially based on the use of the Seiberg-Witten map \cite{Seiberg_1999}, initially introduced in the framework of string theory. The claim underlying the Seiberg-Witten map is that commutative (\textit{i.e.}\ ordinary) gauge theories should be ``gauge equivalent'' to a noncommutative Yang-Mills type theory (viewed as some effective theory of open strings). Roughly speaking, this basically permits one to express a noncommutative theory in terms of a commutative theory with, as a net result however, infinitely many interaction terms stemming from an expansion in the ``noncommutativity parameter'' $\Theta^{\mu\nu}$. This approach, albeit practical, has its inherent limitations stemming from the appearance of an infinite expansion. In particular, renormalizability cannot hold true, at least in the standard sense. Nevertheless, we find instructive to briefly review this approach which has contributed to one step ahead of a better understanding of noncommutative gauge theories.

To briefly illustrate the Seiberg-Witten map, consider the case of a $U(1)$ gauge theory. Then, the Seiberg-Witten map is defined by
\begin{equation}
    \hat{A}_\mu (A_\nu) + \hat{\delta}_{\hat{\lambda}} \hat{A}_\mu (A_\nu)
    = \hat{A}_\mu (A_\nu + \delta_\lambda A_\nu)
    \label{eq:moyal_SW_map}
\end{equation}
where $\hat{A}_\mu $ (resp.\ $A_\mu$) denotes the noncommutative (resp.\ ordinary) gauge field, in the liturgy of the Seiberg-Witten map, and $\delta_{\hat{\lambda}}$ is the infinitesimal noncommutative gauge transformation with parameter $\hat{\lambda}$ given by
\begin{equation}
   \hat{\delta}_{\hat{\lambda}} \hat{A}_\mu
   = \partial_\mu \hat{\lambda} + i [\hat{\lambda}, \hat{A_\mu}]_\theta.
   \label{eq:moyal_gauge_trans_inf}
\end{equation}
The quantity $\hat{\lambda}$ depends on $A_\mu$ and the parameter for the commutative $U(1)$ gauge transformation $\lambda$, namely $\hat{\lambda}=\hat{\lambda}(A_\mu,\lambda)$. This is the relation expressing the ``gauge equivalence''.

Expanding \eqref{eq:moyal_gauge_trans_inf} in powers of $\Theta_{\mu\nu}$, for instance up to the second order in the parameter $\theta$ yields
\begin{equation}
     \hat{\delta}_{\hat{\lambda}} \hat{A}_\mu
     = \partial_\mu \hat{\lambda} - \Theta^{\rho\sigma} \partial_\rho \lambda \partial_\sigma A_\mu + \mathcal{O}(\theta^2),
\end{equation}
which, combined with \eqref{eq:moyal_SW_map}, gives rise to
\begin{align}
    \hat{A}_\mu(A_\nu)
    &= A_\mu + A^\prime_\mu + \mathcal{O}(\theta^2), &
    A^\prime_\mu
    &= - \frac{1}{2} \Theta^{\rho\sigma} A_\rho(\partial_\sigma A_\mu + F_{\sigma\mu}),
\end{align}
$F_{\mu\nu}$ being the ordinary $U(1)$ field strength. Accordingly, the noncommutative field strength takes the form
\begin{equation}
    \hat{F}_{\mu\nu}
    = F_{\mu\nu} + \Theta^{\rho\sigma} (F_{\mu\rho} F_{\nu\sigma} - A_\rho \partial_\sigma F_{\mu\nu}) + \mathcal{O}(\theta^2).
    \label{eq:moyal_SW_map_fs}
\end{equation}

One would proceed similarly whenever a (fermionic) matter field is included so as to obtain a $\theta$-expanded noncommutative QED. Notice that the solutions of the condition of gauge equivalence \eqref{eq:moyal_SW_map} are in general not unique. For a discussion, see \cite{Buric_2007}. This results in some freedom when using the Seiberg-Witten map.

\paragraph{}
It appears that the use of covariant coordinates, such as the one defined by \eqref{eq:moyal_covar_coord}, fits quite well with the Seiberg-Witten approach and proves useful when the above $U(1)$ framework is extended to the case of $SU(N)$ \cite{wess-madore} (see also \cite{Jurco_2000b}). The corresponding starting point, that we merely rephrase from \cite{wess-madore}, is that multiplication of a (covariant) field by a coordinate can in general not be a covariant operation in noncommutative geometry, because the coordinates will not commute with the gauge transformations. The idea is to make the coordinates covariant by adding a gauge potential to them. We refer to \cite{wess-madore} for more technical details.

\paragraph{}
The $\theta$-expanded approach has been subject to numerous works, one part devoted to the exploration of quantum properties, another part investigating some phenomenological aspects.

\paragraph{}
Despite the obstruction to renormalizability indicated at the beginning of this section, one-loop investigations have been performed within the $\theta$-expanded QED as well as in $SU(N)$ extensions. Needless to say, the results are valid at a given order in the expansion parameter, usually the first and second order being accessible. The net result is that the renormalizability is lost, in particular when matter fields are introduced. See \cite{Buric_2007, Carlson_2002, Martin_2007, Bichl_2001, Buric_2006a, Latas_2007, Buric_2008, Tamarit_2009, Martin_2009}. The natural evolution of the studies finally produced various extensions of the standard model, sometimes called generically noncommutative standard model \cite{Tamarit_2009, Calmet_2002, Aschieri_2003, Melic_2005a, Melic_2005b}.

\subsubsection{Modified gauge and curvature}
\label{subsubsec:moyal_modif_gauge_curv}
\paragraph{}
Another approach to noncommutative gauge theory in the semi-classical limit that is, at first order in the deformation parameter $\theta$, was proposed in \cite{Kupriyanov_2020}. The main idea is to add generic terms $\gamma$, $P$ and $R$ to the gauge transformation of the gauge potential $A$ and the expression of the curvature $F$ as follows
\begin{align}
    \delta_f A_\mu 
    &= \gamma_\mu^\nu(A) \partial_\nu(f) + \{ A_a, f\}, &
    F_{\mu\nu}
    &= \tensor{P}{_{\mu\nu}^{\rho\sigma}}(A) \partial_\rho A_\sigma + \tensor{R}{_{\mu\nu}^{\rho\sigma}}(A) \{A_\rho, A_\sigma\}.
\end{align}
To match with the usual commutative limit, these new fields have to satisfy the following condition
\begin{align}
    \gamma_\mu^\nu 
    &= \delta_\mu^\nu + \mathcal{O}(\theta), &
    \tensor{P}{_{\mu\nu}^{\rho\sigma}}
    &= \delta_\mu^\rho \delta_\nu^\sigma - \delta_\mu^\sigma \delta_\nu^\rho + \mathcal{O}(\theta), &
    \tensor{R}{_{\mu\nu}^{\rho\sigma}}
    &= \frac{1}{2} \big( \delta_\mu^\rho \delta_\nu^\sigma - \delta_\mu^\sigma \delta_\nu^\rho \big) + \mathcal{O}(\theta).
\end{align}
Other conditions are required that are the closure of the gauge transformation $[\delta_f, \delta_g] A = \delta_{\{f,g\}} A$ and the covariance of the field strength $\delta_f F = \{F, f\}$. This whole set of conditions corresponds to constraints on $\gamma$, $P$, $R$, to which a solution, if it exists, can give rise to a gauge invariant action.

This model can be enlarged for non-canonical noncommutativity \cite{Kupriyanov_2022}, like Lie algebra type noncommutativity as mentioned in section \ref{subsec:km_other_approach}.

\subsubsection{Phenomenological consequences}
\label{subsubsec:moyal_pheno_cons}
\paragraph{}
Phenomenological works have explored possible new effects stemming from the appearance of new couplings generated by the $\Theta$-expansion, hence considering the resulting gauge models as kind of effective theories with $\Theta^{\mu\nu}$ being a (suitably) small parameter. These have resulted into various bounds on the ``scale of noncommutativity'' defined as
\begin{equation}
    \Lambda_{\mathrm{NC}}
    = \frac{1}{\sqrt{\theta}},
\end{equation}
assuming the convenient parameterization for $\Theta^{\mu\nu}$ given by \eqref{eq:Zebelleparam}. Recall that the mass dimension of $\theta$ is $-2$.

A study of two body scattering processes in linear $e^+-e^-$ colliders involving pairs annihilation, M\"oller-Bhabba scattering and $\gamma \gamma \to \gamma \gamma$ \cite{Hewett_2001} have suggested that a scale $\Lambda_{\mathrm{NC}}\sim\mathcal{O}(1)$ TeV could be accessible in high energy $e^+-e^-$ colliders. The work \cite{Hinchliffe_2001} claims that visible effects could arise for $\Lambda_{\mathrm{NC}} \gtrsim \mathcal{O}(1)$ TeV. The authors of \cite{Schupp_2004} studied the possible consequence of the appearance of Left- and Right-neutrino couplings to the photon, arising through the covariant derivative
\begin{equation}
    D_\mu \psi
    = \partial_\mu \psi - i \sigma e[A_\mu,\psi]_\theta,
\end{equation}
($\sigma$ is some constant) which, at the first order in $\theta$, leads to
\begin{equation}
   D_\mu \psi
   = \partial_\mu \psi + \sigma e \Theta^{\nu\rho} \partial_\nu  A_\mu \partial_\rho \psi + \mathcal{O}(\theta^2)
\end{equation}
resulting in some new $\bar{\nu} \nu \gamma$ coupling in the Lagrangian. By using constraints on additional energy loss in globular stellar clusters, the lower bound on $\Lambda_{\mathrm{NC}}$ is estimated \cite{Schupp_2004} to be $\Lambda_{\mathrm{NC}} \gtrsim \mathcal{O}(100)$ GeV, while in \cite{Minkowski_2003}, the consequences of these new neutrino couplings on the neutrino dipole moment and charge radii are analyzed, leading to an upper bound $\Lambda_{\mathrm{NC}} \lesssim \mathcal{O}(150)$ TeV. This has to be compared with the results of the analysis \cite{Horvat_2009} which uses data from Big Bang Nucleosynthesis, suggesting $\Lambda_{\mathrm{NC}} \gtrsim \mathcal{O}(10^3)$ TeV. These bounds, albeit interesting, are only indicative and would require a complete analysis of the underlying assumptions.

Much stronger limits from ``noncommutative QCD'' have been put forward in \cite{Carlson_2001} due to the appearance of Lorentz-violating operators induced by radiative corrections, suggesting a lower bound $\Lambda_{\mathrm{NC}} \gtrsim \mathcal{O}(10^{17}) $ GeV.

\subsection{Gauge theories on \tops{$\mathbb{R}^{4}_\theta$}{R\^4\_theta} as matrix models: ``Induced'' gauge theory.}
\label{subsec:moyal_induced_gauge}

\subsubsection{General properties.}
\label{subsubsec:moyal_induced_general}
\paragraph{}
Another approach of the noncommutative gauge theories on Moyal spaces has been elaborated a few years after the first investigations of the quantum properties of the simplest noncommutative gauge invariant action \eqref{eq:moyal_YM_action}.

\paragraph{}
This was motivated partly by the difficulties generated by the UV/IR mixing to get a renormalizable theory, partly by the observation that some counterterms involving the coordinate $x^\mu$ (e.g.\ of the form $\sim \tilde{x}^\mu A_\mu$) seemed to appear in one-loop computations around the action \eqref{eq:moyal_YM_action}, suggesting the need to introduce powers of the tensor form/covariant coordinate \eqref{eq:moyal_covar_coord} $ \mathcal{A}_\mu = - i (A_\mu + \Theta^{-1}_{\mu\nu} x^\nu)$, as e.g.\ $\int \dd^4x\  \mathcal{A}_\mu \star_\theta \mathcal{A}^\mu$. The latter term is by itself gauge invariant in view of \eqref{eq:moyal_gauge_trans_cov} and the cyclicity of the integral w.r.t.\ the star-product. This altogether was somewhat reminiscent of a polynomial action in the curvature expressed in terms of $\mathcal{A}_\mu$, keeping in mind $F_{\mu\nu} = [\mathcal{A}_\mu, \mathcal{A}_\nu]_\theta - i \Theta^{-1}_{\mu\nu}$, the second relation of \eqref{eq:moyal_covar_curv}.

\paragraph{}
To investigate this conjecture, two separate ways were followed to compute an effective action depending on the gauge potential, one way based on a loop-computation \cite{wall-wulk1}, \cite{wall-wulk11}, the other one based on a heat kernel expansion \cite{michael-induced}. Both gave the same result. \\
The strategy followed by the authors of \cite{wall-wulk1}, \cite{wall-wulk11} was to couple a complex scalar field theory with a harmonic term $\sim \Omega^2 (\tilde{x}_\mu \phi)^\dag \star_\theta \tilde{x}^\mu \phi$ to an external gauge potential $A_\mu$ by a minimal coupling prescription and to compute the effective action for $A_\mu$ at the one-loop order by integrating over the matter fields. The matter field $\phi$ was assumed to obey the following gauge transformation: $\phi^g = g \star_\theta \phi$, where $g \in \mathcal{U}(\mathbb{R}^4_\theta)$, the unitary gauge group defined in \eqref{eq:moyal_U1}.

This minimal prescription can be conveniently written as
\begin{subequations}
\begin{align}
    \partial_\mu \phi
    &\mapsto \nabla^A_\mu \phi
    = \partial_\mu \phi - i A_\mu \star_\theta \phi,
    \label{eq:moyal_induced_prescrip_conn} \\
    \tilde{x}_\mu \phi
    &\mapsto \tilde{x}_\mu \phi + A_\mu \star_\theta \phi,
    \label{eq:moyal_induced_prescrip_tilde}
\end{align}
    \label{eq:moyal_induced_prescrip}
\end{subequations}
where \eqref{eq:moyal_induced_prescrip_tilde} stems from the relation
\begin{equation}
    \tilde{x}_\mu \phi
    = \tilde{x}_\mu \star_\theta \phi - i \partial_\mu \phi
    = - 2 i \nabla^\xi_\mu \phi + i \partial_\mu \phi
\end{equation}
in which $\xi_\mu = - \frac{1}{2} \tilde{x}_\mu$ was defined in \eqref{eq:moyal_inner_der}. The corresponding action is given by
\begin{equation}
    S(\phi, A)
    = \int \dd^4 x\ \left( 
    |\partial_\mu \phi - i A_\mu \star_\theta \phi|^2
    + \Omega^2 |\tilde{x}_\mu \phi + A_\mu \star_\theta \phi |^2
    + m^2 |\phi|^2
    + \lambda \phi^\dag \star_\theta \phi \star_\theta \phi^\dag \star_\theta \phi \right),
    \label{eq:moyal_induced_act_interm}
\end{equation}
where $\Omega \in [0,1]$, which is gauge invariant thanks to $(\nabla^{A,\xi}_\mu \phi)^g = g \star_\theta \nabla^{A,\xi}_\mu \phi$.

The computation of the gauge invariant effective action formally defined as
\begin{equation}
    e^{-\Gamma(A)}
    = \int D\phi D\phi^\dag \ e^{-S(\phi, A)}
\end{equation}
is standard and reduces at the one-loop order to
\begin{equation}
    e^{-\Gamma_1(A)}
    = \int D\phi D\phi^\dag\ e^{- S_{\mathrm{free}}(\phi)} e^{- S_{\mathrm{int}}(\phi, A)}
\end{equation}
where $S_{\mathrm{free}}(\phi)$ and $S_{\mathrm{int}}(\phi,A)$ denote respectively the quadratic part and the interaction part of $S(\phi,A)$ \eqref{eq:moyal_induced_act_interm}.

The lengthy computation has been carried out in \cite{wall-wulk1}. It appears that the 1-point (tadpole) function is non-zero. It is found to exhibit a quadratic UV divergence (together with subleading logarithmic divergences), namely
\begin{equation}
    \Gamma_1^1
    = - \frac{1}{\epsilon} \frac{\Omega^2}{4 \pi^2 (1 + \Omega^2)^3} \int \dd^4 x\ \tilde{x}^\mu A_\mu + \dots
\end{equation}
The factor $\frac{1}{\epsilon}$ comes from the regulator used in \cite{wall-wulk1} ($\epsilon \to 0$) and the ellipsis stands for the subleading divergences $\sim \log(\epsilon)$ which are as well found to be proportional to $\int d^4x\ \tilde{x}_\mu A_\mu$, again suggesting the occurrence of powers of $\mathcal{A}_\mu$ in the effective action.

\paragraph{}
After collecting all the multi-point contributions, the final action is found to be expressible as \cite{wall-wulk1} (assuming for simplicity $m = 0$)
\begin{align}
\begin{aligned}
   \Gamma_1(A)
   &= \frac{1}{\epsilon} \frac{\Omega^2}{4 \pi^2 (1 + \Omega^2)^3} \int \dd^4x\ (\mathcal{A}_\mu \star_\theta \mathcal{A}^\mu - \frac{1}{4} \tilde{x}^2) 
   \\
   & - \ln(\epsilon) \frac{(1 - \Omega^2)^4}{192 \pi^2 (1 + \Omega^2)^4} \int \dd^4x\ F_{\mu\nu} \star_\theta F^{\mu\nu} 
   \\
   & + \ln(\epsilon) \frac{\Omega^4}{8 \pi^2 (1 + \Omega^2)^4} \Bigg(\int  \dd^4x\ \Big(F_{\mu\nu} \star_\theta F^{\mu\nu} + \{\mathcal{A}_\mu, \mathcal{A}_\nu \}_\theta^2 - \frac{1}{4} (\tilde{x}^2)^2 \Big) \Bigg) + \dots
\end{aligned}
   \label{eq:moyal_ind_onepoint}
\end{align}
where the ellipsis stands for finite contributions.

\paragraph{}
This result has been confirmed by another type of computation done independently in \cite{michael-induced} which was based on heat kernel technics and performed within the matrix base for the Moyal space where it was assumed that the gauge transformation of the scalar field was $\phi^g = g^\dag \star_\theta \phi \star_\theta g$.

\paragraph{}
Putting all together, \eqref{eq:moyal_ind_onepoint} exhibits all the natural $\mathcal{A}_\mu$-depending terms that should presumably be involved in a suitable gauge invariant action depending on $\mathcal{A}_\mu$. Note that these are ``induced'' by summing over field configurations, hence the terminology ``induced gauge theory''.

\paragraph{}
The net conclusion/conjecture of \cite{wall-wulk1}, \cite{michael-induced} is that the following action
\begin{align}
    S
    &= \int \dd^4 x \left( 
    \frac{1}{4} F_{\mu\nu} \star_\theta F^{\mu\nu}
    + \frac{\Omega^2}{4} \{{\cal{A}}_\mu, {\cal{A}}_\nu\}^2_\theta
    + \kappa {\cal{A}}_\mu \star_\theta {\cal{A}}^\mu \right), \label{eq:moyal_ind_matrix_act}
\end{align}
where the mass dimension for the parameters $\Omega$ and $\kappa$ are respectively $0$ and $2$, may be a candidate for a renormalizable gauge theory on Moyal space. The action \eqref{eq:moyal_ind_matrix_act} is obviously gauge invariant and is formallly a (quartic) polynomial in $\mathcal{A}_\mu$, since in particular one has $F_{\mu\nu} = [\mathcal{A}_\mu, \mathcal{A}_\nu]_\theta - i \Theta^{-1}_{\mu\nu}$. Using this latter relation, one can rewrite \eqref{eq:moyal_ind_matrix_act} as
\begin{equation}
    S
    = \int \dd^4 x\ \left(
    - \frac{1}{4} [\mathcal{A}_\mu, \mathcal{A}_\nu]_\theta^2
    + \frac{\Omega^2}{4} \{{\cal{A}}_\mu, {\cal{A}}_\nu\}^2_\theta
    + \kappa {\cal{A}}_\mu \star_\theta {\cal{A}}^\mu \right)
    \label{eq:moyal_ind_matrix_act2}
\end{equation}
where there is an apparent exchange symmetry
\begin{equation}
    i[\mathcal{A}_\mu, \mathcal{A}_\nu]_\theta
    \rightleftharpoons \{{\cal{A}}_\mu, {\cal{A}}_\nu\}_\theta. 
    \label{eq:moyal_ind_matact_symm}  
\end{equation}
Contrary to \eqref{eq:moyal_phi4_act}, \eqref{eq:moyal_ind_matrix_act} is not covariant under the Langmann-Szabo duality. Recall that the Langmann-Szabo duality can be viewed as the symmetry exchange
\begin{equation}
    i [\xi_\mu, \cdot]_\theta
    \rightleftharpoons \{\xi_\mu, \cdot\}_\theta
    \label{eq:moyal_ind_lz}
\end{equation}
of the kinetic part of the renormalizable noncommutative $\phi^4$ model with a harmonic term. Indeed, the latter can be rewritten as 
\begin{equation}
    S_{\mathrm{kin}}(\phi)
    = \int \dd^4 x\ \frac{1}{2} (i [\xi_\mu, \phi]_\theta)^2 + \frac{\Omega^2}{2}\{\xi_\mu, \phi\}_\theta^2,
\end{equation}
resulting from the mere use of algebraic properties of the Moyal algebra, where covariance under the above exchange symmetry is manifest. Finally, the action \eqref{eq:moyal_ind_matrix_act} is invariant under the group of global symmetry of the Moyal space $G = SO(4) \cap Sp(4)$ with action on $A_\mu$ given by $(\mathcal{A}^\Lambda)(x)^\mu = \Lambda^{\mu\nu} \mathcal{A}_\nu (\Lambda^{-1}x)$ for any $\Lambda\in G$.

\paragraph{}
From a comparison of \eqref{eq:moyal_ind_matact_symm} with \eqref{eq:moyal_ind_lz}, it has been tempting to conjecture that \eqref{eq:moyal_ind_matact_symm} may be the gauge theory counterpart of the Langmann-Szabo symmetry.

Besides, by expressing \eqref{eq:moyal_ind_matrix_act} in terms of the gauge potential $A_\mu$, one realizes that the kinetic part for $A_\mu$ involves the operator 
\begin{equation}
    - \partial^2 + \Omega^2 \tilde{x}^2 + 2 \kappa
\end{equation}
whose inverse is nothing but the Mehler kernel. The latter plays an essential role to neutralize the UV/IR mixing in the noncommutative $\phi^4$ model with a harmonic term thanks to its IR damping properties reviewed in the section \ref{subsubsec:moyal_harmonic_term}. Notice, by the way, that a mass term for $A_\mu$ can be present at the tree level in the action without breaking the gauge invariance.

\paragraph{}
These observations have favored the conjecture that the gauge invariant action \eqref{eq:moyal_ind_matrix_act} may be a candidate for a renormalizable gauge theory on Moyal space. To date, this conjecture however has not been proven. It turns out that the action has a complicated vacuum structure which may drastically complicate the achievement of such a proof. The very involved vacuum structure can be already realized from \eqref{eq:moyal_ind_matrix_act} expressed back in terms of the $A_\mu$ field variable. A simple algebraic computation shows that $A_\mu = 0$ is no longer solution of the equation of motion, hence no longer a vacuum.

\paragraph{}
To end up with this section, one can notice that \eqref{eq:moyal_ind_matrix_act} is formally similar to a class of matrix models related in particular to the IIB matrix models or the IKKT model, arising in the literature of string theory \cite{matrix1}. We will come back to this aspect in a while.

\subsubsection{Vacuum configurations.}
\label{subsubsec:moyal_vacuum_conf}
\paragraph{}
In this section, $\mathcal{A}^0_\mu$ denotes generically the field configuration corresponding to a vacuum. To characterize the vacuum configurations \cite{wall-wulk2}, it is convenient to start from the action $S(\mathcal{A})$ \eqref{eq:moyal_ind_matrix_act} and first look for the solutions of the equation of motion. Furthermore, since one will have to expand \eqref{eq:moyal_ind_matrix_act} around a vacuum it is natural to require that 
\begin{equation}
   (\mathcal{A}^{0})^\Lambda(x)_\mu
   = \mathcal{A}_\mu(x),
   \label{eq:moyal_videsym}
\end{equation}
\textit{i.e.}\ the vacuum $\mathcal{A}^0_\mu$ is invariant under the global symmetry group of the action which in $d$-dimensions is $G_d=SO(d)\cap Sp(d)$. This ensures that the expanded action $S(\mathcal{A}^0_\mu + \delta\mathcal{A}_\mu)$ is still invariant under $G_d$ \cite{wall-wulk2}.

From \eqref{eq:moyal_videsym} and the group isomorphisms $G_d \simeq U(\frac{d}{2})$, it follows from the theory of invariants \cite{weylinvariant} that the general expression for the symmetric vacuum $\mathcal{A}^{0}_\mu$ is
\begin{equation}
   \mathcal{A}^{0}_\mu
   = \phi_1 \big(x^2\big) x_\mu + \phi_2 \big(x^2\big) \tilde{x}_\mu
   \label{eq:moyal_vac_conf}
\end{equation}
where $\phi_1$ and $\phi_2$ are two $G_d$-invariant scalar fields which must be combined with the equation of motion for \eqref{eq:moyal_ind_matrix_act}. This latter is given by
\begin{equation}
\begin{aligned}
   0
   = &- 2 (1 - \Omega^2) \mathcal{A}^\nu \star_\theta \mathcal{A}_\mu \star_\theta \mathcal{A}_\nu
   + (1 + \Omega^2) \mathcal{A}_\mu \star_\theta \mathcal{A}^\nu \star_\theta \mathcal{A}_\nu \\
   &+ (1 + \Omega^2) \mathcal{A}^\nu \star_\theta \mathcal{A}_\nu \star_\theta \mathcal{A}_\mu
   + 2 \kappa \mathcal{A}_\mu.
   \label{eq:moyal_eom}
\end{aligned}
\end{equation}
This is a complicated integro-differential equation. Remark that it supports the trivial solution $\mathcal{A}^0_\mu = 0$, which however is not of a great interest since expanding the action around it leads to a non-dynamical (matrix) model.

\paragraph{}
The use of the matrix base, described in the subsection \ref{subsubsec:matrix_base}, fortunately permits one to determine the non-trivial solutions of \eqref{eq:moyal_eom}, as carried out in \cite{wall-wulk2}. A similar method has been used to deal with the scalr theory case \cite{tanas-wal}. It is instructive to sketch the way the equation of motion is solved in the 2-dimensional case, which corresponds to one symplectic pair for $\Theta^{\mu\nu}$.\\
Indeed, first express the equation of motion in terms of a pair of complex covariant coordinates
\begin{align}
    Z(x)
    &= \frac{\mathcal{A}_1(x) + \mathcal{A}_2(x)}{\sqrt{2}}, & 
    Z^\dag(x)
    &= \frac{\mathcal{A}_1(x) - \mathcal{A}_2(x)}{\sqrt{2}}.
\end{align}
Then, ones uses the expression of these covariant coordinates in the matrix base, namely 
\begin{equation}
    Z(x)
    = \sum_{m, n = 0}^\infty Z_{mn} f_{mn}(x)
\end{equation}
(similar for $Z^\dag$) to compute the coordinates (matrix coefficients) of the symmetric vacuum solution \eqref{eq:moyal_vac_conf} in the matrix base. Those are given by
\begin{equation}
    Z_{mn}
    = \frac{1}{2\pi\theta} \int \dd^2x\ Z(x) f_{mn}(x).
\end{equation}
After some algebraic manipulations using polar coordinates, one arrives at
\begin{equation}
    Z_{mn}
    = - i a_m \delta_{m+1, n},
    \label{eq:moyal_grandZ}
\end{equation}
with
\begin{equation}
    a_m
    = \frac{(-1)^{m+1}}{4}c{\sqrt{\frac{m+1}{\theta}}
    \int \dd z\ L^{-1}_{m+1}(z) e^{-\frac{z}{2}}}
    \left( \phi_2 \left(\frac{z \theta}{2}\right) + i \frac{\theta}{2} \phi_1 \left(\frac{z\theta}{2}\right) \right) , 
\end{equation}
where $z = \frac{2 x^2}{\theta}$, $L^n_m$ denotes a generalized Laguerre polynomials. The previous equation combined with the equation of motion, setting $|a_m|^2 = u_{m+1}$ and disregarding the trivial solution yields
\begin{align}
    (3 \Omega^2 - 1) (u_m + u_{m+2}) + 2 (1 + \Omega^2) u_{m+1} + \kappa
    &= 0, &
    m &\in \mathbb{N}, &
    u_0
    &= 0,&
    u_m\ge0
    \label{eq:moyal_norm_seq}
\end{align}
Equation \eqref{eq:moyal_norm_seq} determines the symmetric solutions of the equation of motion which can be classified according to the range of values for $\Omega$ (and $\kappa$) \cite{wall-wulk2}. These however are extrema of the action but still not the possible minima.

\paragraph{}
For a detailed discussion on plausible conditions for these extrema to be actually a minimum, see in \cite{wall-wulk2}. So far, there exists no general property, or theorem, ensuring that a symmetric configuration \eqref{eq:moyal_vac_conf} solving \eqref{eq:moyal_norm_seq} is a minimum of the action so that an explicit check must be done in each case.

This can be rather easily done whenever $\Omega^2 = \frac{1}{3}$ and $\kappa < 0$ which corresponds to 
\begin{equation}
  u_m
  = \sqrt{\frac{-3\kappa}{4}}.
  \label{eq:moyal_norm_min}  
\end{equation}
The corresponding minimum can be written as a function of the coordinates $x^\mu$. It takes the rather complicated expression \cite{wall-wulk2}
\begin{align}
    \mathcal{A}^0_\mu(x)
    &= \sqrt{- 3 \kappa \theta} \left( \frac{e^{\frac{z}{2}} }{{\sqrt{z}}} \int_0^\infty \dd t\ e^{-t}{\sqrt{t}} J_1 \left(2 \sqrt{t z} \right) \sum_{m = 0}^\infty \frac{(-1)^m t^m}{m! \sqrt{m + 1}} \right) \tilde{x}_\mu, &
    z
    &= \frac{2 x^2}{\theta}.
    \label{eq:moyal_vac_min}
\end{align}
A similar process holds for the 4-dimensional case and leads to similar comments and observations with, in addition, many technical complications due to the increase of the Moyal space dimension.

\paragraph{}
The above discussion exhibits clearly the complexity of the vacuum structure of the so-called induced gauge theory. This is the main technical difficulty preventing so far the study of its perturbative renormalizability.

One can notice that the use of the matrix base leads, after expansion around a chosen symmetric vacuum, to a matrix model with, roughly speaking, infinite dimensional matrices, \textit{i.e.}\ operators. Therefore, the use of the matrix base maps in some sense these noncommutative gauge models to matrix models.

\subsubsection{Quantum instability of the vacuum?}
\label{subsubsec:moyal_vac_stability}
\paragraph{}
A first exploration of quantum properties of the matrix model related to a 2-dimensional version of the induced gauge theory, expanded around the particular symmetric vacuum defined by \eqref{eq:moyal_norm_min}, \eqref{eq:moyal_vac_min} has been carried out in \cite{MVW13} (see also \cite{gerwal1}). One of the result is the appearance of a non vanishing 1-point (tadpole) function indicating a quantum vacuum instability. Regardless its gauge theoretic origin, this model is essentially a noncommutative field theory with a particular bounded Jacobi operator as kinetic operator and polynomial interactions after suitable gauge fixing. 

\paragraph{}
The 2-dimensional version of \eqref{eq:moyal_ind_matrix_act} is
\begin{equation}
    S_\Omega[\mathcal{A}]
    = \tr \Big( (1 + \Omega^2) {\cal{A} \cal{A}^\dag \cal{A} \cal{A}^\dag}
    + (3 \Omega^2 - 1) {\cal{A} \cal{A} \cal{A}^\dag \cal{A}^\dag}
    + 2 \kappa {\cal{A} \cal{A}^\dag} \Big)
    \label{eq:moyal_class1},
\end{equation}
where ${\cal{A}} = \frac{\mathcal{A}_1 + i \mathcal{A}_2}{\sqrt{2}}$, ${\cal{A}}^\dag = \frac{{\cal{A}}_1 - i{\cal{A}}_2}{\sqrt{2}}$ and the product is the matrix product, \textit{i.e.}\ the action is implicitly written in the matrix base. Notice in particular that the trace in \eqref{eq:moyal_class1} is the matrix base translation of the Lebesgue integral $\int \dd^2x$. Notice also that the functional action \eqref{eq:moyal_class1} has some similarities with the 6-vertex model. However, the actual analysis and the actual ''identity card'' of the model obviously depends on the choice of a particular vacuum around which the classical theory is expanded.

\paragraph{}
Let us summarize the main steps of the derivation of the gauge fixed action performed in \cite{MVW13}. We set formally ${\cal{A}} = Z + \phi$, ${\cal{A}}^\dag = Z^\dag + \phi^\dag$ into \eqref{eq:moyal_class1}, where $Z$ is a symmetric vacuum configuration, to be chosen in a while, and $\phi$ can be interpreted as a fluctuation around $Z$. One easily obtains an action functional $S[\phi, \phi^\dag]$ invariant under a background transformation related to a nilpotent BRST-like operation, $\delta_Z$, with structure equations:
\begin{align}
    \delta_Z \phi 
    &= - i [Z + \phi, C], &
    \delta_Z \phi^\dag 
    &= - i [Z^\dag + \phi^\dag, C], &
    \delta_Z Z
    &=0, &
    \delta_Z C
    &= i C C.
    \label{eq:moyal_brst_like}
\end{align}
supplemented by
\begin{equation}
    \delta_Z{\bar{C}}=b,\  \delta_Zb=0.
\end{equation}
Here, $C$, ${\bar{C}}$ and $b$ are respectively the ghost, the antighost and the St\"uckelberg field with ghost number equal to $+1$, $-1$ and $0$. As usual, $\delta_Z$ acts as a graded derivation with grading equal to the sum of the form degree and ghost number (modulo 2) and $\delta_Z^2 = 0$. This background symmetry can be fixed by supplementing $S[\phi, \phi^\dag]$ with the gauge-fixing action
\begin{equation}
    S_{\mathrm{GF}}
    = \delta_Z \tr \Big( \bar{C} (\phi - \phi^\dag) \Big)
    = \tr \Big( b (\phi - \phi^\dag) + i \bar{C} [Z - Z^\dag + \phi - \phi^\dag, C] \Big)
    \label{eq:moyal_gauge_fix_phi},
\end{equation}
where $\delta_Z{\bar{C}} = b$, $\delta_Z b = 0$. Upon integrating the $b$ field, the ghost fields decouple.

Introducing now the matrix base, \textit{i.e.}\ setting 
\begin{equation}
    \phi
    = \sum_{m,n} \phi_{mn} f_{mn}(x),   
\end{equation}
the remaining (non-ghost) part of the gauge fixed action reduces to a functional of $\phi$ only given by
\begin{equation}
    S[\Omega; \phi]
    = \sum_{m,n,k,l\in\mathbb{N}} \phi_{mn} \phi_{kl} G_{mn;kl} + S_{int}
    \label{eq:moyal_matrix_act}
\end{equation}
where $S_{int}$ involves the cubic and quartic interactions whose actual form is not of interest here and the kinetic operator takes the complicated expression
\begin{equation}
\begin{aligned}
    G_{mn;kl} 
    =\;& (1 + 5 \Omega^2) \delta_{ml} \delta_{nk} (a_n a_{n+1} + a_n a_{n-1}) \\
    &
    \begin{aligned}
        - (3 \Omega^2 - 1)
        (&\delta_{ml} \delta_{n+1, k-1} a_n a_{n+1} 
        + \delta_{ml} \delta_{n-1, k+1} a_n a_{n-1} \\
        &- 2 \delta_{m, l+1} \delta_{k+1, n} a_n a_l)
    \end{aligned} \\
    &- (1 + \Omega^2)(\delta_{k, n+1} \delta_{m, l+1} a_n a_l + \delta_{n, k+1} \delta_{l, m+1} a_n a_l) + 2 \kappa \delta_{ml} \delta_{nk}
    \label{eq:moyal_kinetic_op}
\end{aligned}
\end{equation}
with the $a_m$'s given by any of the sequences defined in \eqref{eq:moyal_grandZ}-\eqref{eq:moyal_norm_seq}.

\paragraph{}
At this stage, two instructive remarks are in order.

First, note that for $Z = 0$, which corresponds to the trivial vacuum, one can verify that the classical action before the integration of the $b$ field reduces to 
\begin{equation}
    S[0; \phi]
    = \tr \Big( 2 \kappa \phi^\dag \phi + (1 + \Omega^2) \phi \phi^\dag \phi \phi^\dag + (3 \Omega^2 - 1) \phi \phi^\dag \phi^\dag \phi \Big)
\end{equation}
which describes the action of a local matrix model, the so-called 6-vertex model \cite{ginzparg}, solved in \cite{kostov}.

Next, it should be clear from \eqref{eq:moyal_kinetic_op} that the actual expression of the kinetic operator, hence the dynamics, is fixed by the choice of the vacuum.

\paragraph{}
Choosing from now one the symmetric vacuum defined by \eqref{eq:moyal_norm_min}, the kinetic operator \eqref{eq:moyal_kinetic_op} simplifies into
\begin{equation}
    G^{}_{mn; kl}
    = (- \kappa) \big(2 \delta_{ml} \delta_{nk} - \delta_{k, n+1} \delta_{m, l+1} - \delta_{n, k+1} \delta_{l, m+1} \big)
    \label{eq:moyal_kop_symmvac},
\end{equation}
and fulfills the following conservation law for indices 
\begin{equation}
    G^{}_{mn;kl} \ne 0
    \iff m + n = k + l
    \label{eq:moyal_kop_cons}, 
\end{equation}
which is similar to the one obtained from the propagator in the matrix base of the noncommutative $\phi^4$ model with a harmonic term. It implies that $G^{}_{mn;kl}$ depends only on two indices, namely one gets
\begin{equation}
    G_{ml}
    = (-\kappa) (2 \delta_{ml} - \delta_{m, l+1} - \delta_{l, m+1} ).\label{cestjacobi}
\end{equation}
From this, one realizes that $G_{ml}$ is an infinite real symmetric tridiagonal matrix which can thus be related to a Jacobi operator. The inverse of this operator, denoted by $P_{mn;kl}$, verifies
\begin{align}
    \sum_{k,l} G_{mn; kl} P_{lk; sr}
    &= \delta_{mr} \delta_{ns}, &
    \sum_{k,l} P_{nm; lk} G_{kl; rs}
    &= \delta_{mr} \delta_{ns}.
\end{align}
It can be computed by noticing that the kinetic operator \eqref{cestjacobi} defines a bounded Jacobi operator and further using the Favard theorem \cite{gerwal1}, which is a corollary of the spectral theorem, see \cite{gerwal1}. The result is
\begin{align}
   P_{m n; k l} 
   &= \delta_{m+n; k+l} P_{ml}, &
   P_{ml}
   &= \frac{1}{\pi\mu^2} \int_{-1}^1 \dd x\ {\sqrt{\frac{1+x}{1-x}}} U_m(x) U_l(x),
   \label{invers-G}
\end{align}
where $\mu^2 = -\kappa$ and $U_n(x)$ denotes a Chebyshev polynomials of second kind given by
\begin{align}
    U_n(t)
    &:= (n + 1) \ \tensor[_2]{F}{_1}\!\! \left(-n, n+2; \frac{3}{2}; \frac{1-t}{2} \right), &
    n \in \mathbb{N},
    \label{eq:cheb_poly_def}
\end{align}
with $\tensor[_2]{F}{_1}$ being a hypergeometric function.

\paragraph{}
By using the cubic vertex expressed in the matrix base together with the propagator \eqref{invers-G}, a standard albeit cumbersome calculation leads \cite{MVW13} to a non-vanishing 1-point function whose diverging part is given by
\begin{equation}
    \Gamma^1[\phi]
    = \sigma \sum_k (2 P_{k k} - P_{k, k+1}) + \text{finite}
\end{equation}
where $\sigma$ is some unessential constant. A non-zero 1-point (tadpole) function signals that the vacuum considered here is not stable against quantum fluctuations. This unpleasant feature complicates the actual situation of the matrix model description of the gauge theory on the 2-dimensional Moyal space.

\paragraph{}
One could however notice that the present choice of the symmetric vacuum \eqref{eq:moyal_norm_min} leads to a propagator $P_{ml}$ which does not decay at large separation $|m-l|$, indicating a lack of UV decay for the propagator. It turns out that demanding the additional constraint of sufficient UV decay singles out one specific family of vacua among those symmetric vacua classified
in \cite{wall-wulk2}. But this latter does not correspond to some elements of the algebra modeling the Moyal space mostly used in the literature so that some extensions of it (if possible at all) would be needed. Besides, this family of vacua would lead to prohibitively complicated computations.

\paragraph{}
Needless to say, the situation is even more technically complicated in the case of 4-dimensional Moyal space. At the present time, it is not known if or how one can escape a non-vanishing tadpole within the present matrix description.

\newpage
\section{Gauge theories on deformations of \tops{$\mathbb{R}^3$}{R\^3}.}
\label{sec:R3L}

\subsection{From star-products to the matrix base for \tops{$\mathbb{R}^3$}{R\^3}.} \label{subsec:R3L_star_prod}

Deformations of $\mathbb{R}^3$, which we will all denote generically as $\mathbb{R}^3_\lambda$, are quantum spaces whose coordinate algebra obeys the commutation relations of a $\mathfrak{su}(2)$ Lie algebra,
\begin{equation}
    [x^j, x^k]
    = i \tensor{\epsilon}{^{jk}_l} x^l
    \label{eq:r3l_lie_alg}.
\end{equation}
where $\varepsilon$ is the totally antisymmetric Levi-Civita tensor. We will proceed in a way rather similar to the one used in the above sections, except that we will concentrate mainly on the matrix base formulation of $\mathbb{R}^3_\lambda$.\\
There are two reasons for this. First, expressing a noncommutative field theory on $\mathbb{R}^3_\lambda$ within the matrix base leads to easy computations of the quantum (radiative) corrections, while the use of a star-product would result in intractable computations. Next, the use of the matrix base makes apparent the natural decomposition of this quantum space into a direct sum of fuzzy spheres reflecting the fact that the relevant group algebra is the one of the compact Lie group $SU(2)$, for which the Peter-Weyl theorem holds.

\paragraph{}
First, notice that different classes of star-products related to deformations of $\mathbb{R}^3$ have already been presented in the literature. These first appeared within quantization of Poisson manifolds \cite{flato1}, a notion which is relevant here.  Indeed, a common feature of almost all the works dealing with deformations on $\mathbb{R}^3$ in the physics literature is the identification of $\mathbb{R}^3$ with $\mathfrak{su}(2)'$, the dual of $\mathfrak{su}(2)$, which therefore induces the appearance of a Poisson-structure related to $\mathfrak{su}(2)$, hence of a Poisson manifold structure. Recall that the dual of a finite dimensional Lie algebra carries a canonical Poisson manifold structure defined by the Lie bracket.

Related star-products involved in field theory developments are mainly obtained through reduction of star-products related to deformations of $\mathbb{R}^4$. One popular family, obtained from a reduction of the standard (4-dimensional) Moyal product, has been derived in \cite{Pepe-vit}. Another star-product obtained in \cite{algerio} is derived from the (4-dimensional) Wick-Voros product \cite{wick-vor} and exploits an adaptation of the Hopf fibration of the 3-sphere. In this construction, the Hopf map controls the embedding of $\mathbb{R}^3_\lambda$, the deformed $\mathbb{R}^3$, in the deformed $\mathbb{R}^4$, which determines the star-product for $\mathbb{R}^3_\lambda$. This can be viewed as a noncommutative analog of the Kustaanheimo-Stiefel map \cite{kustaan} used in \cite{vitale-one, gervitwal} to obtain a reasonable notion of integration \textit{i.e.}\  a trace functional which, in turn, corresponds to the natural trace for $\mathbb{R}^3_\lambda$. This trace will be introduced below.

\paragraph{}
Note that the star-products obtained above\footnote{These products are in general {\textit{not}} equivalent, hence are corresponding to different deformations of $\mathbb{R}^3$. The equivalence is defined roughly by the existence of a map $T$ such that $T(f) \star T(g) = T(f\ \tilde{\star}\ g)$ in obvious notations. In this respect, Moyal and Wick-Voros products are equivalent.} are not closed w.r.t.\ the related trace functionals. Denoting generically these star-products and traces by $\star_\lambda$ and $\widetilde{\int}$, one would have 
\begin{equation}
    \widetilde{\int}(f \star_\lambda g)(x)
    \ne \widetilde{\int} f(x) g(x),
\end{equation}
contrary to the Moyal product as given by the 2nd relation in \eqref{eq:moyal_prod_closed}. This has motivated the search of other star-products for deformations of $\mathbb{R}^3$ closed w.r.t.\ a trace functional. Keeping in mind that the existence of a star-product closed w.r.t.\ the corresponding trace in the framework of Poisson manifolds is guaranteed by \cite{felder}. Such a star-product closed w.r.t.\ the usual Lebesgue integral $\int d^3x$ has been constructed in \cite{kupri-vit}, modified in \cite{jur-poul-wal17}, \cite{poul-wal17bis} to respect the structures of $*$-algebra which is needed for physical applications \cite{closedproductphy}. This product coincides with the Kontsevich product \cite{Kons1} related to the Poisson manifold dual to $\mathfrak{su}(2)$. It can be expressed as \cite{jur-poul-wal17}
\begin{equation}
    (f \star_\lambda g)(x)
    = \int \frac{\dd^3p}{(2\pi)^3} \frac{\dd^3q}{(2\pi)^3} \mathcal{F}f(p) \mathcal{F}{g}(q) \mathcal{W}(p,q) e^{iB(p,q)x},
    \label{eq:r3l_kont_prod}
\end{equation}
where 
\begin{equation}
    \mathcal{W}(p,q)
    = \frac{|B(p,q)|}{\lambda |p| |q|} \frac{\sin \big(\lambda |p| \big) \sin \big(\lambda |q| \big)}{\sin \big(\lambda |B(p,q)| \big)},
\end{equation}
and $|B(p,q)|$ is an infinite expansion defined by the $SU(2)$ Baker-Campbell-Hausdorff formula
\begin{equation}
    e^{p_j x^j} e^{q_j x^j}
    = e^{B(p,q)_j x^j},
\end{equation}
with
\begin{align}
    B(p, q) &= - B(-q, -p), &
    B(p, 0) &= p.
\end{align}
To obtain \eqref{eq:r3l_kont_prod}, the quantization map $\mathcal{K}$ was used and corresponds to
\begin{align}
    \mathcal{K}
    &= W \circ j^{\frac{1}{2}}(\Delta), &
    j^{\frac{1}{2}}(\Delta)
    &= \frac{\sinh \big(\lambda \sqrt{\Delta} \big)}{\lambda \sqrt{\Delta}},
    \label{eq:r3l_duflo_map}
\end{align}
where $\Delta$ denotes the 3-dimensional Laplacian and $ j^{\frac{1}{2}}(\Delta)$ is the Harish-Chandra map \cite{harish}.

In other words, the product $\star_\lambda$ \eqref{eq:r3l_kont_prod} is obtained from the modified quantization map \eqref{eq:r3l_duflo_map}, which is related to the Duflo quantization map \cite{duflo}. It verifies
\begin{equation}
    \int d^3x\ (f \star_\lambda g)(x)
    = \int d^3x\ f(x) g(x).
    \label{eq:r3l_duflo_map_clos}
\end{equation}
Despite the above rather simple expression for $\star_\lambda$ and in particular the closeness property \eqref{eq:r3l_duflo_map_clos}, the use of this product together with the others mentioned above in noncommutative field theory leads to technical limitations in the computation. These can be avoided by the use of the matrix base to which we turn from now on.

\paragraph{}
The starting point of the construction of the matrix base for $\mathbb{R}^3_\lambda$ follows closely the route of the section \ref{subsec:star_prod_conv_alg}. In particular, recall \cite{harmo-analys} that for a locally compact group $G$ (assumed here to be unimodular) and for any unitary representation $\pi:G\to\mathcal{B}(V_\pi)$ and any $f\in L^1(G)$, there exists a unique operator 
\begin{equation}
    \pi(f): L^1(G) \to \mathcal{B}(V_\pi)
\end{equation}
such that
\begin{equation}
    \langle \pi(f)v, w \rangle_{V_\pi}
    = \int_G \dd x\ f(x) \langle \pi(x)v, w \rangle_{V_\pi}
    \label{eq:reps_conv_alg_Hilbert}
\end{equation}
where $v,w$ are any vectors in $V_\pi$, $\dd x$ is the Haar measure and $\langle \cdot, \cdot \rangle_{V_\pi}$ denotes a Hilbert product in $V_\pi$. Note that \eqref{eq:reps_conv_alg_Hilbert} is nothing but the counterpart of \eqref{eq:reps_conv_alg}.

Now comes a simplification stemming from the fact that the relevant Lie group here is $SU(2)$, in view of \eqref{eq:r3l_lie_alg}, which is compact so that the Peter-Weyl theorem holds true. Indeed, one has the $*$-algebra isomorphism{\footnote{It holds true actually for any {\textit{compact}} group}}
\begin{equation}
    L^2(SU(2))
    \simeq \bigoplus_{m \in \mathbb{N}} \End(V_m)
    \simeq \bigoplus_{m \in \mathbb{N}} \mathbb{M}_m(\mathbb{C}),
    \label{eq:r3l_PW_isomorp}
\end{equation}
where $\End(V_m)$ is the algebra of endomorphisms of the representation space $V_m$ of the unitary irreducible representation of dimension $m$ of $SU(2)$, denoted hereafter by $\pi_m$. $\End(V_m)$ is itself isomorphic to the $m$-dimensional complex-valued matrix algebra $\mathbb{M}_m(\mathbb{C})$. Notice that the product of the algebra structure of the leftmost  (resp.\ rightmost) side corresponds to the convolution product on $SU(2)$ (resp.\ the matrix product).

\paragraph{}
From this isomorphism, it follows that any function $f \in L^2(SU(2))$ is mapped into an infinite sum of operators, namely
\begin{equation}
    f
    \mapsto \sum_{m \in \mathbb{N}} \pi_m({f})
    = \sum_{m \in \mathbb{N}} \int_{SU(2)} \dd x\ f(x) \pi_m(x)
    \label{eq:r3l_Bochner_decomp},
\end{equation}
where the right-hand side of \eqref{eq:r3l_Bochner_decomp} is an operator acting on $\bigoplus_m V_m$. Besides, upon using Schur orthogonality and polarization, one infers that 
\begin{equation}
    \int_{SU(2)} \dd x\ \langle \pi_m(x)u_1, v_1 \rangle \langle \pi_m(x)u_2, v_2 \rangle
    = \frac{1}{m} \langle u_1, u_2 \rangle {\overline{\langle v_1, v_2\rangle}},
\end{equation}
where $\langle \cdot, \cdot \rangle$ denotes the Hilbert product on $V_m$. Then, further introducing the set of functions called the $j, k$-matrix coefficients,
\begin{align}
    \pi^m_{jk}(x)
    &:= \langle \pi_m(x) e^m_j, e^m_k \rangle \in L^2(SU(2)), & 
    1 \leqslant j, k \leqslant m
    \label{eq:r3l_mat_coeff},
\end{align}
for any $x \in SU(2)$ where $\{e_j^m\}$, $1 \leqslant j \leqslant m$ is an orthonormal base of $V_m$. One verifies that
\begin{equation}
    \langle \pi^{m_1}_{jk}, \pi^{m_2}_{ln} \rangle_{L^2}
    = \frac{1}{m_1} \delta_{m_1 m_2} \delta_{jl} \delta_{kn},
    \label{eq:r3l_orth_mat}
\end{equation}
where $\langle f, g \rangle_{L^2} = \int_{SU(2)} \dd x\ \bar{f}(x) g(x)$. Then, it follows that $\{\sqrt{m} \pi^m_{jk},\ 1\leqslant j, k\leqslant m\}$ is an orthonormal basis of $\mathcal{M}_m$, the subspace of $L^2(SU(2))$ spanned by the $\pi^m_{jk}$'s, and that $\mathcal{M}_{m_1}$ and $\mathcal{M}_{m_2}$ are orthogonal for $m_1 \ne m_2$.

Hence, one can write for any function $f \in L^2(SU(2))$, $f(x) = \sum_m \sum_{j, k} f^m_{jk} \pi^m_{jk}(x)$. On the other hand, setting $\hat{f} = \sum_{m \in \mathbb{N}} \int_{SU(2)} \dd x\ f(x) \pi_m(x)$, one has $\langle \hat{f} e_j^m, e_k^m\rangle = \langle f, \pi^{m}_{jk}\rangle_{L^2} = f^m_{jk}$. One can then express $\hat{f}$ \eqref{eq:r3l_Bochner_decomp} as
\begin{equation}
    \hat{f}
    = \sum_m \sum_{j, k} f^m_{jk} e^m_j \otimes e^{m*}_k,
\end{equation}
where $\{e^{m*}_j\}$ is the dual base, \textit{i.e.}\  $ e^{m*}_j (e^m_k) = \delta_{jk}$.

\paragraph{}
We are done! In order to make contact with the notations of the physics literature, simply set
\begin{align}
    m &= 2j+1, &
    j \in \frac{\mathbb{N}}{2},
\end{align}
and
\begin{align}
    e^m_k \otimes e^{m*}_l 
    \to v^j_{kl}
    &:= |jk\rangle \langle jl|, &
    j &\in \frac{\mathbb{N}}{2}, &
    -j \leqslant k, l \leqslant j.
    \label{eq:r3l_phys_not}
\end{align}

\paragraph{}
Let us summarize the above discussion. One easily realizes that for any $j \in \frac{\mathbb{N}}{2}$, $\{v^j_{mn}\}$, $-j\leqslant m,n\leqslant j$ is simply the canonical basis for $\mathbb{M}_{2j+1}(\mathbb{C})$ so that $\mathbb{R}^3_\lambda$ inherits the following orthogonal basis
\begin{align}
    \{v^j_{mn}\}, &&
    -j \leqslant m, n \leqslant j, &&
    j \in \frac{\mathbb{N}}{2}
    \label{eq:r3l_canon_base},
\end{align}
where the $v^j_{mn}$'s are defined in \eqref{eq:r3l_phys_not} and satisfy
\begin{align}
    (v^j_{mn})^\dag = v^j_{nm}, &&
    v^{j_1}_{mn} v^{j_2}_{qp} = \delta^{j_1j_2} \delta_{nq} \ v^{j_1}_{mp}, &&
    -j_1 \leqslant m, n \leqslant j_1, &&
    -j_2 \leqslant p, q \leqslant j_2
    \label{eq:r3l_fusion_rule}
\end{align}
for any $j, j_1, j_2 \in \frac{\mathbb{N}}{2}$. From the above discussion, it follows that $\mathbb{R}^3_\lambda$ can be conveniently represented as
\begin{align}
    \mathbb{R}^3_\lambda 
    = (\bigoplus_{j \in \frac{\mathbb{N}}{2}} \mathbb{M}_{2j+1}(\mathbb{C}), \cdot, \dag )
    \label{eq:r3l_conv_alg},
\end{align}
where $\cdot$ (resp.\ $\dag$) is the usual matrix product (resp.\ hermitian conjugation), consistently with the Peter-Weyl decomposition \eqref{eq:r3l_PW_isomorp}. In other words, keeping in mind that $\mathbb{M}_{2j+1}(\mathbb{C})$ is the algebra of the fuzzy sphere of radius $j$, $\mathbb{R}^3_\lambda$ is nothing but an infinite sum of fuzzy spheres. Accordingly, dropping from now on the hat superscript on any element $\mathbb{R}^3_\lambda$, any $f\in\mathbb{R}^3_\lambda$ has a blockwise expansion of the form
\begin{equation}
    f
    = \sum_{j\in\frac{\mathbb{N}}{2}} \ \sum_{-j\leqslant m, n \leqslant j} f^j_{mn} v^j_{mn},
    \label{eq:r3l_nat_fourier}
\end{equation}
with $f^j_{mn} \in \mathbb{C}$.

\paragraph{}
The orthogonality of \eqref{eq:r3l_canon_base} holds w.r.t.\ the Hilbert product given by $\langle f,g \rangle := \tr(f^\dag g)$ with the trace given by 
\begin{equation}
    \tr(f \star_\lambda g) 
    := 8 \pi \lambda^3 \sum_{j\in\frac{\mathbb{N}}{2}} (2j+1) \tr_j(F^j G^j)
    \label{eq:r3l_trace},
\end{equation}
for any $f,g\in\mathbb{R}^3_\lambda$, where $F^j, G^j \in \mathbb{M}_{2j+1}(\mathbb{C})$ are the matrix arising in the blockwise expansion of $f, g\in\mathbb{R}^3_\lambda$ respectively in the base \eqref{eq:r3l_canon_base} in view of \eqref{eq:r3l_nat_fourier}, explicitly $(F^j)_{mn} = f^j_{mn}$ and similarly for $g$, and $\tr_j$ is the usual trace on $\mathbb{M}_{2j+1}(\mathbb{C})$. The overall factor $8 \pi \lambda^3$ in \eqref{eq:r3l_trace} has been installed for further convenience. Note that $\lambda$ has mass dimension $[\lambda]=-1$.

\paragraph{}
Let us collect usefull algebraic properties. $\mathbb{R}^3_\lambda$ is a unital algebra with unit 
\begin{align}
    \bbone
    = \sum_{j \in \frac{\mathbb{N}}{2}} P_j, &&
    P_j
    = \sum_{m=-j}^j v^j_{mm}
    \label{eq:r3l_unit}
\end{align}
in which $P_j$ is the orthogonal projector on $\mathbb{M}_{2j+1}(\mathbb{C})$, for any $j \in \frac{\mathbb{\mathbb{N}}}{2}$.

Besides, the following relations hold:
\begin{align}
    \tr_j(v^j_{mn}) = \delta_{mn},  &&
    \langle v^{j_1}_{mn}, v^{j_2}_{pq} \rangle
    = 8 \pi \lambda^3(2j_1+1) \delta^{j_1j_2} \delta_{mp} \delta_{nq}.
    \label{eq:r3l_orhto_base_prop}
\end{align}
The center of $\mathbb{R}^3_\lambda$, denoted by $\mathcal{Z}(\mathbb{R}^3_\lambda)$, is the set of the elements of $\mathbb{R}^3_\lambda$ having the following expansion
\begin{equation}
    z 
    = \sum_{j\in\frac{\mathbb{N}}{2}} f(j) P_j
    \label{eq:r3l_center}.
\end{equation}
Using the fact that $f(j)$ can be expressed as a serie depending on $j$, one easily realizes that the center $\mathcal{Z}(\mathbb{R}^3_\lambda)$ is  generated by
\begin{equation}
    x^0
    = \lambda \sum_{j\in\frac{\mathbb{N}}{2}} j P_j
    \label{eq:r3l_center_gen}
\end{equation}
which is often referred in the physics literature as the radius operator. Note that the overall factor $\lambda$ in \eqref{eq:r3l_center_gen} yields $[x^0]=-1$.

\paragraph{}
Finally, from a standard computation, one infers that
\begin{subequations}
\begin{align}
    x^1
    &= \frac{\lambda}{2} \sum_{j, m} \left( \sqrt{(j+m) (j-m+1)} v^j_{m, m-1} + \sqrt{(j-m) (j+m+1)} v^j_{m, m+1} \right),
    \label{eq:r3l_gen_1} \\
    x^2
    &= \frac{\lambda}{2 i} \sum_{j, m} \left( \sqrt{(j+m) (j-m+1)} v^j_{m, m-1} - \sqrt{(j-m) (j+m+1)} v^j_{m,m+1} \right),
    \label{eq:r3l_gen_2} \\
    x^3
    &= \lambda \sum_{j,m} m v^j_{mm},
    \label{eq:r3l_gen_3}
\end{align}
    \label{eq:r3l_gen}
\end{subequations}
from which, by using \eqref{eq:r3l_fusion_rule} one obtains
\begin{align}
    [x^j, x^k]
    &= i \lambda \tensor{\varepsilon}{^{jk}_l} x^l, &
    [x^j, x^0]
    &= 0, &
    j, k, l = 1, 2, 3,
    \label{eq:r3l_com_rel}
\end{align}
and
\begin{equation}
    (x^0)^2 + \lambda x^0
    = \sum_{j = 1}^3 (x^j)^2,
    \label{eq:r3l_com_rel_rad}
\end{equation}
reproducing the popular ``defining relations'' of $\mathbb{R}^3_\lambda$ used in the physics literature, in which the right-hand side of \eqref{eq:r3l_com_rel_rad} is the Casimir operator for $\mathfrak{su}(2)$.

\subsection{Noncommutative differential calculus on \tops{$\mathbb{R}^3_\lambda$}{R\^3\_lambda}.}
\label{subsec:r3l_diff_calc}
\paragraph{}
A convenient differential calculus can be obtained by a mere adaptation of the general framework introduced in the section \ref{sec:nc_diff_calc}, starting from the following Lie algebra of real\footnote{
Recall that a real derivation $D$ verifies $(D(f))^\dag = D(f^\dag)$.
} 
inner derivations of $\mathbb{R}^3_\lambda$ \cite{gervitwal}
\begin{align}
    \mathcal{G} := \{D_\alpha:= i [\theta_\alpha, \cdot]_\lambda\}, &&
    \theta_\alpha := \frac{x_\alpha}{\lambda^2}, &&
    \alpha = 1,2,3,
    \label{eq:r3l_der}
\end{align}
satisfying
\begin{align}
    [D_\alpha, D_\beta] = -\frac{1}{\lambda} \tensor{\epsilon}{_{\alpha\beta}^\gamma} D_\gamma, &&
    \alpha, \beta, \gamma = 1,2,3.
    \label{eq:r3l_der_alg}
\end{align}
According to the section \ref{sec:nc_diff_calc}, the corresponding $\mathbb{N}$-graded differential algebra is given by $(\Omega_\mathcal{G}^\bullet = \oplus_{n\in\mathbb{N}} \Omega^n_\mathcal{G}, \dd, \wedge)$, in which as usual $\Omega^n_\mathcal{G}$ is the space of $n$-linear antisymmetric maps\footnote{
Linearity is w.r.t.\ the elements of $\mathcal{Z}(\mathbb{R}^3_\lambda)$.
}
$\omega: \Omega_\mathcal{G}^n \to \mathbb{R}^3_\lambda$ ($\Omega^0_\mathcal{G} = \mathbb{R}^3_\lambda$), the differential $\dd: \Omega^n_\mathcal{G} \to \Omega^{n+1}_\mathcal{G}$ is defined by 
\begin{equation}
\begin{aligned}
    \dd \omega(X_1, \dots, X_{p+1}) 
    &= \sum_{k=1}^{p+1} (-1)^{k+1} X_k \big( \omega(X_1, \dots, \vee_k, \dots, X_{p+1}) \big) \\
    &+ \sum_{1 \leqslant k < l \leqslant p+1} (-1)^{k+l} \omega([X_k,X_l], \dots, \vee_k, \dots, \vee_l, \dots, X_{p+1}),
\end{aligned}
    \label{eq:r3l_differential}
\end{equation}
for any $\omega\in\Omega^p_\mathcal{G}$, $\rho\in\Omega^q_\mathcal{G}$ and the product $\wedge$ on $\Omega_\mathcal{G}^\bullet$ is
\begin{equation}
\begin{aligned}
    \omega\ \wedge&\ \rho (X_1, \dots, X_{p+q}) \\
    &= \frac{1}{p!q!} \sum_{\sigma\in\mathfrak{S}_{p+q}} \vert\sigma\vert \omega(X_{\sigma(1)}, \dots, X_{\sigma(p)}) \star_\lambda \rho(X_{\sigma(p+1)}, \dots, X_{\sigma(p+q)}),
\end{aligned}
    \label{eq:r3l_calc_diff_prod}
\end{equation}
for any $\omega\in\Omega^p_\mathcal{G}$ and $\rho\in\Omega^q_\mathcal{G}$, where $X_i\in{\cal{G}}$, $\vert\sigma\vert$ is the signature of the permutation $\sigma\in\mathfrak{S}_{p+q}$.

\paragraph{}
Assuming now that the module $\modM$ is one copy of the algebra, \textit{i.e.}\  $\modM = \mathbb{R}^3_\lambda$, and that the hermitian structure is
\begin{equation}
    h(m_1, m_2) = m_1^\dag \star_\lambda m_2, 
\end{equation}
one infers from the section \ref{sec:nc_diff_calc}, equations \eqref{eq:nc_conn_def} and \eqref{eq:nc_conn_form_def}, that the hermitian connection is entirely determined by the 1-form 
\begin{equation}
    A := \nabla(\bbone) \in \Omega^1_{\mathcal{G}}
\end{equation}
while the 2-form curvature is 
\begin{equation}
    F = \dd A + A^2 \in \Omega^2_{\mathcal{G}}, 
\end{equation}
with
\begin{align}
    \nabla_{D_\mu}(f) 
    &:= \nabla_\mu(f) 
    = D_\mu (f) + A_\mu \star_\lambda f, & 
    (A_\mu :=\nabla_\mu(\bbone))
    \label{eq:r3l_conn}
\end{align}
and 
\begin{equation}
    A_\mu^\dag = - A_\mu, 
\end{equation}
for any $f\in\mathbb{R}^3_\lambda$. The curvature reads
\begin{align}
\begin{aligned}
    F(D_\mu, D_\nu) 
    := F_{\mu\nu} 
    &= [\nabla_\mu,\nabla_\nu] - \nabla_{[D_\mu,D_\nu]} & \\
    &= D_\mu (A_\nu) - D_\nu (A_\mu) + [A_\mu,A_\nu]_\lambda + \frac{1}{\lambda} \tensor{\epsilon}{_{\mu\nu}^\gamma} A_\gamma, &
    \mu, \nu = 1,2,3.
\end{aligned}
    \label{eq:r3l_curv}
\end{align}
Using \eqref{eq:nc_conn_gauge}, \eqref{eq:nc_curv_gauge}, one finds that the gauge transformations are
\begin{align}
    A_\mu^g = g^\dag \star_\lambda A_\mu \star_\lambda g + g^\dag \star_\lambda D_\mu (g), &&
    F^g_{\mu\nu} = g^\dag \star_\lambda F_{\mu\nu} \star_\lambda g.
\end{align}
with
\begin{align}
    g := \phi(\bbone), &&
    \phi \in \Aut(\modM, h),
\end{align}%
\vspace{\dimexpr-\abovedisplayskip-\belowdisplayskip-\baselineskip+\jot}%
\begin{align}
    g^\dag \star_\lambda g = g \star_\lambda g^\dag = \bbone 
\end{align}
since the gauge group is the group of the unitary elements of the module $\mathcal{U}(\mathbb{R}^3_\lambda)$ (see section \ref{sec:nc_diff_calc}).

\paragraph{}
As indicated at the end of the section \ref{subsubsec:nc_mod_is_alg} in the case of algebras of {\textit{inner}} derivations, the present framework gives room to the existence of an interesting gauge invariant connection that we now describe.

Indeed, the space $\Omega^1_{\mathcal{G}}$ involves a particular element $\Theta \in \Omega^1_{\mathcal{G}}$ given by
\begin{equation}
    \Theta(D_\mu)
    := \Theta_\mu 
    = -i \theta_\mu
    \label{eq:r3l_inv_conn}
\end{equation}
where $\theta_\mu$ is given by \eqref{eq:r3l_der}. By further making use of \eqref{eq:r3l_differential} and \eqref{eq:r3l_calc_diff_prod}, a simple computation yields
\begin{equation}
    F^{\mathrm{inv}}
    := \dd \Theta + \Theta \star_\lambda \Theta
    = 0
    \label{eq:r3l_inv_curv}.
\end{equation}
Then, introduce the Cartan operations\footnote{Recall that Cartan operations are essential ingredients in the control of the different formulations of equivariant cohomologies relevant to topological field theories \cite{stor-wal} and in the algebraic theory of anomalies \cite{RS}.} describing the action of the Lie algebra of derivations $\mathcal{G}$ on the differential algebra  $\Omega^\bullet_{\mathcal{G}}$, respectively the inner product $\iota$ and Lie derivative $\mathscr{L}$ given by
\begin{align}
    \iota_X 
    &: \Omega^p_{\mathcal{G}} \to \Omega^{p-1}_{\mathcal{G}}, & 
    \iota_X (\omega)(X_1, \dots, X_{p-1})
    &= \omega(X, X_1, \dots, X_{p-1})
    \label{eq:r3l_int_der}, \\
    \mathscr{L}_X
    &: \Omega^p_{\mathcal{G}} \to \Omega^{p}_{\mathcal{G}}, & 
    \mathscr{L}_X
    &= \iota_X \circ \dd + \dd \circ \iota_X,
    \label{eq:r3l_lie_der}
\end{align}
for any $X, X_k \in \mathcal{G}$, $(k = 1, 2, \dots, p-1)$. Recall that $\iota_X$ and $\mathscr{L}_X$ act as derivations with respective degree $-1$ and $0$.

One has the following relation which holds on $\Omega^0_{\mathcal{G}}$,
\begin{equation}
    \dd = - [\Theta, \cdot]_\lambda,
\end{equation}
from which a simple calculation yields
\begin{equation}
    \mathscr{L}_X(\Theta)
    = \iota_X (\dd \Theta) + \dd(\iota_X \Theta)
    = \iota_X (\dd \Theta + \Theta \star_\lambda \Theta)
    = \iota_X (F^{\mathrm{inv}})
    = 0
    \label{invar},
\end{equation}
where the last relation stems from \eqref{eq:r3l_inv_curv}. Hence, $\Theta$ is an invariant form\footnote{
It is however not horizontal since $\iota_X(\Theta) \ne 0$.
}
in the language of Cartan operations.

Besides, a natural noncommutative analog of a symplectic form is defined as a real closed 2-form $\omega$ such that for any $f$ in the algebra, there exists a derivation $\ham(f)$, the analog of Hamiltonian vector field, such that $\omega(X, \ham(f)) = X(f)$ for any derivation $X$ \cite{mdv99}. Then, it can be easily realized that the 2-form
\begin{equation}
    \omega
    := \dd \Theta \in \Omega^2_\mathcal{G}
\end{equation}
can be interpreted as a symplectic form on the algebra $\mathbb{R}^3_\lambda$, where the  Hamiltonian vector field is 
\begin{equation}
    \ham(f) = \mathrm{Ad}_{i f}
\end{equation}
for any $f \in \mathbb{R}^3_\theta$, while the related (real) Poisson bracket is
\begin{equation}
    \left\{f, g\right\}
    := \omega \big(\ham(f), \ham(g)\big)
    = - i \left[f, g\right]
    \label{eq:r3l_poisson_bra}.
\end{equation}

Finally, $\Theta$ defines the 1-form connection for the canonical gauge invariant connection. Setting $\nabla^{\mathrm{inv}}(\bbone) = \Theta$, it is straightforward to verify that it is given by
\begin{equation}
    \nabla^{\mathrm{inv}}(f)
    := \dd f + \Theta \star_\lambda f
    = f \star_\lambda \Theta,
    \label{eq:r3l_inv_form_conn}.
\end{equation}
for any $ f\in\mathbb{R}^3_\lambda$ and one easily verifies that 
\begin{equation}
    \Theta_\mu^g
    = \Theta_\mu
    \label{eq:r3l_inv_theta},
\end{equation}
where $\Theta_\mu$ is defined by \eqref{eq:r3l_inv_conn}. The corresponding curvature is simply given by \eqref{eq:r3l_inv_curv}, so that $\Theta$ defines a flat connection.

\paragraph{}
According to the above discussion and the general definition given in section \ref{sec:nc_diff_calc}, the natural gauge covariant tensor 1-form related to the covariant coordinates is then given by
\begin{equation}
    (\nabla - \nabla^{\mathrm{inv}})(f)
    = (A - \Theta)(f)
    := \mathcal{A}(f)
    \label{eq:r3l_tens_form}
\end{equation}
for any $f\in\mathbb{R}^3_\lambda$ and transform as
\begin{equation}
    \mathcal{A}^g
    = g^\dag \star_\lambda \mathcal{A} \star_\lambda g
    \label{eq:r3l_tens_form_gauge}.
\end{equation}
for any $g \in \mathcal{U}(\mathbb{R}^3_\lambda)$. The covariant coordinates are then
\begin{align}
    \mathcal{A}_\mu 
    = \nabla_\mu - \nabla^{\mathrm{inv}}_\mu 
    = A_\mu + i \theta_\mu, &&
    \mu = 1, 2, 3, 
    \label{eq:r3l_covar_coord}
\end{align}
with 
\begin{align}
    \mathcal{A}_\mu^\dag = - \mathcal{A}_\mu, &&
    \mu = 1, 2, 3.
\end{align}
In terms of the variables $\mathcal{A}_\mu$, the curvature can be written as
\begin{equation} 
    F_{\mu\nu} 
    = [\mathcal{A}_\mu, \mathcal{A}_\nu]_\lambda + \frac{1}{\lambda} \tensor{\epsilon}{_{\mu\nu}^\gamma} \mathcal{A}_\gamma,
    \label{eq:r3l_curv_covar_coord}
\end{equation}
with 
\begin{equation}
    F_{\mu\nu}^g 
    = g^\dag \star_\lambda F_{\mu\nu} \star_\lambda g,
\end{equation}
for any $g \in \mathcal{U}(\mathbb{R}^3_\lambda)$.

\subsection{Gauge theories on \tops{$\mathbb{R}^3_\lambda$}{R\^3\_lambda}.}
\label{subsec:r3l_gauge_th}
\paragraph{}
As it is the case for gauge theories on Moyal spaces, the existence of a covariant coordinate \eqref{eq:r3l_tens_form} yields two different types of gauge theories on $\mathbb{R}^3_\lambda$, depending on the choice of the field variable $\mathcal{A}_\mu$ or $A_\mu$.

\paragraph{}
In the first case, gauge invariant action functionals $S(\mathcal{A})$ merely correspond to matrix models, as we will recall in a while. This observation has been already used within theories on Moyal spaces $\mathbb{R}^4_\theta$ and $\mathbb{R}^2_\theta$ (see section \ref{subsec:moyal_induced_gauge}). Note that the matrix formulation of gauge theories supports actually two interpretations. To illustrate this point, assume that $S(\mathcal{A})$ admits $\mathcal{A}=0$ as vacuum configuration. Then, one could interpret $S(\mathcal{A})$ either as the action describing the dynamics of fluctuations of $\mathcal{A}$ around the vacuum $0$, or alternatively as describing the dynamics of fluctuations of $A$ around the vacuum $\Theta$, in view of \eqref{eq:r3l_tens_form}.

\paragraph{}
The second possibility for which $A_\mu$ is assumed to be the field variable gives rise in some cases to vacuum instabilities. This has been shown in \cite{gervitwal} from the computation of the one-loop tadpole within a gauge theory built in a way to be a formal noncommutative analog of a usual Yang-Mills theory. In this theory, the tadpole does not vanish at the first order in perturbations. Notice that a similar behavior has been exhibited within gauge theories on $\mathbb{R}^2_\theta$ \cite{MVW13}. The actual origin and status of this behavior is not clarified so far.

\paragraph{}
Accordingly, we will first review this latter possibility in section \ref{subsubsec:r3l_yang_mills}. Then, we will mainly focus on the first route based on $\mathcal{A}_\mu$ as the field variable. We will exhibit in particular a class of gauge invariant models \cite{wal-16}, \cite{wal-161} which, after a suitable gauge-fixing, are finite at all orders in perturbation. Finitude is obtained from the combination of the Peter-Weyl decomposition of $\mathbb{R}^3_\lambda$ \eqref{eq:r3l_conv_alg}, the existence of a bound on the absolute value for the propagator together with the addition of a gauge invariant harmonic term in the action. Recall that the addition of a harmonic term was essential to neutralize the UV/IR mixing within scalar field theories on the 4-dimensional Moyal space $\mathbb{R}^4_\theta$ \cite{Grosse:2003aj-pc}
\footnote{
The introduction of the harmonic term increases slightly the decay of the propagator which is sufficient to get rid of the UV/IR mixing effects on the perturbative renormalization.
}
while its occurrence in gauge theories on Moyal spaces was forbidden by gauge invariance. But gauge invariant harmonic terms can be added to any gauge invariant action in the case of $\mathbb{R}^3_\lambda$, as we will show at the beginning of subsection \ref{subsubsec:r3l_matrix_mod}.

\paragraph{}
Before going further, one remark is in order.

We note that the differential calculus presented in subsection \ref{subsec:r3l_diff_calc} which underlies the gauge theory models on $\mathbb{R}^3_\lambda$ in the literature is based on derivations defined in \eqref{eq:r3l_der} which are related to the rotations. This can be easily realized by combining the relation
\begin{equation}
    [x^\mu, f]_\lambda
    = - i \lambda \tensor{\epsilon}{^\mu_\nu^\rho} x^\nu \partial_\rho f
\end{equation}
where $[\cdot, \cdot]_\lambda$ denotes the commutator w.r.t.\ the relevant star-product discussed in section \ref{subsec:R3L_star_prod}, with the derivations defined in \eqref{eq:r3l_der}. This implies that no radial dependence can be accounted for within this framework leading to gauge models with unusual commutative limit, as those reviewed below in sections \ref{subsubsec:r3l_yang_mills} and \ref{subsubsec:r3l_matrix_mod}. Roughly speaking, the resulting models can be viewed as kind of ``rotor models''. For a discussion focused on the derivations  \eqref{eq:r3l_der}, see \cite{PV-corf}. These gauge models however exhibit instructive features at the quantum level, from the viewpoint of noncommutative field theories. Besides, some of them have interesting relationship with models arising either in the string theory context, such as the Alekseev-Recknagel-Schomerus action, which is a particular combination of the Yang-Mills action and the Chern-Simons term on the fuzzy sphere, or with group field theory description of spin foam models for quantum gravity (see subsection \ref{MGMback}).

\subsubsection{Gauge theories on \tops{$\mathbb{R}^3_\lambda$}{R\^3\_lambda} as Yang-Mills type models.}
\label{subsubsec:r3l_yang_mills}
\paragraph{}
One possible way is to try to build on $\mathbb{R}^3_\lambda$ a reasonable noncommutative analog of a Yang-Mills theory, hence massless. For this purpose, it is natural to look for an action which satisfies the following requirements:
\begin{enumerate}[label=(\roman*)]
    \item \label{it:gauge_inv}
    The gauge invariant action functionals are at most quartic in $A_\mu$;
    \item \label{it:no_lin}
    No linear terms in $A_\mu$ are involved;
    \item \label{it:pos_op}
    The kinetic operator is a positive operator (upon gauge fixing).
\end{enumerate}
The requirement \ref{it:gauge_inv} is obviously satisfied by any action of the form
\begin{equation}
    S_{\mathrm{cl}}(A_\mu)
    = \int_{\mathbb{R}^3_\lambda} \mathcal{P}(\mathcal{A}_\mu)   
\end{equation}
where $\mathcal{P}(\mathcal{A}_\mu)$ is a polynomial depending on $\mathcal{A}_\mu$ \eqref{eq:r3l_covar_coord} and involving at most quartic monomials, \textit{i.e.}\ $\sim \mathcal{A}_{\mu_1} \star_\lambda \mathcal{A}_{\mu_2} \star_\lambda \mathcal{A}_{\mu_3} \star_\lambda \mathcal{A}_{\mu_4}$ (a so called quartic ``star-polynomial'' in the physics literature) and $\int_{\mathbb{R}^3_\lambda}$ is nothing but the trace defined by \eqref{eq:r3l_trace}. Recall that the gauge invariance simply stems from the gauge transformation of $\mathcal{A}_\mu$ \eqref{eq:r3l_tens_form_gauge} combined with the cyclicity of the trace.

\paragraph{}
The requirement \ref{it:no_lin} ensures that the classical equations of motion support the solution $A^0_\mu = 0$, which otherwise would imply a non trivial vacuum for the classical action. This was assumed \cite{gervitwal} as an attempt to escape the difficulties linked to a complicated vacuum structure as the one encountered in the gauge theories on Moyal spaces, as presented in section \ref{subsubsec:moyal_vac_stability}.

The construction of a candidate action is a simple matter of algebra. Starting from the following most general expression
\begin{align}
\begin{aligned}
    S_{\mathrm{cl}}(A_\mu)
    = \frac{1}{g^2} \int_{\mathbb{R}^3_\lambda}
    &(\alpha \mathcal{A}_\mu \star_\lambda \mathcal{A}_\nu \star_\lambda \mathcal{A}^\nu \star_\lambda \mathcal{A}^\mu 
    + \beta\mathcal{A}_\mu \star_\lambda \mathcal{A}_\nu \star_\lambda \mathcal{A}^\mu \star_\lambda \mathcal{A}^\nu \\
    &+ \zeta \epsilon^{\mu \nu \rho} \mathcal{A}_\mu \star_\lambda \mathcal{A}_\nu \star_\lambda \mathcal{A}_\rho
    + m \mathcal{A}_\mu \star_\lambda \mathcal{A}^\mu)
\end{aligned}
\end{align}
where $\alpha$, $\beta$, $\zeta$ are real parameters with respective mass dimensions $[\alpha] = [\beta] = 0$, $[\zeta] = 1$, $[m] = 2$ and the coupling constant has mass dimension $[g^2] = 1$. Expanding the fields in the matrix base, introduced in section \ref{subsec:R3L_star_prod}, in which any field has a blockwise expansion of the form \eqref{eq:r3l_nat_fourier}, namely
\begin{equation}
    \mathcal{A}_\mu
    = \sum_{j\in\frac{\mathbb{N}}{2}} \sum_{-j \leqslant m, n\leqslant j} (\mathcal{A}_\mu^j)_{mn} v^j_{mn},
    \label{eq:r3l_covar_coord_mat}
\end{equation}
one easily arrives at
\begin{equation}
\begin{aligned}
   S_{\mathrm{cl}}(A_\mu)
   = \frac{8 \pi \lambda^3}{g^2} \sum_{j \in \frac{\mathbb{N}}{2}} (j+1) \tr_j \Big(
   & \alpha \mathcal{A}^j_\mu \mathcal{A}^j_\nu (\mathcal{A}^j)^\nu (\mathcal{A}^j)^\mu
   + \beta \mathcal{A}^j_\mu \mathcal{A}^j_\nu (\mathcal{A}^j)^\mu (\mathcal{A}^j)^\nu \\
   & +\zeta \epsilon^{\mu \nu \rho} \mathcal{A}^j_\mu \mathcal{A}^j_\nu \mathcal{A}^j_\rho
   + m \mathcal{A}^j_\mu \mathcal{A}^j_\mu \Big),
\end{aligned}
\end{equation}
where, in view of \eqref{eq:r3l_covar_coord_mat}, the product becomes now a matrix product between matrices $\mathcal{A}_\mu^j\in\mathbb{M}_{2j+1}(\mathbb{C})$ with canonical trace $\tr_j$. By merely adjusting the parameters in order to fulfill the above requirements, one finally finds the following family of actions 
\begin{equation}
    S_{\mathrm{cl}}(A_\mu)
    = \frac{8 \pi \lambda^3}{g^2} \sum_{j \in \frac{\mathbb{N}}{2}}(j+1)
    \Big( (F^j_{\mu\nu})^\dag F^j_{\mu\nu}
    +\gamma \big( \epsilon^{\mu \nu \rho} \mathcal{A}^j_\mu \mathcal{A}^j_\nu \mathcal{A}^j_\rho
    + \frac{3}{2\lambda} \mathcal{A}^j_\mu (\mathcal{A}^j)^\mu \big) \Big)  \label{eq:r3l_action}
\end{equation}
where $F^j_{\mu\nu} \in \mathbb{M}_{2j+1}(\mathbb{C})$, for any $j \in \frac{\mathbb{N}}{2}$, stems from the expansion of $F_{\mu\nu}$ \eqref{eq:r3l_curv} in the matrix base defined in section \ref{subsec:R3L_star_prod} and $\gamma$ is a real parameter with mass dimension $[\gamma] = 1$. Recall that in \eqref{eq:r3l_action}, it is understood that $\mathcal{A}$ is expressed in terms of $A$ through \eqref{eq:r3l_covar_coord}.

\paragraph{}
Upon rewriting \eqref{eq:r3l_action} as
\begin{equation}
    S_{\mathrm{cl}}(A_\mu)=\sum_{j\in\frac{\mathbb{N}}{2}}S^j_{\mathrm{cl}}(A_\mu)
\end{equation}
one easily realizes that $S_{\mathrm{cl}}(A_\mu)$ is the sum of actions $S^j_{\mathrm{cl}}(A_\mu)$ describing Yang-Mills-Chern-Simons actions on fuzzy spheres \cite{madore91} $\mathbb{S}^j \simeq \mathbb{M}_{2j+1}(\mathbb{C})$. This can be traced back to the particular structure of $\mathbb{R}^3_\lambda$ which reduces to a direct sum of fuzzy spheres stemming from the Peter-Weyl decomposition \eqref{eq:r3l_PW_isomorp}. For a review on fuzzy objects, see e.g.\ \cite{bal2005}.

\paragraph{}
It appears that each $S^j_{\mathrm{cl}}(A_\mu)$ is formally similar to the action for the Alekseev-Recknagel-Schomerus gauge action on the fuzzy sphere which describes the low-energy effective action at the leading order of the string tension of a brane on the sphere $\mathbb{S}^3$ with non-vanishing Neveu-Schwarz 3-form field strength. Hence, this observation exhibits the link existing between the classical action for a noncommutative gauge theory on $\mathbb{R}^3_\lambda$ described as a ``noncommutative Yang-Mills theory'' and the above model describing a particular brane dynamics.

\paragraph{}
The investigation of the quantum properties is technically not easy. The use of the matrix base however leads in some instance to tractable computations. A convenient simplification arises whenever the theory is massless \cite{gervitwal} which occurs for a particular value of the parameters \cite{gervitwal} at which the computation of the propagator for $A_\mu$ becomes tractable in a Feynman-Landau type gauge $D_\mu (A^\mu) = 0$. It permits one to perform a first exploration of the perturbative properties of these noncommutative gauge theories. A convenient BRST gauge-fixing of \eqref{eq:r3l_action} yields the gauge-fixed action
\begin{equation}
\begin{aligned}
    S
    =&\ S_{\mathrm{cl}}(A_\mu) + s \widetilde{\tr}(\overline{C}(D_\mu A_\mu) + \xi\overline{C}b) \\
    =&\ \widetilde{\tr} \Big(A_\mu \Big[ - 2 \delta_{\mu\nu} D^2 + (2 + \frac{1}{4\xi}) D_\mu D_\nu \Big] (A_\nu)
    + 4 i D_\mu (A_\nu) [A_\mu, A_\nu] \\
    &- i \frac{4}{3\lambda} \epsilon^{\mu\nu\rho} A_\mu A_\nu A_\rho
    - 2 (A_\mu A^\mu)^2 + 2 A_\mu A_\nu A^\mu A^\nu \Big) \\
    &+ \widetilde{\tr} \big(- \overline{C} D^2 (C) + i \overline{C} D_\mu [A_\mu, C] \big),
\end{aligned}
    \label{eq:r3l_action_calc}
\end{equation}
where we have set, to simplify the notations,
\begin{equation}
    \widetilde{\tr}
    := \frac{8 \pi \lambda^3}{g^2} \sum_j(j+1) \tr_j, \label{tildelatrace}
\end{equation}
and the nilpotent BRST operation $s$ is defined by 
\begin{align}
    s A_\mu 
    = D_\mu(C) - i [A_\mu, C] 
    = [\mathcal{A}_\mu, C], &&
    s C = i C C, &&
    s \overline{C} = b,  &&
    s b = 0,
\end{align}
where $C$, $\overline{C}$, $b$ are respectively the antighost, ghost and St\"uckelberg field with respective ghost numbers $1$, $-1$ and $0$. The choice of the gauge parameter $\xi=-\frac{1}{8}$ leads to a diagonal kinetic term, a choice which will be assumed in the rest of this section \ref{subsubsec:r3l_yang_mills}.

\paragraph{} 
The propagator for $A_\mu$ can be computed by expressing, in the matrix basis of section \ref{subsec:R3L_star_prod}, the quadratic part of $S$ \eqref{eq:r3l_action_calc}, which takes the generic form
\begin{equation}
    S_2(A_\mu)
    = 2 \sum_{j_1,j_2} \sum_{m_1, \tilde m_1} \sum_{m_2, \tilde m_2} 
    (A_\mu)^{j_1}_{m_1 \tilde m_1} (\Delta)^{j_1, j_2}_{m_1 \tilde m_1; m_2 \tilde m_2} (A_\mu)^{j_2}_{m_2 \tilde m_2},
    \label{eq:r3l_quadrat_matrix}
\end{equation}
where the kinetic operator in the matrix base denoted by $(\Delta)^{j_1, j_2}_{m_1 \tilde m_1; m_2 \tilde m_2}$ has a complicated expression whose explicit form is not of interest here. It verifies however the relation $\Delta^{j_1 j_2}_{m n; k l} = \Delta^{j_1 j_2}_{k l; m n}$ together with the indices conservation law 
\begin{align}
    \Delta^{j_1 j_2}_{m n; k l} \ne 0
    \implies j_1 = j_2, &&
    m + k = n + l,
    \label{eq:r3l_consindice}
\end{align}
and can be shown to be positive so that it fulfils the requirement \ref{it:pos_op} given at the beginning of this section. The propagator $P^{j_1 j_2}_{m n; k l}$ is defined as the inverse of $\Delta^{j_1 j_2}_{m n; k l}$ by
\begin{align}
    \sum_{k, l = - j_2}^{j_2} \Delta^{j_1 j_2}_{m n; l k} P^{j_2  j_3}_{k l; r s} 
    = \delta^{j_1 j_3} \delta_{m s} \delta_{n r}, &&
    \sum_{m, n = -j_2}^{j_2} P^{j_1 j_2}_{r s; m n} \Delta^{j_2  j_3}_{n m; k l}
    = \delta^{j_1 j_3} \delta_{r l} \delta_{s k},
    \label{eq:r3l_def_prop}
\end{align}
which implies
\begin{align}
    P^{j_1 j_2}_{m n; k l} \ne 0
    \implies j_1 = j_2, &&
    m + k = n + l
    \label{eq:r3l_considice_prop}.
\end{align}
It can be expressed as \cite{gervitwal}
\begin{equation}
    P^{j_1 j_2}_{m n; p q}
    = \frac{g^2}{8 \pi \lambda} \delta^{j_1 j_2} \frac{1}{(j_1 + 1) (2j_1 + 1)}
    \sum_{l = 0}^{2 j_1} \sum_{k = - l}^l \frac{1}{l (l + 1)} (Y^{j_1 \dag}_{l  k})_{n m} (Y^{j_2}_{l k})_{q p},
    \label{eq:r3l_complex_prop}
\end{equation}
where $Y^j_{l k}$, $l \in \mathbb{N}$, $0 \leqslant l \leqslant 2j$, $- l \leqslant  k \leqslant l$, denotes the Fuzzy Spherical Harmonics which have been introduced in the context of the fuzzy spheres. For more technical details, see e.g.\ \cite{dasetal}.

Alternatively, one may introduce the Wigner 3-j symbols to express \eqref{eq:r3l_complex_prop}, namely
\begin{equation}
    P^{j j}_{m n; p q}
    = \frac{g^2}{8 \pi \lambda(j + 1)}
    \sum_{l, k}(-1)^{-(n + p)} \frac{2 l + 1}{l (l + 1)}
    \wign{j}{m}{j}{-n}{l}{k} \wign{j}{q}{j}{-p}{l}{k}.
    \label{eq:r3l_simpler_prop}
\end{equation}

One can observe that \eqref{eq:r3l_complex_prop} and \eqref{eq:r3l_simpler_prop} are singular whenever $l = 0$, corresponding to an IR singularity, as it can be expected for a massless gauge theory. Similar comments obviously apply to the ghost propagator, in view of \eqref{eq:r3l_action_calc} and $\xi = - \frac{1}{8}$.

\paragraph{}
The computation of the 1-point (tadpole) function for the $A_\mu$ at the one-loop order is cumbersome. This has been carried out in \cite{gervitwal}. It is found that the gauge and ghost contributions to the 1-point function for the component $A_3$ of the gauge potential vanish separately. This is a direct consequence of the index conservation law \eqref{eq:r3l_considice_prop} combined with algebraic properties of the Fuzzy Spherical Harmonics of the Wigner 3j symbols.

Besides, it is found that the tadpole for the $A_1$ and $A_2$ components are non vanishing. Setting $A_\pm = A_1 \pm i A_2$, the corresponding contribution to the one-loop effective action denoted by $\Gamma^1(A_\mu)$ can be cast into the form
\begin{equation}
   \Gamma^1(A_\mu)
   = \frac{4 \pi \lambda^2}{g^2}
   \sum_{j \in \frac{\mathbb{N}}{2}} \sum_{ - j \leqslant n, p \leqslant j} \!\! 
   \Pi(j, n) \big(\delta_{n + 1, p}(A_-)^j_{n p} - \delta_{n, p + 1} (A_+)^j_{n p} \big),
   \label{eq:r3l_effective_lin}
\end{equation}
with
\begin{equation}
    \Pi(j, n) 
    = 2 \sum_{m = -j}^j (j + 1) \big(F^j(m) P^{jj}_{m+1 n+1; n m} - F^j(n) P^{jj}_{m n; n m} \big)
    \label{eq:r3l_goldstone_op}
\end{equation}
and
\begin{equation}
    F^j(m)
    = [(j + m + 1) (j - m)]^{\frac{1}{2}}.
\end{equation}
This result needs to be commented which leads to some conclusions.

\paragraph{}
First, it appears that no IR singularity is present in \eqref{eq:r3l_goldstone_op}, stemming from the fact that the potentially IR singular part of \eqref{eq:r3l_goldstone_op} is given by \cite{gervitwal}
\begin{equation}
    \Pi(j, n)_{\mathrm{IR}}
    \sim \lim_{\varepsilon \to 0} \frac{1}{\varepsilon} \sum_m 
    \big( F^j(m) \delta_{nm} - F^j(n) \delta_{nm} \big)
    = 0
\end{equation}
implying that $\Pi(j, n)$ is IR finite. Besides, the UV behavior of the tadpole is controlled by its large $j$ limit. One finds \cite{gervitwal}
\begin{equation}
    \lim_{j \to \infty} \Pi(j, -j)
    \sim - \lambda \sqrt{j} \log (j).
    \label{eq:r3l_lim_eff}
\end{equation}
indicating that the 1-point function is UV diverging.

\paragraph{}
Next, the occurrence of non-vanishing tadpole indicates that the classical vacuum configuration becomes unstable under quantum fluctuations. This can be traced back to the fact that the term linear in $A_\mu$ induced by the quantum (one-loop) fluctuations in the effective action obviously prevents the classical vacuum configuration $A_\mu = 0$ to be an extremal point of the effective action. Thus, moving ahead consistently into the perturbative expansion would need to tune the vacuum at each order, \textit{i.e.}\ performing an expansion of the field variable around the right vacuum at each order of perturbation.

Note that the appearance of a non-zero tadpole at the one-loop order in other noncommutative gauge theories has been already pointed out in section \ref{sec:moyal}, in particular in a family of gauge matrix models on $\mathbb{R}^2_\theta$ or in models attempting to implement an IR damping to neutralize the UV/IR mixing as shown in section \ref{subsubsec:moyal_harmonic_term}. So far, it seems to be a specific one-loop feature of various gauge theory models on different quantum spaces. In the same way, we will again encounter a non-vanishing tadpole for gauge models on the $\kappa$-Minkowski space (see subsection \ref{tadpole}).

\paragraph{}
Let us summarize and list the different implications of this feature.

A non-vanishing tadpole is equivalent to a non-zero vacuum expectation value for the gauge potential, \textit{i.e.}\ $\langle A_\mu \rangle \ne 0$ which may signal Lorentz (rotational in the present case) symmetry breaking. Besides, the effective action is no longer BRST invariant and it seems unlikely possible to balance its gauge variation by another variation of some higher order terms which signals that the classical symmetry is broken.

As far as we know, no solid study has been achieved in a possible elimination or clever exploitation of the appearance of non-zero tadpoles, apart from observing that some show up in some gauge models.

\subsubsection{Gauge theories on \tops{$\mathbb{R}^3_\lambda$}{R\^3\_lambda} as matrix models.}
\label{subsubsec:r3l_matrix_mod}
\paragraph{}
In this section, we choose the field variable to be the covariant coordinate $\mathcal{A}_\mu$ defined in \eqref{eq:r3l_covar_coord} which leads to the matrix model formulation of gauge theories on $\mathbb{R}^3_\lambda$, as done in section \ref{subsec:moyal_induced_gauge} for $\mathbb{R}^4_\theta$.

\paragraph{}
First, observe that gauge invariant harmonic terms can be added to any gauge invariant action on $\mathbb{R}^3_\lambda$ built from the differential calculus described in section \ref{subsec:r3l_diff_calc}. Indeed, one observes that the gauge invariant object $\Theta_\mu \Theta^\mu$ verifies
\begin{equation}
    \Theta_\mu \Theta^\mu \in \mathcal{Z}(\mathbb{R}^3_\lambda).
    \label{eq:r3l_harm_center}
\end{equation} 
This can be readily obtained from \eqref{eq:r3l_com_rel}, \eqref{eq:r3l_com_rel_rad}, \eqref{eq:r3l_der} and \eqref{eq:r3l_inv_conn}, keeping in mind that the center of $\mathbb{R}^3_\lambda$, $\mathcal{Z}(\mathbb{R}^3_\lambda)$, is generated by $x^0$. One therefore concludes that any polynomial $P(\mathcal{A})$ satisfies the relation
\begin{equation}
    \tr(P(\mathcal{A}) \Theta_\mu \Theta^\mu)^g
    = \tr(P(\mathcal{A}) \Theta_\mu \Theta^\mu)
\label{eq:r3l_harm_poly}
\end{equation}
which holds since one has \eqref{eq:r3l_tens_form_gauge}. The equation \eqref{eq:r3l_harm_poly} expresses gauge invariance. To see that, simply combine the gauge invariance of $\Theta_\mu$ with \eqref{eq:r3l_tens_form_gauge} and \eqref{eq:r3l_harm_center}, and make use of the cyclicity of the trace. Setting for convenience 
\begin{equation}
    \mathcal{A}_\mu = i \Phi_\mu,
\end{equation}
and keeping in mind \eqref{eq:r3l_harm_poly} and \eqref{eq:r3l_der}, one finally concludes that gauge invariant harmonic terms 
\begin{equation}
    \sim \tr(x^2 \Phi_\mu \Phi^\mu)
\end{equation}
are allowed in any gauge invariant classical action on $\mathbb{R}^3_\lambda$.

\paragraph{}
Families of gauge invariant classical actions can be built from the trace of any gauge-covariant polynomial functional in the covariant coordinates $\mathcal{A}_\mu$, \textit{i.e.}\ 
\begin{equation}
    S_{\mathrm{inv}}(\mathcal{A}_\mu)
    = \tr \big( P(\mathcal{A}_\mu) \big).
\end{equation}

Natural requirements for the gauge invariant action are, as in section \ref{subsubsec:r3l_yang_mills}:
\begin{enumerate}[label=(\roman*)]
    \item $P(\mathcal{A}_\mu)$ is at most quartic in $\mathcal{A}_\mu$,
    \item $P(\mathcal{A}_\mu)$ does not involve linear term in $\mathcal{A}_\mu$ (not tadpole at the classical order),
    \item the kinetic operator is positive.
\end{enumerate}

Standard calculations single out \cite{wal-16} a familly of gauge invariant (positive) actions fulfilling the above requirements and parametrized as
\begin{equation}
    S_\Omega 
    = \frac{1}{g^2} \tr \Big( 
    \big( F_{\mu\nu} - \frac{i}{\lambda} \tensor{\epsilon}{_{\mu\nu}^\rho} \Phi_\rho \big)^\dag \big(F^{\mu\nu} - \frac{i}{\lambda} \tensor{\epsilon}{^{\mu\nu\rho}} \Phi_\rho \big)
    + \Omega \left\{\Phi_\mu, \Phi_\nu \right\}^2 
    + (M + \mu x^2) \Phi_\mu \Phi^\mu \Big),
    \label{eq:r3l_zeta_0}
\end{equation}
in which $\Omega \geqslant 0$, $mu > 0$, $M > 0$ and the parameters have the following mass dimensions: $[\Omega] = 0$, $[\mu] = 4$, $[M] = 2$, $[g^2] = 1$, assuming that the relevant dimension for $\mathbb{R}^3_\lambda$ is the ``engineering dimension'' $d = 3$. By using the equation of motion for \eqref{eq:r3l_zeta_0} given by
\begin{equation}
    4 (\Omega + 1) \big(\Phi_\rho \Phi_\mu \Phi^\mu + \Phi_\mu \Phi^\mu \Phi_\rho \big)
    + 8 (\Omega - 1) \Phi_\mu \Phi_\rho \Phi^\mu
    + 2 (M + \mu x^2) \Phi_\rho 
    = 0,
    \label{eq:r3l_eqn_motion}
\end{equation}
one easily realizes that $\Phi_\rho = 0$ is the absolute minimum of \eqref{eq:r3l_zeta_0}.

\paragraph{}
Combining \eqref{eq:r3l_zeta_0} with the expression for the curvature in terms of $\mathcal{A}_\mu $ \eqref{eq:r3l_curv_covar_coord}, the classical gauge invariant action can be rewritten as
\begin{equation}
    S_{\mathrm{cl}}[\Phi]
    = \frac{1}{g^2} \tr \Big( 
    [\Phi_\mu, \Phi_\nu]^2 + \Omega \{\Phi_\mu, \Phi_\nu\}^2 + (M + \mu x^2) \Phi_\mu\Phi^\mu \Big),
    \label{eq:r3l_class_act}
\end{equation}
with $\{a, b\} := a b + b a$ which, keeping in mind \eqref{eq:r3l_nat_fourier}, clearly exhibits the structure of a matrix model. Notice that the action \eqref{eq:r3l_class_act} can be expressed as
\begin{equation}
    S_{\mathrm{cl}}[\Phi] 
    = \sum_{j \in \frac{\mathbb{N}}{2}} S^{(j)} [\Phi],
    \label{eq:r3l_action_decomp}
\end{equation}
as a direct consequence of the Peter-Weyl decomposition of $\mathbb{R}^3_\lambda$ \eqref{eq:r3l_conv_alg}, which can be easily verified by a direct computation using \eqref{eq:r3l_nat_fourier} and \eqref{eq:r3l_trace}. In \eqref{eq:r3l_action_decomp}, each term $S^{(j)}[\Phi]$ simply describes a scalar action on the fuzzy sphere $\mathbb{S}^j\simeq \mathbb{M}_{2j+1}(\mathbb{C})$.

\paragraph{}
It turns out that the family of classical gauge invariant actions \eqref{eq:r3l_class_act}, \eqref{eq:r3l_action_decomp} leads to all orders finite gauge theories \cite{wal-16}, among which one is even exactly solvable \cite{wal-161}. To see that, one starts by fixing the gauge. This can be most conveniently done by using the gauge
\begin{equation} 
    \Phi_3 = \theta_3.
\end{equation}
The corresponding BRST symmetry is defined by the following nilpotent operation\footnote{
In by now standard notations, $C$ (resp.\ ${\bar{C}}$) is the ghost (resp.\ antighost) field with ghost number $+1$ (resp.\ $-1$) and $b$ is the St\"uckelberg field with ghost number $0$. Recall that the Slavnov operation $s$ acts as an antiderivation w.r.t.\ the grading defined by (the sum of) the ghost number and the degree of forms (modulo 2).
}
$s$
\begin{align}
    s\Phi_\mu = i [C, \Phi_\mu], &&
    s C = i C C, &&
    s {\bar{C}} = b, &&
    s b = 0.
    \label{eq:r3l_brs_op}
\end{align}
We then supplement \eqref{eq:r3l_class_act} with the $s$-exact gauge-fixing action given by 
\begin{equation}
    S_{\phi \pi}
    = s \tr \big({\bar{C}} (\Phi_3 - \theta_3) \big) 
    = \tr \big(b (\Phi_3 - \theta_3) - i {\bar{C}} [C, \Phi_3] \big),
    \label{eq:r3l_gauge_fix} 
\end{equation}
and further integrating over the St\"uckelberg field $b$ yields the gauge-fixed action 
\begin{equation}
    S^f_\Omega 
    = \frac{2}{g^2} \tr\big( \Phi \mathcal{Q} \Phi^\dag + \Phi^\dag \mathcal{Q}\Phi \big)
    + \frac{16}{g^2} \tr \big( (\Omega+1) \Phi \Phi^\dag \Phi \Phi^\dag + (3 \Omega - 1) \Phi \Phi \Phi^\dag \Phi^\dag \big),
    \label{eq:r3l_quasilsz}
\end{equation}
where we have defined for further convenience
\begin{align}
    \Phi := \frac{1}{2}(\Phi_1 + i \Phi_2), && 
    \Phi^\dag := \frac{1}{2}(\Phi_1 - i \Phi_2),
\end{align}
and the self-adjoint kinetic operator is given by
\begin{equation}
    \mathcal{Q}
    = M \bbone + \mu L(x^2) + 8 \Omega L(\theta_3^2) + 4 i (\Omega - 1) L(\theta_3) D_3
    \label{eq:r3l_kin_op}.
\end{equation}
in which $L(.)$ denotes the left multiplication.

\paragraph{}
One observes that the expression of the gauge fixed action \eqref{eq:r3l_quasilsz} is close to the one describing the family of complex LSZ models \cite{LSZ}, up to the kinetic operators which are somewhat different. Recall that these models are noncommutative complex scalar field theories defined on Moyal spaces, some of which can be solved exactly \cite{LSZ} thanks to the existence of a duality connecting space coordinates with momenta \cite{LZ} which occurs whenever a harmonic term is introduced. This latter can be physically interpreted as a fixed magnetic background.

Whenever $\Omega=1/3$, the quartic interaction depends only on $\Phi\Phi^\dag$ and the action is {\textit{formally}} similar to the action for an exactly solvable LSZ model, again up to differences in the respective kinetic operators. It appears that this gauge theory model is also exactly solvable as shown in \cite{wal-161}. Indeed, the corresponding partition function is expressible as a product of factors labeled by $j\in\frac{\mathbb{N}}{2}$. Each factor is nothing but a $\tau$-function for a 2-dimensional Toda hierarchy and can be actually interpreted as the partition function for the reduction of the gauge-fixed theory on the matrix algebra $\mathbb{M}_{2j+1}(\mathbb{C})$, \textit{i.e.}\ on a fuzzy sphere of radius $j$.

\paragraph{}
The salient common property of the family of gauge theory models described by \eqref{eq:r3l_quasilsz} for arbitrary $\Omega\ge0$ is that they are finite to all orders in perturbations. The lengthy proof is rather technical and is detailed in \cite{wal-16}. It is however instructive to give below sketchy elements of it in the case $\Omega=1$, the extension of the case of arbitrary $\Omega$ being an easy task. It turns out that the finitude of the gauge theories \eqref{eq:r3l_quasilsz} results from the combination of three special features:
\begin{enumerate}[label=(\roman*)]
    \item a sufficient rapid decay of the propagator at large $j$ implying that the correlations at large separation indices decay as well,
    \item the role of a UV and IR cut-off played by $j$, the radius of the fuzzy sphere components,
    \item the existence of an upper bound for the propagator that depends only on the cut-off.
\end{enumerate}

For $\Omega=1$, one uses the matrix base expansion of the fields to write, as usual in matrix models, the kinetic term as
\begin{equation}
    \frac{2}{g^2} \tr_j \big( \Phi \mathcal{Q} \phi^\dag + \Phi^\dag \mathcal{Q} \Phi \big)
    := \sum_{\mu = 1}^2 \sum_{m, n, k, l} (\Phi^\mu)^j_{mn} (\Phi^\mu)^j_{kl} (\mathcal{Q})^j_{m n;k l}
    \label{eq:r3l_qj_def}
\end{equation}
where
\begin{equation}
    \mathcal{Q}^{j_1 j_2}_{m n;k l} 
    := \frac{8 \pi \lambda^3}{g^2} w(j_1) \big( M + \mu \lambda^2 j_1 (j_1 + 1) + \frac{4}{\lambda^2} (k^2 + l^2) \big) \delta^{j_1j_2} \delta_{ml} \delta_{nk}
\end{equation}
and $w(j)=2j+1$. The propagator, obtained from $\sum_{j_2,k,l} \mathcal{Q}^{j_1j_2}_{mn;lk} P^{j_2j_3}_{kl;rs} = \delta^{j_1j_3} \delta_{ms} \delta_{nr}$, together with $\sum_{j_2,n,m} P^{j_1j_2}_{rs;mn} Q^{j_2j_3}_{nm;kl} = \delta_{j_1j_3} \delta_{rl} \delta_{sk}$ is positive and can be expressed easily as
\begin{equation}
    P^{j_1j_2}_{mn;kl} 
    = \frac{g^2}{8 \pi \lambda^3} \frac{1}{w(j_1) (M + \lambda^2 \mu  j_1 (j_1 + 1) + \frac{4}{\lambda^2}(k^2 + l^2))} \delta^{j_1 j_2} \delta_{ml} \delta_{nk},  \label{eq:r3l_prop_mat_mod}
\end{equation}
and, more importantly, satisfies the following estimate \cite{wal-16}
\begin{equation}
    0 
    \leqslant P^{j_1 j_2}_{mn; kl}
    \leqslant \frac{\Pi(M, j_1)}{w(j_1)} \delta_{j_1 j_2} \delta_{ml} \delta_{nk}, \label{eq:r3l_envelop_model}
\end{equation}
for any $j_1,j_2\in\frac{\mathbb{N}}{2}$, $- j_1 \leqslant m, n, k, l \leqslant j_1$, where
\begin{equation}
    \Pi(M,j)
    := \frac{g^2}{8 \pi \lambda^3} \frac{1}{(M + \lambda^2 \mu j(j + 1))}.
    \label{eq:r3l_prop_bound}
\end{equation}
The right-hand side of \eqref{eq:r3l_prop_bound} is nothing but the propagator of the gauge theories which would have been obtained in the gauge $\Phi_3=0$, called the ``truncated model'' in \cite{wal-16}. It turns out that these latter gauge-fixed theories are related to a particular class of noncommutative field theories on $\mathbb{R}^3_\lambda$ shown to be finite to all orders in perturbation in \cite{vit-wal-12}. Equation \eqref{eq:r3l_prop_bound} thus means that the propagator \eqref{eq:r3l_prop_mat_mod} for the gauge theories \eqref{eq:r3l_quasilsz} decays faster than the one for a finite theory, which intuitively suggests that \eqref{eq:r3l_quasilsz} are also finite. This is actually what happens.

\paragraph{}
Consider first the ``truncated model''. Within this gauge model, one first observes that the amplitude for any ribbon diagrams depends only on $j \in \frac{\mathbb{N}}{2}$ and that, at fixed $j$, any ribbon loop contributes at most to a factor $(2j + 1)^x$, $x \leqslant 2$. Recall that any ribbon in such a diagram is made of 2 lines, each line carrying 2 bounded indices $m, n = -j, \dots, j$ in view of the general structure for the propagator given by the left-hand side of \eqref{eq:r3l_prop_mat_mod}. Observe that there is a ``conservation of the indices'' along each line, as it can be seen by observing the delta function in the expression of the propagator \eqref{eq:r3l_prop_mat_mod}, each delta defining the indices affected to a line.

% Utiliser le footnote pour expliquer les diagrames a rubans, faire un diagrame pour le progateur (5.99)
% Diagramatic representation of the propagator

Now, consider the amplitude $\mathbb{A}^j(\mathcal{D})$ for a general ribbon diagram $\mathcal{D}$\footnote{A ribbon diagram built from the quartic vertices is characterized by 4 positive integers $(V,I,F,B)$, respectively the number of vertices, of internal ribbons, of faces, of boundaries. Recall that $F$ is obtained by closing the external lines of a diagram and counting the number of closed {\textit{single}} lines, while $B$ is equal to the number of closed lines with external legs. The number of ribbon loops is $\mathcal{L} = F - B$. Such a diagram can be drawn on a Riemann surface of genus $g$ with $2 - 2g = V - I + F$.}. Notice that the superscript $j$ in $\mathbb{A}^j(\mathcal{D})$ indicates that the amplitude is related to the dynamics of the $j$-th component of the fields $\Phi$ and $\Phi^\dag$, stemming from their decomposition in the matrix basis for $\mathbb{R}^3_\lambda$. Correlations between field components of different $j$'s vanish. This is almost obvious from the decompositions \eqref{eq:r3l_nat_fourier}, \eqref{eq:r3l_orhto_base_prop} and \eqref{eq:r3l_action_decomp}. Plainly, such an amplitude is finite for finite $j$, while from the above discussion and making use of \eqref{eq:r3l_envelop_model}, it is not difficult to show that 
\begin{equation}
    {\mathbb{A}}^j(\mathcal{D})
    \leqslant K \frac{w(j)^{V - I} (2j + 1)^{2(F - B)}} {(j^2 + \frac{M}{\lambda \mu^2})^I}
    \label{eq:r3l_estim_amp},
\end{equation}
where $K$ is some positive constant (recall that $w(j)=2j+1$). One therefore concludes that the right-hand side of \eqref{eq:r3l_estim_amp} is finite whenever $j\to\infty$, provided
\begin{equation}
    \omega(\mathcal{D})
    = I + 2 B + 2 (2g - 2) + V 
    \geqslant 0,
    \label{eq:r3l_power_count}
\end{equation}
which thus gives rise to a power counting formula. This is always verified for any $g \geqslant 1$, while, for $g = 0$, the only possibly troublesome case occurs when $V = 1$, which corresponds to the 2-point function. But simple topological considerations for the ribbon diagrams related to the planar and non-planar contributions to the 2-point function yields respectively $B = 2$ and $B = 1$ so that \eqref{eq:r3l_power_count} is still verified.

Finally, one can prove \cite{wal-16} that the amplitude $\mathfrak{A}^j({\mathcal{D}})$ for a ribbon diagram $\mathcal{D}$ computed from the gauge-fixed model \eqref{eq:r3l_quasilsz} when $\Omega = 1$ satisfies the estimate
\begin{equation}
    \vert \mathfrak{A}^j({\mathcal{D}}) \vert
    \leqslant \vert \mathbb{A}^j({\mathcal{D}}) \vert
    < \infty,
\end{equation}
since ${\mathbb{A}}^j(\mathcal{D}) < \infty$. This implies that $\mathfrak{A}^j({\mathcal{D}})$ is finite for any value of $j$. Hence, all the ribbon amplitudes stemming from the action \eqref{eq:r3l_quasilsz} when $\Omega = 1$ are finite, \textit{i.e.}\ \eqref{eq:r3l_quasilsz} at $\Omega = 1$ is perturbatively finite to all orders. The extension of this result for arbitrary values of $\Omega$ has been achieved in \cite{wal-16}.

\subsection{Relations with models of brane dynamics and group field theory.}\label{MGMback}
\paragraph{}
It was noticed in the area of string theory that the presence of a background $B$-field generates a noncommutative structure on the world volume of branes, as exemplified for branes on a 2-torus \cite{Douglas_1998} and further developed in \cite{Cheung_1998, Schomerus_1999, Seiberg_1999} which basically reflects itself in the description of branes dynamics thought as (effective) noncommutative field theories.
 
\paragraph{}
In this vein, the authors of \cite{Steinacker_2004} gave convincing arguments that the dynamics of open strings moving in a curved space with the metric of a 3-sphere in the presence of a non-vanishing Neveu-Schwarz $B$-field and with $D$-branes is equivalent, at the leading order in the string tension, to a gauge theory on a {\textit{fuzzy sphere}} which involves in particular a Chern-Simons term. The generic form of the corresponding action is given by (adapting the notations of \cite{Steinacker_2004})
\begin{equation}
    S^j_{\mathrm{ARS}}
    = \tr_j \left( F^j_{\mu\nu} (F^j)^{\mu\nu} \right)
    + \tr_j \left(\epsilon^{\mu\nu\rho}\mathcal{A}^j_\mu \mathcal{A}^j_\nu \mathcal{A}^j_\rho + \frac{1}{R^2} \mathcal{A}^j_\mu (\mathcal{A}^j)^\mu \right) \label{ARS-action}
\end{equation}
up to unessential terms, where the fields are all defined on a fuzzy sphere $\mathbb{S}^j\simeq\mathbb{M}_{2j+1}(\mathbb{C})$, indicated by the superscript $j$ on the fields, $\tr_j$ is defined in \eqref{tildelatrace}. The sum over $j$ is not present since only one given fuzzy sphere is relevant here. $R$ denotes the radius of the 3-sphere.

\paragraph{}
It can be easily realized that \eqref{ARS-action} belongs to the family of actions defined in \eqref{eq:r3l_action} when this latter is restricted to one component $j$, \textit{i.e.}\ {\textit{when projected on a single fuzzy sphere}} $\mathbb{S}^j$, with similar restriction applying to the gauge transformations. This observation exhibits clearly the link between the family of so-called Yang-Mills type gauge theories on $\mathbb{R}^3_\lambda$ reviewed in section \ref{subsubsec:r3l_yang_mills} and the above effective model for brane dynamics.

\paragraph{}
The classical study performed in \cite{Steinacker_2004} was supplemented by an analysis of some 1-loop perturbative properties in \cite{Castro_Villarreal_2005} (see also \cite{Steinacker_2004}) in a Feynman-type gauge, e.g.\ similar to the one used in \eqref{eq:r3l_action_calc}. This gauge was limited however to a restricted range of parameters defining the action, where it was claimed that the gauge theory suffers from UV/IR mixing. This was further conjectured to extend to any values for the parameters. Besides, a non-vanishing tadpole was also found.

So far, the link between the above gauge theory model on a fuzzy sphere describing a brane dynamics and the gauge theory of section \ref{subsubsec:r3l_yang_mills} has not been fully explored. One may conjecture that the action \eqref{eq:r3l_action} exhibits UV/IR mixing as a mere extrapolation of the claim of \cite{Castro_Villarreal_2005}. The actual relevance (if any) of \ref{eq:r3l_action} for some brane dynamics description is not known.

Besides, one may notice that the family of actions \eqref{eq:r3l_class_act} considered in section \ref{subsubsec:r3l_matrix_mod} is also related (but not similar) to \eqref{ARS-action}. This is apparent by observing that \eqref{eq:r3l_class_act} can be expressed \cite{wal-16} as an action of the form
\begin{align}
\begin{aligned}
   S_{\mathrm{cl}}(\mathcal{A})
   &= \widetilde{\tr}(F_{\mu\nu} F^{\mu\nu})
   + \widetilde{\tr} \Big(\zeta \epsilon^{\mu\nu\rho} \mathcal{A}_\mu \mathcal{A}_\nu \mathcal{A}_\rho + M \mathcal{A}_\mu \mathcal{A}^\mu + \mu x_\mu x^\mu \mathcal{A}_\nu \mathcal{A}^\nu + \Omega \{\mathcal{A}_\mu, \mathcal{A}_\nu\}^2 \Big) \\
   &= \sum_{j\in\frac{\mathbb{N}}{2}} \Big( S^j_{\mathrm{ARS}} + \tr_j \big( \mu x^2 \mathcal{A}^j_\mu (\mathcal{A}^j)^\mu + \Omega\{\mathcal{A}^j_\mu,  \mathcal{A}^j_\nu\}^2 \big) \Big),
\end{aligned}
   \label{autreformule}
\end{align}
where $M$, $\zeta$ and $\Omega$ are parameters, the $x^\mu$'s denote the coordinate functions of $\mathbb{R}^3_\lambda$ and again $\mathcal{A}^j_\mu$ denotes the component $j$ in the matrix base expansion of $\mathcal{A}_\mu$ (see subsection \ref{subsec:R3L_star_prod}). Hence, \eqref{eq:r3l_class_act} is built from the sum of the above actions $S^j_{ARS}$ augmented by a harmonic-type term similar to the harmonic terms introduced in field theories on Moyal spaces to neutralize the UV/IR mixing (see subsection \ref{subsubsec:moyal_harmonic_term}) plus the square of the anticommutator of two $\mathcal{A}_\mu$'s, as the one appearing in the ``induced'' gauge theory on $\mathbb{R}^4_\theta$ \eqref{eq:moyal_ind_matrix_act} (see subsection \ref{subsubsec:moyal_induced_general}). In view of the all order finitude of \eqref{eq:r3l_class_act}, \eqref{autreformule} reviewed in section \ref{subsubsec:r3l_matrix_mod}, it seems reasonable to conjecture that these two terms likely neutralize the UV/IR mixing in the gauge theory. Note that the presence of the term $\sim\sum_jx^2\mathcal{A}^j_\mu (\mathcal{A}^j)^\mu$ is essential to insure finitude \cite{wal-16}.

\paragraph{}
Gauge theories on $\mathbb{R}^3_\lambda$ and more generally field theories on $\mathbb{R}^3_\lambda$ exhibit interesting relationship with group field theory models. Group field theory appeared in the context of quantum gravity and can be roughly viewed as a kind of generalization of matrix models. Group field theories aim at modeling quantum gravity from a combinatoric of non-local quantum field theories on group manifolds, mostly linked to the rotation or Lorentz group. In some sense, they can be viewed as quantum field theories {\textit{of}} space-time instead {\textit{on}} space-time and stand at the crossroad of loop quantum gravity and spinfoam models. For reviews, see for instance \cite{Oriti_2011, Oriti_2013, Oriti_2009} and references therein.

\paragraph{}
Recall that the typical form of the action for a $d$-dimensional group field theory model is given by
\begin{align}
\begin{aligned}
    S_{\mathrm{GFT}}
    =&\ \frac{1}{2} \int_{G} \dd g_i\ \dd h_i\ 
    \phi(g_i) K(g_i h^{-1}_i) \phi(h_i) \\
    &+ \frac{\lambda}{(d+1)!} \int_G \dd g_{ij}\ 
    V(g_{ij} g^{-1}_{ji}) \phi(g_{1j}) \cdots \phi(g_{(d+1)j}),
\end{aligned}
    \label{eq:r3l_GFT_action}
\end{align}
where $\phi$ is a complex-valued field depending on $d$ elements of a Lie group $G$, $\phi:G^d\mapsto\mathbb{C}$, with Haar measure $\dd g$ (with the notation $\phi(g_i) := \phi(g_1, g_2, \dots, g_d)$) and $K$ and $V$ are respectively the kinetic operator and vertex function. This latter describes the interaction of $(d-1)$ simplices forming a $d$ simplex by gluing them along their $(d-2)$ faces. The field $\phi$ is (often) required to satisfy the right-invariance condition $\phi(g_i h) = \phi(g_i)$ for any $h\in G$, possibly supplemented with a reality condition.

\paragraph{}
The relationship between group field theory, noncommutative field theories on some deformations of $\mathbb{R}^3$ and matrix models has been underlined for some time. This has been first shown by considering matter coupled to 3-dimensional quantum gravity \cite{Freidel_2006a, Freidel_2006b} which, upon integrating out the gravity fluctuations, yields a noncommutative field theory on a deformation of $\mathbb{R}^3$, interpreted as the effective theory describing the dynamics of matter. The key-point underlying the construction is to interpret the group manifold as the momentum space which is then related to the configuration space by a ``noncommutative'' group Fourier transform. Of course, the noncommutativity of the group transfers to the configuration space which thus becomes noncommutative. In addition, the effective theory is found to be invariant under a deformed action of the Poincar\'e group. 

To give a flavor of the above analysis, one first observes that the quantum gravity amplitudes are given by the evaluation of the diagrams generated by a noncommutative (scalar) field theory with cubic interaction, defined on a deformed (noncommutative) $\mathbb{R}^3 \simeq \mathfrak{su}(2)$. The corresponding action is given by \cite{Freidel_2006a, Freidel_2006b}
\begin{equation}
    S
    = \frac{1}{8\pi\kappa^3} \int{\dd^3x} 
    \left( \frac{1}{2}\partial_\mu \phi \star \partial_\mu \phi
    - \frac{1}{2} \frac{\sin^2(m \kappa)}{\kappa^2} \phi \star \phi
    + \frac{\lambda}{3!} \phi \star \phi \star \phi \right),
    \label{eq:r3l_freidel_action}
\end{equation}
where $m$ stands for the mass of the field $\phi$ and $\kappa$ here is related to the Newton constant $G_N$ by $\kappa = 4 \pi G_N$ and thus has the dimension of a length. The star-product in \eqref{eq:r3l_freidel_action} can be defined through the product of the following exponential quantities
\begin{equation}
    e^{\frac{1}{2\kappa} \tr(g_1 x)} \star e^{\frac{1}{2\kappa} \tr(g_2 x)}
    = e^{\frac{1}{2\kappa} \tr(g_1 g_2 x)}
    \label{eq:r3l_star_expo}
\end{equation}
for any $g_1, g_2 \in SU(2)$ with $x = x^\mu \sigma_\mu$ where $\sigma_\mu$ are hermitian Pauli matrices. Then, defining the ``momentum''
\begin{equation}
    P_\mu = \frac{1}{2 i \kappa} \tr(g \sigma_\mu),
\end{equation}
the equation \eqref{eq:r3l_star_expo} becomes
\begin{equation}
    e^{i \vec{P_1} \vec{x}} \star e^{i \vec{P_2} \vec{x}}
    = e^{i(\vec{P_1} \oplus \vec{P_2}) \vec{x}},
\end{equation}
with the deformed composition of momenta
\begin{equation}
    \vec{P_1} \oplus \vec{P_2}
    = \vec{P_1} \sqrt{1 - \kappa^{2} P_2^2} + \vec{P_2} \sqrt{1 - \kappa^{2} P_1^2} - \kappa \vec{P_1} \wedge \vec{P_2}   
\end{equation}
where $\wedge$ denotes the 3-dimensional vector cross product. Note that this star-product is related to the noncommutative Lie algebra of coordinates $[x^\mu, x^\nu] = i \kappa \tensor{\epsilon}{^{\mu\nu}_\rho} x^\rho$ together with deformed phase space as $[x^\mu, P_\nu] = i \delta^\mu_\nu \sqrt{1 - \kappa^2 P^2} - i \kappa \tensor{\epsilon}{^\mu_\nu^\rho}P_\rho$, usually interpreted as signaling that the momentum space is curved.

\paragraph{}
The link with a group field theory formulation is obtained by defining a group Fourier transform as first proposed in \cite{Freidel_2006a, Freidel_2006b} and further investigated in \cite{Freidel_2008, Joung_2009}. From \cite{Freidel_2006a, Freidel_2006b}, it is defined as
\begin{equation}
   \phi(x)
   = \int_{SU(2)} \dd g\ \mathfrak{F}(\phi)(g) e^{\frac{\kappa}{2} \tr(g x)}
   := \int_{SU(2)} \dd g\ \varphi(g) e^{\frac{\kappa}{2} \tr(g x)}
   \label{eq:r3l_freidel_fourier}.
\end{equation}
where $\mathfrak{F}$ is the group Fourier transform. One can verify that the usual convolution product on $SU(2)$, which has its usual expression given by $(\varphi \ \hat{\circ}\ \psi)(g) = \int_{SU(2)} \dd h\ \varphi(g h^{-1})\psi(h)$
for any $\varphi, \psi$ in the convolution algebra of $SU(2)$, is dual to the star-product defined by \eqref{eq:r3l_star_expo}, since one has
\begin{equation}
    (\phi \star \psi)(x)
    = \int_{SU(2)} \dd g\ \big(\mathfrak{F}(\phi) \ \hat{\circ}\ \mathfrak{F}(\psi) \big)(g) e^{\frac{\kappa}{2} \tr(g x)}.
\end{equation}
Now, by using the group Fourier transform \eqref{eq:r3l_freidel_fourier}, the action \eqref{eq:r3l_freidel_action} can be expressed as
\begin{align}
\begin{aligned}
    S
    =&\ \frac{1}{2} \int_{SU(2)} \dd g\ \left( P^2(g) - \frac{\sin^2(\kappa  m)}{\kappa^2} \right) \varphi(g) \varphi(g^{-1}) \\
    &+ \frac{\lambda}{3!} \int_{SU(2)} \dd g_1\ \dd g_2\ \dd g_3\ \delta(g_1 g_2 g_3) \varphi(g_1) \varphi(g_2) \varphi(g_3),
\end{aligned}
    \label{eq:r3l_GFT_2d}
\end{align}
which in view of \eqref{eq:r3l_GFT_action} is the action of a (2-dimensional) group field theory. The link with the matrix model formulation goes as follows. Use the Peter-Weyl theorem to expand $\varphi(g)$ as $\varphi(g)=\sum_{j\in\frac{\mathbb{N}}{2}}(2j+1)\varphi^j_{mn}\pi^{mn}(g)$ as in section \ref{subsec:R3L_star_prod}. Combine it with \eqref{eq:r3l_GFT_2d} yields
\begin{equation}
    S
    = \frac{1}{2} \sum_{j\in\frac{\mathbb{N}}{2}} (2j+1)
    \left( \sum_{-j \leqslant m,n \leqslant j}\!\!  \varphi^j_{mn} \varphi^j_{nm}
    + \frac{\lambda}{3!} \sum_{m,n,p} \varphi^j_{mn} \varphi^j_{np} \varphi^j_{pm} \right),
    \label{eq:r3l_GFT_matrix}
\end{equation}
which describes a tower of uncoupled matrix models, each one being defined on a fuzzy sphere $\mathbb{S}^j$. This provides a description of the dynamics of the matter field after integrating out the gravitational sector, which is alternative to the description by the noncommutative field theory \eqref{eq:r3l_freidel_action}.

Note that the latter action is found to be invariant under the deformed action of
the quantum double of $SU(2)$ \cite{Freidel_2006a, Freidel_2006b} which already has some flavor of the actual $\kappa$-Poincar\'e invariance considered in the section \ref{sec:kappa}.

\paragraph{}
The initial studies summarized above have been followed by many works, see e.g.\ \cite{Oriti_2011, Oriti_2013, Oriti_2009}. While the above discussion shows that the classes of gauge theories on $\mathbb{R}^3_\lambda$ share obviously similar features with the noncommutative or matrix model representations of group field theory models, it is so far unknown if (some of) these gauge theories may represent specific types of group field theory models.

\newpage
\section{Gauge theories on \tops{$\kappa$}{kappa}-Minkowski space-time}
\label{sec:kappa}
\paragraph{}
We are interested now in gauge theories defined on the $\kappa$-Minkowski space-time $\mathcal{M}_{\kappa}$, the deformation of the Minkowski space. For a nice review of historical constructions and phenomenological interest of the $\kappa$-Minkowski space, see \cite{Lukierski_2017} and references therein.

\paragraph{}
It turns out that the $\kappa$-Minkowski space is rigidly linked to a deformation of the Poincar\'{e} algebra, called the $\kappa$-Poincar\'e algebra which can be viewed as the quantum (i.e. noncommutative) analog of its algebra of symmetries.\\
The first deformation of the Poincar\'{e} algebra was done in the pioneering work  \cite{Lukierski_1991} and is now known as the $\kappa$-Poincar\'{e} algebra $\mathcal{P}_\kappa$. The latter may be defined as a Hopf algebra generated by $(P_\mu)_{\mu = 0, \dots, d}$ ,the deformed translations, $(M_j)_{j = 1, \dots, d}$, the deformed rotations and $(N_j)_{j = 1, \dots, d}$, the deformed boosts. Note that this set of generators is sometimes referred as the classical basis. For more details on Hopf algebra see subsection \ref{subsubsec:star_prod_hopf_alg}. \\
An alternative basis, so-called Majid-Ruegg basis \cite{MR1994}, consist in trading $P_0$ for $\mathcal{E} = e^{-\frac{P_0}{\kappa}}$. The Hopf algebra structure of $\mathcal{P}_\kappa$ is as follows
\begin{subequations}
\begin{align}
		[M_j,M_k] &= i \tensor{\epsilon}{_{jk}^l} M_l, & 
		[M_j,N_k] &= i\tensor{\epsilon}{_{jk}^l} N_l, & 
		[N_j,N_k] &= -i\tensor{\epsilon}{_{jk}^l} M_l, \\
		[M_j,P_k] &= i\tensor{\epsilon}{_{jk}^l} P_l, &
		[P_j,\mathcal{E}] &= [M_j,\mathcal{E}]=0, &
		[P_j, P_k] &= 0.
		\label{eq:kM_kP_Hopf_alg_alg_tran}
\end{align}%
    \vspace{\dimexpr-\abovedisplayskip-\baselineskip+\jot}%
\begin{align}
    [N_j,\mathcal{E}] &= -\dfrac{i}{\kappa}P_j\mathcal{E}, &
    [N_j,P_k] = - \frac{i}{2} \delta_{jk} \left( \kappa(1-\mathcal{E}^2) + \frac{1}{\kappa} P_l P^l \right) + \frac{i}{\kappa} P_j P_k,
\end{align}%
    %\vspace{\dimexpr-\abovedisplayskip-\belowdisplayskip-\baselineskip+\jot}%
\begin{align}
    \Delta P_0 &= P_0 \otimes 1 + 1 \otimes P_0, &
	\Delta P_j &= P_j \otimes 1 + \mathcal{E} \otimes P_j, 
	\label{eq:kM_kP_Hopf_alg_coalg_tran} \\
	\Delta \mathcal{E} &= \mathcal{E} \otimes \mathcal{E}, &
	\Delta M_j &= M_j \otimes 1 + 1 \otimes M_j,
\end{align}%
    \vspace{\dimexpr-\abovedisplayskip-\baselineskip+\jot}%
\begin{align}
    \Delta N_j = N_j\otimes 1 + \mathcal{E}\otimes N_j - \frac{1}{\kappa} \tensor{\epsilon}{_j^{kl}} P_k \otimes M_l,
\end{align}%
    %\vspace{\dimexpr-\abovedisplayskip-\belowdisplayskip-\baselineskip+\jot}%
\begin{align}
    \varepsilon(P_0) = \varepsilon (P_j) = \varepsilon(M_j) = \varepsilon(N_j) = 0, &&
    \varepsilon(\mathcal{E})=1,
\end{align}%
    %\vspace{\dimexpr-\abovedisplayskip-\belowdisplayskip-\baselineskip+\jot}%
\begin{align}
		S(P_0)&=-P_0, &
		S(\mathcal{E}) &= \mathcal{E}^{-1}, &
		S(P_j) &= -\mathcal{E}^{-1}P_j,
\end{align}%
    \vspace{\dimexpr-\abovedisplayskip-\baselineskip+\jot}%
\begin{align}
		S(M_j) &= -M_j, &
		S(N_j) &= -\mathcal{E}^{-1}(N_j-\dfrac{1}{\kappa} \tensor{\epsilon}{_j^{kl}} P_k M_l).
\end{align}%
    %\vspace{\dimexpr-\abovedisplayskip-\baselineskip+\jot}%
    \label{eq:kM_kP_Hopf_alg}
\end{subequations}
The parameter $\kappa$ is the deformation parameter that has mass dimension $1$. It is often associated to the Planck mass.

\paragraph{}
In \cite{MR1994}, the authors exhibited a bicrossproduct structure on $\kappa$-Poincar\'{e} and used it to define $\kappa$-Minkowski via duality. Explicitly, one observes in \eqref{eq:kM_kP_Hopf_alg} that $\mathcal{P}_\kappa$ contains the deformed translations $\mathcal{T}_\kappa$, generated by the $P_\mu$'s, as a sub-Hopf algebra. Moreover, setting $P_\mu = 0$ one obtains the Hopf algebra $U\mathfrak{so}(1,d)$ so that $\mathcal{T}_\kappa$ projects onto $U\mathfrak{so}(1,d)$. This allows to built an action $\actl: \mathcal{T}_\kappa \otimes U\mathfrak{so}(1,d) \to U\mathfrak{so}(1,d)$ and a coaction $\coactr: U\mathfrak{so}(1,d) \to U\mathfrak{so}(1,d) \otimes \mathcal{T}_\kappa$, so that $\kappa$-Poincar\'{e} can be defined as the bicrossproduct $\mathcal{P}_\kappa = \mathcal{T}_\kappa \bicros U\mathfrak{so}(1,d)$.

\paragraph{}
One then defines $\kappa$-Minkowski as the dual Hopf algebra of the deformed translations, that is $\mathcal{M}_\kappa = \mathcal{T}_\kappa'$. For more details on Hopf algebra duality, see subsection \ref{subsubsec:star_prod_hopf_alg}. Note that $\mathcal{T}_\kappa$ is not cocommutative, in view of  \eqref{eq:kM_kP_Hopf_alg_coalg_tran}, due to the factor $\mathcal{E}$. This implies, by duality, that $\mathcal{M}_\kappa$ is noncommutative, and its noncommutativity involves $\mathcal{E}$. This duality allows to compute the full Hopf algebra structure of $\mathcal{M}_\kappa$ thanks to \eqref{eq:hopf_alg_dual_alg}, \eqref{eq:hopf_alg_dual_coalg} and \eqref{eq:hopf_alg_dual_ant}. One obtains that $\mathcal{M}_\kappa$ is generated by $(x^\mu)_{\mu = 0, \dots, d}$ that satisfies
\begin{subequations}
\begin{align}
    [x^0, x^j]_\kappa &= \frac{i}{\kappa} x^j, &
    [x^j, x^k]_\kappa &= 0, 
    \label{eq:kM_kM_Hopf_alg_alg}\\
    \Delta(x^\mu) &= x^\mu \otimes 1 + 1 \otimes x^\mu, &
    S(x^\mu) &= - x^\mu,
\end{align}
    \label{eq:kM_kM_Hopf_alg}
\end{subequations}
where we denoted, $[f, g]_\kappa = f \star_\kappa g - g \star_\kappa f$ and $\star_\kappa$ is the associative product of $\mathcal{M}_\kappa$.

In terms of symmetries, the dual structure allows $\mathcal{T}_\kappa$ to act on $\mathcal{M}_\kappa$ through a natural action. The bicrossproduct structure also allow to dualize the coaction $\coactr$ to action of $U\mathfrak{so}(1, d)$ on $\mathcal{M}_\kappa$. After computation, one finds that for any $f \in \mathcal{M}_\kappa$,
\begin{subequations}
\begin{align}
    (P_\mu \actl f)(x) &= -i \partial_\mu f(x), &
    (\mathcal{E} \actl f)(x) &= f(x^0 + \frac{i}{\kappa}, \vec{x}), \\
    (M_j \actl f)(x) &= \big( \tensor{\epsilon}{_{jk}^l} x^k P_l \actl f \big) (x), &&
\end{align}%
    \vspace{\dimexpr-\abovedisplayskip-\baselineskip+\jot}%
\begin{align}
    (N_j \actl f)(x)
    &= \left( \Big( \frac{1}{2}x^j \big( \kappa (1 - \mathcal{E}^2) + \frac{1}{\kappa} P_lP^l \big) + x^0 P_j - \frac{i}{\kappa} x^k P_k P_j  \Big) \actl f \right)(x).
\end{align}
    \label{eq:kM_kP_action}
\end{subequations}

\paragraph{}
Generalised $\kappa$-deformation of the Minkowski space were latter considered \cite{Lukierski_2002}. The Lie algebra part of $\kappa$-Minkowski \eqref{eq:kM_kM_Hopf_alg_alg} is replaced by
\begin{align}
    [x^\mu, x^\nu]_\kappa = \frac{i}{\kappa} (a^\mu x^\nu - x^\mu a^\nu),
    \label{eq:kM_gen_kM_alg}
\end{align}
where $a^\mu$ is a constant $4$-vector of Minkowski space. Up to a noramlization factor, one is reduced to three choice for $a$, that are time-like vector $a_\mu a^\mu = -1$, light-like vector $a_\mu a^\mu = 0$ and space-like vector $a_\mu a^\mu = +1$. One recovers \eqref{eq:kM_kM_Hopf_alg_alg} from \eqref{eq:kM_gen_kM_alg} by taking time-like $\kappa$-Minkowski space, e.g.\ $a = (1, 0, \dots ,0)$.

\paragraph{}
The fact that $\kappa$-deformation offers a phenomenological framework makes it interesting to study. The main aspect of the former lies in the fact that $\kappa$-Poincar\'{e} realises a doubly special relativity \cite{Lukierski_2003}. Doubly special relativity (see \cite{Kowalski_Glikman_2005} for a review) is a framework for possible testable effects of quantum gravity in which a minimal length is added to the axioms of special relativity, that is of order $\frac{1}{\kappa}$. This additional axiom imposes to work in a noncommutative space. Other testable frameworks arise in $\kappa$-deformation, like deformed dispersion relation or relative locality. For a recent review on quantum gravity phenomenology see \cite{Addazi_2022}.

\subsection{Gauge theories with \tops{$\kappa$}{kappa}-Lorentz invariant calculus}
\label{subsec:kM_kL_gauge_th}
\paragraph{}
In this section, we follow the works \cite{DJMTWW2004a, DJMTWW2004b, DMT2004, DMMW2004, DJM2005}. The authors considered the $\kappa$-Poincar\'{e} algebra \eqref{eq:kM_kP_Hopf_alg} but redefined the translation generators to have an undeformed algebraic sector. They also built a $\kappa$-Lorentz invariant $d+1$-dimensional differential calculus through the notion of derivative-valued forms. The integral is defined as in the commutative case, supplemented with a non-constant volume form to have the cyclicity property with the star-product \eqref{eq:kM_trace_prop}. Finally, the gauge theory is built through the Seiberg-Witten map \eqref{eq:moyal_SW_map}.

\subsubsection{Star-product and symmetries}
\label{subsubsec:kM_kL_star_prod_and_symm}
\paragraph{}
The star-product of \cite{DJMTWW2004a, DJMTWW2004b} on $\kappa$-Minkowski is obtained through the Campbell-Baker-Hausdorff-quantization method. The star-product chosen to realize the commutation relations of $\mathcal{M}_{\kappa}$ is the symmetric star-product defined by
\begin{align}
\begin{aligned}
    \left(f \star_\kappa g\right)\left(z\right)
    &= \lim\limits_{ \substack{x \to z \\ y \to z} } \exp \left(
    z^{j}\frac{\partial}{\partial x^{j}}
        \left(\frac{\frac{\partial}{\partial x^{0}} + \frac{\partial}{\partial y^{0}}}{\frac{\partial}{\partial x^{0}}}
        e^{-\frac{i}{\kappa}\frac{\partial}{\partial y^{0}}}
        \frac{1 - e^{-\frac{i}{\kappa}\frac{\partial}{\partial x^{0}}}}
        {1 - e^{-\frac{i}{\kappa}\left(\frac{\partial}{\partial x^{0}} + \frac{\partial}{\partial y^{0}}\right)}} - 1\right) \right.\\
    &\left.+ z^{j}\frac{\partial}{\partial y^{j}}
        \left(\frac{\frac{\partial}{\partial x^{0}} + \frac{\partial}{\partial y^{0}}}{\frac{\partial}{\partial y^{0}}}
        \frac{1 - e^{-\frac{i}{\kappa}\frac{\partial}{\partial y^{0}}}}
        {1 - e^{-\frac{i}{\kappa}\left(\frac{\partial}{\partial x^{0}} + \frac{\partial}{\partial y^{0}}\right)}} - 1\right)\right)
    f \left( x \right) g \left( y \right).
\end{aligned}
    \label{eq:kM_star_prod_1}
\end{align}
Up to the first order in $\frac{1}{\kappa}$, this product writes
\begin{align}
    (f \star_\kappa g) (x) 
    &= f (x) g(x) 
    + \frac{i}{2\kappa}x^{j}\Big(\partial_{0} f(x) \partial_{j} g(x) 
    - \partial_{j} f(x) \partial_{0} g(x) \Big) 
    + \mathcal{O}\left(\frac{1}{\kappa^{2}}\right),  
\end{align}
where the notation $\partial_\mu = \frac{\partial}{\partial x^\mu}$ is used.

\paragraph{}
The first symmetry considered over this space correspond to the deformed algebra of rotations. It is often called the $\kappa$-Lorentz algebra and denoted $\mathfrak{so}_\kappa(1, d)$. It corresponds to the counterpart of $U\mathfrak{so}(1, d)$ above. The following requirements are imposed on those rotations:
\begin{enumerate}[label = (\roman*)]
    \item they form a module over $\mathcal{M}_\kappa$ and are consistent with \eqref{eq:kM_kM_Hopf_alg_alg},
    \item they form a close algebra,
    \item they are deformations of the usual rotations $\mathfrak{so}(1, d)$,
    \item $[M_j, x^\mu]$ and $[N_j, x^\nu]$ are at most linear in the rotations.
\end{enumerate}
These conditions have a unique solution \cite{DMT2004}, which algebraic sector correspond to the undeformed rotations $\mathfrak{so}(1, d)$ and action on $\kappa$-Minkowski is deformed to
\begin{align}
\begin{aligned}
    M_j f(x)
    &= \tensor{\epsilon}{_{jk}^l} x^k \partial_l f(x), \\
    N_j f(x)
    &= \left( x^j\partial_0 - x^0\partial_j + x^j \partial_\mu \partial^\mu \frac{e^{\frac{i}{\kappa}\partial_0} -1}{2\partial_0} - x^\mu \partial_\mu \partial_j \frac{e^{\frac{i}{\kappa}\partial_0} -1 - \frac{i}{\kappa}\partial_0}{ \frac{i}{\kappa} \partial_0^2} \right) f(x).
\end{aligned}
    \label{eq:kM_kL_action}
\end{align}

\paragraph{}
The next symmetry considered is the deformed translation defined through the derivations over the algebra $\mathcal{M}_\kappa$. These derivations, noted $\partial^\star_\mu$, form the counterpart of $\mathcal{T}_\kappa$ above. The following requirements are imposed on those derivations:
\begin{enumerate}[label = (\roman*)]
    \item they form a module over $\mathcal{M}_\kappa$ and are consistent with \eqref{eq:kM_kM_Hopf_alg_alg},
    \item they form a commutative algebra, \textit{i.e.}\ $[\partial^\star_\mu, \partial^\star_\nu] = 0$,
    \item they are deformations of the usual derivatives, \textit{i.e.}\  $[\partial^\star_\mu, x^\nu] = \delta_\mu^\nu + \mathcal{O}(\frac{1}{\kappa})$,
    \item $[\partial^\star_\mu, x^\nu]$ is at most linear in the derivations,
    \item the derivations form a module of $\mathfrak{so}_\kappa(1, d)$.
\end{enumerate}
These conditions have only three isomorphic solutions \cite{DMT2004}, which can be reduced to
\begin{align}
    \partial^\star_{0} f(x)
    &= \partial_0 f(x), &
    \partial^\star_j f(x)
    &= \frac{e^{\frac{i}{\kappa}\partial_{0}} - 1}{\frac{i}{\kappa} \partial_{0}} \partial_{j} f(x).
    \label{eq:kM_kL_der_action}
\end{align}
The operator $\partial^\star_{0}$ has a trivial coproduct, contrary to $\partial^\star_j$ which has a twisted one. Therefore, $\partial^\star_0$ obeys a regular Leibniz rule with respect to the star-product \eqref{eq:kM_star_prod_1}, but $\partial^\star_j$ satisfies a twisted Leibniz rule.

\paragraph{}
Mixing these two algebra of symmetries gives back the $\kappa$-Poincar\'{e} algebra  \eqref{eq:kM_kP_Hopf_alg}, where $\partial^\star_\mu$ correspond to $P_\mu$. However, its action on the $\kappa$-Minkowski space, corresponding to equations \eqref{eq:kM_kL_action} and \eqref{eq:kM_kL_der_action}, are different from the bicrossproduct construction ones \eqref{eq:kM_kP_action}.

\paragraph{}
The authors of \cite{DJMTWW2004a,DJMTWW2004b} wanted an undeformed algebraic sector of the $\kappa$-Poincar\'{e} algebra. Therefore, they defined the components of the Dirac operator $(D_{\mu})_{0\leqslant \mu\leqslant d}$ by
\begin{align}
    D_{0} 
    &= \kappa \sin \left(\frac{1}{\kappa} \partial^\star_{0} \right) - \frac{i\kappa}{2} \partial^\star_k \partial^{\star k} e^{-\frac{i}{\kappa} \partial^\star_0}, &
    D_{j}
    &= \partial^\star_{j} e^{-\frac{i}{\kappa} \partial^\star_0}.
\end{align}
Considering now the algebra generated by $D_\mu$, $M_j$ and $N_j$, one have a Hopf algebra with the Poincar\'{e} commutation relations. However, this implies a highly non-linear coalgebra sector, and so strongly deformed Leibniz rules. The action of the Dirac operator on the $\kappa$-Minkowski space writes
\begin{align}
\begin{aligned}
    D_{0} f(x) 
    &= \left( \kappa \sin \left(\frac{1}{\kappa} \partial_{0} \right) - \frac{1}{\frac{i}{\kappa}\partial^{2}_{0}} \left( \cos \left(\frac{1}{\kappa} \partial_{0} \right) - 1 \right) \partial^{j} \partial_{j} \right) f(x),\\
    D_{j} f(x) 
    &= \frac{1 - e^{-\frac{i}{\kappa} \partial_{0}} } {\frac{i}{\kappa} \partial_{0}} \partial_{j} f(x).
\end{aligned}
\end{align}

\subsubsection{Differential calculus and integration}
\label{subsubsec:kM_kL_diff_calc_int}
\paragraph{}
The discussion on how to define a differential calculus on $\kappa$-Minkowski is undertook in \cite{DMT2004}. This is first done by defining a deformed de Rham differential operator $\dd$. The latter operator is required to satisfy the following requirements
\begin{enumerate}[label = (\roman*)]
    \item \label{it:kM_kL_diff_nilpot}
    the differential must be nilpotent, \textit{i.e.}\ $\dd^2 = 0$,
    \item \label{it:kM_kL_diff_1_form}
    the application of $\dd$ to coordinates gives a $1$-form: $[\dd, x^\mu] = \xi^\mu$,
    \item \label{it:kM_kL_diff_inv}
    the differential is invariant under $\mathfrak{so}_\kappa(1,d)$, \textit{i.e.}\ $[M_j, \dd] = 0$ and $[N_j, \dd] = 0$,
    \item \label{it:kM_kL_diff_Leib}
    it satisfies undeformed Leibniz rule for the point-wise product: $\dd(f g) = \dd(f) g + f \dd(g)$.
\end{enumerate}
From the antsatz $\dd = \xi^\mu D_\mu$, \ref{it:kM_kL_diff_1_form} and \ref{it:kM_kL_diff_inv} are automatically satisfied and \ref{it:kM_kL_diff_nilpot} is achieved when $[\xi^\mu, \partial_\nu] = 0$ and $\{\xi^\mu, \xi^\nu\} = 0$. The condition \ref{it:kM_kL_diff_Leib} is a consequence of \ref{it:kM_kL_diff_1_form}.

\paragraph{}
An important point to note is that, adding as a requirement the commutator $[\xi^\mu, x^\nu]$ is closed in the space of $1$-forms, imposes the differential calculus to be at least $d+2$-dimensional. The fact that the differential calculus needs one supplementary dimension to fulfill symmetry condition is not new. Indeed, it was shown in \cite{Sitarz_1995} that a bicovariant, Lorentz-invariant differential calculus on $3+1$-dimensional $\kappa$-Minkowski need at least to be $5$-dimensional. This is, however, done with full consistency with respect to the commutative limit since, in this context, one of the base $1$-form vanishes when $\kappa \to +\infty$.

\paragraph{}
The path chosen by \cite{DMT2004} is to require that the differential calculus is still $d+1$-dimensional. This is done by promoted $[\xi^\mu, x^\nu]$ to be derivative-valued and so to relax the closeness hypothesis. This path leads to cumbersome expressions for $\xi^\mu$.

\paragraph{}
Finally, an integral is needed to have an action functional. The integral is defined as a linear map satisfying the trace property with respect to the star-product, \textit{i.e.}\ 
\begin{equation}
    \int f \star_\kappa g = \int g \star_\kappa f.
    \label{eq:kM_trace_prop}
\end{equation}
This trace property \eqref{eq:kM_trace_prop} is not fulfilled for the usual Lebesgue measure. A solution consists of writing the integral with a non-trivial volume element as
\begin{equation}
    \int f = \int\left( \dd^{d+1} x\; \mu(x) \right) f(x).
    \label{eq:kM_kL_measure}
\end{equation}
The trace property is achieved if
\begin{align}
    \partial_{0} \mu(x) &= 0, &
    x^{j}\partial_{j} \mu (x) &= - d\ \mu(x).
\end{align}
where $d$ is the space dimension of $\kappa$-Minkowski. One can note that it is independent of $x^0$ and of the deformation parameter $\kappa$, therefore it does not vanish at the commutative limit. Furthermore, it is singular at $0$ and is not uniquely defined.

These issues are solved by requiring that the computation of physical quantities is independent of the choice of this function. Thus, the explicit form of $\mu$ is not necessary.

\subsubsection{Action functional and gauge transformation}
\label{subsubsec:kM_kL_action_and_gauge}
\paragraph{}
In \cite{DMMW2004}, the gauge theory, for an arbitrary finitely generated gauge group, is built through the use of the Seiberg-Witten map \eqref{eq:moyal_SW_map}. It further contains the notion of covariant derivative, which is built as $\kappa$-Lorentz invariant. Moreover, the $U(1)$ gauge theory is considered in \cite{DJM2005}.

\paragraph{}
The components $(\mathcal{D}_{\mu})_{0\leqslant \mu\leqslant d}$ of a covariant derivative are defined thanks to the components of the Dirac operator defined above:
\begin{equation}
    \mathcal{D}_{\mu} = D_{\mu} - iA_{\mu}
\end{equation}
where $A_{\mu}$ is the gauge potential. This covariant derivative is both gauge and $\kappa$-Lorentz invariant.

The curvature tensor is defined by
\begin{equation}
    \mathcal{F}_{\mu\nu} = i [\mathcal{D}_{\mu}, \mathcal{D}_{\nu} ].
\end{equation}
It transforms covariantly because $\mathcal{D}$ does. This curvature is derivative-valued and so expresses, in the basis of the $\mathcal{D}_\mu$, as
\begin{align}
    \mathcal{F}_{\mu\nu} 
    = F_{\mu\nu} + T_{\mu\nu}^\rho \mathcal{D}_\rho + \cdots + T_{\mu\nu}^{\rho_1 \cdots \rho_n} \mathcal{D}_{\rho_1} \cdots \mathcal{D}_{\rho_n}.
\end{align}
The first term, $F$ correspond to the noncommutative field strength which goes to the usual field strength at the commutative limit. The second term $T_{\mu\nu}^\rho$ correspond to a torsion-like term.

\paragraph{}
The $U(1)$ gauge invariant action is introduced as
\begin{align}
    S = - \frac{1}{4} \int \dd^{d+1}x\ \mu(x)\ X_2 \star_\kappa F_{\mu\nu} \star_\kappa F^{\mu\nu},
\end{align}
where $X_2$ is added to compensate for $\mu$ at the commutative limit. $X_2$ is further required to be gauge covariant so that $S$ is gauge invariant using \eqref{eq:kM_trace_prop}. At first order in $\kappa$, one has $X_2 = \mu^{-1}(1 - \frac{d}{\kappa} A_0^0)$, where $A^0_\mu$ refers to the commutative gauge potential.

The action obtained at first order is invariant under the commutative $U(1)$ gauge and the undeformed Poincar\'{e} transformations.

\subsection{Gauge theories on \tops{$\kappa$}{kappa}-Minkowski via twists}
\label{subsec:kM_twist}
\paragraph{}
In this subsection, we follow the works \cite{DJ2011, DJP2014}. The authors use Drinfel'd twists to deform the Poincar\'{e} Hopf algebra and extract the space-time geometry. The gauge theory is then built with the Seiberg-Witten map \eqref{eq:moyal_SW_map}.

\subsubsection{Star-product from twists}
\label{subsubsec:km_twist_star_prod}
\paragraph{}
We briefly recall key elements of the twist deformation construction applied to the case of the $\kappa$-Minkowski space. For more details about this construction see subsection \ref{subsec:star_prod_twists}. Let $\mathcal{M}$ be the Minkowski space and $\mathfrak{iso}(1,d)$ the Poincar\'{e} Lie algebra. One constructs the canonical Poincar\'{e} Hopf algebra by considering the Hopf algebra of the universal envelopping algebra $U \mathfrak{iso}(1,d)$, see subsection \ref{subsubsec:star_prod_eg} for more details. One then deforms the Poincar\'{e} Hopf algebra through a twist element $\twiF_\kappa \in U \mathfrak{iso}(1,d) \otimes U\mathfrak{iso}(1,d)$. 

The above discussion led to the deformed Poincar\'{e} algebra, which is the algebra of symmetries leaving $\mathcal{M}$ invariant. In order to keep $\mathcal{M}$ invariant under the deformed symmetries, one needs to change its product denoted by $\cdot$ into a noncommutative star-product $\star_\kappa$ defined by \eqref{eq:star_prod_twist}. In the context of the Minkowski space, this writes
\begin{equation}
    f \star_\kappa g = \cdot \circ \twiF_\kappa^{-1}(f \otimes g)
    \label{eq:km_star_prod_twist_general}
\end{equation}
where $\cdot: \mathcal{M} \otimes \mathcal{M} \to \mathcal{M}$ is the usual pointwise product of function on the Minkowski space, defined by $\cdot(f \otimes g) = f \cdot g$.

\paragraph{}
The twist formalism provides a straightforward way to define an undeformed differential calculus, as the usual differential $\dd$ already satisfies requirements \eqref{eq:kM_star_calc_diff}. More details about twisted differential calculus are given in subsection \ref{subsec:diff_calc_twist}. The latter differential calculus has the following elementary properties
\begin{subequations}
\begin{align}
    \dd(f \star_\kappa g) &= (\dd f) \star_\kappa g + f \star_\kappa (\dd g), \\
    \dd^{2} &= 0,\\
    \dd f &= (\partial_\mu f) \dd x^\mu 
    = (\partial^\star_\mu f) \star_\kappa \dd x^\mu,
    \label{eq:kM_star_calc_diff_star_der}
\end{align}
    \label{eq:kM_star_calc_diff}
\end{subequations}
where the last equation \eqref{eq:kM_star_calc_diff_star_der} defines the so-called $\star$-derivatives $\partial^\star$, which is only valid if a basis of $1$-forms $\dd x^\mu$ exists.

Moreover, one defines the deformed product of forms $\wedge_\kappa$ by twisting the undeformed one $\wedge$ as in \eqref{eq:wedge_prod_twist}. The previous base $\dd x^\mu$ then allows to decompose forms as $\omega = \omega_{\mu_1 \cdots \mu_n} \star_\kappa \dd x^{\mu_1} \wedge_\kappa \dots \wedge_\kappa \dd x^{\mu_n}$, for $\omega \in \Omega^n(\mathcal{M}_\kappa)$. Furthermore, one can show that a deformed product of $1$-forms of the basis is akin to the undeformed one and is anti-symmetric, namely
\begin{equation}
    \dd x^\mu \wedge_\kappa \dd x^\nu 
    = \dd x^\mu \wedge \dd x^\nu
    = -\dd x^\nu \wedge \dd x^\mu 
    = -\dd x^\nu \wedge_\kappa \dd x^\mu.
\end{equation}
Since basis $1$-forms anticommute, the volume form remains undeformed
\begin{equation}
    \dd^{d+1}_\kappa x 
    := \dd x^0\wedge_\kappa \dd x^1\wedge_\kappa \cdots \wedge_\kappa \dd x^{d} 
    = \dd x^0 \wedge \dd x^1 \wedge \cdots \wedge \dd x^{d} 
    = \dd^{d+1} x.
\end{equation}
Finally, there is a natural notion of trace, that is the one of the commutative case.

\paragraph{}
In \cite{DJ2011}, the authors consider the following so-called Abelian twist
\begin{align}
    \twiF^{\mathrm{A}}_\kappa
    &= e^{-\frac{i}{2}\Theta^{ab} X_a \otimes X_b}
    = e^{-\frac{i}{2\kappa}(\partial_0 \otimes x^j\partial_j - x^j\partial_j \otimes \partial_0)}.
\end{align}
where $a, b = 0, 1$ and the $X_a$ are commuting vector fields, here chosen to be $X_0 = \partial_0$ and $X_1 = x^j\partial_j$. $\Theta$ is the constant antisymmetric deformation matrix here given by
\begin{equation}
    \Theta 
    = \begin{pmatrix}
     0 &  \frac{1}{\kappa} \\ 
    -\frac{1}{\kappa} &  0
    \end{pmatrix}
\end{equation}
where $\kappa$ is the deformation parameter. This twist generates through \eqref{eq:km_star_prod_twist_general}, the algebra $\mathcal{M}_\kappa$ with the star-product \eqref{eq:kM_star_prod_1}. First derivations of the $\kappa$-Minkowski space through twist deformation were made in \cite{Bu_2008, Meljanac_2007}.

However, one should note that $X_1$ corresponds to the space dilatation generator and is therefore not in the universal enveloping algebra of the Poincar\'{e} algebra. In other words, $\twiF^\mathrm{A}_\kappa \; \slashed{\in}\; U\mathfrak{iso}(1,d) \otimes U\mathfrak{iso}(1,d)$. The Poincar\'{e} algebra has to be enlarged to the inhomogeneous general linear algebra $\mathfrak{igl}(1,d)$, which contains dilatations. Therefore, one has $\twiF^\mathrm{A}_\kappa \in U\mathfrak{igl}(1,d) \otimes U\mathfrak{igl}(1,d)$. This means that the $\kappa$-Minkowski space studied is not only invariant under the deformed Poincar\'{e} algebra, but under the deformed $\mathfrak{igl}(1,d)$ algebra.

\paragraph{}
The previous observation is even deeper since there exists a no-go theorem \cite{Borowiec_2014} stating that, in dimension $2$ and $4$, the $\kappa$-Poincar\'{e} algebra cannot be recovered by a twist deformation. In other words, there are no twist in $U\mathfrak{iso}(1,3) \otimes U\mathfrak{iso}(1,3)$ that allows to reconstruct the $\kappa$-Minkowski space.

\paragraph{}
In the case of the Abelian twist, the integral of a top degree form is cyclic
\begin{equation}
    \int \omega_1 \wedge_\kappa \omega_2 
    = (-1)^{d_1\cdot d_2} \int \omega_2 \wedge_\kappa \omega_1,
    \label{eq:kM_cyclic_trace}
\end{equation}
with $d_{j}=\mathrm{deg}(\omega_{j})$ and  $d_1 + d_2 = d+1$.

\paragraph{}
In \cite{DJP2014}, the authors proposed to allow only one more degree of freedom instead of infinitely many. This is done by deforming the Poincar\'{e}-Weyl algebra $\mathfrak{iwso}(1,d)$. One goes from the Poincar\'{e} algebra to the Poincar\'{e}-Weyl algebra by adding the dilatation generator. They consider the Jordanian twist 
\begin{align}
    \twiF^{\mathrm{J}}_\kappa
    = e^{-i D \otimes \sigma}
    \in U\mathfrak{iwso}(1,d) \otimes U\mathfrak{iwso}(1,d), &&
    \sigma \in \ln \left( 1 + \frac{1}{\kappa} P_0 \right)
\end{align}
where $D = -i x^\mu \partial_\mu$ is the dilatation generator and $P_0 = -i \partial_0$ is the time translation generator. This twist was first derived in \cite{Borowiec_2009}. The price to pay here is the loss of the cyclicity of the twist, so we are sent back to the problems presented in the previous section \ref{subsec:kM_kL_gauge_th}.

\subsubsection{Action functional from Hodge duality}
\label{subsec:kM_action_and_Hodge}
\paragraph{}
In this subsection \ref{subsec:kM_action_and_Hodge}, the notations of the Seiberg-Witten map are used (see subsection \ref{subsec:moyal_theta_exp} for more details). Explicitly, fields with a hat, e.g.\ $\hat{F}$, correspond to noncommutative fields and their commutative counterpart will be noted without hat, e.g.\ $F$.

\paragraph{}
The gauge field $\hat{A}$ and the curvature $\hat{F}$ are introduced through the Seiberg-Witten map \eqref{eq:moyal_SW_map} and \eqref{eq:moyal_SW_map_fs} respectively. Then, one can decompose them on the basis of $1$-forms as
\begin{align}
    \hat{A} = \hat{A}_\mu \star_\kappa \dd x^\mu, &&
    \hat{F} = \frac{1}{2} \hat{F}_{\mu\nu} \star_\kappa {\dd}x^\mu \wedge_\kappa {\dd}x^\nu.
\end{align}
The action functional considered in \cite{DJ2011} is
\begin{equation}
    S = \frac{1}{2} \int \hat{F} \wedge_\kappa (\sstar \hat{F}),
    \label{eq:kM_twist_action}
\end{equation}
where $\sstar$ is the Hodge operator, defined as in the undeformed case, that is $(\sstar \hat{F})_{\mu\nu} = \frac{1}{2} \epsilon_{\mu\nu\lambda\sigma} \hat{F}^{\lambda\sigma}$, where $\epsilon$ is the fully antisymmetric Levi-Civita symbol. This definition has the advantage of having a straightforwardly nice commutative limit. However, it appears that such an action is not gauge invariant. This issue can be traced back to the fact that $\sstar \hat{F}$ does not transform covariantly.

\paragraph{}
A discussion in \cite{DJ2011}, pursued in \cite{DJP2014}, is undertook to address the problem of definition of a Hodge duality. The authors put forward three different methods to solve this issue, all involving new forms. These new fields do not change the number of degrees of freedom due to the Seiberg-Witten map, but they can introduce additional covariant terms in the expanded actions provided one discusses the freedom of the Seiberg-Witten map. These additional terms could be used to render some nice properties of the theory, like renormalizability.

\paragraph{}
The first method is to introduce a $2$-form $\hat{Z}$ and write the action \eqref{eq:kM_twist_action} as $S = \frac{1}{2} \int \hat{F} \wedge_\kappa \hat{Z}$. To have a good commutative limit, $\hat{Z}$ must satisfy $\hat{Z} = \sstar F + \mathcal{O}(\frac{1}{\kappa})$, where $F$ is the commutative field strength. This new action writes at first order in $\kappa$
\begin{equation}
    S = - \frac{1}{4} \int \dd^{d+1} x\ F_{\mu\nu} F^{\mu\nu} 
    - \frac{1}{2} \tensor{C}{^{\sigma\rho}_\lambda} x^\lambda F^{\mu\nu} F_{\mu\nu} F_{\sigma\rho}
    + 2 \tensor{C}{^{\sigma\rho}_\lambda} x^\lambda F^{\mu\nu} F_{\mu\rho} F_{\nu\sigma}
    \label{eq:kM_action_twist_meth1}
\end{equation}
which is invariant under commutative $U(1)$ gauge transformations. $C$ is defined as the structure constant of the $\kappa$-Minkowski Lie algebra \eqref{eq:kM_kM_Hopf_alg_alg}, that is $[x^\mu, x^\nu] = \tensor{C}{^{\mu\nu}_\lambda} x^\lambda$. Here, the covariant additional terms are proportional to $x$ and has not yet been interpreted geometrically.

\paragraph{}
The issue with the Hodge gauge transformation arises because functions does not commute with base forms $\dd x^\mu$, explicitly, for $f \in \mathcal{M}_\kappa$, $f \star_\kappa \dd x^\mu \neq \dd x^\mu \star_\kappa f$. The second method consists in considering the basis of $1$-forms $\vartheta$ which commutes with functions, $f \star_\kappa \vartheta^\mu = \vartheta^\mu \star_\kappa f$. One should note that this change of basis is twist dependant.

Such a transformation usually turns the flat integration measure to a curved one. Therefore, this results in changing the deformed Hodge definition itself. This is done by introducing a form $\hat{G}$ so that $\sstar \hat{F} = \frac{1}{2} \epsilon_{\mu\nu\lambda\sigma} \hat{G}^{\mu\rho\nu\tau} \star_\kappa \hat{F}_{\rho\tau} \star_\kappa \vartheta^\lambda \wedge_\kappa \vartheta^\sigma$ transforms covariantly. This $\hat{G}$ can be seen as the deformed metric part of the Hodge operator. Therefore, to recover a good commutative limit, one needs $\hat{G}^{\mu\rho\nu\tau} = \sqrt{-g} g^{\mu\rho} g^{\nu\tau} + \mathcal{O}(\frac{1}{\kappa})$.

Throught this procedure the action \eqref{eq:kM_twist_action} is invariant under the noncommutative $U(1)$ gauge transformation. Moreover, it yields also \eqref{eq:kM_action_twist_meth1} at first order if $\kappa$ (when going back to the $\dd x^\mu$ basis).

\paragraph{}
The third method avoids using the Hodge dual and defines the action as
\begin{equation}
    S = \frac{1}{2} \int \epsilon_{\mu\nu\rho\sigma} \left(
    \frac{1}{2} \big( \hat{f}^{\mu\nu} \star_\kappa \hat{F} + \hat{F} \star_\kappa \hat{f}^{\mu\nu} 
    - \frac{1}{12} \hat{f}^{\lambda\tau} \star_\kappa \hat{f}_{\lambda\tau} \star_\kappa \hat{e}^{\mu} \wedge_\kappa \hat{e}^\nu \right)
    \wedge_\kappa \hat{e}^\rho \wedge_\kappa \hat{e}^\sigma
    \label{eq:kM_twist_action_meth3}
\end{equation}
rather then \eqref{eq:kM_twist_action}. In the previous equation $\hat{e}$ is the noncommutative tetrad and $\hat{f}$ is a new field which is determined by the equations of motion. Both fields transforms covariantly and are defined through the Seiberg-Witten map. This method was first introduced in \cite{Aschieri_2012b} for matter coupled to gravity.

The action \eqref{eq:kM_twist_action_meth3} is also invariant under the noncommutative $U(1)$ gauge transformation and yields \eqref{eq:kM_action_twist_meth1} at first order in $\kappa$.

\paragraph{}
The analysis of the action \eqref{eq:kM_twist_action} made in \cite{DJ2011} leads to modified dispersion relation for plane wave solution with birefringenceless superluminal or supraluminal solutions, depending of the type (time-like, light-like or space-like) of the $\kappa$-Minkowski space.

However, the previous analysis, reviewed in subsection \ref{subsec:kM_kL_gauge_th}, had no superluminal or supraluminal solutions. The authors of \cite{DJ2011} put forward that this divergence could come from the enlarged space of symmetries in the twisted model which is not $\kappa$-Poincar\'{e} but rather the deformation of $\mathfrak{igl}(1,3)$.

\subsection{\tops{$\kappa$}{kappa}-Poincar\'{e} invariant gauge theories}
\label{subsec:kM_kP_invariant_gauge}
\paragraph{}
It is useful to list the five main assumptions underlying the construction of $\kappa$-Poincar\'e-invariant gauge theories on $\kappa$-Minkowski spaces as proposed in \cite{MW2020a}, \cite{MW2020b}.
\begin{enumerate}
    \item The action is a quadratic polynomial functional in the curvature and is both invariant under $\mathcal{P}_\kappa$ and the noncommutative gauge symmetry, assumed for simplicity to be a noncommutative analog of a $U(1)$ symmetry.
    \item The commutative limit (\textit{i.e.}\ the limit $\kappa\to\infty$) of the classical action coincides with the action describing an ordinary gauge theory.
    \item The connection is described by a twisted version of a noncommutative connection defined on a right module $\mathbb{E}$ over $\mathcal{M}_\kappa$.
    \item The module $\mathbb{E}$ is assumed to be a copy of $\mathcal{M}_\kappa$, \textit{i.e.}\ $\mathbb{E}\simeq\mathcal{M}_\kappa$ and is promoted to the status of hermitian module. Thus, it is endowed with an hermitian structure.
    \item The action of $\mathcal{M}_\kappa$ on $\mathbb{E}$, defined as a linear map $\Phi:\mathbb{E}\otimes\mathcal{M}_\kappa\to\mathbb{E}$, is twisted by an automorphim of $\mathcal{M}_\kappa$. Namely, one has:
    \begin{align}
        \Phi(m\otimes f) &:= m \bullet f = m \star_\kappa \sigma(f), &
        \sigma &\in \mathrm{Aut}(\mathcal{M}_\kappa)
    \label{zeaction},
    \end{align}
    for any $m\in\mathbb{E}$ and $f\in\mathcal{M}_\kappa$.
\end{enumerate}

\paragraph{}
Before going into details, it is instructive to comment these assumptions and summarize the main features of these $\kappa$-Poincar\'e invariant gauge theories. The two first assumptions are clearly motivated by reasonable physical considerations. In particular, $\kappa$-Poincar\'e invariance is a rather natural assumption together with the need to recover a usual gauge theory, here QED, at the commutative limit. The price to be paid for the $\kappa$-Poincar\'e invariance requirement is the loss of cyclicity of the trace in the action which becomes twisted.

The third and fourth assumptions permit one to introduce a noncommutative analog of hermitian connection which gives rise in \cite{MW2020a}, \cite{MW2020b} to a noncommutative analog of a Yang-Mills theory. Besides, the choice $\mathbb{E}\simeq\mathcal{M}_\kappa$ selects a noncommutative analog of a $U(1)$ gauge theory for simplicity. The introduction of twisted structures is a rather natural solution to compensate for the effects of the loss of cyclicity of the twisted trace just mentioned above, without which the gauge invariance of the action could not be achieved. It turns out that this latter can be achieved only whenever the dimension of the $\kappa$-Minkowski space-time is equal to 5. Therefore, a salient prediction from this noncommutative gauge theory model is the existence of one extra dimension.

The last assumption corresponds to the freedom in choosing the way the algebra modeling $\mathcal{M}_\kappa$ acts on the module. This is controlled by the choice of the automorphism $\sigma$ in \eqref{zeaction}. In fact, as shown in \cite{module-paper}, changing $\sigma$ alters only the hermiticity condition ruling the gauge potential $A_\mu$ while keeping the above value of the dimension of $\mathcal{M}_\kappa$ unchanged, as well as the general features of the corresponding gauge theories. Thus, in the following we will restrict ourselves to $\sigma = \id$, hence considering untwisted action as  $m \bullet f = m \star_\kappa f$.

\subsubsection{A star-product from the affine group algebra.}
\label{subsubsec:kM_star_prod_affine}
\paragraph{}
A convenient star-product for $\mathcal{M}_\kappa$ can be obtained from a simple application of the scheme given in subsection \ref{subsec:star_prod_conv_alg}. Note that this star-product has been used in \cite{DS, PW2018, PW2019} to construct and explore quantum properties of scalar field theories on $\mathcal{M}_\kappa$ invariant under the action of $\kappa$-Poincar\'e. We summarize the main steps leading to its construction together with the natural involution coming along with it, which both reflect the algebraic objects defining the convolution algebra of the group relevant to $\kappa$-Minkowski space.

\paragraph{}
Start from the Lie algebra of coordinates $\mathfrak{g}$ given by the commutation relations
\begin{align}
    [x^0, x^j]
    = \frac{i}{\kappa} x^j, &&
    [x^j, x^k] = 0, &&
    j, k = 1, 2, \dots, d. 
\end{align}
The corresponding Lie group is the affine group \cite{khalil} $\mathcal{G} := \mathbb{R} \ltimes_{{\phi}} \mathbb{R}^{d}$ where $\phi: \mathbb{R} \to \text{Aut}(\mathbb{R}^{d})$, given below. $\mathcal{G}$ is not unimodular. The left- and right-invariant Haar measures, respectively $\dd \mu$ and $\dd \nu$, are related by $\dd \nu(s) = \Delta_{\mathcal{G}}(s)\ \dd \mu(s)$, for any $s \in \mathcal{G}$, where $\Delta_{\mathcal{G}}: \mathcal{G} \to \mathbb{R}^+/\{0\}$ is the modular function, a continuous group homomorphism.

\paragraph{}
Next, consider the convolution product and involution in $L^1(\mathcal{G})$ expressed in terms of the right-invariant Haar measure given for any $F, G \in L^1(\mathcal{G})$, by
\begin{align}
    (F \tcvp G)(t) 
    = \int_{\mathcal{G}} \dd \nu(s)\ F(ts^{-1})G(s), && 
     F^*(t) 
     = \overline{F}(t^{-1}) \Delta_{\mathcal{G}}(t)
    \label{decadix}
\end{align}
where $\overline{F}$ is the complex conjugation of $F$. Now, use the group law for $\mathcal{G}$ given by
\begin{subequations}
\begin{align}
    W(p_0, \vec{p}) W(q_0, \vec{q}) 
    &= W(p_0 + q_0, \vec{p} + {e^{-p_0/\kappa} \vec{q}}), &
    \id_{{\mathcal{G}}} &= W(0, 0),\\
    W^{-1}(p_0, \vec{p}) &= W(- p_0, -e^{p_0/\kappa} \vec{p} ),
\end{align}
\end{subequations}
with $\vec{p}, \vec{q} \in \mathbb{R}^{d}$. The modular function expresses as ${\Delta_{\mathcal{G}} (p_0, \vec{p}) = e^{d\, p_0/\kappa}}$ and one defines $F(W) = F(p_0, \vec{p}) := \mathcal{F}f(p_0, \vec{p})$ for any $F \in L^1(\mathcal{G})$, where $\mathcal{F}$ denotes the Fourier transform. These expressions, combined with \eqref{decadix} yield
\begin{subequations}
\begin{align}
    (F \tcvp G)(p_0, \vec{p})
    &= \int_{\mathbb{R}^{d+1}} \dd q_0 \dd^d \vec{q}\  F(p_0 - q_0, \vec{p} - e^{(q_0 - p_0) / \kappa} \vec{q}) G(q_0, \vec{q})
    \label{convolbis} \\
    F^*(p_0, \vec{p}) &= e^{d\, p_0/\kappa} \overline{F} ( - p_0, -e^{p_0/\kappa} \vec{p}).
    \label{involbis},
\end{align}
    \label{eq:kM_kP_conv_inv_bis}
\end{subequations}
Furthermore, one has $\dd \nu(W) = \dd p_0 \dd^d \vec{p}$ showing that the right-invariant Haar measure is the usual Lebesgue measure. Then, following subsection \ref{subsec:star_prod_conv_alg}, we combine the bounded non-degenerate ${}^*$-representation of $L^1(\mathcal{G})$, $\pi: L^1(\mathcal{G}) \to \mathcal{B}(\mathcal{H})$, $\pi(F) = \int_{\mathcal{G}} \dd \nu(s) F(s) \pi_U(s)$ where $\pi_U: \mathcal{G} \to \mathcal{B}(\mathcal{H})$ is a unitary representation of $\mathcal{G}$. The representation $\pi$ satisfies $ \pi (F \tcvp G) = \pi(F) \pi(G)$, $\pi(F)^\dag = \pi(F^*)$. Now, from the definition of the Weyl quantization map \eqref{eq:weyl_map}, $Q(f)=\pi(\mathcal{F}f)$ which must obey
\begin{align}
    Q(f \star_\kappa g) &= Q(f) Q(g), & 
    (Q(f))^* &= Q(f^\dag)
    \label{propQ},
\end{align}
one deduces
\begin{align}
    f \star_\kappa g &= \mathcal{F}^{-1}(\mathcal{F}f\circ\mathcal{F}g), &
    f^\dag &= \mathcal{F}^{-1}(\mathcal{F}(f)^*)
    \label{starinvol-def},
\end{align}
which, by simply using \eqref{eq:kM_kP_conv_inv_bis} yield \cite{DS, PW2018}
\begin{align}
    (f \star_\kappa g)(x) &= \int \frac{\dd p_0}{2 \pi} \dd y^0\ e^{-i y^0 p_0} f(x^0 + y^0, \vec{x}) g(x^0, e^{-p_0/\kappa} \vec{x}), 
    \label{star-kappa}\\
    f^\dag(x) &= \int \frac{\dd p_0}{2\pi} \dd y^0\ e^{-i y^0 p_0} \overline{f} (x^0 + y^0, e^{-p_0/\kappa} \vec{x})
    \label{invol-kappa},
\end{align}
thus defining the star-product for $\kappa$-Minkowski space and the corresponding involution. Before going further, some comments are in order.

Note that the quantization map $Q$ is defined upon identifying functions on $\mathcal{G}$ with functions on $\mathbb{R}^{d+1}$. Besides, functions involved in the convolution product and involution map defined above are interpreted as Fourier transforms of functions of space-time coordinates, hence the occurrence of $\mathcal{F}f$ in the definition of $Q$.

\paragraph{}
For a detailed mathematical characterization of the relevant algebra of functions on which \eqref{star-kappa} and \eqref{invol-kappa} can be defined, see \cite{DS, PW2018}. A careful characterization of a multiplier algebra, involving coordinate functions, constants, etc ..., convenient for physical applications has been nicely carried out in \cite{DS}. For our present purpose, it will be sufficient to know that such a ``sufficiently large'' algebra of functions exists. We denote it by $\mathcal{M}_\kappa$.

\paragraph{}
Two salient properties must be outlined, which will be needed in the rest of this section.

First, the Lebesgue integral (\textit{i.e.}\ the right-invariant measure) defines a twisted trace w.r.t.\ the star-product \eqref{star-kappa}, which is apparent in
\begin{equation}
    \int \dd^{d+1}x\ (f \star_\kappa g)(x) = \int \dd^{d+1}x\ \big( (\mathcal{E}^d \actl g) \star_\kappa  f \big)(x),
    \label{twisted-trace}
\end{equation}
for any $f,g \in \mathcal{M}_\kappa$. Hence the cyclicity of the trace w.r.t.\ the start-product is lost but is replaced by a new property. In fact, any functional of the form $\varphi(f) = \int \dd^{d+1}x\ f$ defines a KMS weight \cite{kuster} for the modular group of automorphisms $\sigma_t(f) = e^{i t d P_0/\kappa} \actl f$. For a discussion on the relationship with the Tomita-Takesaki modular theory \cite{takesaki} and possible physical implications including the possibility to define a global time observer independent in the spirit of past works \cite{connrov}, see \cite{PW2018}. In the sequel, the automorphism defined by $\mathcal{E}^{d} \actl$ in equation \eqref{twisted-trace} is called  ``modular twist''. Observe that this latter depends on the spacial dimension $d$ of the $\kappa$-Minkowski space.

Next, the KMS weight is invariant under the action of the $\kappa$-Poincar\'e algebra $\mathcal{P}_\kappa$ since \cite{DS}:
\begin{equation}
    h \blacktriangleright S 
    = \int \dd^{d+1}x\ h \actl \mathcal{L} 
    = \epsilon(h) S
    \label{poinca-invar},
\end{equation}
for any $h \in \mathcal{P}_\kappa$ where $\epsilon$ is the co-unit of $\mathcal{P}_\kappa$, implying that {\textit{any action involving the Lebesgue integral is $\kappa$-Poincar\'e invariant}}. The $\kappa$-Poincar\'e invariance of the action is a physically natural requirement: the $\kappa$-Poincar\'e symmetry is assumed to be the relevant symmetry at an energy scale close to $\kappa$, which reduces to the usual Poincar\'e symmetry at low energy scale for which the Minkowski space becomes relevant.

In the following, we denote by $\langle\cdot, \cdot\rangle$ the following Hilbert product on $\mathcal{M}_\kappa$ defined by
\begin{equation}
    \langle f,g \rangle := \int \dd^{d+1}x\ (f^\dag \star_\kappa g)(x)
    \label{hilbert-prod}
\end{equation}
for any $f, g \in \mathcal{M}_\kappa$. A useful formula, valid for any $f \in \mathcal{M}_\kappa$ is
\begin{equation}
    \int \dd^{d+1}x\ (f \star_\kappa g^\dag)(x)
    = \int \dd^{d+1}x\ f(x){\bar{g}}(x),
    \label{algeb-1}
\end{equation}

\subsubsection{Reconciling twisted trace and gauge invariance}
\label{subsubsec:kM_kP_gauge_inv}
\paragraph{}
It is known that the loss of cyclicity of the trace generates a problem to achieve the gauge invariance of an action of the form $S \sim \int \dd^{d+1}x\ P(F_{\mu\nu})$
where $P$ is a polynomial depending on the curvature $F_{\mu\nu}$. The latter curvature arises from the use of a, by now, standard framework involving a standard noncommutative differential calculus together with the associated notion of noncommutative connection and curvature, as reviewed in section \ref{sec:nc_diff_calc}.

To illustrate this point, note that such a framework would produce, at the end of the day, a curvature with gauge transformations of the form
\begin{equation}
    F_{\mu\nu}^g = g^\dag \star_\kappa F_{\mu\nu} \star_\kappa g
    \label{untwis-gauge}
\end{equation}
for any $g \in \mathcal{U} = \{g \in \mathbb{E},\ g^\dag \star_\kappa g = g \star_\kappa g^\dag = \mathds{1}\}$. $\mathcal{U}$ is the gauge group generated by the unitary elements of the module $\mathbb{E}$, on which module the connection is defined. The module will be assumed to be a copy of $\mathcal{M}_\kappa$, \textit{i.e.}\ $\mathbb{E} \simeq \mathcal{M}_\kappa$, which is (almost) always assumed in the literature. Then, the action of $\mathcal{U}$ on $\langle F,F\rangle = \int \dd^{d+1}x\  F_{\mu\nu}^\dag \star_\kappa F_{\mu\nu}$ would produce
\begin{equation}
    \langle F^g, F^g\rangle 
    = \int \dd^{d+1}x\ {(\mathcal{E}^{d}(g) \star_\kappa g^\dag)} \star_\kappa F_{\mu\nu}^\dag \star_\kappa F_{\mu\nu},
    \label{zepb}.
\end{equation}
so that $  \langle F^g,F^g\rangle\ne \langle F,F\rangle$ unless the prefactor $\mathcal{E}^{d}(g) \star_\kappa g^\dag$ is eliminated.

\paragraph{}
One apparent solution would be to impose $\mathcal{E}^{d}(g) \star_\kappa g^\dag = \mathds{1}$, thus restricting {\textit{formally}} the gauge group to
\begin{equation}
    \mathcal{U}_\Delta 
    = \{ g\in\mathcal{U}, \mathcal{E}^{d}(g) = g \}
    \subset \mathcal{U}. 
\end{equation}
However, the condition $\mathcal{E}^{d}(g)=g$ imposes very (actually too) strong restrictions on the possible functions $g \in \mathbb{E}$ which, as discussed in \cite{MW2020a}, \cite{module-paper} reduces to constants. Indeed, the constraint implies $g(x^0 + \frac{i}{\kappa}, \vec{x}) = g(x^0, \vec{x})$ so that $g$ is periodic in $x_0$. As discussed in \cite{MW2020a, module-paper}, $g$ must be an entire function. But by the Liouville theorem, any entire periodic function is constant.

\paragraph{}
Another solution to escape \eqref{zepb} is the use of a suitably twisted differential, together with a related notion of twisted connection and curvature, leading to a twisted version of \eqref{untwis-gauge} able to compensate for the factor $\mathcal{E}^{d}(g) \star_\kappa g^\dag$ in \eqref{zepb}. This is the way followed in \cite{MW2020a}, \cite{MW2020b}, reviewed in the following subsections.

\paragraph{}
Twisted structures appeared in twisted spectral triples, naturally related to twisted differential calculi, and used in various contexts, see e.g.\ \cite{como-1, martinet, Devastato-Martinetti, Filaci-Martinetti}. To give a flavor of these objects, a twisted spectral triple with Dirac operator $D$ satisfies the condition that for any $f$ in the algebra $\algA$ of the triple, $[D,f]_\rho := Df-\rho(f)D$ is a bounded operator on the Hilbert space $\mathcal{H}$ defining the triple. Here, $\rho \in \mathrm{Aut}(\algA)$ is the twist which is a {\textit{regular automorphism}}, namely
\begin{equation}
    \rho(f)^\dag = \rho^{-1}(f^\dag)
    \label{regular}
\end{equation}
for any $f\in\algA$. For an untwisted spectral triple $(\algA, D, \mathcal{H})$, one would have instead that $[D,f]= Df-fD$ is bounded. The twisted commutator $[D,f]_\rho$ acts as a twisted derivation on $\algA$ as
\begin{equation}
    \delta_\rho(f g)
    := [D,fg]_\rho 
    = [D, f]_\rho g + \rho(f)[D, g]_\rho 
    = \delta_\rho(f) g + \rho(f) \delta_\rho(g)
\end{equation}
for any $f,g\in\algA$. It can be further extended to a derivation in the set of 1-forms defined by $\Omega^1_D=\{\omega=\sum_i f_i[D,k_i],\ f_i,k_i\in\algA \}$, up to technical conditions to be satisfied.

\paragraph{}
As far as gauge theories on $\mathcal{M}_\kappa$ are concerned, one convenient way to obtain twisted differential calculi is to use twisted derivations for constructing derivation-based differential calculi in the spirit of subsection \ref{subsec:nc_diff_calc_der}. A twisted derivation is a linear map $X: \algA \to \algA$ for some algebra $\algA$, to be identified in a while with the algebra $\mathcal{M}_\kappa$, satisfying a twisted Leibnitz rule
\begin{equation}
    X(a \star b) = X(a) \star \alpha(b) + \beta(a)\star X(b),\label{leibnitz-twisted}
\end{equation}
for any $a, b \in \algA$, where $\star$ is its associative product. In \eqref{leibnitz-twisted}, $\alpha, \beta \in \text{Aut}(\algA)$. Twisted derivations have been considered from various viewpoints in algebra, see e.g.\ \cite{orebis, Hom-lie}. 

\subsubsection{Twisted differential calculus, curvature and connection.}
\label{subsubsec:kM_kP_twist_diff_calc}
\paragraph{}
Owing to the duality between deformed translations $\mathcal{T}_\kappa \subset \mathcal{P}_\kappa$ and $\mathcal{M}_\kappa$ and to the requirement that the commutative limit of a reasonable action for a gauge theory on $\mathcal{M}_\kappa$ must describe a usual (commutative) gauge theory, it is natural to look for twisted derivations as elements in $\mathcal{T}_\kappa$.

As shown in \cite{MW2020a, MW2020b}, there exists a unique Abelian Lie algebra of twisted derivations of the Hopf subalgebra $\mathcal{T}_\kappa$ fulfilling this requirement. It is defined by:
\begin{equation}
    \mathfrak{D}_\gamma
    = \big\{ X_\mu: \mathcal{M}_\kappa \to \mathcal{M}_\kappa,\  
    X_0 = \kappa \mathcal{E}^\gamma(1 - \mathcal{E}),\ 
    X_j = \mathcal{E}^\gamma P_j,\  
    j = 1, 2, \dots, d \big\},
    \label{tausig-famil}
\end{equation}
where $\gamma\in\mathbb{R}$ and $(d+1)$ denotes the dimension of the $\kappa$-Minkowski space. $\mathfrak{D}_\gamma$ satisfies $[X_\mu,X_\nu]:=X_\mu X_\nu-X_\nu X_\mu=0$ and
\begin{equation}
    X_\mu(f \star_\kappa h) 
    = X_\mu(f) \star_\kappa \mathcal{E}^\gamma(h) + \mathcal{E}^{1 + \gamma}(f) \star_\kappa X_\mu(h),
    \label{tausigleibniz}
\end{equation}
and one has $(X \actr z)(f) := X(f) \star_\kappa z = z \star_\kappa X(f) = (z \actl X)(f)$, for any $f \in \mathcal{M}_\kappa$ and any $z \in \mathcal{Z}(\mathcal{M}_\kappa)$, the center of $\mathcal{M}_\kappa$. Therefore, $\mathfrak{D}_\gamma$ inherits a structure of $\mathcal{Z}(\mathcal{M}_\kappa)$-bimodule. Note that the above derivations are {\textit{not}} real derivations, since one can easily verify that for any $f\in\mathcal{M}_\kappa$
\begin{equation}
    (X_\mu(f))^\dag = -\mathcal{E}^{-2\gamma-1} (X_\mu(f^\dag)).
    \label{xpasreel}
\end{equation}

\paragraph{}
Using $\mathfrak{D}_\gamma$, a mere adaptation of the derivation-based differential calculus of subsection \ref{subsec:nc_diff_calc_der} to incorporate twists leads to a family of twisted differential calculus indexed by a real parameter $\gamma$ \cite{MW2020a}. The differential calculus based on the Lie algebra of twisted derivations $\mathfrak{D}_\gamma$, correspond to the differential calculus \eqref{eq:ncdc_set_forms} replacing $\Der(\algA)$ by $\mathfrak{D}_\gamma$. Explicitly, this calculus is fully characterized by the following graded differential algebra
\begin{equation}
    \left( \Omega^\bullet = \bigoplus_{n=0}^{d+1} \Omega^n(\mathfrak{D}_\gamma),
    \wedge, 
    {\dd} \right),
    \label{diff-algebra}
\end{equation}
where $\Omega^0(\mathfrak{D}_\gamma) = \mathcal{M}_\kappa$, $\Omega^n(\mathfrak{D}_\gamma)$ is the space of $n$-linear antisymmetric forms. These $n$-forms correspond to $\alpha: \mathfrak{D}^{n}_\gamma \to \mathcal{M}_\kappa$ satisfying $\alpha(X_1, X_2, \cdots, X_n) \in \mathcal{M}_\kappa$ and $\alpha(X_1, X_2, \cdots, X_n \actr z) = \alpha(X_1, X_2, \dots, X_n) \star_\kappa z$, for any $z \in \mathcal{Z}(\mathcal{M}_\kappa)$ and any $X_1, \dots, X_n \in \mathfrak{D}_\gamma$. Note that the linearity holds w.r.t.\  $\mathcal{Z}(\mathcal{M}_\kappa)$.

The associative product $\wedge_\kappa: \Omega^m(\mathfrak{D}_\gamma) \otimes \Omega^n(\mathfrak{D}_\gamma) \to \Omega^{m+n}(\mathfrak{D}_\gamma)$ is given by \eqref{eq:form_prod} and the differential $\dd: \Omega^n(\mathfrak{D}_\gamma) \to \Omega^{n+1}(\mathfrak{D}_\gamma)$ is defined by \eqref{eq:koszul}. As in \eqref{eq:d2_0}, one can check that $\dd$ satisfies $\dd^2=0$. However, the Leibniz rule \eqref{eq:ncdc_leibniz} is now twisted
\begin{equation}
    \dd (\omega \wedge_\kappa \eta) 
    = \dd \omega \wedge_\kappa \mathcal{E}^\gamma(\eta) + (-1)^{|\omega|} \mathcal{E}^{1 + \gamma}(\omega) \wedge_\kappa \dd \eta,
    \label{leibniz-form}
\end{equation}
where $|\omega|$ is the degree of $\omega$.

\paragraph{}
A notion of twisted connection on a right module over the algebra can then be defined  \cite{MW2020a, MW2020b} which extends the notion of noncommutative connection on a right module over the algebra as reviewed in subsection \ref{subsec:nc_conn_right}\footnote{
Note that other extensions of this notion of connection related to differential calculus based on $\epsilon$-derivations, which generalize graded derivations, have been considered in \cite{epsilon-stuff}.
}.

For $\mathbb{E}$ assumed to be a copy of $\mathcal{M}_\kappa$, it is defined as a map $\nabla_{X_\mu}:\mathbb{E}\to\mathbb{E}$ for any $X_\mu\in\mathfrak{D}_\gamma$ such that
\begin{subequations}
\begin{align}
    \nabla_{X_{\mu}}(m \star_\kappa  a) 
    &= \nabla_{X_{\mu}}(m) \star_\kappa \mathcal{E}^{\gamma}(a) + \mathcal{E}^{\gamma + 1}(m) \star X_{\mu}(a), \\
    \nabla_{X_\mu \star_\kappa z + X^\prime_\mu}(m)
    &= \nabla_{X_\mu}(m) \star_\kappa z + \nabla_{X^\prime_\mu}(m),
\end{align}
    \label{twist-conn}
\end{subequations}
for any $m\in\mathbb{E}\simeq\mathcal{M}_\kappa$, $z\in\mathcal{Z}(\mathcal{M}_\kappa)$ and $a\in\mathcal{M}_\kappa$. One can note that the $\caZ(\mathcal{M}_\kappa)$-linearity \eqref{eq:nc_conn_def_lin} is untouched but the Leibniz rule \eqref{eq:nc_conn_def_lin} is now twisted.

The curvature map $\mathcal{F}(X_\mu, X_\nu) := \mathcal{F}_{\mu\nu}$, $\mathcal{F}_{\mu\nu}: \mathbb{E} \to \mathbb{E}$, fulfilling $\mathcal{F}_{\mu\nu}(m \star_\kappa a) = \mathcal{F}_{\mu\nu}(m) \star_\kappa a$ as a morphism of module, is defined by
\begin{equation}
    \mathcal{F}_{\mu\nu} 
    = \mathcal{E}^{1-\gamma} (\nabla_\mu \mathcal{E}^{-1-\gamma} \nabla_\nu - \nabla_\nu \mathcal{E}^{-1-\gamma} \nabla_\mu).
    \label{twist-curvat}
\end{equation}
This definition correspond to a twisted version of \eqref{eq:nc_curv_def}. To make contact with the notations used in the physics literature, one defines
\begin{align}
  \nabla_\mu &:= \nabla_{X_\mu}, & 
  A_\mu &:= \nabla_{\mu}(\mathds{1}), &
  \mathcal{F}_{\mu\nu}(\mathds{1}) &:= F_{\mu\nu}, 
  \label{lesdefs}
\end{align}
which gives rise to the following expression for the curvature
\begin{equation}
    F_{\mu\nu}
    = \mathcal{E}^{-2\gamma} (X_\mu A_\nu - X_\nu A_\mu)
    + \mathcal{E}^{1-\gamma}(A_\mu) \star_\kappa \mathcal{E}^{-\gamma}(A_\nu)
    - \mathcal{E}^{1-\gamma}(A_\nu) \star_\kappa \mathcal{E}^{-\gamma}(A_\mu).
    \label{zecourbure}
\end{equation}

\paragraph{}
The map $\nabla_{X_\mu}$ \eqref{twist-conn} defined above extends to a map $\nabla: \mathbb{E}\to\mathbb{E} \otimes \Omega^1(\mathfrak{D}_\gamma)$ with
\begin{align}
    \nabla(a) &= A \star_\kappa \mathcal{E}^{\gamma}(a) + 1 \otimes \dd a, &
    A &\in \Omega^1(\mathfrak{D}_\gamma)
    \label{nabla-diff}
\end{align}
with $A(X_\mu)=A_\mu$ while the map $\mathcal{F}_{\mu\nu}: \mathbb{E} \to \mathbb{E}$ \eqref{twist-curvat} extends to $F: \mathbb{E} \to \mathbb{E} \otimes \Omega^2(\mathfrak{D}_\gamma)$ such that
\begin{equation}
    F = \mathcal{E}^{-2\gamma} (\dd A) + \mathcal{E}^{-\gamma} \big( \mathcal{E}(A) \wedge_\kappa A \big)
    \label{curvat-diff},
\end{equation}
and one has the following Bianchi identity:
\begin{equation}
    \dd F = \mathcal{E}^{1+\gamma}(F) \wedge_\kappa  A - \mathcal{E}^{2}(A) \wedge_\kappa \mathcal{E}^{\gamma}(F)
    \label{bianchi}.
\end{equation}

\paragraph{}
To introduce the gauge transformations, $\mathbb{E}$ is endowed with a hermitian structure
\begin{equation}
    h(m_1,m_2)=m_1^\dag \star_\kappa m_2.
    \label{gabuzo},
\end{equation}
and thus becomes a hermitian module. Recall that a hermitian structure is defined by \eqref{eq:nc_hermit_struct}.

The gauge group is then defined as the set of automorphisms of $\mathbb{E}$ compatible with the hermitian structure \eqref{gabuzo}. It is entirely determined by the group of unitary elements of $\mathbb{E}$ given by
\begin{equation}
    \mathcal{U}
    = \{g \in \mathbb{E},\ g^\dag \star_\kappa g = g \star_\kappa g^\dag = \mathds{1} \}
    \label{huron}
\end{equation}
The twisted gauge transformations are \cite{MW2020a}
\begin{equation}
    \nabla_{{\mu}}^g(a) = \mathcal{E}^{\gamma+1} (g^\dag) \star_\kappa \nabla_{{\mu}}(g \star_\kappa a),
    \label{gaugebitwist1}
\end{equation}
so that 
\begin{align}
    A_\mu^g 
    &= \mathcal{E}^{\gamma+1}(g^\dag) \star_\kappa A_\mu \star_\kappa \mathcal{E}^{\gamma}(g)
    + \mathcal{E}^{\gamma+1}(g^\dag) \star_\kappa X_\mu(g), &
    F_{\mu\nu}^g 
    &= \mathcal{E}^2(g^\dag) \star_\kappa F_{\mu\nu} \star_\kappa g,
\label{amugtwist}
\end{align}
which extend to the $1$-form connection and $2$-form curvature as
\begin{align}
    A^g 
    = \mathcal{E}^{\gamma+1}(g^\dag) \wedge_\kappa A \star_\kappa \mathcal{E}^{\gamma}(g)
    + \mathcal{E}^{\gamma+1}(g^\dag) \wedge_\kappa \dd g, &&
    F^g 
    = \mathcal{E}^2(g^\dag) \wedge_\kappa F \wedge_\kappa g,
    \label{zegaugetransfo}
\end{align}
for any $g\in\mathcal{U}$ and $a\in\mathcal{M}_\kappa$. Observe that the gauge transformations for the curvature are independent on $\gamma$. Moreover, $A_\mu$ \eqref{lesdefs} satisfies
\begin{equation}
    A_\mu = \mathcal{E}^{2\gamma+1} (A_\mu^\dag)
    \label{hermit1}
\end{equation}
instead of the usual relation $A_\mu = A_\mu^\dag$, reflecting the existence of a twisted condition defining twisted hermitian connections
\begin{equation}
    h(\mathcal{E}^{-1} \nabla_{X_{\mu}}(m_1), m_2) 
    + h(\mathcal{E}^{-1}(m_1), \nabla_{X_{\mu}}(m_2)) 
    = X_\mu h(m_1,m_2)
    \label{cledag}
\end{equation}
for any $X_\mu\in\mathfrak{D}_\gamma$, $m_1,m_2\in\mathbb{E}$.

\subsubsection{From classical gauge invariant action to BRST gauge-fixing}
\label{subsubsec:kM_kP_BRST}
\paragraph{}
A natural action candidate for a $\kappa$-Poincar\'e invariant gauge theory on $\mathcal{M}_\kappa$ is given by
\begin{equation}
    S_\kappa
    = \int \dd^{d+1}x\ F^{\mu\nu} \star_\kappa F_{\mu\nu}^\dag.
    \label{zelagrangien}
\end{equation}
Note that $S_\kappa$ is real thanks to \eqref{hilbert-prod} while $\kappa$-Poincar\'e invariance stems from \eqref{poinca-invar}. To control the gauge invariance under $\mathcal{U}$ \eqref{huron}, one computes
\begin{align}
    \int \dd^{d+1}x\ (F^{\mu\nu})^g \star_\kappa (F_{\mu\nu}^g)^\dag
    &= \int \dd^{d+1}x\ \big({\mathcal{E}^{d-2}(g) \star_\kappa \mathcal{E}^2(g^\dag)} \big) \star_\kappa F^{\mu\nu} \star_\kappa F_{\mu\nu}^\dag
    \label{computation}
\end{align}
where we used the twisted trace formula \eqref{twisted-trace} and \eqref{zegaugetransfo}. From \eqref{computation}, one concludes that $S_\kappa$ is gauge invariant under the $\mathcal{U}$ transformations if and only if
\begin{equation}
    \mathcal{E}^{d-2}(g) \star_\kappa \mathcal{E}^2(g^\dag) = 1
\end{equation}
which is verified for
\begin{equation}
    d + 1 = 5.
\end{equation}
Hence, a salient prediction from this noncommutative gauge theory model is the existence of one extra dimension. More formally, $\kappa$-Poincar\'e and gauge invariance can only co-exist in five dimensions.

\paragraph{}
The gauge-fixing can be achieved by exploiting a suitable BRST symmetry. While a BRST symmetry, leaving invariant the classical action and formally related to the gauge group $\mathcal{U}$, can be constructed, the fact that the differential calculus is twisted slightly affects the algebraic structures underlying the standard BRST framework. This will be commented in a while. For a mathematical analysis, see \cite{MW2021}.

To simplify the discussion, set $\gamma=0$ from now on. It is first convenient to define the following ``infinitesimal'' gauge transformation
\begin{align}
    \delta_\omega A_\mu
    = X_\mu(\omega)
    + A_\mu \star_\kappa \omega
    - \mathcal{E}(\omega) \star_\kappa A_\mu, &&
    \delta_\omega A
    = \dd\omega + A \star_\kappa \omega - \mathcal{E}(\omega) \star_\kappa A
\end{align}
where $A$ is the $1$-form connection, which satisfies the commutation relation given by 
$\left[\delta_{\omega_1}, \delta_{\omega_2}\right]
:= \delta_{\omega_1} \delta_{\omega_2} 
- \delta_{\omega_2} \delta_{\omega_1}
= \delta_{\omega_1 \star_\kappa \omega_2 - \omega_2 \star_\kappa \omega_1}
= \delta_{[\omega_1,\omega_2]_\kappa}$ so that the set of $\delta_\omega$ transformations has a Lie algebra structure and 
\begin{equation}
    \delta_\omega S_{\kappa} = 0.
\end{equation}
Then, by introducing the ghost field $C$, one can define the BRST operation $s_{0}$, with $s_0^2=0$, by
\begin{align}
    s_{0} A
    = - \dd C - A \wedge_\kappa C - \mathcal{E}(C) \wedge_\kappa A, &&
    s_{0}C
    = - C \wedge_\kappa C.
    \label{equat-struct}
\end{align}
acting as usual on products of forms as $s_{0}(\rho \wedge_\kappa \eta) = s_{0}(\rho) \wedge_\kappa \eta + (-1)^{|\rho|}\rho \wedge_\kappa s_{0}(\eta)$ and anticommuting with $\dd$, namely $s_{0}\dd + \dd s_{0} = 0$.
This yields
\begin{equation}
    s_{0} F 
    = F \wedge_\kappa C - \mathcal{E}^2(C) \wedge_\kappa F,
\end{equation}
where $F$ is the curvature 2-form and one can verify that
\begin{equation}
    s_{0} S_{\kappa} = 0.
\end{equation}
Observe that the BRST operation $s_{0}$ is {\textit{formally}} similar to the BRST operation for a commutative (non-Abelian) Yang-Mills theory, up to a twist when operating on $A$ and $F$. This reflects merely the fact that the present twisted BRST symmetry is related to a twisted gauge transformation $\delta_{\omega}$.

\paragraph{}
Choosing the gauge $X_\mu(A^\mu) = 0$, the BRST gauge-fixed action takes the form
\begin{align}
\begin{aligned}
    S_{\mathrm{gf}} 
    &= S_{\kappa} + s_{0} \int \dd^5x \left( \overline{C}^\dag \star_\kappa \mathcal{E}^{-4} \big(X_\mu (A_\mu) \big) \right)\\
    &= S_{\kappa} + \int \dd^5x\ \left(
    b.X_\mu(A^\mu) + \overline{C}X^\mu X_\mu(C)
    - \overline{C}X_\mu \big( \mathcal{E}(C) \star_\kappa A^\mu 
    - A^\mu \star_\kappa C \big) \right).
\end{aligned}
    \label{eq:kM_kP_gf_action}
\end{align}
Here, $\overline{C}$ and $b$ are respectivly the antighost and St\"uckelberg fields
with respective ghost numbers $-1$ and $0$, and
\begin{align}
    s_0\overline{C} = b, &&
    s_0 b = 0.
\end{align}
It appears that perturbative computations can be rather easily done from this action, thanks in particular to the expression for the star-product which yields rather simple algebraic expressions. This has been exemplified in \cite{MW2021} and will be reviewed in the next subsection.

\paragraph{}
One remark is in order. The structure equations \eqref{equat-struct} do not fit within the standard BRST framework in that they do not stem from a ``horizontality condition''. The complete mathematical analysis is given in \cite{MW2021}.

To give a flavor of the relevant algebraic framework, one would have expected to obtain \eqref{equat-struct} from the expansion in ghost number of the constraint $\widetilde{F}=F$, where $F$ is given by \eqref{curvat-diff} and $\widetilde{F}$ is obtained from $F$ through the replacement $\dd \to \widetilde{\dd} = \dd + s_0$. This can be already suspected by noticing that $\widetilde{\dd}$ mixes $s_0$, an untwisted derivation, and $\dd$ which is twisted. As shown in \cite{MW2021}, there exists another BRST operation $s_1$ with $s_1^2=0$, defined by
\begin{align}
        s_{1}A = \dd C - \mathcal{E}(C) \wedge_\kappa A - \mathcal{E}(A) \wedge_\kappa C,\ \ 
        s_{1}C = -\mathcal{E}(C) \wedge_\kappa C
    \label{BRS-algebraic}
\end{align}
but however $s_1 S_\kappa \ne 0$. In \eqref{BRS-algebraic}, $s_1$ obeys the same twisted Leibniz rule as $\dd$ and can be obtained from
\begin{equation}
   \widehat{F} = F 
\end{equation}
where $\widehat{F}$ is obtained from $F$ throught the replacement $\dd \to \widehat{\dd}_1 = \dd + s_1$.

Although $s_1$ is no longer a symmetry of $S_\kappa$, it can be shown \cite{MW2021} to be rigidly linked to $s_0$, the actual BRST symmetry of $S_\kappa$ by a continuous transformation. The actual role of $s_1$ in the quantum theory aspects (if any) is presently unknown.

\subsubsection{Some physical properties}
\label{tadpole}
\paragraph{}
One physical prediction of $\kappa$-Poincar\'e invariant gauge theory on $\kappa$-Minkowski space is the existence of one extra dimension. Related phenomenological properties have been discussed in \cite{MW2020b} within models with ``universal extra dimension'' (UED) \cite{HP2007, DKM2010}, assuming a simple compactification scheme on the simple orbifold $\mathbb{S}^1/\mathbb{Z}_2$. Combining the scaling relation $ M_P^2=\frac{\kappa^3}{\mu}$ ($M_P$ is the 4-dimensional Planck mass), which holds true generically in these models, with the experimental constraint from LHC data on the size $\mu^{-1}$ of the extra dimension, $\mu\gtrsim\mathcal{O}(1-5)\ \text{TeV}$, one obtains
\begin{equation}
    \kappa \gtrsim \mathcal{O}(10^{13})\ \text{GeV},
    \label{kappaconserv1}
\end{equation}
where the 4-dimensional Planck mass $M_{\mathrm{P}}\sim \mathcal{O}(10^{19})$ GeV and $\kappa$ in \eqref{kappaconserv1} must be identified with the 5-dimensional bulk Planck mass.

This bound can be improved \cite{MW2020a} by using the observational constraints from gamma ray bursts (GRB) \cite{Addazi_2022} on the flight-time of cosmological photons. In the present gauge theory, the in-vacuo dispersion relation for the 4-dimensional photon, which is identified to the zero-mode of the 5-dimensional $A_\mu$ field in the compactification scheme, is entirely fixed by the noncommutative differential calculus together with the kinetic operator for $A_\mu$. By expanding in inverse powers of $\kappa$, one finds
\begin{equation}
    E^2 - |\vec{p}|^2 - \frac{1}{\kappa} E^3 + \mathcal{O} \left(\frac{1}{\kappa^2} \right)
    = 0
    \label{dispersion}.
\end{equation}
which combined with recent data from GRB yields
\begin{equation}
    \kappa \gtrsim \mathcal{O}(10^{17}-10^{18})\ \text{GeV}.
    \label{kappa-mgm}
\end{equation}
Combining this latter constraint to the above scaling relation gives
\begin{equation}
    \mu \gtrsim \mathcal{O}(10^{13}-10^{16})\ \text{GeV},
\end{equation}
thus corresponding to a very small size for the extra dimension within UED models.

\paragraph{}
The BRST framework of \cite{MW2021} can serve to explore perturbative properties of the gauge theory model described by $S_\kappa$ \eqref{zelagrangien}. This machinery has been used in \cite{HMW2022b} to explore one-loop properties, showing in particular that a non-vanishing one-point function (tadpole) shows up at this order. The corresponding contribution to the one-loop effective action takes the form
\begin{equation}
    \Gamma^1(A) = \int \dd^5 x\ K(\kappa) A_0(x),
    \label{gamma1}
\end{equation}
where $K(\kappa)$ is a gauge dependant diverging integral (to be regularized). Note that the computation leading to \eqref{gamma1} has been carried out either with a kind of noncommutative generalization of the Lorentz gauge, \textit{i.e.}\ $X^\mu(A_\mu) = 0$ or with an interpolating gauge $A_0 = \lambda$.

\paragraph{}
The appearance of the non-zero term $\Gamma^1(A)$ in the 1-loop
effective action implies that the classical vacuum of the theory is not stable against quantum fluctuations, signaling a gauge (BRST) symmetry breaking. Note that the fact that $\langle A_\mu\rangle\ne0$, stemming from \eqref{gamma1}, suggests that the Lorentz invariance is broken by radiative corrections.

\subsection{Other approaches}
\label{subsec:km_other_approach}
\paragraph{}
Several other attempts of defining a gauge theory on $\kappa$-Minkowki space have been proposed and may lead to fruitful developments in the future.

In \cite{KFS2013}, considering more generally a Lie type noncommutative space, the authors propose to use the group of translations in Fourier space, which is commutative, contrary to the space itself. They define a noncommutative gauge theory which requires the introduction of an extra gauge field component. No quantum observable quantity is exhibited though.

Then, let us cite \cite{KV2020, KKV2021}, in which the authors modified the gauge transformation of the gauge potential and the definition of the curvature to incorporate noncommutative gauge transformations. This modifications are introduced in subsection \ref{subsubsec:moyal_modif_gauge_curv}. This approach assumes slowly varying fields, which allows to approximate the $\star$-commutator of functions by the Poisson bracket, and so to work at first order in $\kappa$. However, so far, this attempt led to introduce again a measure function \eqref{eq:kM_kL_measure}, with all the difficulties mentioned subsection \ref{subsec:kM_kL_gauge_th}. Two interpretations of this measure function are proposed. It could be an artefact of the curvature of space-time, or it could be used to modify the original bracket defining $\mathcal{M}_{\kappa}$.

To finish, let us mention a few promising works. First, in \cite{PV2015}, the authors defined a new star-product realizing $\kappa$-Minkowski commutation relations from Wick-Voros star-product and showed it can be obtained from a Jordanian twist. Second, in \cite{LSV2022} the authors studied a variation of $\kappa$-Minkowski space, the so called $\rho$-Minkowski space. However, no gauge theory has been developed yet in any of these two attempts.

\newpage
\section{Gravity on quantum spaces: beyond Yang-Mills}
\label{sec:beyond_YM}
\paragraph{}
The first step to describe the gravity dynamics in noncommutative spaces is to extend consistently its structures and dynamical fields. It appears that generalizing the notion of metric or tetrad is not that straightforward and is certainly not unique. This leads to several approaches aiming to generalize gravity to quantum spaces. The purpose of this section is to review the approaches that have, as far as we know, been undertaken so far. For an earlier review covering a part of the above approaches, see \cite{Muller_Hoissen_2008}.

Note that the spectral approach of Connes will not be discussed here. For physically relevant studies on the spectral action related to the standard model and gravity see \cite{Chamseddine_1997, Chamseddine_2007, Dabrowski_2022}. For interesting adaptations to the noncommutative torus, see \cite{Connes_2014, Fathizadeh_2013, Fathizadeh_2015, Lesch_2016}. For earlier studies, see \cite{Chamseddine_1993, Landi_1994, Sitarz_1994}. Modified Heisenberg uncertainty relations related to volume quantization are investigated in \cite{Chamseddine_2015}.

\paragraph{}
Recall that a ``noncommutative space-time manifold'' is modeled by a noncommutative algebra $\algA$ (e.g.\ an associative algebra of smooth functions). The derivations of $\algA$ are thus natural noncommutative analogs of the usual vector fields which act as derivations on the commutative algebra of functions representing a commutative space-time manifold. Let us denote the set of derivations of $\algA$ by $\mathrm{Der}(\algA)$. Besides, a natural noncommutative analog of the notion of a vector bundle, which should enter the noncommutative extension of connection and gauge transformations, is provided by the notion of (projective) module over $\algA$, denoted by $\modM$, in the spirit of the Serre-Swan theorem. For more mathematical details, see section \ref{sec:nc_diff_calc}.

\paragraph{}
In a commutative setting, a linear connection is a connection on the tangent bundle. As a consequence, this generalizes to a connection on $\mathrm{Der}(\algA)$. Accordingly, one wants to promote $\mathrm{Der}(\algA)$ as a module over the algebra, which is a natural extension. There is a canonical action from $\algA$ on $\mathrm{Der}(\algA)$ defined by
\begin{equation}
    (X\actr a)(b) = X(b) \star a, 
\end{equation}
for $X\in\mathrm{Der}(\algA)$, $a,b\in\algA$. However, for $\actr$ to be an action, $X\actr a$ must be a derivation (that is, the action must be consistent with the module structure), so that it must satisfy the Leibniz rule. Therefore,  for $b,c\in\algA$,
\begin{align*}
	(X\actr a)(b \star c) 
	&= X(b \star c) \star a 
	= X(b) \star c \star a + b \star X(c) \star a \\ 
	&= X(b) \star c \star a + b \star (X \actr a)(c)
\end{align*}
is a derivation if and only if $a$ and $c$ commutes. As $c$ can be set arbitrarily, we need $a\in\mathcal{Z}(\algA)$. Hence, $\mathrm{Der}(\algA)$ is not a $\algA$-module, but rather a $\mathcal{Z}(\algA)$-module.

As this feature does not quite correspond to the commutative case, some authors consider it as the main obstacle to define a linear connection on a noncommutative algebra. However, two possible ways out, respectively based on central bimodules (see subsection \ref{subsec:central_bimodules}) and braided geometry (see subsection \ref{subsec:braided_geometry}), attempt to overcome this problem.

Note that several authors have described connections with the help of covariant derivatives, that are connections on the set of forms, noted $\Omega(\algA)$. This avoids the above problem as the set of forms is a  $\algA$-module. The central bimodule approach (see subsection \ref{subsec:central_bimodules}) actually studies the correspondence between linear connections and covariant derivatives in the context of bimodule having the supplementary property of being central.

\paragraph{}
In general relativity, the gravitational field is rigidely linked to the metric and all the geometry relies on this field. The noncommutative generalization of the metric is still in debate in the scientific literature. It may become complex, see for example \cite{Chamseddine_2001a}. But complex metrics yields physical inconsistencies. The latter happens when the antisymmetric part of the metric does not vanish. In the case of a complex metric, this antisymmetric part may correspond to the imaginary part of the metric because of the hermiticity condition $g(X,Y) = \overline{g(Y,X)}$. Furthermore, there are several ways to define it: bilinear or sesquilinear, symmetric or hermitian, invertible or nondegenerate, real or not, on derivations or on forms, \textit{etc}...

These latter issues also happen when defining the metric by replacing usual products with star-products $\star$ in the equation
\begin{align}
    g_{\mu\nu} = \eta_{IJ}\tensor{\vartheta}{^I_\mu} \tensor{\vartheta}{^J_\nu},
    \label{eq:tetrad_def}
\end{align}
where $g$ is the base space metric, $\eta$ is the Minkowski metric and $\vartheta$ is the inverse tetrad\footnote{
To make the denomination clear, the inverse tetrad, also called inverse vierbein (in $4$ dimensions), inverse vielbein (in any dimensions) or co-frame field, $\vartheta$ are $1$-forms dual to the vector fields called tetrad $e$, also named vierbein (in $4$ dimensions), vielbein (in any dimensions) or frame field. Even if tetrad and vierbein refers explicitly to $4$ dimensions, they are also used regardless the dimension.
}. This method has been widely used in the approach detailed in section \ref{subsec:star_prod_inc}. Then, the antisymmetric part of the metric $g_{[\mu\nu]} = \eta_{IJ} [\tensor{\vartheta}{^I_\mu}, \tensor{\vartheta}{^J_\nu}]_\star$ is not vanishing since the star-product $\star$ is no longer commutative.

A discussion on the structure of a noncommutative version of the metric is given in \cite{Moffat_2000}. Gravity with a complex hermitian metric has negative energy, free ghosts, and free tachyons. The weak field theory is further unstable. Gravity with a complex symmetric metric also contains negative energy ghost states. Several ways out have been proposed. \cite{Chamseddine_2001a} considered only the symmetric part of the metric in the related action. The work \cite{Moffat_2000} took a hyperbolic complex metric, \textit{i.e.}\  $g_{\mu\nu} = g_{\{\mu\nu\}} + i\gamma g_{[\mu\nu]}$ with $\gamma^2=-1$. \cite{Nishino_2002} considered only global Lorentz symmetry instead of local one, as it is done in the context of teleparallel gravity (see details in subsection \ref{subsubsec:telepar_grav}). These numerous problems even motivated some authors \cite{Chamseddine_2003} to develop a metric free approach by considering the tetrad as the gravitational field.

\paragraph{}
One remark is in order. Most of the works dealing with gravity on quantum spaces have been mostly algebraic. They left aside topology. One should stress that some topology should appear at some stage in order to build a reasonable action. So far, this hampers physical discussions in many of the reviewed studies.

\subsection{Central bimodules}
\label{subsec:central_bimodules}
\paragraph{}
The work \cite{Dubois_Violette_1996} defines a notion of central bimodule. Basically, central bimodules are bimodules for which left and right actions coincide for elements of the center.

Connections, curvatures and torsion on such central bimodules are defined in the usual way with the additional feature that one can define a connection on the dual module. Details on this construction are collected in subsection \ref{subsec:nc_central_bimod}. Useful properties are summarized in the rest of this subsection.

\paragraph{}
Consider a central bimodule $\modM$ over an algebra $\algA$. The dual bimodule $\modM'^{{}_\algA}$ is defined as the set of homomorphisms of bimodule from $\modM$ to $\algA$. Recall that a homomorphism of bimodule is a linear map that satisfies \eqref{eq:bimod_homomorp_def}.\\
Now, if $\modM$ is a $\algA$-bimodule then $\modM'^{{}_\algA}$ can be granted a $\mathcal{Z}(\algA)$-module structure. Conversely, if $\modM$ is a $\mathcal{Z}(\algA)$-bimodule then $\modM'^{{}_\algA}$ can be granted a $\algA$-module structure. From any of the two choices, starting from a connection $\nabla$ on $\modM$, one can define a unique dual connection $\nabla'$ on $\modM'^{{}_\algA}$ given by \eqref{eq:dual_conn_def}.

\paragraph{}
Whenever $\modM = \Omega^1(\algA)$, one has $\modM'^{{}_\algA} = \mathrm{Der}(\algA)$. Then, starting from a connection over the $\algA$-bimodule $\Omega^1(\algA)$, one can define a unique connection on $\mathrm{Der}(\algA)$ which is a linear connection. Furthermore, the authors of \cite{Dubois_Violette_1996} define a notion of (pseudo)-Riemannian metric and a notion of metric compatibility, based on the duality. This, together with a torsion free condition, allows one to define a unique Levi-Civita connection on $\mathrm{Der}(\algA)$, which can be given by an explicit expression.

\paragraph{}
Such a notion of linear connection was already used in \cite{Mourad_1995}. In the commutative case, one has $\nabla(f \omega) = \nabla(\omega f)$, for $f\in\mathcal{C^\infty(M)}$ and $\omega\in\Omega^1(\mathcal{M})$. Therefore, to get a proper commutative limit, \cite{Madore_1997, Mourad_1995} introduce a twist $\tau$ in the definition of the linear connection through
\begin{align}
	\nabla_X(f \actl \omega) 
	&= X(f) \actl \omega + f \actl \nabla_X(\omega), &
	\nabla_X(\omega \actr f) 
	&= \tau(\omega \actr X(f)) + \nabla_X(\omega) \actr f,
\label{eq:bimodule_conn}
\end{align}
for any $f\in \algA$, $X\in\mathrm{Der}(\algA)$ and $\omega\in\Omega^1(\algA)$, in which the twist is defined by 
\begin{equation}
\tau(\omega \actr f) = f \actl \omega,     
\end{equation}
for any $\omega \in \Omega^1(\algA)$, $f\in\algA$.

\paragraph{}
Note that the above duality also occurs for right (or left) modules, and one could build similarly a notion of dual connection. However, \cite{Dubois_Violette_1996} put forward that the set of 1-forms is itself a bimodule. Moreover, only the central bimodules canonically reduce to the modules when $\algA$ is commutative.

\paragraph{}
Connections on bimodules have been applied to many noncommutative spaces like the algebra of matrix valued functions $\mathcal{C^\infty(M)}\otimes \mathbb{M}_N(\mathbb{C})$ \cite{Madore_1995, Dabrowski_1996} or the quantum plane $xy=qyx$ \cite{Dabrowski_1996, Dubois_Violette_1995, Georgelin_1996} and also used in general settings like the tame differential calculus (see section \ref{subsec:tame_diff_calc}), quantum principal fiber bundles (see section \ref{subsec:quantum_prin_bunde}) or fuzzy spaces (see section \ref{subsec:fuzzy_spaces}).

\subsection{Tame differential calculus}
\label{subsec:tame_diff_calc}
\paragraph{}
A close albeit somewhat different approach is the use of tame differential calculus on a {\textit{centered}} bimodule\footnote{
This is not to be confused with a central bimodule as in section \ref{subsec:central_bimodules}!} in \cite{Bhowmick_2020a, Bhowmick_2020b}. The set of differential 1-forms is used to define connection, torsion, curvature and metric, rather than from the set of derivations. A Koszul formula is derived, from which a Levi-Civita connection follows.

\paragraph{}
Consider a (noncommutative) algebra $\algA$. Here, the differential calculus $(\Omega^\bullet(\algA),\dd)$ is taken to be the universal differential calculus over $\algA$. A connection is defined as a linear map $\nabla : \Omega^1(\algA) \to \Omega^1(\algA) \otimes_\algA \Omega^1(\algA)$ satisfying the Leibniz rule 
\begin{equation}
    \nabla(\omega \actr a) = \nabla(\omega) \actr a + \omega \otimes_\algA \dd a
\end{equation}
for $\omega\in\Omega^1(\algA)$ and $a\in\algA$. The corresponding torsion is then
\begin{equation}
  T_\nabla = \wedge \circ \nabla + \dd : \Omega^1(\algA) \to \Omega^2(\algA).
\end{equation}
$\Omega^1(\algA)$ is a $\algA$-bimodule which is further assumed to be projective and finitely generated. This latter assumption guarantees the existence of connections on the module.

\paragraph{}
This approach relies on two important structures that are centered bimodule and tame differential calculus.

\paragraph{}
First, given a $\algA$-bimodule $\modM$, its center is defined as
\begin{equation}
 \mathcal{Z}(\modM) = \{m\in\modM, m \actr a = a \actl m, \forall a\in\algA \}.  
\end{equation}
Now, $\modM$ is called centered if $\mathcal{Z}(\modM)$ is right $\algA$-total, that is, if the right $\algA$-linear span of $\mathcal{Z}(\modM)$ equals $\algA$. Observe that this condition is stronger than for the central bimodule of section \ref{subsec:central_bimodules}, as in the former case, any element of $\modM$ must commute with the whole $\mathcal{Z}(\algA)$ and not with $\algA$. Moreover, a centered bimodule is central.

\paragraph{}
Next, a differential calculus $(\Omega^\bullet(\algA),\dd)$ over an algebra $\algA$ is called tame if it verifies the following conditions:

\begin{enumerate}[label = (\roman*)]
	\item \label{it:tame_diff_calc_forms}
	The space of 1-forms is given by $\Omega^1(\algA) = \mathcal{Z}(\Omega^1(\algA)) \otimes_{\mathcal{Z}(\algA)} \algA$.
	\item \label{it:tame_diff_calc_modules}
	The following short exact sequence of right $\algA$-module splits :
	\begin{align*}
		0 
		\to \mathrm{Ker}(\wedge)
		\to \Omega^1(\algA)\otimes_\algA\Omega^1(\algA)
		\to \mathrm{Ran}(\wedge) = \Omega^2(\algA)
		\to 0
	\end{align*}
	\item \label{it:tame_diff_calc_symm}
	Define the following object $\sigma = 2P_{\mathrm{sym}}-1$, where $P_{\mathrm{sym}}$ is the idempotent in the set $\Hom_\algA(\Omega^1(\algA) \otimes_\algA \Omega^1(\algA), \Omega^1(\algA) \otimes_\algA \Omega^1(\algA))$ with image $\mathrm{Ker}(\wedge)$ and kernel $\mathrm{Ker}(\sigma) $ being the complement of $\mathrm{Ker}(\wedge)$ in $\Omega^1(\algA)\otimes_\algA \Omega^1(\algA)$. Then, $\sigma(\omega \otimes_\algA \eta) = \eta \otimes_\algA \omega$, for any $\omega, \eta\in \mathcal{Z}(\Omega^1(\algA))$.
\end{enumerate}
Let us comment on this definition. It follows that the condition \ref{it:tame_diff_calc_forms} implies that the set of $1$-forms $\Omega^1(\algA)$ of a tame differential calculus is automatically a centered bimodule. If follows from \ref{it:tame_diff_calc_modules} that $\Omega^1(\algA)$ admits a torsionless connection, which is the Grassmann connection. Finally, the last condition \ref{it:tame_diff_calc_symm} is present to define a proper symmetry condition for the metric.

\paragraph{}
As far as the metric is concerned, its extension to noncommutative space can be done either by considering a sesquilinear form on the module of 1-forms, as it was made in section \ref{subsec:central_bimodules}, or by considering a complex bilinear form on the module of 1-forms. The latter option is chosen by the authors of \cite{Bhowmick_2020a, Bhowmick_2020b} as a way to avoid $*$-structures. However, the two choices are not equivalent and the uniqueness of Levi-Civita connections may be lost in this formalism using the first option.\\
A pseudo-Riemannian metric $\mathrm{g} \in \Hom_\algA(\Omega^1(\algA)\otimes_\algA\Omega^1(\algA), \algA)$ satisfies the following conditions:
\begin{enumerate}[label = (\roman*)]
	\item \label{it:tame_metric_symm}
	the ``symmetricity'' condition $\mathrm{g} \circ \sigma = \mathrm{g}$, for $\sigma$ defined above,
	\item \label{it:tame_metric_nondegen}
	the map $\Omega^1(\algA) \to (\Omega^1(\algA))'^{{}_\algA}$, $\omega \mapsto \mathrm{g} (\omega \otimes_\algA \cdot)$ is an isomorphism of right $\algA$-modules, where $(\Omega^1(\algA))'^{{}_\algA} = \Hom_\algA(\Omega^1(\algA),\algA)$. It is further said to be bilinear if it is a $\algA$-bimodule map. 
\end{enumerate}
The symmetry condition \ref{it:tame_metric_symm} implies that for $\omega$ or $\eta$ either belonging to $\mathcal{Z}(\Omega^1(\algA))$, then $\mathrm{g} (\omega \otimes_\algA \eta) = \mathrm{g} (\eta \otimes_\algA \omega)$.

\paragraph{}
Finally, a connection $\nabla$ on $\Omega^1(\algA)$ is said to be compatible with a metric $\mathrm{g}$ on $\mathcal{Z}(\Omega^1(\algA))$ if one has
\begin{align}
	(g \otimes_\algA \mathrm{id}) 
	\big( \sigma_{23}(\nabla(\omega)\otimes_\algA \eta) +
	\omega \otimes_\algA \eta \big) 
	&= \dd g(\omega \otimes_{\mathcal{Z}(\algA)} \eta),
	& \omega,\eta\in \mathcal{Z}(\Omega^1(\algA)),
	\label{eq:metric_comp_tame_diff_calc}
\end{align}
where 
\begin{equation}
   \sigma_{23}(\omega \otimes_\algA \eta \otimes_\algA \rho) = \omega \otimes_\algA \rho \otimes_\algA \eta,  
\end{equation}
for any $\omega,\eta,\rho\in \Omega^1(\algA)$.

\paragraph{}
The authors of \cite{Bhowmick_2020a, Bhowmick_2020b} proved that the tame differential calculus is a proper setting for the existence and uniqueness of a Levi-Civita connection, that is, a torsionless connection on $\Omega^1(\algA)$ which is compatible with a pseudo-Riemannian bilinear metric. They also show that this connexion is a bimodule connection in the sense of \eqref{eq:bimodule_conn} over $\Omega^1(\algA)$ with the canonical choice for the twist $\tau=\sigma$ the symmetry operator defined above.

\paragraph{}
Another interesting feature of their model is that the set of vector fields is not defined as $\mathrm{Der}(\algA)$, or $(\Omega^1(\algA))'^{{}_\algA}$, but through the metric, which connects 1-forms to vector fields.\\
Given a tame differential $(\Omega^1(\algA),\dd)$ over an algebra $\algA$ and a pseudo-Riemannian bilinear metric $g$, one defines the set of vector fields as
\begin{align*}
	\mathfrak{X}(\algA) = \big\{ g(\omega \otimes_\algA \cdot), \;\omega\in\Omega^1(\algA) \big\} 
	\; \subseteq \; (\Omega^1(\algA))'^{{}_\algA}.
	\label{eq:vector_fields_tame}
\end{align*}
Several important results follow from this definition.

First, $\mathfrak{X}(\algA) = \mathcal{Z}(\Omega^1(\algA)'^{{}_\algA})$. Then, $\mathfrak{X}(\algA)$ is a $\mathcal{Z}(\algA)$-submodule of $\Omega^1(\algA)'^{{}_\algA}$ which is right $\algA$-total in $\Omega^1(\algA)'^{{}_\algA}$. Moreover, $\mathfrak{X}(\algA)$ is a Lie subalgebra of $\mathrm{Der}(\algA)$, via its action on $\algA$ defined by $\delta_X(a) = X(\dd a)$, for $X\in\mathfrak{X}(\algA)$ and $a\in\algA$. It is shown that for a differential calculus over derivations that is tame, then $\mathfrak{X}(\algA)$ is isomorphic to $\mathrm{Der}(\algA)$.\\

Notice that such a space allows one to define a covariant derivative. Let $(\Omega^1(\algA),\dd)$ be a tame differential calculus over an algebra $\algA$ and $\nabla$ a connection on $\Omega^1(\algA)$. Given $X,Y\in\mathfrak{X}(\algA)$, we define $\nabla_X(Y) \in \Omega^1(\algA)'^{{}_\algA}$ by
\begin{align}
	\nabla_Y(X)(\omega) 
	&= \delta_Y(X(\omega)) - (X\otimes_\algA Y)(\nabla(\omega)),
	& \omega\in\Omega^1(\algA).
	\label{eq:covariant_der_tame}
\end{align}

It is instructive to observe that this formula is akin to \eqref{eq:dual_conn_def}. Through the covariant derivative formulation, usual definition of torsionless, metric compatibility and Koszul formula for the Levi-Civita are recovered.\\
Note that this approach is related to an earlier work \cite{Bhowmick_2019}.

\subsection{Pseudo-Riemannian calculi}
\label{subsec:pseudo_riemann_calc}
\paragraph{}
In the setting of derivation-based differential calculus, one could be tempted to consider the vector fields to be (a sub-Lie algebra of) the derivations as both match in the commutative limit from an algebraic point of view, but this might not be the only possible case. Furthermore, in a noncommutative setting, nothing ensures that the tangent bundle is generalized by the set of derivations.

In the approach of pseudo-Riemannian calculi, the generalization of the tangent bundle is considered to be a general module $\modM$, which is however linked to the hermitian derivations by the existence of a frame field.

\paragraph{}
Inspired by \cite{Rosenberg_2013}, which was revised in \cite{Peterka_2017}, \cite{Arnlind_2017} defines a version of pseudo-Riemannian calculi of modules over noncommutative algebras. The aim of this calculus is to study the properties of the noncommutative space through its generalized tangent bundle, considered here to be a general module over the algebra.

The noncommutative space is modelled as a $*$-algebra $\algA$ and the vector bundle is generalized as a right $\algA$-module $\modM$. In this particular case, $\modM$ is considered to be the space of generalized vector fields. One constructs the metric $h$ on $\modM$ as a non-degenerate hermitian form on $\modM$. The hermitian structure is akin to \eqref{eq:nc_hermit_struct}, introduced in the framework of derivation-based differential calculus and the constraint of non-degeneracy states that if, for any $m_1 \in \modM$, one has $h(m_1, m_2) = 0$, then this implies $m_2 = 0$.

The module $\modM$ is however linked to the hermitian derivations over the algebra $\Der^*(\algA)$ by the existence of a $\mathbb{R}$-linear map $e: \Der^*(\algA) \to \modM$. This map is the generalization of the soldering form. Recall that the soldering form is an isomorphism between the tangent bundle and a vector bundle and is here akin to a frame field. Furthermore, the real metric calculus over $\algA$ is defined through two assumptions:
\begin{enumerate}[label = (\roman*)]
	\item \label{it:prc_met_calc_gen}
	The image $e\big( \Der^*(\algA) \big) = \modM_e$ generates $\modM$ as an $\algA$-module.
	\item \label{it:prc_met_calc_real}
	The metric $h$ is real on $\modM_e$, namely 
\begin{equation}
 h \big(e(X), e(Y) \big)^* = h \big(e(X), e(Y) \big),   
\end{equation}	
for any $e(X), e(Y) \in \modM_e$, .
\end{enumerate}
The assumption \ref{it:prc_met_calc_gen} allows $e$ to be surjective in some sense. This implies that the whole structure of $\modM$ can be characterized thanks to $\Der^*(\algA)$ and $e$. The assumption \ref{it:prc_met_calc_real} allows the metric $h$ to be symmetric on $\modM_e$, which is a physically motivated requirement.

\paragraph{}
Now the connection is defined as for \eqref{eq:nc_conn_def}, except that the definition uses $\Der^*(\algA)$ (instead of the full $\Der(\algA)$) and that $\caZ(\algA)$-linearity \eqref{eq:nc_conn_def_lin} is replaced by $\mathbb{C}$-linearity since, in general, $\Der^*(\algA)$ is not closed under the action of $\caZ(\algA)$.

Furthermore, a connection $\nabla$ is said to be real if for any $X,Y\in\Der^*(\algA)$, one has
\begin{equation}
    h(\nabla(X), Y) = h(Y, \nabla(X))^*. 
\end{equation}
For real connections, two additional definitions can be conveniently introduced. First, the metric compatibility is defined on $\modM_e$ and $\Der^*(\algA)$ through \eqref{eq:nc_hermit_conn}. Then, one defines a torsion-free condition by the requirement that for any $X, Y \in \Der^*(\algA)$, 
\begin{equation}
    \nabla_X (e(Y)) - \nabla_Y (e(X)) - e([X, Y]) = 0, 
\end{equation}
which is the usual definition of the torsion on $\modM_e$.

\paragraph{}
The above set-up corresponds to what the authors call a pseudo-Riemannian calculus. They show that, within this framework, the Levi-Civita connection is unique whenever it exists (which may not be the case). Its existence is achieved when $\modM$ is a free module, that is generated by a finite number of elements.

\paragraph{}
The curvature of a connection $\nabla$ is defined as usual through \eqref{eq:nc_curv_def}, but again on $\Der^*(\algA)$. It is said to be real if for any $X,Y\in\Der^*(\algA)$, 
\begin{equation}
  h \big( \nabla \circ \nabla(e(X)), e(Y) \big) = h \big( e(Y), \nabla \circ \nabla(e(X)) \big).  
\end{equation}
The inverse of the metric is not unique but one can define a pseudo-inverse, which is unique up to the choice of an element of $\modM$. Therefore, there are several possible scalar curvatures given a curvature. However, for a given pseudo-inverse metric, the scalar curvature is unique.

Note that the above set-up was further applied to a non-compact space, namely the noncommutative cylinder \cite{Arnlind_2019}.

\subsection{Quantum tangent space}
\label{subsec:quantum_tangent_space}
\paragraph{}
The notion of duality used for the central bimodules approach (see section \ref{subsec:central_bimodules}) applies as well to modules over an algebra and one considers the derivations as the dual of the 1-forms in this spirit.

However, this is not the only way to set-up a noncommutative analog of vector field using the differential forms. Indeed, as described in section 14.1 of \cite{Klimyk_1997}, for a given left-invariant first order differential calculus $(\Omega^1(\algA),\dd)$ on a Hopf algebra $(\algA, \mu_\star, 1, \Delta, \varepsilon, S)$, one can define the quantum tangent space of this quantum group via its dual Hopf algebra $\algA'$ (see section 1.2.8 in \cite{Klimyk_1997}).

\paragraph{}
More precisely, recall that the first order differential calculus is called left-invariant if there is a linear map $\Delta_L: \Omega^1(\algA) \to \algA \otimes \Omega^1(\algA)$, called the left coaction, which satisfies 
\begin{align}
	\Delta_L(a \star \dd b) 
	&= \Delta(a) \star (\mathrm{id} \otimes \dd) \Delta(b),
	& a,b\in\algA.
\label{eq:left_covariant_fodc}
\end{align}
The space of left invariant elements, that is $\omega\in\Omega^1(\algA)$ such that $\Delta_L(\omega) = 1 \otimes \omega$, is denoted $\tensor[_{\inv}]{\Omega^1(\algA}{})$. The unique projection on left invariant elements $P_L: \Omega^1(\algA) \to \tensor[_{\inv}]{\Omega^1(\algA)}{}$ is given by
\begin{equation}
    P_L(\omega) 
    = \mu_\star \circ (S\otimes \mathrm{id})\circ \Delta_L(\omega) 
    = \sum S(\omega_{(-1)})\omega_{(0)},    
\end{equation}
for $\omega\in\Omega^1(\algA)$. Here, we use the Sweedler notations\footnote{The Sweedler notations write $\Delta_L(\omega) = \sum \omega_{(-1)} \otimes \omega_{(0)}$.}.

This allows us to define the vertical space $\mathcal{V}_\Omega = \{ a\in\mathrm{Ker}(\varepsilon), P_L(\dd a) = 0\}$.

\paragraph{}
The quantum tangent space to $\Omega^1(\algA)$ is then
\begin{align}
	T_\Omega \algA = \{ X\in\algA', \ X(1) = 0 \ \textnormal{and} \ X(a)=0 \ \forall a\in \mathcal{V}_\Omega \} \subset \algA'.
\label{eq:quantum_tangent_space}
\end{align}
The condition $X(1)=0$ makes us consider the tangent space at the identity while the condition $X(a) = 0$ for any $a\in \mathcal{V}_\Omega$, ensures that the space is actually tangent.\\

The duality can be further enlightened by the fact that there exists a unique bilinear form $\langle \cdot, \cdot \rangle : T_\Omega \algA \times \Omega^1(\algA) \to \algA$ given by 
\begin{equation}
 \langle X, a\dd b\rangle = \varepsilon(a)X(b),    
\end{equation}
for any $X\in T_\Omega \algA$ and $a,b\in\algA$. This bilinear form is even a non-degenerate dual pairing between $\tensor[_{\inv}]{\Omega^1(\algA)}{}$ and $T_\Omega \algA$. Then, one can define a pairing between $\algA$ and $T_\Omega \algA$ via
\begin{align}
	\langle X, P_L(\dd a) \rangle 
	&= X(a),
	& X\in T_\Omega \algA, a\in\algA.
\label{eq:dual_paring_quantum_tangent_space}
\end{align}
Therefore, this corresponds to the usual pairing between $\algA$ and $\algA'$.

From this duality, one defines the left action of $T_\Omega \algA$ on $\algA$ : 
\begin{equation}
 X \actl a = \big( (\mathrm{id} \otimes \langle X, \cdot \rangle) \circ \Delta \big) (a) = a_{(1)} \langle X, a_{(2)} \rangle.  
\end{equation}
Now let $\{X_j\}_{j= 1, \dots, n}$ be a basis of $T_\Omega \algA$ and  $\{\omega_j\}_{j= 1, \dots, n}$ the dual basis of $\tensor[_{\inv}]{\Omega^1(\algA)}{}$ with respect to the dual pairing. Then, there exists a family of functions $\{\sigma^j_k : \algA \to \algA\}_{j,k= 1, \dots, n}$ such that
\begin{align}
	X_j \actl (ab) 
	&= X_k \actl (a) \sigma^k_j \actl (b) + a X_j \actl (b)
	& a,b\in\algA, j= 1, \dots, n.
\label{eq:quantum_tangent_twisted_leibniz}
\end{align}
Therefore, one concludes that the quantum tangent space corresponds to twisted derivations. These twists correspond to transition functions between right $\actr_\algA$ and left $\actl_\algA$ actions of $\algA$ on $\Omega^1(\algA)$. Explicitly, 
\begin{equation}
    \omega_j \actr_\algA a = \big((\mathrm{id} \otimes \sigma_j^k) \circ \Delta \big) (a) \actl_\algA \omega_k,  
\end{equation}
for $j,k= 1, \dots, n$. They further satisfy
\begin{align*}
    \sigma^j_k(ab) 
    &= \sigma^j_l(a)\sigma^l_k(b), &
    \sigma^j_k(1) 
    &= \delta^j_k, &
    a,b\in\algA, j,k= 1, \dots, n.
\end{align*}

\paragraph{}
Finally, one can define a connection $\nabla$ on a right $\algA$-module $\modM$ as satisfying the following Leibniz rule  
\begin{align}
	\nabla_{X_j} \actl (ma) 
	&= \nabla_{X_k} \actl (m) \sigma^k_j \actl (a) + m X_j \actl (a), &
	m\in\modM, a\in\algA, j= 1, \dots, n.
\end{align}

\paragraph{}
This framework was applied in \cite{Arnlind_2020, Arnlind_2022} to the quantum sphere to define a Levi-Civita connection. Metric compatibility is introduced through a (sesquilinear) invertible hermitian form $\mathrm{g}(\cdot, \cdot)$. Invertibility of $\mathrm{g}$ is defined as having an induced map $\mathrm{g}^\flat : \modM \to \modM'$, defined as $\mathrm{g}^\flat(m) = \mathrm{g}(m,\cdot)$, for $m\in\modM$, which is invertible.

\subsection{Braided geometry}
\label{subsec:braided_geometry}
\paragraph{}
Another recent approach is the one explored in \cite{Aschieri_2020} (and references therein). This study considers a less wider class of algebra which are $\algH$-module braided symmetric, or braided commutative, algebras $\algA$, for $\algH$ being a triangular Hopf algebra.

The former is built on another formalism based on the twist deformation of the product and all other structures that follow: the Lie algebra of vector fields, the forms, the wedge product... Even if the basic idea is somehow different, this formalism has shown some similarities with the star-product incorporation method detailed in section \ref{subsec:star_prod_inc}. 

Both approaches are reviewed below and a discussion about their differences is added. Finally, the application of the braided geometry to $L_\infty$-algebras is presented.

\subsubsection{The new settings: braidings and symmetry}
\label{subsubsec:braid_braid}
\paragraph{}
The base space is encoded as an $\algH$-module algebra $\algA$, for $\algH$ a triangular Hopf algebra which encodes its symmetry. Recall that a Hopf algebra $(\algH, m, 1, \Delta, \varepsilon, S)$, which definition is recalled in section \ref{subsec:star_prod_twists}, is called quasi-triangular if it possesses an invertible element $\mathcal{R} \in \algH \otimes \algH$, called the $\mathcal{R}$-matrix, such that
\begin{subequations}
	\begin{align}
		(\Delta \otimes \mathrm{id}) (\mathcal{R}) 
		= \mathcal{R}_{13}&\mathcal{R}_{23}, \qquad 
		(\mathrm{id} \otimes \Delta) (\mathcal{R}) 
		= \mathcal{R}_{13}\mathcal{R}_{12}, 
		\label{eq:quasitri_Hopf_alg_1} \\
		\tau \circ \Delta(h) 
		&= \mathcal{R} \Delta(h) \mathcal{R}^{-1}, 
		\quad  h \in \algH.
		\label{eq:quasitri_Hopf_alg_quasicocommu}
	\end{align}
	\label{eq:quasitri_Hopf_alg}
\end{subequations}
where $\tau(h_1 \otimes h_2) = h_2 \otimes h_1$. We used here the usual notations for the $\mathcal{R}$-matrix. Explicitly, for $\mathcal{R}= \sum \mathcal{R}_{(1)} \otimes \mathcal{R}_{(2)}$ in the Sweedler notations, the notations used is
\begin{align}
	\mathcal{R}_{jk} 
	&= \sum 1 \otimes \cdots \otimes 1 \otimes \overset{(j)}{\mathcal{R}_{(1)}} \otimes 1 \otimes \cdots \otimes 1 \otimes \overset{(k)}{\mathcal{R}_{(2)}} \otimes 1 \otimes \cdots \otimes 1, &
	j < k = 1, \dots, n,
	\label{eq:R-matrix_notation}
\end{align}
as being an element of $\algH^{\otimes n} = \algH \otimes \cdots \otimes \algH$.The notion of quasi-triangular Hopf algebra was built to have an almost cocommutative algebra. In other words, the cocommutative condition writes $\tau \circ \Delta = \Delta$ and \eqref{eq:quasitri_Hopf_alg_quasicocommu} is a deformation of it. A quasi-triangular Hopf algebra is further said to be triangular if one has  $\mathcal{R}_{21}\mathcal{R} = 1 \otimes 1$, or equivalently $\tau(\mathcal{R}^{-1}) = \mathcal{R}$. Triangular Hopf algebras are even closer to being cocommutative, as the previous relations give constraints on the inverse $\mathcal{R}$-matrix.\\
Finally, an $\algH$-module algebra $\algA$ is both an algebra and a $\algH$-module such that the product and the action are compatible. Explicitly, for a right module (this is similar for a left module)
\begin{align}
    h \actl (a \star b)
    &= \sum (h_{(1)} \actl a) \star (h_{(2)} \actl b)
    \label{eq:module_alg}
\end{align}
using Sweedler notations $\Delta(h) = \sum h_{(1)} \otimes h_{(2)}$.\\
Besides, an $\algH$-module algebra $\algA$, with $\algH$ a triangular Hopf algebra, is called braided symmetric, or braided commutative, if
\begin{align}
	a \star b 
	&= (\mathcal{R}_1^{-1} \actl b) \star (\mathcal{R}_2^{-1} \actl a), &
	& a,b\in\algA,
\label{eq:braided_symm}
\end{align}
where $\mathcal{R} = \mathcal{R}_1 \otimes \mathcal{R}_2 \in \algH \otimes \algH$ is the $\mathcal{R}$-matrix of $\algH$. In the same way \eqref{eq:quasitri_Hopf_alg_quasicocommu} is the deformation of the cocommutative condition of $\algH$, the equation \eqref{eq:braided_symm} is the deformation of the commutative condition of $\algA$.

\paragraph{}
The idea underlying the use of such a structure comes from the attempt of \cite{Weber_2020} to build a noncommutative Cartan calculus that escapes from central bimodules, and so from the center of the algebra, but that sticks to derivation-based differential calculus.

In this framework, the author considers the set of braided derivatives of $\algA$, noted $\mathrm{Der}_\mathcal{R}(\algA)$ that is the space of linear operators from $\algA$ to itself satisfying the following braided Leibniz rule
\begin{align}
	X(a \star b) 
	&= X(a) \star b + (\mathcal{R}_1^{-1}\actl a) \star (\mathcal{R}_2^{-1}\actl X)(b), &
	& X \in \mathrm{Der}_\mathcal{R}(\algA), a,b\in\algA.
\label{eq:braided_Leibniz_rule}
\end{align}
The action of $\algH$ on endomorphisms of $\algA$ is given by the adjoint action. This is a braided Lie algebra with the braided commutator $[X,Y]_\mathcal{R} = XY - (\mathcal{R}_1^{-1}\actl X)(\mathcal{R}_2^{-1}\actl Y)$, for $ X, Y \in \mathrm{Der}_\mathcal{R}(\algA)$. But braided derivations become also a $\algA$-bimodule for actions
\begin{align}
	(a \actl_\algA X)(b) 
	&= a \star X(b) &
	(X \actr_\algA a)(b) 
	&= X(\mathcal{R}_1^{-1}\actl b) \star (\mathcal{R}_2^{-1}\actl a), &
	& X\in\mathrm{Der}_\mathcal{R}(\algA), a,b \in \algA
\label{eq:braided_action}
\end{align}
where we noted $\actr_\algA$ and $\actl_\algA$ the left and right actions of $\algA$ on $\mathrm{Der}_\mathcal{R}(\algA)$, to distinguish from the action $\actl$ of $\algH$ on $\algA$. This can be proven using both the braided symmetry \eqref{eq:braided_symm} and the braided Leibniz rule \eqref{eq:braided_Leibniz_rule}. We refer to \cite{Weber_2020_PhD} for a proof of the above statement (see Lemma 4.2.1) and also for other properties of such structures.

\paragraph{}
The author of \cite{Aschieri_2020} then defines a connection on $\algH$-equivariant braided symmetric $\algA$-bimodule, like $\mathrm{Der}_\mathcal{R}(\algA)$. As emphasised through \eqref{eq:braided_action}, left and right actions play different roles, with one acting canonically and the other acting as a braided deformation. This deformed symmetry forces the author to define two kinds of connections (right and left connections) through the usual formulas. A right (resp.\ left) connection is then a braided left (resp.\ right) connection. It follows that notions of left and right curvatures and torsions are defined through the usual formulas for the corresponding connections. A notion of covariant derivative, based on Cartan operators, is also defined and Cartan formulas for curvature and torsion are obtained.\\
The notion of pseudo-Riemannian metric is defined together with a (left or right) Levi-Civita connection over $\mathrm{Der}_\mathcal{R}(\algA)$. This connection is shown to be unique thanks to a braided commutative version of the Koszul formula.

\subsubsection{The previous setting: deformed vector fields}
\label{subsubsec:braid_deformed_vf}
\paragraph{}
This formalism generalizes the one studied previously in \cite{Aschieri_2006} of Drinfeld twists applied to the canonical Hopf algebra of the universal enveloping algebra of vector fields. Such a framework was first used on the Moyal space \cite{Aschieri_2005}.

In the spirit of section \ref{sec:star_prod}, we denote in this section the classical (undeformed) algebra of function by $\mathcal{A}$ with the product $\cdot$, and its quantum deformation by $\algA$ with product $\star$.

Here, we recall some facts about twists. The full treatment was made in subsection \ref{subsec:star_prod_twists}. Given a Hopf algebra $(\algH, m, 1, \Delta, \varepsilon, S)$, a Drinfeld twist $\twiF \in \algH \otimes \algH$ is an invertible element that satisfies
\begin{align}
	\twiF_{12}(\Delta\otimes \mathrm{id}) \twiF &= \twiF_{23}(\mathrm{id} \otimes \Delta) \twiF, &
	(\varepsilon \otimes \mathrm{id}) \twiF &= (\mathrm{id} \otimes \varepsilon)\twiF = 1,
\label{eq:drinfeld_twist}
\end{align}
where\footnote{
Those notations are similar to the $\mathcal{R}$-matrix one, see \eqref{eq:R-matrix_notation}.
} 
$\twiF_{12} = \twiF\otimes 1$ and $\twiF_{23} = 1\otimes\twiF$.

To make contact with what we saw above, we now define a new Hopf algebra, noted $\algH^\twiF$, with same elements, product and counit but with deformed coproduct and antipode through
\begin{align}
	\Delta^\twiF &= \twiF \Delta \twiF^{-1} &
	S^\twiF &= \chi S \chi^{-1},
\label{eq:coprod_antipode_drinfeld}
\end{align}
with $\chi = \twiF_1 S(\twiF_2)$. This new Hopf algebra is triangular with $\mathcal{R}$-matrix $\mathcal{R} = \twiF_{21}\twiF^{-1}$. Moreover, for a given $\algH$-module algebra $\mathcal{A}$, one can define a $\algH^\twiF$-module algebra $\algA$ with a noncommutative (star-)product
\begin{align}
	a \star b 
	&= \twiF^{-1}_1(a) \cdot \twiF^{-1}_2(b), &
	a,b\in\algA,
\label{eq:star_prod_drinfeld_twist}
\end{align}
with $\twiF^{-1} = \twiF^{-1}_1 \otimes \twiF^{-1}_2 \in \algH \otimes \algH$.

\paragraph{}
Finally, the contact with physics is made through a choice of particular $\algH$ and $\mathcal{A}$. Considering a smooth manifold, its vector fields form a Lie algebra, noted $\Xi$ in \cite{Aschieri_2006}. We consider $\mathcal{A} = U\Xi$, the enveloping algebra of $\Xi$ and $\algH = U\Xi$, the canonical Hopf algebra of $U\Xi$. The star-product and the Hopf algebra $\algA = U\Xi_\star$ are built as above. It appears that the Lie algebra of $U\Xi$ generates a braided Lie algebra $U\Xi_\star$, with the bracket $[\cdot,\cdot]_\twiF$ defined as the braided commutator given above. The space of deformed vector fields $\Xi_\star$ is then defined as having similar generators to $\Xi$ but with the bracket $[\cdot,\cdot]_\star$.\\

The set of $1$-forms is defined through a twisted tensor product $\otimes_\algA = \otimes_\mathcal{A} \circ \twiF^{-1}$ and the whole set of forms $\Omega^\bullet(\algA)$ is defined through a twisted wedge product between forms $\wedge_\star = \wedge_\mathcal{A} \circ \twiF^{-1}$ (see section \ref{subsec:diff_calc_twist} for more details). A definition of (braided) connection on $\Xi_\star$ is set together with its torsion and curvature. A $\star$-metric is defined as a symmetric element of $\Omega^\bullet(\algA) \otimes_\algA \Omega^\bullet(\algA)$, where symmetricity is ruled by braided transposition given in \eqref{eq:braided_symm}.

\paragraph{}
The formulation of gauge theories in such a model is studied in \cite{Wess_2006}. In \cite{Kobakhidze_2010}, the author proposed to twist both local Lorentz invariance (through the tetrad formalism) and coordinate transformation invariance with this setting to have a covariant theory.

\paragraph{}
Despite the fact that this framework was developed earlier, recent results are obtained by making use of it as in \cite{Aschieri_2021a}. The authors considered a cosmological space-time having a deformed Poincar\'{e}-Weyl symmetry. The deformation is encoded in a Jordanian twist 
\begin{align}
	\twiF &= \exp (-iD\otimes \sigma), &
	\sigma &= \ln \left(1+\frac{1}{\kappa}P_0 \right)
\label{eq:kappa_jordan_twist}
\end{align}
where $D = -ix^\mu \partial_\mu$ and $P_0=-i\partial_0$ are the generators of dilations and time translations respectively. The star-product is then defined \eqref{eq:star_prod_drinfeld_twist} and its set of forms through the twisted tensor product defined above. However, the metric is not defined as reviewed above. Instead, a deformed Hodge operator is defined over $\Omega^\bullet(\algA)$ through 
\begin{equation}
    \sstar^\twiF(\omega) = \twiF^{-1}_1(\sstar) \twiF^{-1}_2(\omega),   
\end{equation}
where $\omega\in\Omega^\bullet(\algA)$ and the action of $\twiF^{-1}_1$ on the usual Hodge operator $\sstar$ is the adjoint action. 

This definition gives the freedom of choosing a metric for the usual Hodge operator. One can then deform any classical metric into what the authors called its $\kappa$-deformation. Choosing the metric to be Minkowski, one recovers the $\kappa$-Minkowski space as constructed in subsection \ref{subsec:kM_twist}, with star-product \eqref{eq:kM_star_prod_1}. In \cite{Aschieri_2021a}, the Friedman-Lemaître-Robertson-Walker metric is chosen.

The author uses this $\kappa$-deformed Hodge operator to compute, in vacua, the dispersion relation of a scalar field and observes an energy dependant light speed and a new expression for the time lag between a low and a high energetic photons. However, the background space-time is still considered classical so that there are no quantum interactions between photons and the space-time in their model. They also put forward that their method can be implemented in the study of a much wider class of classical and quantum spaces.

\paragraph{}
Drinfeld twists were also extended to fuzzy spaces \cite{Kurkcuoglu_2006}. To do so, the star-product was simplified and a so called pseudo-twist was defined.

\subsubsection{Braided \tops{$L_\infty$}{L\_infinity}-algebras}
\label{subsubsec:braided_l_infty_alg}
\paragraph{}
The gravity theory formulated in $L_\infty$-algebras is also quantized using the framework of braided geometry described above \cite{Ciric_2021}. First, Drinfeld twists are applied on a natural Hopf algebra structure given to the universal enveloping Lie algebra of vector fields. Then,  $L_\infty$-algebras are twisted via a twist of their brackets and a $h$-adic version of their Lie algebras.

The gauge transformations are thus braided and a covariant action under left and right braided gauge transformations is derived. 

\paragraph{}
The theory is explicitly computed for the $L_\infty$ version of Einstein-Cartan action. Recall that the Einstein-Cartan action writes
\begin{equation}
	S_{\mathrm{EC}}(\vartheta, \omega) = \frac{c^4}{32\pi G} \int \epsilon_{IJKL}\ \vartheta^{I} \wedge \vartheta^{J} \wedge \left(R^{KL}(\omega) - \frac{\Lambda}{6} \vartheta^{K} \wedge \vartheta^{L} \right).
\label{eq:einstein_cartan_action}
\end{equation}
with $\vartheta$ the inverse tetrad, $R$ the curvature of the spin-connection $\omega$, $G$ the gravitational constant, $c$ the speed of light and $\Lambda$ the cosmological constant.

The curvature and torsion of both translation connection $e$, called $F^\star_e$, and spin connection $\omega$, called $F^\star_\omega$ are computed. The torsion equation states $F^\star_\omega=0$ and the curvature equation states $F^\star_e=0$.

In the commutative case of $L_\infty$-algebras, see \cite{Ciric_2020}, the torsion equation reduces to the torsion free condition and the curvature equation reduces to Einstein equation. These equations appear to be much more complex in their braided version as the torsion equation involves more braided covariant forms than just the left and right torsions. The curvature equation also surprisingly contains new terms that are not braided covariant on their own. The full equation remains braided gauge invariant though.\\
Former equations show that braided geometry imposes a non-trivial, even non-braided covariant, mixing of the gauge fields $e$ and $\omega$.

\paragraph{}
Finally, the commutative limit corresponds to the case where there is no twist, that is $\mathcal{R} = 1\otimes 1$. In this case, the action reduces to an $L_\infty$ version of the Einstein-Cartan action \eqref{eq:einstein_cartan_action}. 

\paragraph{}
This setting has also a predecessor \cite{Barnes_2016} which studies differential geometry on noncomutative and nonassociative spaces arising from quantization of manifolds via cochain twist deformations.

\subsection{Quantum principal fiber bundle}
\label{subsec:quantum_prin_bunde}
\paragraph{}
One of the earliest attempts to develop a noncommutative version of gauge theory was done in \cite{Brzezinski_1993} through the notion of quantum principal fiber bundle, also called Hopf-Galois extension. This object aims at generalizing the notion of principal fiber bundles when the structure group is a quantum group (\textit{i.e.}\  a Hopf algebra) and so it uses algebraic versions of all elements of principal bundles. Note that other proposals to describe the quantum version of a principal bundle exist in the literature, see e.g.\ \cite{Pflaum_1994, Durdevic_1996}.

\paragraph{}
A quantum principal fiber bundle is defined as an algebraic dual version of a classical fiber bundle. More precisely\footnote{Notations we choose here are different from the ones in \cite{Brzezinski_1993} for consistency with the global notations of this work.}, 
$\algH\to \mathcal{P}\to \algA$ is said to be a quantum principal fiber bundle with universal differential calculus, structure group $\algH$ and base space $\algA$ if
\begin{enumerate}[label = (\roman*)]
	\item \label{it:qpfb_hopf}
	$(\algH, m, 1, \Delta, \varepsilon, S)$ is a Hopf algebra,
	\item \label{it:qpfb_coaction}
	$(\mathcal{P},\Delta_R)$ is a right $\algH$-comodule algebra,
	\item \label{it:qpfb_inv}
	$\algA=\{p\in\mathcal{P}, \; \Delta_R(p) = p\otimes 1 \}$,
	\item \label{it:qpfb_freeness}
	$(m\otimes \mathrm{id})\circ (\mathrm{id} \otimes \Delta_R):\mathcal{P\otimes P\to P\otimes H}$ is a surjection (freeness condition),
	\item \label{it:qpfb_exactness}
	$\mathrm{Ker}({}^{v})=\Omega^1(\mathcal{P})_{\mathrm{hor}}$ (exactness condition).
\end{enumerate}

There are several ingredients to define here. 

First, $\algH$ is the quantum version of the structure group. Then, the smooth manifold bundle is generalized by an algebra $\mathcal{P}$ and the smooth manifold constituting the base space is generalized by an algebra $\algA$. 1-forms are considered instead of vector fields.

Besides, the coaction $\Delta_R:\mathcal{P\to P\otimes H}$ defined in condition \ref{it:qpfb_coaction} is a dual version of the right action of the structure group on the principal bundle. The freedom condition \ref{it:qpfb_freeness} of this coaction is the dual version of the freedom condition of the usual action that states the induced right action is an inclusion. 

Moreover, the algebra $\algA$ is chosen to be the subalgebra of $\mathcal{P}$ invariant under the coaction $\Delta_R$ (condition \ref{it:qpfb_inv}) to mimic smooth functions on the base space manifold. 

Finally, there is a canonical inclusion $j:\algA\hookrightarrow \mathcal{P}$ which corresponds to the canonical projection in the classical case. This allows to define horizontal $1$-forms as 
\begin{equation}
    \Omega^1(\mathcal{P})_\mathrm{hor} 
    = \mathcal{P}j(\Omega^1(\algA))\mathcal{P}
    \subseteq \Omega^1(\mathcal{P}).
\end{equation}
The map ${}^{v}$ realizes an extension of the vertical component of a vector field (also called fundamental vector field) for the action $\Delta_R$ through 
\begin{align*}
	{}^{v} = (m_\mathcal{P} \otimes \mathrm{id}) \circ (\mathrm{id} \otimes \Delta_R) \big|_{\mathrm{Ker}(m_\mathcal{P})} : \Omega^1(\mathcal{P}) \to \mathcal{P} \otimes \algH,
\end{align*}
with $m_\mathcal{P}$ the product in $\mathcal{P}$. Thus, the exactness condition \ref{it:qpfb_exactness} imposes that the two definitions of horizontal and vertical 1-forms coincide.

\paragraph{}
From the above considerations, one can define the left sub-$\mathcal{P}$-module of vertical forms as 
\begin{equation}
    \Omega^1(\mathcal{P}) 
    = \Omega^1(\mathcal{P})_\mathrm{hor} \oplus \Omega^1(\mathcal{P})_\mathrm{ver},
\end{equation}
or equivalently, 
\begin{equation}
    \Omega^1(\mathcal{P})_\mathrm{ver}
    = {}^{v}(\Omega^1(\mathcal{P})).   
\end{equation}
Then, the projection $\Pi : \Omega^1(\mathcal{P}) \to \Omega^1(\mathcal{P})_\mathrm{ver}$, given the right invariance property $\Delta_R\Pi = (\Pi \otimes \mathrm{id})\Delta_R$, is a connection on $\mathcal{H\to P\to A}$. Moreover, every connection $\Pi$ can be associated with a connection form $\nabla : \algH\to \Omega^1(\mathcal{P})$. Conversely, for a connection $1$-form $\nabla$, $\Pi = m_\mathcal{P}\circ(\mathrm{id}\otimes\nabla)\circ {}^{v}$ is a connection.

\paragraph{}
This notion was improved in \cite{Hajac_1996}, through the notion of strong connection, to better fit with the classical definition. From the latter notion, strong tensorial forms were also defined together with a (global) curvature form satisfying a quantum version of classical properties that the curvature automatically satisfies.

\paragraph{}
The notion of trivial quantum bundle is defined in a straightforward manner in order to recover a trivial bundle when considering the algebra of smooth functions. Let us make explicit the gauge transformations on these bundles. 

Let $\algH \to \mathcal{P} \to \algA$ be a quantum principal bundle. It is said to be trivial if there exists a trivialization $\phi : \algH \hookrightarrow \mathcal{P}$, that is a convolution invertible map that satisfies $\Delta_R\circ\phi = (\phi\otimes \mathrm{id})\circ \Delta$ and $\phi(1) = 1$. 

The convolution is taken here in the sense of convolution between an algebra and a coalgebra. Explicitly, let $(\algA_1,m_1,\eta_1)$ be an algebra and $(\algA_2,\Delta_2,\varepsilon_2)$ be a coalgebra, then one defines for any $f_1,f_2\in\mathrm{Lin}(\algA_2,\algA_1)$, $f_1 \ \tcvp\  f_2 = m_1 \circ (f_1 \otimes f_2) \circ \Delta_2$. The neutral element of $\tcvp$ is $\eta_1 \circ \varepsilon_2$ and $f\in\mathrm{Lin}(\algA_2,\algA_1)$ is said to be convolution invertible if there exists a map $f^{-1}\in\mathrm{Lin}(\algA_2,\algA_1)$ such that $f \ \tcvp\  f^{-1} = f^{-1} \ \tcvp\  f = \eta_1 \circ \varepsilon_2$.

Let $g: \algH \to \mathcal{P}$ be a convolution invertible linear map such that $g(1)=1$, then $\phi^g = (j\circ g)\ \tcvp\  \phi$ is a gauge transformation of $\phi$. This gauge transformation can be interpreted as a change of local coordinates in $\mathcal{P}$.

Finally, the locally trivial quantum bundle is defined to be a collection of trivial bundles pasted together via gauge transformations, just as in the classical case.

\paragraph{}
Note that all these notions are defined for the universal differential calculus over $\mathcal{P}$. However, they can be defined for other differential calculi like the bicovariant one. In such a case, additional constraints are put in some of the above definitions to make these objects compatible with the bicovariant structure.

\paragraph{}
This topic has exhibited many interesting features. However, since we are only interested in gravity theories in a noncommutative framework, we will not review the works dealing with quantum principal bundles.

\subsubsection{Frame resolution}
\label{subsubsec:frame_resolution}
\paragraph{}
The study of gravity within this setting rapidly turned to Riemannian geometry \cite{Majid_1997}. The idea lies in a new formalism of classical Riemannian geometry based on frame resolution, which we now recall.

Let $G\to\mathcal{P\to M}$ be a principal fiber bundle over a manifold $\mathcal{M}$, and $V$ a representation of $G$, \textit{i.e.}\  a left $G$-module. Let $\mathcal{E = P} \times_G V$ be an associated vector bundle, and $e \in \Omega_{\mathrm{tens}}^1(\mathcal{P},V)$, a tensorial 1-form, that is a $G$-equivariant and horizontal form. $(\mathcal{P},G,V,e)$ is said to be a frame resolution over the tangent bundle if $\Omega_{\mathrm{tens}}^1(\mathcal{P},V)$ is isomorphic to bundle maps $T\mathcal{M\to E}$.

This definition comes from the example of the frame bundle $\mathcal{P}=L\mathcal{M}$, $G=GL_d(\mathbb{R})$ and $V=\mathbb{R}^d$, for a $d$-dimensional manifold $\mathcal{M}$. In such a case, the tensorial $1$-form $e\in\Omega^1(L\mathcal{M}, \mathbb{R}^d)$ completely determines the frame bundle. This is the precise reason why only $e$ and not all $\Omega_{\mathrm{tens}}^1(\mathcal{P},V)$ is a datum of the frame resolution.

Dualizing this definition allows one to define a frame resolution over 1-forms. Explicitly, $(\mathcal{P},G,V,e)$ is said to be a frame resolution over the 1-forms if $\Omega_{\mathrm{tens}}^1(\mathcal{P},V')$ is isomorphic to bundle maps $\mathcal{E}'\to T'\mathcal{M}$, where $V'$ is the dual of $V$, $\mathcal{E}'=\mathcal{P}\times_G V'$ and $T'\mathcal{M}$ is the cotangent bundle over $\mathcal{M}$. Furthermore, one can enlarge this definition to more general cases through isomorphisms
\begin{align*}
	\Omega^k(\mathcal{M}) \otimes_{\mathcal{C^\infty(M)}} \Omega^1(\mathcal{M}) &\simeq \Omega_{\mathrm{tens}}^k(\mathcal{P}, V'), &
	\Omega^k(\mathcal{M}) \otimes_{\mathcal{C^\infty(M)}} T\mathcal{M} &\simeq \Omega_{\mathrm{tens}}^k(\mathcal{P}, V), &
	k \in \mathbb{N}_0.
\end{align*}

Through these isomorphisms, a connection on $\Omega^1(\mathcal{M})$, its torsion and its curvature are defined in $\Omega_{\mathrm{tens}}$ with the isomorphisms given above. Moreover, the metric $\mathrm{g}: T\mathcal{M} \to \Omega^1(\mathcal{M})$ is isomorphic to $\gamma\in\Omega^1_{\mathrm{tens}}(\mathcal{P},V')$. Furthermore, given $(\mathcal{P}, G, V, e)$ a frame resolution then $\mathrm{g}$ is non-degenerate if and only if $(\mathcal{P},G,V',e')$ is itself a frame resolution, called the dual frame resolution or coframe resolution.

This key point allows metric compatibility to be akin to co-torsion freedom, and the co-torsion corresponds here to the covariant derivative of $\gamma$. Therefore, a Levi-Civita connection is equivalent within the frame resolution set-up to a torsionless, cotorsionless condition on the frame resolution.

In the case of a trivial bundle, $e$ is the tetrad (called $V$-bein in the paper), and $e'$ corresponds to the inverse tetrad (called $V$-cobein in the paper).

\paragraph{}
The quantum version of the above frame resolution is now reviewed.

A frame resolution of $(\algA, \Omega^1(\algA))$ is a quantum principal bundle $\algH \to \mathcal{P} \to \algA$, a right $\algH$-comodule $V$ and an $\algH$-equivariant map $e:V\to \mathcal{P}\Omega^1(\algA)$ such that the induced left $\algA$-module map is an isomorphism $s_e : \mathcal{E} \to \Omega^1(\algA)$ by applying $e$ and multiplying by an element in $\mathcal{P}$, $\mathcal{E}$ being an $\algH$-equivariant element of $\mathcal{P}\otimes V$ which encodes a quantum version of vector fields.

For the dual version of the frame resolution, one needs to define a left action on the opposite algebra $\algH^{\mathrm{op}}$, dual to the original right action $\Delta_R$. The coframe resolution then relies on $V'$ and $e':V'\to \Omega^1(\mathcal{A)P}$, giving an isomorphism $\mathcal{E}'\simeq \Omega^1(\algA)$. The metric corresponding to this frame resolution is then given by
\begin{align}
	g = e'(u^j) \otimes_\mathcal{P} e(u_j) 
	\;\in \Omega^1(\algA)\otimes_\algA \Omega^1(\algA),
\label{eq:metric_frame_resolution_qpfb}
\end{align}
for $\{u^j\}_j$ a basis of $V$ and $\{u_j\}_j$ a basis of $V'$. This constitutes the quantum self-dual version of Riemannian geometry.

One can also define a strong connection in this framework followed by its torsion, cotorsion and curvature.

\paragraph{}
Note that one of the missing key elements of the previous model is the lack of some topological structure. This is introduced via two concepts. First, the notions of involution $*$ and related $*$ objects are brought by the formalism of connection on a central bimodule developed in section \ref{subsec:central_bimodules}. Next, the connection compatibility is introduced through bar categories. Those were made by \cite{Beggs_2009}.

The authors further applied this theory to $1+1$-dimensional $\kappa$-Minkowski \cite{Beggs_2014}, by studying all central quantum-symmetric metrics. For $d>2$, they show that there are only timeless metrics. The $d+1$-dimensional case is extrapolated considering its $1+1$-dimensional rotational symmetric version. The differential calculus considered is the most general one compatible with the coordinate commutation relations. Their classical limit leads to a wider structure than the usual gravity on Minkowski space and generates non-trivial geodesics. Furthermore, a full noncommutative theory is elaborated and leads to a unique pair of Levi-Civita connections, one vanishing at the commutative limit. The Ricci scalar is also shown to be non-zero, with corrections starting at the second order, and with a singularity at $r=0$, $r$ being the radial coordinates. 

\paragraph{}
The above idea was also used in \cite{Borowiec_2016}, for $3+1$-dimensional $\kappa$-Minkowski where the basic element is not the central metric but central coframe field. Asking for this field to be central almost fixes it and forces it to be static, \textit{i.e.}\  time-independent. The differential calculus is similar to the previous one and the metric is recovered here through \eqref{eq:tetrad_def}. Solutions with free parameters are obtained, again extending the commutative limit for different types of matter.

\paragraph{}
This quantum principal fiber bundle based on Hopf-Galois extension was shown to be compatible with Drinfeld twists \cite{Aschieri_2016a, Aschieri_2016b}. One can either twist the structure group $\algH$ of the bundle and the automorphism group of the principal bundle independently but in a compatible way, or twist the all Hopf-Galois extension structure.

\paragraph{}
More recently, this framework was enlarged to non-affine bases with a sheaf theoretic approach \cite{Aschieri_2021b}, based on the formulation of \cite{Pflaum_1994}. This definition is built on cleft locality, that is on having a finite cover of open cleft extensions. This condition is akin to being locally trivial. Differential calculus on the sheaf trivialization can also be constructed \cite{Aschieri_2021c} using a differential calculus on the total space which is covariant under the structure group.

\subsubsection{Principal comodule algebras}
\label{subsubsec:prin_comod_alg}
\paragraph{}
The study of quantum principal fiber bundles turned to a general formulation  expressed through principal comodule algebras, also called $C$-Galois extensions \cite{Brzezinski_1996, Brzezinski_1998}.

Let $\algH$ be a Hopf algebra with bijective antipode and $\mathcal{P}$ a right $\algH$-comodule algebra with coaction $\Delta_\algH: \mathcal{P} \to \mathcal{P} \otimes \algH$, multiplication $m : \mathcal{P} \otimes \mathcal{P} \to \mathcal{P}$ and unit $\eta:\mathbb{C} \to \mathcal{P}$. The map $\ell : \algH \to \mathcal{P\otimes P}$ is called a strong connection if 
\begin{subequations}
\begin{align}
	\ell(1) 
	&= 1\otimes 1 &
	m \circ \ell 
	&= \eta \circ \varepsilon \\
	(\ell \otimes \mathrm{id}) \circ \Delta 
	&= (\mathrm{id} \otimes \Delta_\algH) \circ \ell &
	(S \otimes \ell) \circ \Delta
	&= (\tau \otimes \mathrm{id}) \circ (\Delta_\algH \otimes \mathrm{id}) \circ \ell,
\end{align}
\end{subequations}
where $\tau : \mathcal{P} \otimes \algH \to \algH \otimes \mathcal{P}$ is the flip map. If an $\algH$-comodule algebra $\mathcal{P}$ admits a strong connection, then it is called a principal $\algH$-comodule algebra.

\paragraph{}
Through this definition, the principal comodule algebras turn out to be rigidely linked with the notion of strong connection \cite{Hajac_1996} presented above \cite{Brzezinski_2012}.

In terms of geometry, the interpretation is similar to the quantum principal fiber bundle presented above. Explicitly, $\algH$ plays the role of the structure group, $\mathcal{P}$ corresponds to the principal bundle and the base space is the subalgebra of the coaction invariants. 

\paragraph{}
The notion of gauge transformation is also included in this formalism \cite{Brzezinski_1998} and meets the one of the quantum principal bundle. It consists of convolution invertible linear maps that satisfy an algebraic version of the verticality condition.

\subsection{Fuzzy spaces}
\label{subsec:fuzzy_spaces}
\paragraph{}
A fuzzy space $\algA$ refers to a subclass of quantum spaces, \textit{i.e.}\ of noncommutative algebra, that is generated by noncommutative ``coordinates'' $q^\mu$. They satisfy the relations
\begin{align*}
	[q^\mu, q^\nu]
	&= i \kbar q^{\mu\nu},
	& [q^{\mu_1}, q^{\mu_2 \cdots \mu_p}] 
	= i \kbar q^{\mu_1 \cdots \mu_p}.
\end{align*}
The $q^{\mu_1 \cdots \mu_p}$ are defined recursively with mass unit $p-2$ and must satisfy the Jacobi identity. $\kbar$ is a fundamental area scale, presumably linked to the squared Planck scale.

In the commutative limit, \textit{i.e.}\  when $\kbar \to 0$, $q^\mu$ matches the commutative coordinates $x^\mu$, up to some renormalization factors. The $q^{\mu_1 \cdots \mu_p}$ then goes to elements $x^{\mu_1 \cdots \mu_p}$. Because of Lorentz invariance, $x^{\mu_1 \cdots \mu_p}$ for $p\geqslant 3$ are functions of $x^\mu$ and $x^{\mu\nu}$ and only $4$ of the $6$ elements $x^{\mu\nu}$ are independent. This implies that extra dimensions appear in the commutative limit except if the $q^{\mu\nu}$'s are chosen nilpotent.

\paragraph{}
The study of such fuzzy spaces is motivated by quantum gravity considerations. For details about the physical implications of such a model see \cite{Madore_1997, Madore_1998, Madore_1999}. The main idea the authors put forward is that these noncommutative spaces introduce a natural UV cut-off in the theory due to the fuzziness of the coordinates. Contrary to discrete structures, fuzzy spaces allow to preserve Lorentz covariance while introducing this cut-off. The latter is (expected to be) useful for the regularization of the perturbative expansion of quantum field theories. Still, the space-time structure is expected to be dynamically determined by the gravity. Therefore, a gravity theory on fuzzy spaces could match with an old idea in which the gravitational field enforces the UV regularization of the theory. One can illustrate this idea through the following diagram
\begin{align*}
	\text{Gravity}
	\longrightarrow \text{Fuzzy space-time}
	\longrightarrow \text{Cut-off}
	\longrightarrow \text{Regularized quantum fields}.
\end{align*}
Finally, one could close the diagram by arguing that gravity is determined by the matter content, \textit{i.e.}\  the quantum fields, through Einstein equation.

\paragraph{}
Derivations on the algebra $\algA$ are considered to be finitely generated, that is generated by a finite number, denoted by $N$, of elements $e_I$. It is the analogue of the frame fields. Moreover, they are supposed to be inner, that is there exists $N$ antihermitian elements of $\algA$, noted $\lambda_I$, such that $e_I(f) = [\lambda_I, f]$, for $f\in\algA$. The dimension $N$ may be different from the space-time dimension $d+1$, but one should have $N \geqslant d+1$. However, in the case $N > d+1$, the interpretation of these extra dimensions remains unclear.

One then defines the differential calculus first through the differential $\dd$ by Koszul formula \eqref{eq:koszul}, which writes for first degree forms $\dd f(e_I) = e_I(f) = [\lambda_I, f]$. Then, the existence of $N$ forms $\vartheta^I$, dual to $e_I$, generating the 1-forms $\Omega^1(\algA)$ is hypothesized. The duality condition writes $\vartheta^I(e_J) = \delta^I_J$. It is the analogue of the coframe fields. The previous definition has the straightforward consequence that $\Omega^1(\algA)$ is generated by only one element $\vartheta = \lambda_I \vartheta^I$ as 
\begin{align*}
	\dd f 
	&= [\vartheta, f], &
	f \in \algA.
\end{align*}
The wedge product is then defined through the central elements $\tensor{P}{_{IJ}^{KL}}$ by $\vartheta^I \wedge \vartheta^J = \tensor{P}{_{IJ}^{KL}} (\vartheta^K \otimes \vartheta^L)$. This element must satisfy $\tensor{P}{_{IJ}^{KL}} = \frac{1}{2}(\delta_I^K \delta_J^L - \delta_I^L \delta_J^K) + \mathcal{O}(\kbar)$ to recover a good commutative limit. One can go further and define a constraint on this setting through
\begin{align*}
	2 \lambda_K \star \lambda_L \star \tensor{P}{_{KL}^{IJ}} - \lambda_K \star \tensor{F}{^K_{IJ}} - K_{IJ}
	= 0,
\end{align*}
with $\dd \vartheta + \vartheta \wedge \vartheta = - \frac{1}{2} K_{IJ} \vartheta^I \wedge \vartheta^K$, $\dd \vartheta^I = - \frac{1}{2} \tensor{C}{^I_{JK}} \vartheta^J \wedge \vartheta^K$ and $\tensor{F}{^I_{JK}} = \tensor{C}{^I_{JK}} + 2\lambda_L \star \tensor{P}{^{(IL)}_{JK}}$.

\paragraph{}
Once 1-forms are constructed, one defines a connection $\nabla:\Omega^1(\algA) \to \Omega^1(\algA) \otimes_\algA \Omega^1(\algA)$ through the Leibniz rule \eqref{eq:nc_conn_form_def}. It is thus determined by its components
\begin{align*}
	\nabla(\vartheta^I) 
	= - \omega^I_{JK} \vartheta^J \otimes \vartheta^K.
\end{align*}
It can be promoted to a connection on higher order forms through a graded Leibniz rule. This allows to define the curvature $2$-form $\Omega$ and so the Riemann tensor $R$ through 
\begin{align*}
	\nabla^2(\vartheta^I)
	&= - \tensor{\Omega}{^I_J} \otimes \vartheta^J &
	\tensor{\Omega}{^I_J}
	&= \frac{1}{2} \tensor{R}{^I_{JKL}} \vartheta^K \wedge \vartheta^L.
\end{align*}
Finally, the torsion is defined as $T = \dd - \wedge \circ \nabla$. One defines a metric as a bimodule map $g : \Omega^1(\algA) \otimes \Omega^1(\algA) \to \algA$ such that $g(\vartheta^I \otimes \vartheta^J) = \eta^{IJ}1$, where $\eta$ is the Minkowski metric.

\paragraph{}
The interpretation of $N$ has led to different theories of gravity using this formalism. 

First, $N$ may refer to extra dimensions of space-time. Contrary to usual Kaluza-Klein theories, the extra dimensions are here ``noncommutative'' and have a finite number of modes. This may be done by considering the almost commutative algebra $\algA = \mathcal{C^\infty(M)} \otimes \mathbb{M}_N(\mathbb{C})$, where $\mathcal{M}$ is a $d+1$-dimensional space-time manifold \cite{Madore_1997, Madore_2015}. Then $N$ is here linked to the dimension of the matrix algebra that encodes the extra dimensions. Note that this model is suitably fitted for matrix algebras as they have only inner derivations and are finitely generated. Moreover, such a model involves a natural trace that is defined by both the integral on $\mathcal{M}$ and the trace of matrices and so enables to define an action.

The dimension of the phase-space can also be an interpretation of $N$. In this case, the $\lambda_I$ are interpreted as momenta (and noted $p$) \cite{Maceda_2004, Buric_2005, Buric_2006b}.

Eventually, $N$ may be interpreted as the dimension of the flat manifold that embeds the space-time manifold \cite{Madore_1998}.

\subsection{Star-product incorporation}
\label{subsec:star_prod_inc}
\paragraph{}
Another widely used idea was to consider different formulations of gravity but replacing usual products by a noncommutative product $\star$. This idea emerged in the early 2000 and has triggered a lot of interest. At that time, the physics community was enthusiastic about the Moyal space-time and the promises of the Seiberg-Witten map. Therefore, this formulation of noncommutative gravity follows a general procedure: Consider an action formulation of gravity, replace usual the product by Moyal star-product, compute the curvature and express noncommutative quantities with commutative quantities using the Seiberg-Witten map.

\paragraph{}
This procedure shares some similarities with the old setting of braided geometry (see section \ref{subsec:braided_geometry}), but it is sometimes not emphasised or rigorously enlightened.

The link between Moyal space and gravity was first made within string theory and a whole domain of the literature focuses on the two as evoked in the review \cite{Szabo_2006}.

\paragraph{}
Concerning gauge theories, the star-product incorporation faces the difficulty of inconsistent gauge transformations. It can be formulated as part of a wider problem due to the noncommutativity of the base space.

We consider $\mathfrak{g}$ to be the Lie algebra of a Lie group $G$, with a matrix representation, and $A$ to be a $\mathfrak{g}$-valued connection. Then, the noncommutative analogue of $A$ is an element of $\mathfrak{g} \otimes \algA$, that is a matrix with coefficients in $\algA$. However, since $\algA$ is noncommutative, the Lie algebra closure rules are likely to be broken. Explicitly, for $\mathfrak{a} \otimes f, \mathfrak{b} \otimes g \in \mathfrak{g} \otimes \algA$, one has
\begin{align*}
	[\mathfrak{a} \otimes f, \mathfrak{b} \otimes g]
	&= [\mathfrak{a}, \mathfrak{b}]_\mathfrak{g} \otimes fg + \mathfrak{ab} \otimes [f,g]_\star.
\end{align*}
The first term is stable in $\mathfrak{g} \otimes \algA$ but the second is not and does not vanish as the star-product is not commutative. 

The usual solution to this issue is to consider that the connection takes values in $U(\mathfrak{g}) \otimes \algA$, where $U(\mathfrak{g})$ is the universal enveloping algebra of $\mathfrak{g}$. Then, conditions are imposed on $A$ to recover a connection that takes values in $\mathfrak{g} \otimes \algA$. For the most used case of $G = U(N)$, and so $\mathfrak{g} = \mathfrak{u}(N)$, $A$ takes values in $\mathbb{M}_N(\mathbb{C}) \otimes \algA$. To recover $U(N)$ from $\mathbb{M}_N(\mathbb{C})$, a hermiticity condition is imposed on $A$ using the involution of $\algA$.

\paragraph{}
The star-product incorporation scheme faces unsolved difficulties. In particular, going from commutative quantities to noncommutative ones raises the question of terms ordering before trading the usual commutative product for the star-product.

Furthermore, the combination of the Moyal product and the Seiberg-Witten map has shown to lead to very cumbersome computations, even when considering only the lowest order of non-vanishing corrections of the commutative quantities. Thus, the analysis cannot usually be performed beyond the computation of the curvature.

Finally, some inconsistencies have been put forward like in \cite{Aschieri_2011} where the Seiberg-Witten map does not ``commute'' with the computational procedure to derive the equations of motion, as explained below.

\subsubsection{\tops{$U(N)$}{U(N)} gauges theories}
\label{subsubsec:star_prod_inc_Un_gauges}
\paragraph{}
In $1+1$ dimensions, \cite{Cacciatori_2002a} used a $U(1,1)$ gauge theory of topological gravity as a starting point to build a noncommutative model. The usual product is replaced by the Moyal product and diffeomorphism invariance is replaced by an invariance under deformed diffeomorphisms. Finally, the notion of metric is defined as in the commutative case, via \eqref{eq:tetrad_def}, and some noncommutative analogues of classical metrics are studied. A similar study was performed in $2+1$ dimensions \cite{Cacciatori_2002b} and in $4+2$ dimensions \cite{Deliduman_2006}. An approach similar to \cite{Cacciatori_2002a} was considered in \cite{Balachandran_2006} with a deformed Poincar\'{e} algebra.

This was further claimed to be a low-energy limit of a bosonic string theory \cite{Alvarez_2006}. However, the deformed diffeomorphism symmetry was not recovered.

\paragraph{}
Chern-Simons supergravity in odd dimensions was also studied within this context \cite{Castellani_2013}, especially the $5$-dimensional case with gauge symmetry $SU(2,2)$ \cite{Aschieri_2014}. See also \cite{Dordevic_2022}.

\paragraph{}
The author of \cite{Chamseddine_2003} proposed a noncommutative version of gravity based on a $U(2,2)$ gauge gravity that breaks to $U(1,1)\times U(1,1)$. The breaking is performed through constraints that can lead either to a topological Gauss-Bonnet term, a conformal gravity or Einstein gravity. The action formulation is metric-free. The noncommutative version is implemented by replacing the usual product by the Moyal product. However, in this noncommutative version, the constraints cannot be tuned to recover Einstein gravity.

In \cite{Cardella_2003}, a similar approach is considered but exploits the classical breaking of $U(2,2)$ into $SO(1,3)$ to analogously break ``noncommutative'' $U(2,2)$ into a noncommutative analogue of $SO(1,3)$. The commutative limit of $SO(1,3)$ gauge theory is recovered.

\subsubsection{Gauge theories on real gauge groups}
\label{subsubsec:star_prod_inc_real_gauges}
\paragraph{}
As far as the noncommutative versions of special orthogonal and spin gauge theories are concerned, it appears that the procedure is quite similar as for $U(N)$ gauges. It relies on considering subgroups of $U(N)$ that are stable by an anti-automorphism (an automorphism that reverses the order of product factors), which supplements the involution in $\algA$. Then, a hermiticity condition for the latter anti-automorphism is imposed, in addition to the hermiticity condition for the involution of $\algA$. This anti-automorphism plays the role of the transposition in usual $SO$ and $Sp$ groups.

\paragraph{}
For canonical noncommutativity $[x^\mu,x^\nu] = i\Theta^{\mu\nu}$, \cite{Bonora_2000} uses $\Theta$ as a dynamical variable and generates the anti-automorphism $f(x,\Theta) \mapsto f(x,-\Theta)$, for $f\in\algA$. The hermiticity condition for the latter anti-automorphism is imposed via $A(x,\Theta) = - \tensor[^t]{A}{}(x,-\Theta)$, where ${}^t$ is the transposition. These conditions imply that the components of $A$ occurring in its expansion in power series of $\Theta$, are all real. This setting raises the issue of the low energy limit as $\Theta$ must be a constant within this limit. \cite{Bars_2001} exhibited other possible anti-automorphisms for $O(N)$ and $Sp(2N)$. Finally, \cite{Jurco_2000a} generalized this procedure for an arbitrary finitely generated Lie algebra $\mathfrak{g}$. The main idea is that any $U(\mathfrak{g})$-valued connection can be expressed by a $\mathfrak{g}$-valued connection via a Seiberg-Witten map.\\

Using this last method, \cite{Vacaru_2001} proposed a noncommutative version of a metric-affine gauge gravity. Metric-affine gravity is a generalization of the Poincar\'{e} gauge gravity that relies on gauging the affine group and uses symmetry breaking to recover the $SO(1,3)$ gauge version of gravity with the inverse tetrad arising as the Goldstone boson (see for example \cite{Sardanashvily_2016}). Also with the latter method, \cite{Dimitrijevic_2012, Dimitrijevic_2014} used a noncommutative version of $SO(2,3)$ to deform Stelle-West gravity \cite{Stelle_1979, Stelle_1980}. The Stelle-West gravity is a $SO(1,4)$ gauge version of gravity with an extra scalar field of the MacDoweel-Mansouri type, and so Einstein gravity is recovered via a process of symmetry breaking due to gauge fixing of the scalar field. The authors obtained second order corrections to the Einstein-Hilbert action \eqref{eq:einstein_hilbert_tetrad_action}. Recall that the Einstein-Hilbert action writes in the tetrad formalism
\begin{equation}
	S_{\mathrm{EH}}(e) 
	= \frac{c^4}{16\pi G} \int \dd^{d+1}{x}\; |\det(e)| \big( \tensor{e}{_I^\mu} \tensor{e}{_J^\nu} \tensor{R}{_{\mu\nu}^{IJ}} - (d-1)\Lambda \big),
    \label{eq:einstein_hilbert_tetrad_action}
\end{equation}
where $e$ is the tetrad, $R$ the Riemann tensor and $\tensor{e}{_I^\mu} \tensor{e}{_J^\nu} \tensor{R}{_{\mu\nu}^{IJ}}$ corresponds to the Ricci scalar in tetrad formalism, $\Lambda$ the cosmological constant, $G$ the gravitational constant and $c$ the speed of light. First, the two models, one obtained by performing the symmetry breaking first and then the ($\star$ product) deformation and the other one obtained by doing the opposite, do not lead to the same results so that deformation and symmetry breaking do not commute. Furthermore, cosmological constant-like terms arise and are position-dependant in some regimes.\\
A recent review \cite{Dimitrijevic_2022}, by some of the latter authors, was written on this topic, together with deformed Chern-Simons (super)gravity discussed in section \ref{subsubsec:star_prod_inc_Un_gauges}.

\paragraph{}
Finally, \cite{Moffat_2000} used a complex version of the tetrad based on  the gauge group for complex symmetric metric $CSO(1,3)$. Then, the Moyal star-product was incorporated in \eqref{eq:tetrad_def} to have a complex symmetric metric. Gravity with a complex symmetric metric also contains negative energy ghost states that vanish in the case of an hyperbolic symmetric metric.

\cite{Chamseddine_2001b} used a noncommutative version of $SO(1,4)$ which is broken into a noncommutative version of $ISO(1,3)$. The noncommutative $SO(1,4)$ is defined here as $U(1,4)$ plus a reality condition. The Einstein-Hilbert action \eqref{eq:einstein_hilbert_tetrad_action} with the Moyal product is considered and a torsionless condition is imposed by hand. The expansion of the action up to second order in $\Theta$ has been undertaken but not completed as the computations are very involved.

\subsubsection{\tops{$SL_2(\mathbb{C})$}{SL\_2(C)} gauge}
\label{subsubsec:star_prod_inc_sl_gauges}
\paragraph{}
Starting from the $SL_2(\mathbb{C})$ gauge version of gravity developed in \cite{Isham_1972}, the author of \cite{Chamseddine_2004} introduced a complexified tetrad. This turns the theory into a bigravity-like theory with a massless graviton (the real part of the metric) and a massive graviton (the imaginary part of the metric) coupled to a scalar field. Then, the author incorporated the Moyal star-product. This turns the theory into a $GL_2(\mathbb{C})$ gauge theory. Indeed, one has
\begin{align}
	[\xi, A]_\star = [\xi^aT^a, A^bT^b]_\star = \frac{1}{2} \{\xi^a,A^b\}_\star [T^a, T^b] +  \frac{1}{2} [\xi^a,A^b]_\star \{T^a, T^b\}
	\label{eq:nc_gauge_trans}
\end{align}
where $A$ is the gauge potential, $\xi$ is the transformation parameter, $\{T^a\}_a$ is the set of generators of the algebra that generates the transformation and $\{\cdot, \cdot\}$ is the anticommutator. Therefore, due to the noncommutativity, the term $[\xi^a,A^b]_\star$ does not vanish, imposing that anticommutators of generating elements of the gauge group must now be taken into account. For $SL_2(\mathbb{C})$, these generators can be taken as $\gamma^{ab} = \gamma^a \gamma^b$, which points towards the spinor representation. Anticommutators then lead to add the identity matrix and $\gamma_5$ so that the gauge group turns to $GL_2(\mathbb{C})$. An analysis up to first order in $\theta$, the noncommutative parameter of the Moyal space, is then undertaken but not completed due to the complexity of the computations.

\paragraph{}
This setting was also used in \cite{Aschieri_2009} to make a $GL_2(\mathbb{C})$ $\star$-gauge and $\star$-diffeomorphism invariant action extending the Einstein-Cartan \eqref{eq:einstein_cartan_action} one. The latter paper relies on the formulation of braided wedge product developed in section \ref{subsec:braided_geometry}. Still, this model generates a mismatch between getting the solution of the equations of motions then using the Seiberg-Witten map, and doing the converse \cite{Aschieri_2011}. Therefore, to avoid using this map, \cite{DiGrezia_2014} used this formalism to deform parallelizable space-time with (anti-)self-dual connection. The author obtained a partial differential equation giving the solutions of field equations and applied it to deform the Kasner metric.

\paragraph{}
Another construction based on the former setting was done by \cite{Miao_2010} with a deformation parameter $\theta$ that is position-dependant and considering the symplectic manifold generated by the first order in $\theta$. The symplectic structure considered is similar to \cite{Chaichian_2010} (see section \ref{subsubsec:telepar_grav}), but the symplectic form is built order by order. Within this setting, the first order correction to the curvature does not vanish, but only depends on derivatives of $\theta$, so that it vanishes for a constant deformation.

\paragraph{}
However, other constructions make the $SL_2(\mathbb{C})$ gauge arise. Starting from an $SO(1,3)$ invariant action of topological gravity\footnote{This action writes
	\begin{equation}
		S = \int_{\mathcal{M}} R(\omega) \wedge R(\omega),
		\label{eq:action_topo_grav}
	\end{equation}
where $\mathcal{M}$ is a $3+1$-dimensional manifold and $R$ is the curvature of the spin connection $\omega$.} and using the decomposition $SO(1,3) = SL_2(\mathbb{C}) \otimes SL_2(\mathbb{C})$, \cite{Garcia_2003a} made a noncommutative topological gravity theory with star-product incorporation.

The same authors further used the self-dual Pleba\'{n}ski \cite{Plebanski_1977} $SL_2(\mathbb{C})$ gauge version of gravity to extract a noncommutative version of it \cite{Garcia_2003b}. The previous formalism was adapted to twist the deformation of Moyal braided geometry (see details in section \ref{subsec:braided_geometry}) to include the deformed diffeomorphism invariance \cite{Estrada_2008}. The deformed diffeomorphisms correspond to diffeomorphisms for the deformed (Moyal) product.

\paragraph{}
A $3$-dimensional version of gravity with $GL_2(\mathbb{C})$ gauge group was performed on canonical noncommutativity, like $\mathbb{R}^3_\lambda$, \cite{Banados_2001, Chatzista_2018}.

\subsubsection{Other star-products}
\label{subsubsec:other_star_prod}
\paragraph{}
A deformation of the Einstein-Hilbert action up to the second order in the deformation parameter was considered in \cite{Dobrski_2011}. Instead of using the Moyal product, the author used the Fedosov product of endomorphisms on a symplectic manifold. The Moyal product can still be reconstructed through an explicit isomorphism. Even if the deformation is not unique, it was found that, up to the second order, all the information on the deformation is gathered in the metric for all the deformations considered. Finally, the deformed actions are (passive) diffeomorphism invariant since correction terms only involve covariant quantities.

\subsubsection{Teleparallel gravity}
\label{subsubsec:telepar_grav}
\paragraph{}
Teleparallel gravity is a formulation of gravity where local Lorentz symmetry is frozen down to global Lorentz symmetry, and thus describes general relativity macroscopically. One can also define teleparallel gravity as the study of gravity on curvature-free manifolds. For a theory of complex gravity, which may be needed for noncommutative gravity, this feature allows to get rid of negative energy ghosts since the local Lorentz symmetry does not hold anymore. Moreover, the noncommutative trait of space automatically breaks the local Lorentz symmetry so that teleparallel gravity may be a good candidate for a noncommutative version of gravity.

\paragraph{}
The link between teleparallel gravity and canonical noncommutative gauge theory was first put forward in \cite{Langmann_2001}. Starting from a $U(1)$ noncommutative Yang-Mills-like gauge theory on flat $\mathbb{R}^{2n}$, the authors constructed a teleparallel gravity theory on $\mathbb{R}^n$ with infinitesimal diffeomorphism invariance, that is, under change of basis of the tangent space. 

\paragraph{}
\cite{Nishino_2002} used teleparallel gravity with a global $U(1,3)$ gauge symmetry. A constrained Lagrangian is considered so that no negative energy ghosts appear, the irrelevant complex fields do not contribute up to quadratic order and the linearized field equations obtained from the Einstein-Hilbert action \eqref{eq:einstein_hilbert_tetrad_action} are recovered. Finally, noncommutative features appear by replacing the usual product by the Moyal product.

\paragraph{}
A somehow close approach to emergent gravity (see section \ref{subsec:emergent_grav}) is performed in \cite{Chaichian_2010}. The commutator of space-time coordinates is assumed to be non-constant, in a spirit similar to \eqref{eq:nc_dynamical_moyal}. The lowest order in the deformation parameter $\theta$ is considered so that the space-time becomes a Poisson manifold (see \eqref{eq:poisson_moyal}), and so $\theta$ must then fulfil a Jacobi identity. The Poisson manifold can be turned into a symplectic manifold considering $\theta^{-1}$ to be non-degenerate and closed. The connection and curvature are defined in this geometry by defining the Poisson bracket for forms. In their work, the authors considered a curvature-free symplectic manifold and so handled teleparallel gravity. A noncommutative Yang-Mills gauge theory is studied in this symplectic structure.

\subsection{Emergent gravity}
\label{subsec:emergent_grav}
\paragraph{}
Based on a matrix model of $U(N)$ gauge theory on the Moyal space, the author of \cite{Steinacker_2007} developed an appealing theory of emergent gravity. In the usual framework of matrix models, $SU(N)$ gauge symmetry cannot be incorporated because of relation \eqref{eq:nc_gauge_trans} and the fact that $\mathfrak{su}(N)$ is not closed under anticommutation (see equation \eqref{eq:uN_lie_alg_comm}) but $\mathfrak{u}(N)$ is. Thus, the split $U(N) \simeq SU(N) \times U(1)$ is used to recover the usual Yang-Mills theory in $SU(N)$ gauges. However, the remaining $U(1)$ gauge is known to be pathological as it is responsible for the IR divergence in the UV/IR mixing (see subsection \ref{subsubsec:moyal_simplest_YM} for more details). However, this $U(1)$ part can be interpreted as a coupling between the $SU(N)$ gauge field and an emerging gravitational field.

\paragraph{}
In this subsection \ref{subsec:emergent_grav}, the notations of the Seiberg-Witten map are used (see subsection \ref{subsec:moyal_theta_exp} for more details). Explicitly, fields with a hat, e.g.\ $\hat{F}$, correspond to noncommutative fields and their commutative counterpart will be noted without hat, e.g.\ $F$.

\subsubsection{Gauge theory on the Moyal space}
\paragraph{}
We recall here some important features of the matrix model on the Moyal space. The construction of this gauge theory is covered in more details in section \ref{subsec:moyal_YM}.

\paragraph{}
A $U(1)$ gauge theory arises whenever the module is taken to be a copy of the algebra, as detailed in section \ref{subsubsec:nc_mod_is_alg}. In this case, given a connection $\nabla$, one can define the hermitian connection form $\hat{A}_\mu = \nabla_{\partial_\mu}(1)$ and the associate curvature form $i \hat{F}_{\mu\nu} = \partial_\mu \hat{A}_\nu - \partial_\nu \hat{A}_\mu - i [\hat{A}_\mu, \hat{A}_\nu]_\theta$.

However, due to the inner property \eqref{eq:moyal_inner_der} of the derivations, there exists a gauge invariant connection \eqref{eq:moyal_inv_conn} $\nabla^{\inv}_{\partial_\mu}(f) = i f \star_\theta \Theta^{-1}_{\mu\nu}x^\nu$, where $\star_\theta$ is the Moyal product \eqref{eq:moyal_prod}. The associated curvature writes \eqref{eq:Moyal_curv_inv_gauge} $i \hat{F}^{\inv}_{\mu\nu} = \Theta^{-1}_{\mu\nu}$. One then defines the covariant coordinate as the difference of the two connections \eqref{eq:moyal_covar_coord}, that is
\begin{align}
	\hat{\mathcal{A}}_\mu 
	&= - i (\hat{A}_\mu + \Theta^{-1}_{\mu\nu}x^\nu).
	\label{eq:covar_coord_u1}
\end{align}
Then one has \eqref{eq:courb-moyal}
\begin{align}
	\hat{F}_{\mu\nu}
	&= [\hat{\mathcal{A}}_\mu, \hat{\mathcal{A}}_\nu]_\theta - i \Theta^{-1}_{\mu\nu}.
	\label{eq:curv_covar_coord_u1}
\end{align}

The corresponding action writes
\begin{align}
	S 
	&= \int \dd^{2d}x\; \hat{F}_{\mu\nu} \star_\theta \hat{F}^{\mu\nu}
	\label{eq:action_u1}
\end{align}
where $\hat{F}^{\mu\nu} = (g^{\mathrm{B}})^{\mu\sigma} (g^{\mathrm{B}})^{\nu\tau} \hat{F}_{\sigma\tau}$ and $g^\mathrm{B}$ is a background metric \eqref{eq:background_metric}. Note that \cite{Steinacker_2007} considered the equivalent action with curvature form $\hat{F} - \hat{F}^{\inv}$, so that it writes 
\begin{align}
	S 
	&= \int \dd^{2d}x\; [\hat{\mathcal{A}}_\mu, \hat{\mathcal{A}}_\nu]_\theta \star_\theta [\hat{\mathcal{A}}^\mu, \hat{\mathcal{A}}^\nu]_\theta.
	\label{eq:action_covar_coord_u1}
\end{align}

\paragraph{}
The generalization of the previous setting to the $U(N)$ case is made by considering the module to be $N$ copies of the algebra (see section \ref{subsubsec:nc_mod_is_N_alg}). One defines a connection form as previously with $\hat{A}_\mu \in \mathbb{R}^{2d}_\theta \otimes \mathbb{M}_N(\mathbb{C})$. Thus, the connection can be seen as a matrix with elements in $\mathbb{R}^{2d}_\theta$. Note that $\hat{A}$ is required to be anti-hermitian so that $\hat{A}_\mu \in \mathbb{R}^{2d}_\theta \otimes \mathfrak{u}(N)$. One can then make contact with a usual non-Abelian gauge theory considering $\{T^a\}_{a = 0, \dots, N^2-1}$ to be the $N^2$ generators of $\mathfrak{u}(N)$ \eqref{eq:uN_lie_alg_tr} and writing $\hat{A}_\mu = \hat{A}_\mu^a T^a$. The curvature form is then defined as $\hat{F}_{\mu\nu} = \partial_\mu \hat{A}_\nu - \partial_\nu \hat{A}_\mu - i [\hat{A}_\mu, \hat{A}_\nu]_\Theta$. One can write the latter in the $\mathfrak{u}(N)$ basis as such $\hat{F} = \hat{F}^a T^a$, with $\hat{F}^a_{\mu\nu} = \partial_\mu \hat{A}^a_\nu - \partial_\nu \hat{A}^a_\mu + \frac{1}{2} f^{abc} \{\hat{A}^b_\mu, \hat{A}^c_\nu\}_\theta - \frac{i}{2} d^{abc} [\hat{A}^b_\mu, \hat{A}^c_\nu]_\theta$ \eqref{eq:moyal_uN_curv}.

As before, one can build a gauge invariant connection $\nabla^{\inv}_{\partial_\mu}(f^je_j) = i (\partial_\mu(f^j) - \Theta^{-1}_{\mu\nu}x^\nu \star_\theta f^j)e_j$, where $f^j\in\mathbb{R}^{2d}_\theta$ and $e_j = (0, \dots, 0, \overset{(j)}{1}, 0, \dots, 0)$, for any $j= 1, \dots, N$.  One then defines the covariant coordinate as the difference of the two connections, that is
\begin{align}
	\hat{\mathcal{A}}_\mu 
	&= - i (\hat{A}_\mu + \Theta^{-1}_{\mu\nu}x^\nu \bbone_N).
	\label{eq:covar_coord_un}
\end{align}
Note that $\hat{\mathcal{A}}$ could be written in the $\{T^a\}_a$ basis.

Similarly as before, the action writes \eqref{eq:Moyal_action_un}
\begin{align}
	S 
	&= \int \dd^{2d}x\; \tr_{\mathfrak{u}(N)} (\hat{F}_{\mu\nu} \star_\theta \hat{F}^{\mu\nu}),
	\label{eq:action_un}
\end{align}
where once again, indices have been raised thanks to the background matrix $g^\mathrm{B}$. In \cite{Steinacker_2007}, the equivalent action with curvature form $\hat{F} - \hat{F}^{\inv}$ is considered, so that it writes 
\begin{align}
	S 
	&= \int \dd^{2d}x\; \tr_{\mathfrak{u}(N)} \big( [\hat{\mathcal{A}}_\mu, \hat{\mathcal{A}}_\nu]_\theta \star_\theta [\hat{\mathcal{A}}^\mu, \hat{\mathcal{A}}^\nu]_\theta \big).
	\label{eq:action_covar_coord_un}
\end{align}

The equations of motions are found to be 
\begin{align}
	[ \hat{\mathcal{A}}_\mu, [\hat{\mathcal{A}}^\mu, \hat{\mathcal{A}}^\nu]_\theta ]_\theta
	&= 0
	\label{eq:eom_un}
\end{align}
which has a particular solution given by $\hat{\mathcal{A}}^\mu = x^\mu \bbone_N$.

\paragraph{}
Given a star-product, one can construct an effective field theory by considering only the lowest order in the deformation parameter, \textit{i.e.}\  the lowest order in $\Theta$. When doing so, the bracket becomes
\begin{align}
	[f, g]_\theta
	&= \Theta^{\mu\nu} \partial_\mu f \partial_\nu g + \mathcal{O}(\Theta^3)
	= i\{f, g\}_\theta^1 + \mathcal{O}(\Theta^3), &
	f,g \in\mathbb{R}^{2d}_\theta
	\label{eq:poisson_moyal}
\end{align}
and then turns the space-time into a Poisson manifold with Poisson bracket $\{\cdot, \cdot\}_\theta^1$.

\subsubsection{Gravity emergence}
\paragraph{}
In the setting of \cite{Steinacker_2007}, weak deformations are considered so that everything can be associated with the lowest power of $\Theta$ in its power expansion. Weak deformations also allow to make use of the Seiberg-Witten map to express quantities with the usual commutative connection form $A$.

\paragraph{}
In this setting, gravity emerges from two processes. First, considering only the effective theory of the $U(N)$ theory around a specific vacuum leads to the appearance of a dynamical metric. This metric arises thanks to the $U(1)$ gauge field part of the full $U(N)$ gauge field, while the remaining $SU(N)$ gauge field corresponds to the commutative Yang-Mills ones. However, the action \eqref{eq:action_covar_coord_un} does not contain an Einstein-Hilbert-like term. The latter appears upon quantization of the previous action.

\paragraph{}
Consider a perturbation around the vacuum $\hat{\mathcal{A}}_{\mathrm{vac}}^\mu = x^\mu \bbone_N$, given by 
\begin{align}
	\hat{\mathcal{A}}^\mu 
	&= x^\mu \bbone_N + \alpha^\mu(x).
	\label{eq:gauge_pot_vac_pert}
\end{align}
Noting that $\alpha$ is $\mathfrak{su}(N)$-valued and that $T^0 \propto \bbone_N$ \eqref{eq:uN_lie_alg_0}, one can reconstruct the full trace of the gauge potential $\hat{\mathcal{A}}^\mu = \big(x^\mu + (\alpha^0)^\mu \big) \bbone_N + (\alpha^a)^\mu(x) T^a$, where the sum over $a$ does not contain $0$ anymore. The new $\mathfrak{u}(1)$ part of the gauge potential $y^\mu = x^\mu + (\alpha^0)^\mu$ will be considered as the coordinates of the noncommutative space and satisfies the noncommutativity condition
\begin{align}
	[y^\mu, y^\nu]_\theta
	&= \tilde{\Theta}^{\mu\nu}(y).
	\label{eq:nc_dynamical_moyal}
\end{align}
Therefore, one interprets $\alpha^0$ as the fluctuation of the quantum space: it determines the Poisson structure \eqref{eq:poisson_moyal} of the dynamical $\tilde{\Theta}$ and the metric $g$. The latter metric arises when writing the action \eqref{eq:action_covar_coord_un} with gauge potential \eqref{eq:gauge_pot_vac_pert} and using the Seiberg-Witten map in the weak deformation approximation. Such an action writes
\begin{align}
	S
	&= \int \dd^{2d}y\; \rho(y)  \tr_{\mathfrak{u}(N)} \big(4\eta(y)\bbone_N - g^{\mu\lambda} g^{\nu\sigma} F_{\mu\nu} F_{\lambda\sigma} \big) +  2 \eta(y) \tr_{\mathfrak{u}(N)} ( F \wedge F)
	\label{eq:effective_action_gauge}
\end{align}
with the metric $g$ given by
\begin{align}
	g^{\mu\nu}(y)
	&= - \tilde{\Theta}^{\mu\lambda}(y) \tilde{\Theta}^{\nu\sigma}(y) g^\mathrm{B}_{\lambda\sigma},
	\label{eq:effective_metric}
\end{align}
where $\eta(y) = \frac{1}{4} g^{\mu\nu} g^\mathrm{B}_{\mu\nu}$, and $\rho(y) = \sqrt{\det(\tilde{\Theta}^{-1})} =  (\det(g)\det(g^\mathrm{B}))^{\frac{1}{4}}$ is the symplectic volume element of the Poisson manifold. Here $g^\mathrm{B}$ correspond to the background metric defined in \eqref{eq:background_metric}. The commutative field strength $F$ corresponds to the field strength of $A$.

This action \eqref{eq:effective_action_gauge} corresponds to the usual Yang-Mills action in curved space with metric $g$ and extra densities $\eta(y)$ and $\rho(y)$. Both make the background metric $g^\mathrm{B}$ appear explicitly.

\paragraph{}
Interesting features arise. From $U(N)$ gauge theory the coupling between the $SU(N)$ and $U(1)$ parts is essential since it generates the coupling between the gauge potential and gravitational field. The $\eta(y)\rho(y)$ term is not a cosmological constant term, but leads to the vacuum equations of motion of gravity $g^{\mu\lambda}(y) \partial_\lambda \Theta^{-1}_{\mu\nu}(y) = 0$. Those are derived as the equations of motion of the $U(1)$ degree of freedom. It implies that the curvature of $g$ should vanish when considering the dynamical metric to be a perturbation of the flat background: $g = g^\mathrm{B} + h$.

\paragraph{}
To fully incorporate gravity, the action above needs to involve an Einstein-Hilbert-like term. Recall that the Einstein-Hilbert action writes
\begin{align}
    S_{\mathrm{EH}}(g)
    &= \frac{c^4}{16 \pi G} \int \dd^{d+1}x \sqrt{-\det(g)} \big(R(g) - (d-1)\Lambda \big),
    \label{eq:einstein_hilbert}
\end{align}
where $R$ is the Ricci scalar and $\Lambda$ the cosmological constant. However, there is no need to add an extra-term in the action as the Einstein-Hilbert action can arise from quantization. The induced gravity was first proposed by \cite{Sakharov_1967} and later applied to quantization \cite{Visser_2002}.

A linearized gravitational field is considered, namely $g = g^\mathrm{B} + h$. It is shown that even if a $U(1)$ gauge field cannot reproduce all the off-shell degrees of freedom of the graviton, it still gathers the two polarizations of the on-shell graviton.

\paragraph{}
In such formalism, the action for bosons is directly derived in \cite{Steinacker_2007}. The fermionic action was derived in \cite{Klammer_2008}.

\subsubsection{Other approaches}
\paragraph{}
The authors of \cite{Rivelles_2003} had already hinted a correspondence between noncommutative field theory and gravity. Through \eqref{eq:moyal_inner_der}, a translation in the Moyal space is similar to a gauge transformation, a feature that was only associated to general relativity. The previous authors considered an Abelian noncommutative gauge theory given by \eqref{eq:action_u1} and used the Seiberg-Witten map to link the noncommutative field strength $\hat{F}$ to its commutative counterpart $F$. To first order in $\theta$, the action \eqref{eq:action_u1} reduces to commutative Abelian gauge theory coupled to a metric field given by
\begin{align}
	g^{\mu\nu} 
	&= \eta^{\mu\nu} + \Theta^{\mu\rho} \tensor{F}{_\rho^\nu} + \Theta^{\nu\rho} \tensor{F}{_\rho^\mu} + \frac{1}{2} \eta^{\mu\nu} \Theta^{\rho\sigma} F_{\rho\sigma},
	\label{eq:emergent_metric}
\end{align}
where $\eta$ is the Minkowski metric. The same procedure was elaborated for real (uncharged) scalar fields and complex (charged) scalars. The same conclusion was found except that the charged scalar had the extra term in \eqref{eq:emergent_metric} multiplied by $\frac{1}{2}$: The gravitational coupling thus depends on the charge of the field. The different dispersion relations of the three fields were also studied.

This setting was enhanced to a non-Abelian case and using a broader Seiberg-Witten map, as explained in the review \cite{Yang_2007}. Considering the action \eqref{eq:action_un}, the author make use of the so called exact Seiberg-Witten map $\hat{F}_{\mu\nu} = \frac{1}{1 + \Theta^{\rho\sigma} F_{\rho\sigma}} F_{\mu\nu}$. At first order in $\theta$, this action reduces to the Yang-Mills action in curved space-time with metric \eqref{eq:emergent_metric}.

\paragraph{}
In the context of teleparallel gravity, \cite{Cortese_2010} introduced gravity in the noncommutative Yang-Mills Lagrangian by adding a term of the form $\int \dd^4 x\; A_\mu \star_\theta \xi^\nu \star_\theta \big( -\frac{1}{4} \delta^\mu_\nu F_{\rho\sigma} \star_\theta F^{\rho\sigma} + F_{\nu\rho} \star_\theta F^{\mu\rho} \big)$. This new term changes the Yang-Mills action into the one in curved space-time with (co)-frame fields depending explicitly on $\xi$ and $A$. The Seiberg-Witten map at first order in the deformation parameter is used.

\subsection{Quantum Poincar\'{e} gauge theory}
\label{subsec:quantum_poinca_gauge}
\paragraph{}
This approach relies on a quantum version of Poincar\'e gauge gravity. The Poincar\'e gauge gravity is a gauge theory based on gauging the Poincar\'e group and produces a spin-torsion theory, also called $U_4$ theory or Einstein-Cartan(-Sciama-Kibble) theory, that is a theory of gravity with torsion. The main idea is to gauge a quantum version of the Poincar\'e group, explicitly $ISO_q(1,3)$

The formulation of the Poincar\'{e} gauge theory is based on the introduction of the tetrad and the spin connection which correspond respectively to the connection on the translations and Lorentz subgroups of the full Poincar\'{e} group. For early developments of this theory see \cite{Utiyama_1956, Kibble_1961, Hehl_1976} and for reviews see \cite{Ivanenko_1983, Blagojevic_2013}. The authors of \cite{Castellani_1994} formulated a similar gauge version of the gravity using a $q$-deformation of the Poincar\'{e} group, noted $ISO_q(1,3)$, instead of the classical one.

\paragraph{}
A bicovariant differential calculus is set on $ISO_q(1,3)$ and left-invariant 1-forms are defined through a dual version of the $q$-Lie algebra generators. The exterior derivative is defined on the $q$-Lie algebra and extended to the space of forms via the deformed Cartan-Maurer equation. This sets the global differential calculus on the $q$-Lie algebra $ISO_q(1,3)$ and allows the curvature to be defined as usual. The tetrad and spin connections are then defined as the 1-forms associated to the deformed translations and Lorentz generators respectively. Finally, the Einstein-Cartan Lagrangian \eqref{eq:einstein_cartan_action} is considered.

The torsion-free and Einstein equations are found to have similar expressions to the classical ones but with $q$-commuting fields instead of usual complex numbers (called c-numbers in the paper). This allows for a straightforward classical limit when $q\to 1$.

\paragraph{}
This approach was used in \cite{Bimonte_1998a} with a $q$-deformed Einstein-Cartan action. The authors defined the Christoffel symbols $\Gamma$ through the vanishing covariant derivative of the tetrad 
\begin{align}
	\partial_\mu \tensor{e}{_I^\nu} + \omega_{I \mu}^J \tensor{e}{_J^\nu} + \tensor{e}{_I^\rho} \Gamma^\nu_{\mu\rho} 
	&= 0, &
	I,\mu,\nu = 0, ..., d
    \label{eq:tetrad_connection_def}
\end{align}
and introduced a metric using $q$-deformed version of \eqref{eq:tetrad_def}. The authors introduced matter to get the full $q$-deformed Einstein's equation (without cosmological constant). The classical limit of the deformed action and its Hamiltonian formalism of Ashtekar variables is recovered.

A following paper of the same authors \cite{Bimonte_1998b} addressed other aspects of this approach: the link between $q$-fields and c-numbers and a reality condition. The latter applies to the tetrad and spin connections for two kinds of ``reality'', explicitly $|q|=1$ or $q\in\mathbb{R}$, and are consistent with their commutation relations. For both reality conditions, the metric and the action are real. Moreover, a physical interpretation of $q$-fields is unknown so that physics is encoded in c-numbers. As the metric and the action are not c-numbers here, the authors turn some quantities into dimensionful ones, especially the Newton constant, which now have non-trivial commutation relations. Thus, dimensionless ratios of quantities are made c-numbers. This allows to recover the full physical equivalence of the metric theory of gravity of Einstein and this metric theory, in the limit $q\to 1$.

\subsection{Gravity on almost-commutative spaces}
\label{subsec:almost_comm_space}
\paragraph{}
An almost commutative space corresponds to the product $\mathcal{M} \times \mathbb{Z}_N$, where $\mathcal{M}$ is a smooth manifold and $\mathbb{Z}_N$ is a discrete set of $N$ points. On such a space, the authors of \cite{Coquereaux_1991} built a Yang-Mills model with $2\times2$ matrix-valued gauge fields. They obtained two copies of Yang-Mills theory with the gauge fields being the diagonal elements of the $2\times2$ matrix, whereas the off-diagonal elements correspond to the Higgs fields. 

Given two matrix-valued forms $\rho = (\rho_{jk})_{j, k = 1,2}$ and $\eta = (\eta_{jk})_{j, k = 1, 2}$, one can define their wedge product and differential as
\begin{subequations}
\begin{align}
	\rho \wedge \eta
	&= \begin{pmatrix}
		\rho_{11} \wedge \eta_{11} + (-1)^{\mathrm{deg}(\rho_{12})} \rho_{12} \wedge \eta_{21} &
		\rho_{12} \wedge \eta_{22} + (-1)^{\mathrm{deg}(\rho_{11})} \rho_{11} \wedge \eta_{12} \\
		\rho_{21} \wedge \eta_{11} + (-1)^{\mathrm{deg}(\rho_{22})} \rho_{22} \wedge \eta_{21} &
		\rho_{22} \wedge \eta_{22} + (-1)^{\mathrm{deg}(\rho_{21})} \rho_{21} \wedge \eta_{12}
	\end{pmatrix}, 
	\label{eq:matrix-valued_wedge} \\
	\dd \rho
	&= \begin{pmatrix}
		\dd \rho_{11} + \rho_{12} + \rho_{21} &
		- \dd \rho_{12} - \rho_{11} + \rho_{22} \\
		- \dd \rho_{21} + \rho_{11} - \rho_{22} &
		\dd \rho_{22} + \rho_{12} + \rho_{21}
	\end{pmatrix}
	\label{eq:matrix-valued_diff}
\end{align}
	\label{eq:matrix-valued_wedge_diff}
\end{subequations}

Following the same procedure, the authors of \cite{Mohammedi_1992} introduced a matrix-valued spin-connection and a matrix-valued vierbein. The corresponding matrix-valued curvature, torsion and gauge transformation are defined. The latter are found to be diagonal. A matrix-valued Einstein-Hilbert action is considered. The latter contains two different universes, in which the spin-connection and the vierbein correspond to the diagonal terms of their respective matrices, both being coupled to one another. Finally, the off-diagonal vierbein is determined by the equations of motion and the one of the spin connection corresponds to a scalar field that couples to gravity in a non-trivial way.

\newpage
\section{Conclusion}
\label{sec:conc}
\paragraph{}
Gauge theories on Moyal spaces have been the subject of many efforts, focused in particular on the investigation of their quantum properties and their perturbative renormalisability. Besides, some interesting adaptation to ``noncommutative versions'' of the standard model have been proposed together with phenomenological estimates which would come out if these gauge structures would happen to be physically relevant. Despite many efforts, no convincing solution able to cure or render harmless the UV/IR mixing has been found so far. Although a family of gauge theory models on the 4-d Moyal space, expressed as matrix models, has been conjectured to be renormalisable, no one has actually proven or infirm this conjecture. The structure of the vacuum configurations complicates drastically its analysis. To date, renormalisabilty issue within gauge theories on Moyal spaces remains an open question.

\paragraph{}
Gauge theories on $\mathbb{R}^3_\lambda$ exhibit some relationships with a class of brane models as well as with group field theory models. In that latter case, however, group field theory models can be represented as noncommutative (scalar) field theories on $\mathbb{R}^3_\lambda$ and one does not know if some gauge theories could come into play. The particular structure of $\mathbb{R}^3_\lambda$, which basically stems from the compacity of $SU(2)$, the Lie group related to its algebra of coordinates, renders the formulation of these noncommutative gauge theories in terms of matrix models very natural. Roughly speaking, one finally ends up with an infinite sum of models on fuzzy spheres.

Among these gauge models on $\mathbb{R}^3_\lambda$, a few families of gauge theories which are perturbatively finite to all orders and even solvable for some of them, have been constructed. Note however that these finite theories resort on a particular noncommutative differential calculus whose commutative limit is not the usual de Rham calculus. No radial dependence is present. Accordingly, the commutative limit of these gauge theories leads to kind of rotor models. It appears that other types of differential calculus with a usual commutative limit have been poorly explored. An interesting point to investigate would be the construction of gauge theories based on calculus whose structure should change the form and properties of the kinetic operator involved in the action, which may possibly alter its one-loop properties.

\paragraph{}
The case of field theories on $\kappa$-Minkowski space-time is expected to be the most physically interesting. Amazingly, there are relatively few works dealing with the construction of a gauge theory on this quantum space, compared to the numerous contributions devoted to scalar field theories. It turns out that the construction of a gauge invariant action is complicated by technical obstacles. Up to now, three types of  gauge action have been proposed during the two past decades. Two of them make use of the Seiberg-Witten map and are based, either on a $\kappa$-Lorentz invariant differential calculus or a calculus obtained by using a twist, in the Drinfeld twists vein (with corresponding star-product). They seem difficult to handle for some exploration of one-loop radiative corrections, in part due to the complicated expressions for the star-products. The third type of action is obtained by requiring additional $\kappa$-Poincar\'e invariance of the gauge action and manage the twisted trace which necessarily appears thanks to the use of a twisted differential calculus. Perturbative computations can be done as the star-product has a rather simple expression, compared to those used in the theories above. However, this gauge theory exists only in 5 dimensions. Besides, a vacuum instability has been evidenced. To date, it is not know if the theories suffer from UV/IR mixing.

\paragraph{}
In some sense, the above gauge theories can be viewed as noncommutative generalisation of Yang-Mills theories as they are all built from noncommutative connection on a right (or left) module which are designed to extend Yang-Mills connections. Numerous attempts aiming to incorporate gravity to noncommutative spaces have been proposed.

One of the main debate in such theories is about the definition of the ``noncommutative metric'' as different approaches lead to different definitions. Furthermore, in some cases, the commutative limit of such a metric gives rise to unphysical effects. Note that the metric free approach, through the tetrad, has also some loopholes.

Most of the approaches to noncommutative gravity remain algebraic, meaning that no action functional has been explored. Note however that two of the discussed settings considers action functionals, but both remains at the semi-classical level.

\section*{Acknowledgements}
\paragraph{}
J.-C.\ W thanks G.\ Amelino-Camelia, J.\ Barett, D.\ Blaschke, A.\ Connes, L.\ Dabrowski, F.\ D'Andrea, L.\ Freidel, J.\ Gracia-Bondia, T.\ Juri\'{c}, G.\ Landi, F.\ Lizzi, P.\ Martinetti, S.\ Silverstrov, A.\ Sitarz, H.\ Steinacker, D.\ Vassilevich, P.\ Vitale for discussions and exchanges during the past decade. K.\ H thanks M.\ Dimitrievi\'{c}-Ciri\'{c}, S.\ Majid, P.\ Martinetti, A.\ Sitarz, P.\ Vitale for enlightening discussions and comments. We thanks the Action CA18108 QG-MM ``Quantum Gravity phenomenology in the multi-messenger approach'' from the European Cooperation in Science and Technology (COST).

%-- BIBLIOGRAPHY ------------------------------------------------------------%

%----------------------------------------------------------------
%---------------------- End of the document ---------------------
\end{document}